%% file: draft_JP.tex
\documentclass[preprint,3p]{elsarticle}
\usepackage{mymacros}
\usepackage{amssymb}
\usepackage{amsthm}
\usepackage[mathlines]{lineno}
\usepackage{graphicx}
\usepackage{units}
\usepackage{url}
\usepackage{amsmath}
\usepackage{amsfonts}
\usepackage{bm}
\usepackage{textcomp}
\usepackage{subfigure}
\usepackage{multicol}
\usepackage{multirow}
\usepackage{verbatim}
\usepackage{rotating}
\usepackage[colorlinks,linkcolor=red,citecolor=red]{hyperref}
\usepackage{float}
\usepackage{epsfig}
\usepackage{dcolumn}
\usepackage{bm}
\usepackage{color}
\usepackage{pstricks}
\usepackage{pst-node}
\usepackage{times}
\usepackage{indentfirst}
\usepackage[english]{babel}

\biboptions{numbers,sort&compress}

\journal{Physics Letters B}

\begin{document}

\begin{frontmatter}

  \title{{\bf \boldmath Determination of spin and parity of $D^{*}_{(s)}$ mesons}}

  \author{\input{authors.tex}}

  \begin{abstract}
The spin and parity of the charmed mesons $D_{s}^{*+}$, $D^{*0}$ and
$D^{*+}$ are determined for the first time to be $J^P=1^{-}$ with
significances greater than 10$\sigma$ over other hypotheses of $2^{+}$
and $3^{-}$, using an $e^+e^-$ collision data sample with an integrated luminosity of 3.19 fb$^{-1}$ collected by the BESIII detector at a center-of-mass energy of 4.178 GeV.
Different spin-parity hypotheses for $D_{s}^{*+}$, $D^{*0}$, and $D^{*+}$ mesons are tested via a helicity amplitude analysis of the processes $\ee\to D^{*+}_{s}D^{-}_{s}$, $D^{*0}D^{0}$ and $D^{*+}D^{-}$, with $D^{*+}_{s}\to D^{+}_{s} \gamma$, $D^{*0}\to D^{0}\pi^{0}$, and $D^{*+}\to D^{+}\pi^{0}$.
The results confirm the quark model predictions.
    \\
    \\
    \text{Keywords:~~charmed meson, spin and parity, BESIII}
  \end{abstract}

\end{frontmatter}


\begin{multicols}{2}
\section{Introduction}
Charmed-meson spectroscopy provides a powerful tool to achieve a better understanding of the strong interaction. 
Following the experimental discovery of the $D^+$-meson~\cite{Peruzzi:1976sv}, first predictions of the charmed meson spectrum from the quark model emerged in the 1980s~\cite{Godfrey:1985xj}. Although the quark model has successfully predicted masses and spin-parity assignments, experimental confirmations are in many places still missing. Notably, 
the supposed assignment of $D_{s}^{*+}$, $D^{*0}$ and $D^{*+}$ as states with quantum numbers $J^P=1^-$~\cite{ParticleDataGroup:2020ssz,Gell-Mann:1964ewy,Gaillard:1974mw,DeRujula:1975qlm,Glashow:1970gm} has not yet been confirmed by any experiment.
The Particle Data Group (PDG) labels $J^P$ of the $D_{s}^{*+}$ meson as `unknown'~\cite{ParticleDataGroup:2020ssz}, 
the only experimental measurement~\cite{Nguyen:1977kk} found a spin of $1$ for the $D^{*0}$, ruling out spin $0$.  The observed decay modes $D^{*}_{(s)} \to D_{(s)} \pi$ rule out unnatural spin-parity assignments, \emph{i.e.} $1^+$, $2^-$, or $3^+$.  The hypotheses $1^-$, $2^{+}$, and $3^{-}$ remain to be tested. 

Over the past decade, charmed-meson spectroscopy has undergone a resurgence due to the discovery of numerous excited charm and charm-strange meson states by the BaBar~\cite{BaBar:2014omp,BaBar:2010zpy,BaBar:2006gme,BaBar:2009rro,BaBar:2014jjr}, Belle~\cite{Belle:2007hht,BaBar:2014omp}, CLEO~\cite{CLEO:2003ggt} and LHCb~\cite{Chen:2021ftn} collaborations. 
The observation of these new resonances has provided essential knowledge about the radial excitations $(2S)$ and orbital excitations with angular momenta $L=1$ and $2$. The knowledge of spin and parity of the $D_{(s)}^{*}$ is vital to determine the quantum numbers of higher excited $D_{(s)}^{**}$ states which are reconstructed via $D_{(s)}^{*}$ mesons.
Recently, the first candidates for a hidden-charm tetraquark with strangeness, $Z_{cs}(3985)^{+}$~\cite{BESIII:2020qkh} and $Z_{cs}(3985)^{0}$~\cite{BESIII:2022qzr}  were observed by BESIII in their decays to the final states $(D^{-}_{s}D^{*0} +D^{*-}_{s}D^0)$ and $(D^{+}_{s}D^{*-} +D^{*+}_{s}D^-)$. 
Hence, the spin and parity of the $D^{*0}$ and $D^{*+}_{s}$ are essential to determine the quantum numbers of $Z_{cs}(3985)^{+}$.

In this Letter, the spin-parity quantum numbers of $D_{s}^{*+}, D^{*0}$ , and $D^{*+}$ mesons are determined using a data set with an integrated luminosity of 3.19~fb$^{-1}$ collected with the BESIII detector at the center-of-mass (CM) energy of 4.178$\gev$.
The spin-parity hypotheses $1^-$, $2^+$, and $3^-$ are tested by means of the helicity amplitude analysis of $\ee\to D^{*+}_{s}D^{-}_{s}$, $D^{*0}\bar{D}^{0}$, $D^{*+}D^{-}$, with $D^{*+}_{s}\to D^{+}_{s} \gamma$, $D^{*0}\to D^{0}\pi^{0}$, and $D^{*+}\to D^{+}\pi^{0}$. Throughout this letter, charge conjugation is implied, unless explicitly stated otherwise.
A partial reconstruction technique is adopted, namely, one $D_{(s)}$ and $\pi^0$($\gamma$) particles are detected to identify the $D_{(s)}^{*}$ decay, and the other $D_{(s)}$ in the event is undetected. The $D_{(s)}$ candidates are reconstructed in the hadronic decay modes $D_s^{+} \rightarrow K_S^0 K^+$, $D^0 \rightarrow K^-\pi^+$, and $D^+ \rightarrow K^-\pi^+\pi^+$.  
The intermediate states $K_S^0$ and $\pi^0$ are reconstructed via their decays to $\pi^+\pi^-$ and $\gamma\gamma$, respectively.

\section{BESIII Detector and Monte Carlo Simulation}
\label{mc}
The BESIII detector~\cite{BESIII:2009fln} records symmetric $e^+e^-$ collisions provided by the BEPCII storage ring~\cite{Yu:2016cof}, which operates in the CM energy range from 2.0~$\gev$  to 4.946~$\gev$, where BESIII has collected large data samples~\cite{BESIII:2020nme}.
The cylindrical core of the BESIII detector covers 93\% of the full solid angle and consists of a helium-based multilayer drift chamber~(MDC), a time-of-flight system~(TOF) using plastic scintillators in the central region (barrel) and multi-gap RPCs in the end caps, and a CsI(Tl) electromagnetic calorimeter~(EMC), which are all enclosed in a superconducting solenoidal magnet providing a 1.0~T magnetic field.
The solenoid is supported by an octagonal flux-return yoke instrumented with resistive plate counter muon identification modules interleaved with steel.
The charged-particle momentum resolution at $1\gevc$ is $0.5\%$, and the specific ionization energy loss (d$E/$d$x$) resolution is $6\%$ for electrons from Bhabha scattering.
The EMC measures photon energies with a resolution of $2.5\%$ ($5\%$) at $1\gev$ in the barrel (end cap) region.
The time resolution in the TOF barrel region is 68~ps, while that in the end cap region is 60~ps~\cite{etof}.

Simulated data samples are produced with a {\sc GEANT-4}~\cite{GEANT4:2002zbu} based Monte Carlo (MC) package,
which includes the geometric description~\cite{Huang:2022wuo} of the BESIII detector and the detector response, are used to determine detection efficiencies and to estimate backgrounds. 
The simulation includes the beam energy spread and initial state radiation in the $e^+e^-$ annihilations modeled with the event generator {\sc kkmc} ~\cite{ref:kkmc}. 
An MC sample of inclusive decays, forty times larger than the data set, includes the production of open charm processes, the initial state radiation production of vector charmonium(-like) states, and the continuum processes.
The known decay modes are modeled with {\sc evtgen}~\cite{ref:evtgen} using branching fractions taken from the PDG~\cite{ParticleDataGroup:2020ssz}, and the remaining unknown charmonium decays are modeled with \mbox{\sc lundcharm}~\cite{ref:lundcharm}.
Final state radiation~(FSR) from charged final state particles is incorporated using {\sc photos}~\cite{photos}.
MC samples of $e^+e^-\to \pi^0(\gamma)D_{(s)}D_{(s)}$, simulated following a 3-body phase space (PHSP) distribution, are used to determine the selection efficiency. In these simulations, one of the $D_{(s)}$ mesons decays inclusively, based on the branching fractions listed in PDG, and the second one decays to the signal final states. 
\section{Event selection}
\label{select}
To select the signal candidates of the analyzed processes $e^+e^- \rightarrow D_s^{*+}D_s^{-}\rightarrow \gamma D_s^+D_s^-$, $e^+e^- \rightarrow D^{*0}\bar{D}^{0}\rightarrow \pi^0 D^0\bar{D}^0$, and $e^+e^- \rightarrow D^{*+}D^{-}\rightarrow \pi^0 D^+D^-$, the following selection criteria are implemented. The decay modes of $D_{(s)}$ used for reconstruction are $D_s^+\rightarrow K_S^0K^+$,$D^0\rightarrow K^-\pi^+$ and $D^+\rightarrow K^-\pi^+\pi^+$.

The distance of closest approach of each charged track to the $\ee$ interaction point (IP) is required to be within 10~cm along the beam direction and within 1~cm in the plane perpendicular to the beam direction, except for the tracks from $K_S^0$ decays.
The polar angle $\theta$ between the direction of a charged track and  the positron beam must satisfy $|\!\cos\theta|<0.93$ for an effective measurement in the active volume of the MDC.
The d$E/$d$x$ information recorded by the MDC and the time-of-flight information measured by the TOF are used to identify particles by calculating the probabilities $\mathcal{P}$ for various particle hypotheses.
Charged kaons and pions are identified requiring $\mathcal{P}(K)>\mathcal{P}(\pi)$ and $\mathcal{P}(\pi)>\mathcal{P}(K)$, respectively.

Shower clusters in the EMC without associated charged tracks are identified as photon candidates if the deposited energy is greater than 25 MeV in the barrel region ($|\!\cos\theta|<0.80$ ) or greater than 50 MeV in the endcap region ($0.86 <|\!\cos\theta|< 0.92$).
To suppress background from electronic noise and coincidental EMC showers, the difference between the event start time and the EMC signal is required to be smaller than $700$~ns.
The $\pi^0$ candidates are reconstructed from photon pairs with an invariant mass $M(\gamma\gamma)$ within~[0.115, 0.150] GeV/$c^2$.

The $K_S^0$ candidates are reconstructed from two oppositely charged tracks, without a particle identification (PID) requirement.
These tracks are required to originate at a common decay vertex (with a vertex $\chi^2$ less than 100) lying within a distance of 20 cm from the interaction point along the beam direction.
Furthermore, the decay vertex is required to be separated from the IP by a distance of at least twice the fitted vertex resolution.
The invariant mass of $\pi^+\pi^-$ pairs, $M(\pi^+\pi^-)$, is required to be in the range [0.487, 0.511] GeV/$c^2$.

The purity of the selected sample is improved by various constraints listed in Table~\ref{tab:masswindow}, involving 
the invariant mass of the reconstructed $D_{(s)}$ meson $M(D_{(s)})$ and the energy difference $\Delta E$ between the total energy of the $\pi^0$($\gamma)D_{(s)}D_{(s)}$ system in the CM frame and the CM energy $E_{0}$.
The applied constraints correspond to three times the resolution of the respective observables:
\begin{eqnarray}
\Delta E = (E_{D_{(s)}}+E_{\pi^0(\gamma)}+E_{\rm{rec}}) - E_{0},
\label{eq:deltaE}
\end{eqnarray}
where $E_{D_{(s)}}$ and $E_{\pi^0({\gamma})}$ are the energies of reconstructed $D_{(s)}$ and $\pi^0$($\gamma$) from $D_{(s)}^{*}$, respectively, and
$E_{\rm{rec}}$ is the energy of the recoil side, defined as
\begin{eqnarray}
E_{\rm{rec}}=\sqrt{\left|-(\vec{p}_{D_{(s)}}+\vec{p}_{\pi^0(\gamma)})\right|^2c^2+m_{D_{(s)}}^2c^4},
\label{eq:recE}
\end{eqnarray}
where $\vec{p}_{D_{(s)}}$ is the total momentum of the reconstructed $D_{(s)}$ meson, $\vec{p}_{\pi^0(\gamma)}$ is the momentum of the $\pi^0$($\gamma$) from $D_{(s)}^{*}$, and $m_{D_{(s)}}$ is the known mass of the $D_{(s)}$ meson~\cite{ParticleDataGroup:2020ssz}.
For each decay mode with multiple $D_{(s)}\pi^0(\gamma)$ candidates in an event, the one with the minimal $|\Delta E|$ is selected.

In order to suppress background contributions and to improve the momentum resolution, a kinematic fit is performed for the three processes. In the case of the $\gamma D_s^+D_s^-$ decay mode, the invariant mass of $K_{S}^{0}K^+$ is constrained to the known $D_{s}^{+}$ mass, and the recoil mass of $D_{s}^{+}\gamma$ is constrained to the known $D_s^-$ mass.
After the kinematic fit, the four-momenta of all final state particles are updated for further analyses.

\begin{figure*}[!hbt]
\centering
\subfigure[]{\includegraphics[width=0.68\columnwidth]{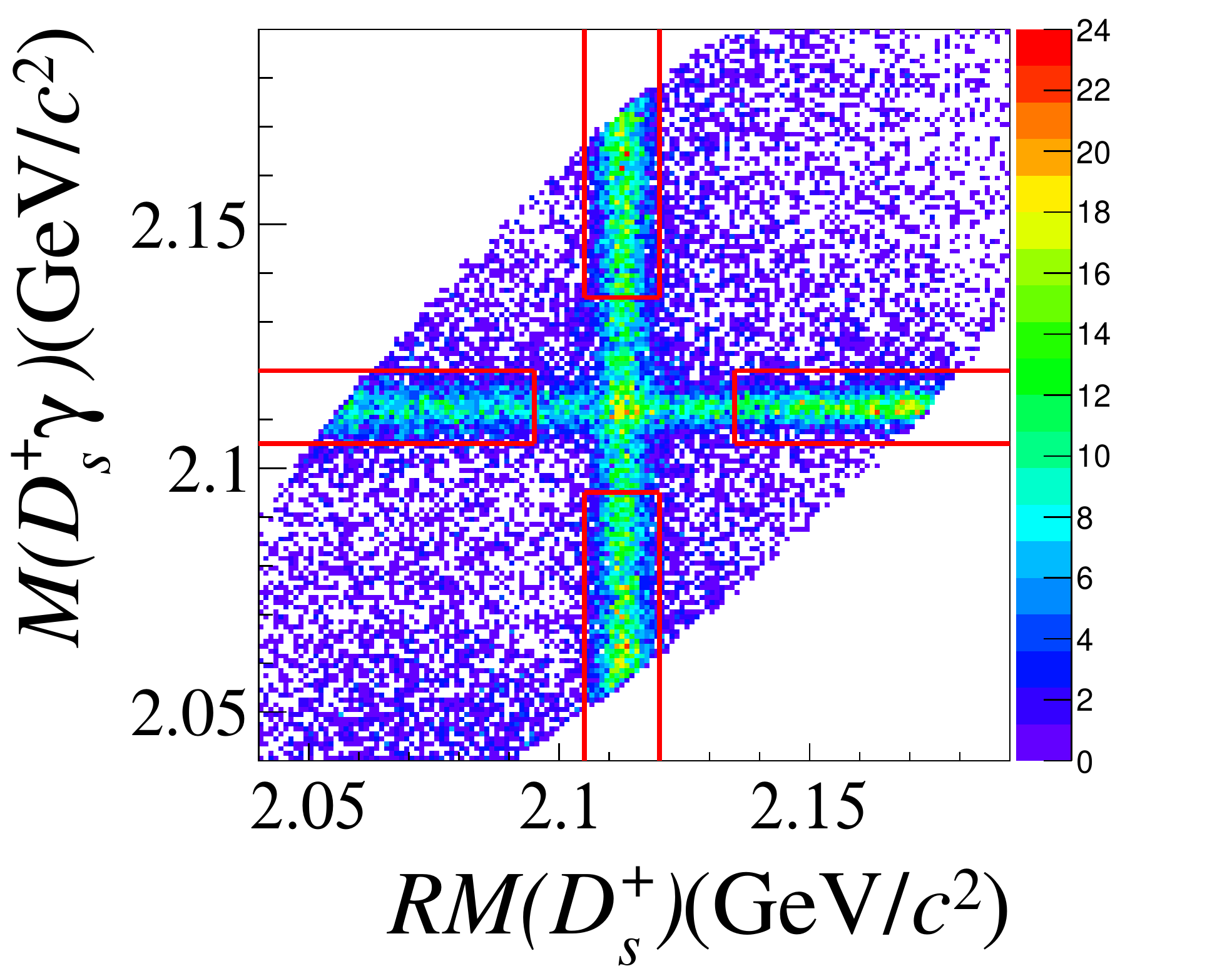}}
\subfigure[]{\includegraphics[width=0.68\columnwidth]{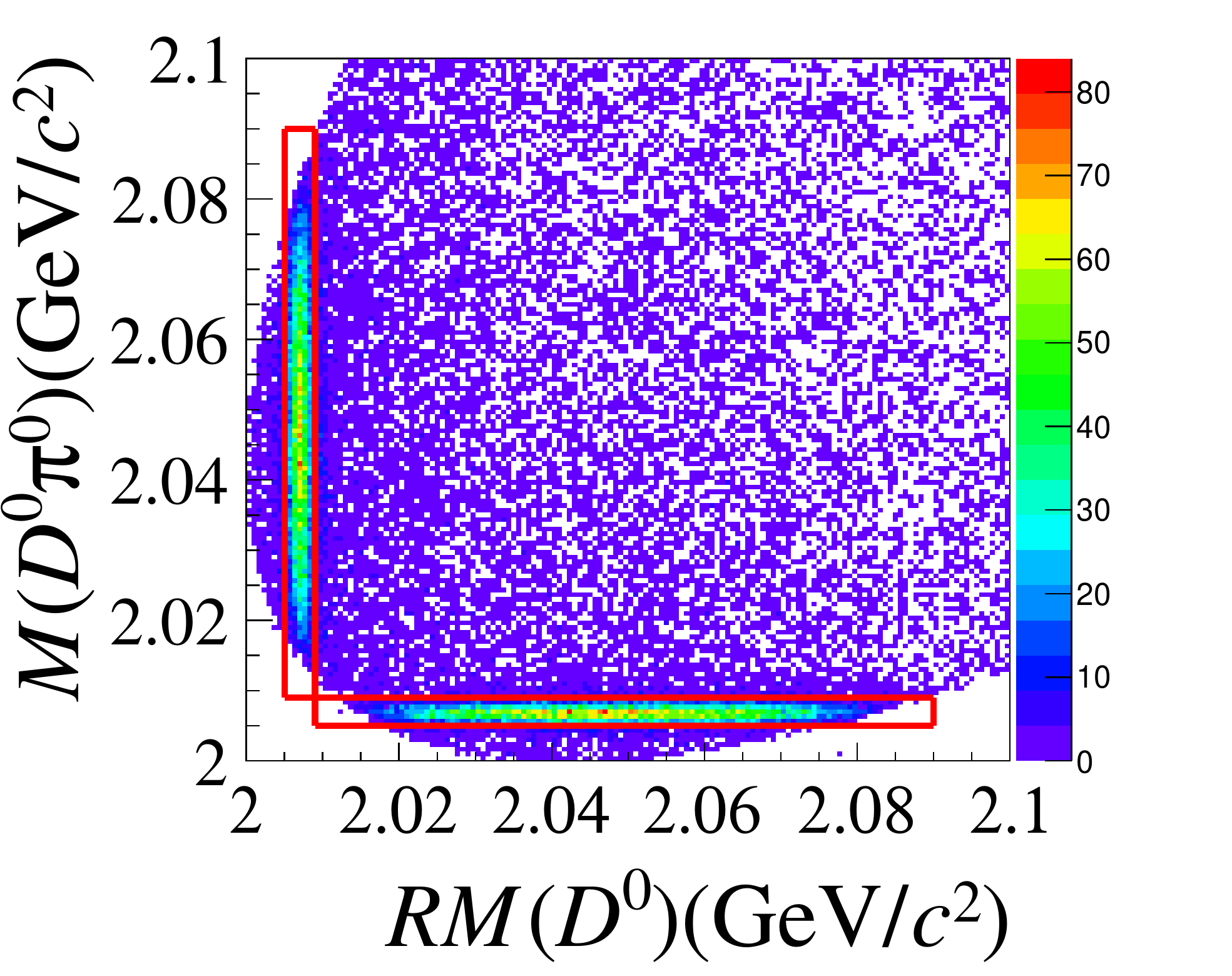}}
\subfigure[]{\includegraphics[width=0.68\columnwidth]{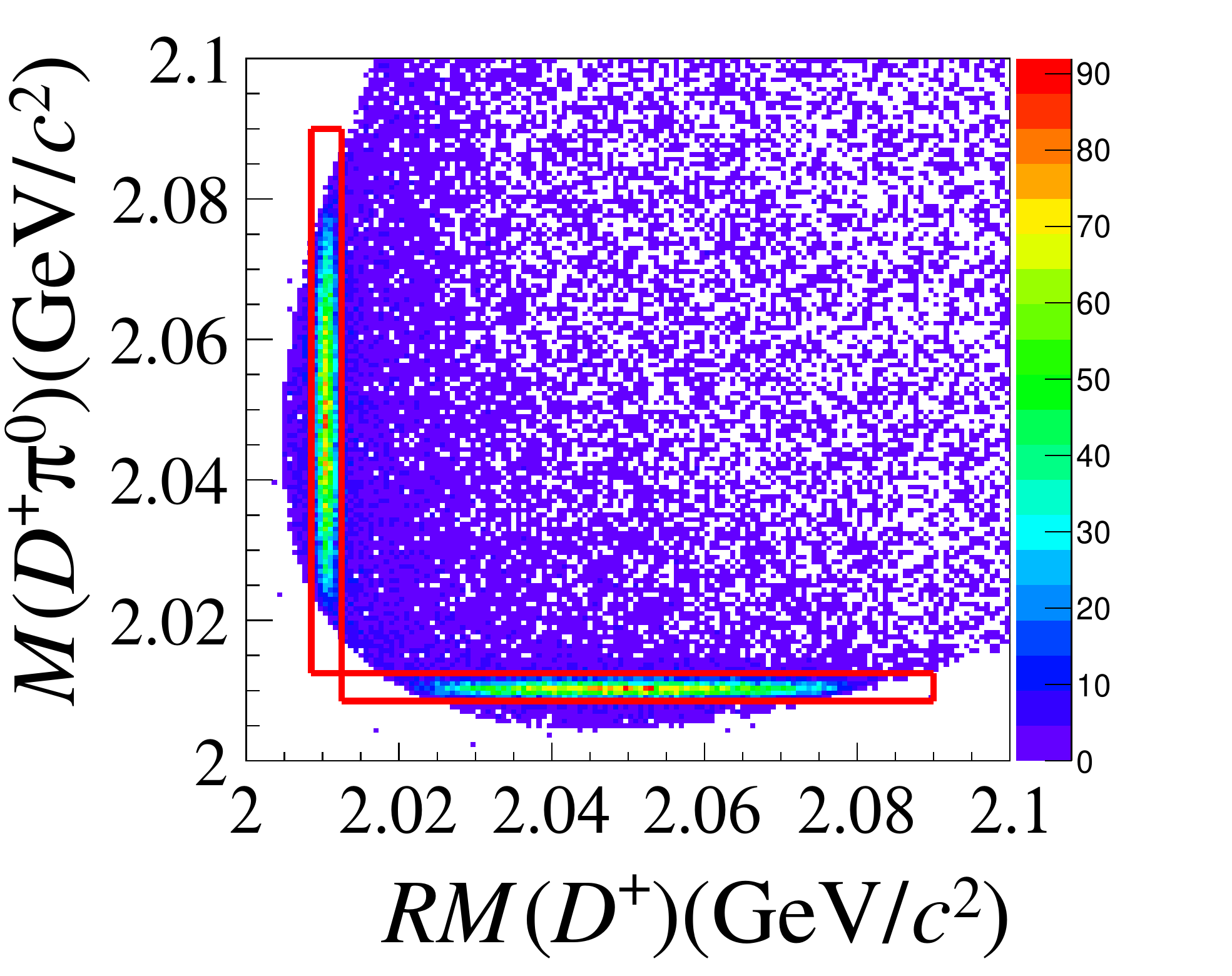}}
\caption{Two-dimensional distributions of $M(D_{(s)}\pi^0(\gamma))$ and  $RM(D_{(s)})$, where the red rectangle denotes the signal region. }
\label{fig:crossCut}
\end{figure*}

Figure~\ref{fig:crossCut} shows two-dimensional distribution of the $D^{*}_{(s)}$ invariant mass $M(D_{(s)}\pi^0(\gamma))$ and the $D_{(s)}$ recoil mass $RM(D_{(s)})$, where two structures are evident.
Taking Fig.~\ref{fig:crossCut}(a) as an example, the horizontal band represents the $e^+e^- \rightarrow D_s^{*+}D_s^{-}$ process, while the vertical band represents the $e^+e^- \rightarrow D_s^{*-}D_s^{+}$ process.
For the further analysis, only events located in regions of the horizontal or vertical bands defined in Table~\ref{tab:masswindow} are retained.
In addition, events in the region common to the horizontal and vertical bands are rejected, since it is impossible to determine whether they stem from the $\ee\to D_s^{*+}D_s^{-}$ or the $\ee \to D_s^{+}D_s^{*-}$ process.
Here, the horizontal band is defined as $D_s^{*+}$-tag sample and the vertical band is defined as $D_s^+$-recoil sample.
The background events are studied with the inclusive MC, and the background contamination is determined to be less than 8\%.
These two data samples are used to perform the helicity amplitude analysis.
Similar selection procedures are also applied to $D^{*0}$ and $D^{*+}$.
\begin{table*}[!htbp]
   \caption{Requirements of $\Delta E$, $M(D_{(s)})$, $M(D_{(s)}\pi^0(\gamma)$ and $RM(D_{(s)})$ for each data sample.}
   \begin{center}
   \begin{tabular}{c|c|c|c|c}
	\hline \hline
Data sample & $\Delta E (\mev)$ & $M(D_{(s)}) (\mevcc)$ & $M(D_{(s)}\pi^0(\gamma)) (\mevcc)$ & $RM(D_{(s)}) (\mevcc)$\\ \hline
$D_s^{*+}$-tag  &\multirow{2}{*}{($-$20, 30)}&\multirow{2}{*}{(1950, 1990)}& (2105, 2120)      &$\notin$(2095, 2135)   \\ \cline{1-1}  \cline{4-5}
$D_s^+$-recoil  &&&$\notin$(2095, 2135)&(2105, 2120)\\  \hline
$D^{*0}$-tag    &\multirow{2}{*}{($-$30, 30)}&\multirow{2}{*}{(1840, 1890)}&    (2005, 2009)    & (2010, 2090)  \\      \cline{1-1}  \cline{4-5}
$D^0$-recoil    &&&  (2010, 2090)& (2005, 2009) \\   \hline
$D^{*+}$-tag    &\multirow{2}{*}{($-$20, 20)} &\multirow{2}{*}{(1850, 1890)}&  (2008.5, 2012.5)    &  (2010, 2090)\\      \cline{1-1}  \cline{4-5}
$D^+$-recoil    &&& (2010, 2090)&  (2008.5, 2012.5)\\
	\hline \hline
   \end{tabular}
    \end{center}
 \label{tab:masswindow}
\end{table*}

\begin{figure*}[!htb]
 \centering
\includegraphics[width=4.5in,angle=0]{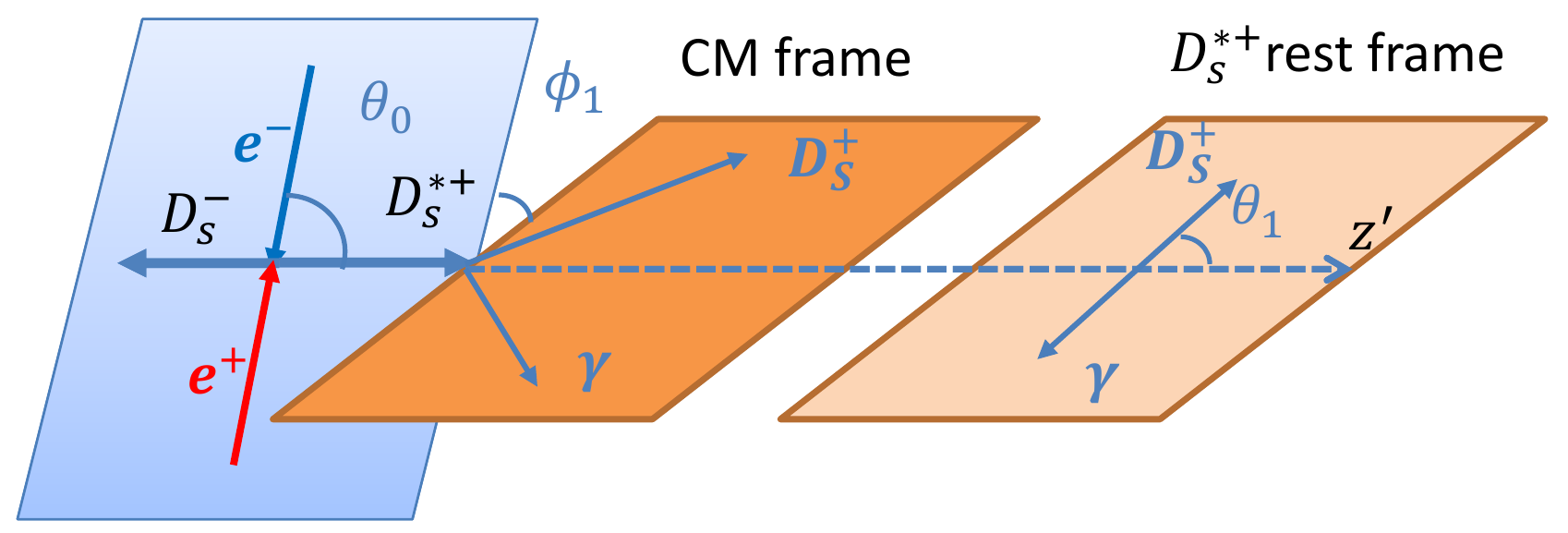}
\put(-280,90){\rotatebox{90}}
\caption{Definition of the helicity frame for $\ee\to\gamma^* \to D_s^{*+}D_s^-, ~D_s^{*+}\to \gamma D_s^+$.}
\label{fig:helicity}
\end{figure*}

\section{Formalism}
To examine the implications of the spin and parity of the $D_{s}^{*+}, D^{*0}$, and $D^{*+}$, the helicity formalism~\cite{Jacob:1959at,Chung:1971ri} is applied to the analysis of the joint angular distribution for charmed mesons and their daughter particles.
Figure~\ref{fig:helicity} shows the helicity frame for the $\ee\to D^{*+}_{s}D^{-}_{s}$ process.
The helicity angle $\theta_0$ is defined as the angle between the $D^{*+}_s$ momentum and the $e^+$ beam axis in the CM system.
The helicity angles $\theta_1$ and $\phi_1$ are related to the decay $D^{*+}_s\to D^{+}_s\gamma$. The angle between the $D^+_s$ momentum in the $D^{*+}_s$ rest frame and the $D^{*+}_s$ momentum in the CM frame is defined as $\theta_1$, the angle between the $D^{*+}_s$ production and decay planes is $\phi_1$.

For the two-body decay, $A(J, m)\to B(s, \lambda)C(\sigma, \nu)$, the helicity amplitude $F^J_{\lambda,\nu}$ is related to the covariant amplitude via~\cite{Chung:1997jn,Chung:1993da,Chung:2007nn}
\begin{footnotesize}
\begin{equation}
\label{chung_forma}
F^J_{\lambda,\nu} =\sum_{LS}\sqrt{\frac{2L+1}{ 2J+1}} \langle L0,S\delta|J,\delta\rangle\langle s\sigma,\lambda-\nu|S,\delta\rangle g_{LS}r^L{B_L(r)\over B_L(r_0)}, \\
\end{equation}
\end{footnotesize}where $\delta=\lambda-\nu$, $g_{LS}$ is the $LS$-coupling constant, $S$ the total spin, $L$ the orbital angular momentum, $r$ the magnitude of momentum difference between the two final state particles, and $B_L$ the Blatt-Weisskopf factor \cite{Santopinto:2016pkp}.

The process $\ee$ $\to$ $\gamma^*$ $\to$ $D_s^{*+}$($\lambda_R$)$D_s^-$, $~D_s^{*+}\to \gamma(\lambda_1) D_s^+$ is discussed for illustration. The $D^{*+}_s$ meson is assigned  spin $J$ and helicity $\lambda_R$, the decay photon and the virtual photon $\gamma^*$ have helicities $\lambda_1$ and $m$, respectively. Then the joint amplitude is
\begin{equation}
\begin{aligned}
  A(m,\lambda_1,\vec\omega)&=\sum_{\lambda_R} F_{\lambda_R,0}D^{1*}_{m,\lambda_R}(\phi_{0},\theta_{0},0)\\
                           & BW(m_{12})F^{J}_{0,\lambda_1}D^{J*}_{\lambda_R,-\lambda_1}(\phi_{1}, \theta_{1},0),
\end{aligned}
\end{equation}
where $\vec\omega=(\theta_0,\phi_0,\theta_1,\phi_1,m_{12})$, $D^J_{m,\lambda}(\phi,\theta,0)$ is the Wigner-$D$ function, $BW$ denotes the Breit-Wigner function, $m_{12}$ is the $\gamma D^+_s$  invariant mass, and $F_{\lambda_R,0}$ and $F^J_{0,\lambda_1}$ are the helicity amplitudes of the two sequential decays.
The decay amplitude for the $D^{*0}$ and $D^{*+}$ mesons can be constructed analogously. Then the differential cross section of the sequential decay via $D_s^*$ is 
\begin{equation}
\mathcal{W}^{J^P} (\theta_0,\theta_1,\phi_1,m_{12})= 
\overline{\sum_{m,\lambda_1}}|A(m,\lambda_1,\vec\omega)|^2,
\end{equation}
with summation over $m=\pm1$ and $\lambda_1=\pm1$. 

\section{Significance estimation}

The spin-parity hypotheses are tested using the likelihood function,
\begin{equation}
\mathcal{L}^{J^P}=\prod_{i=1}^N \frac{1}{\mathcal{C}}\mathcal{W}^{J^P}(\theta^i_0,\theta^i_1,\phi^i_1,m_{12}),
\end{equation}
where $N$ is the number of selected events, ($\theta^i_0,\theta^i_1,\phi^i_1$) are the helicity angles for the $i$-th event describing the $D^*_{(s)}$ decay, and $\mathcal{W}^{J^P}$ is the differential cross section of the sequential decay.
The normalization factor $\mathcal{C} = \int \mathcal{W}^J(\theta^i_0,\theta^i_1,\phi^i_1,m_{12}) \, \rm{d}cos\theta_0 \, \rm{d}cos\theta_{1} \, \rm{d}\phi_1 \, \rm{d}m_{12}$ is numerically evaluated using a phase-space MC sample, but including only accepted events in order to also include efficiency effects.  

We perform an unbinned maximum likelihood fit to the angular distribution of the selected events in the signal region.
The background contributions are subtracted from the log-likelihood values based on selected events in the inclusive MC falling inside the signal region.
The {\sc minuit} package \cite{James:1994vla} is used to minimize the negative net log-likelihood defined by
\begin{equation}
S=-\ln\mathcal{L}^{J^P} = -\alpha[\ln\mathcal{L}^{J^P}(N_s)-\omega_{\rm{bkg}}\ln\mathcal{L}^{J^P}(N_b)],
\end{equation}
where $N_s$ ($N_b$) is the relative ratio of the number of events in data and MC samples. The background weight, $\omega_{\rm{bkg}} = 0.025$, is the 
ratio of the integrated luminosities of the data and the MC sample.
To achieve an unbiased uncertainty estimation, the normalization factor derived in Ref.~\cite{Langenbruch:2019nwe} is taken into account, expressed as
\begin{equation}
\alpha= \frac{N_s-\omega_{\rm{bkg}}N_{b}}{N_{s}+\omega_{\rm{bkg}}^{2}N_{b}}.
\end{equation}
\begin{figure*}[!hbt]
\centering
\includegraphics[width=0.68\columnwidth]{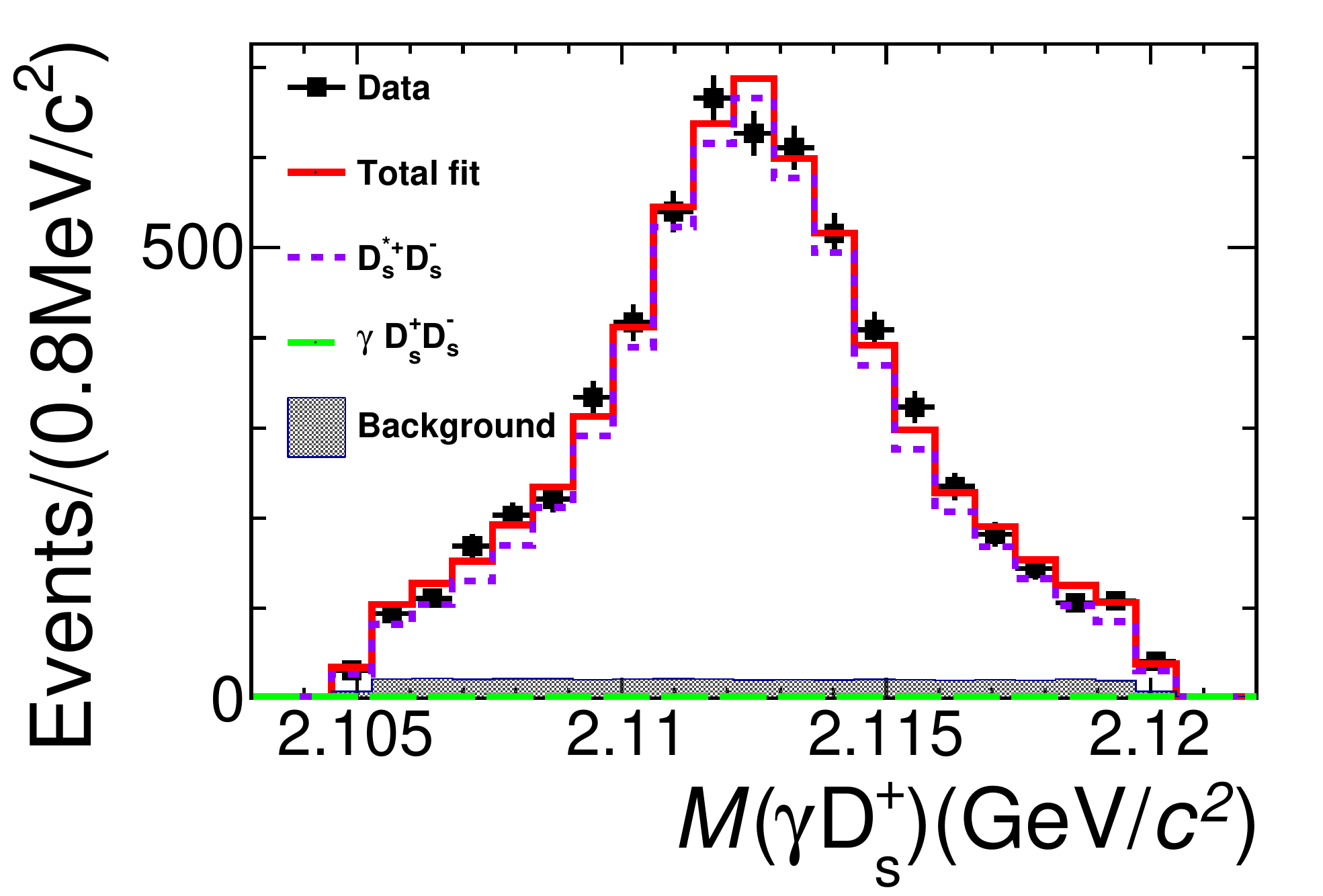}
\includegraphics[width=0.68\columnwidth]{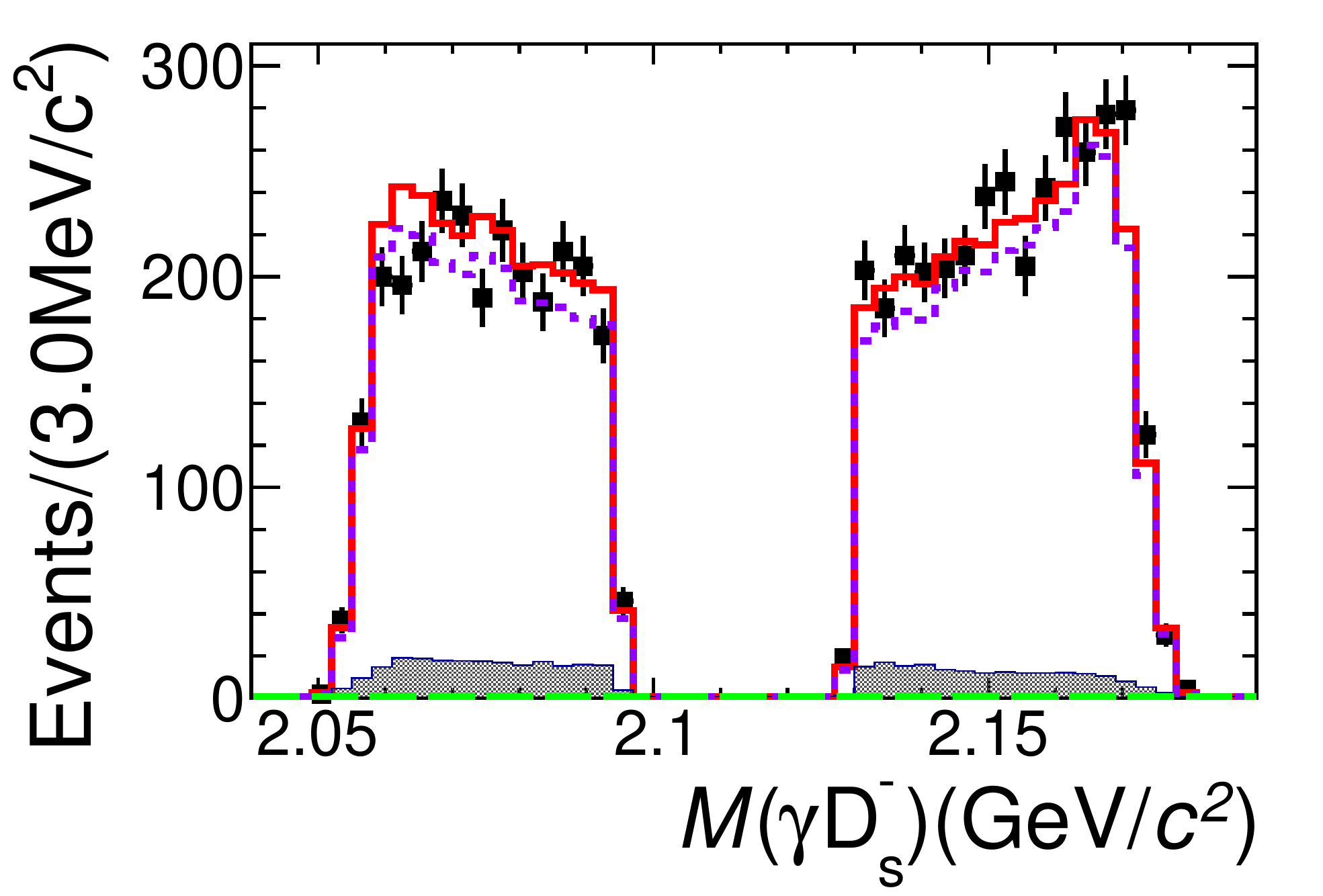}
\includegraphics[width=0.68\columnwidth]{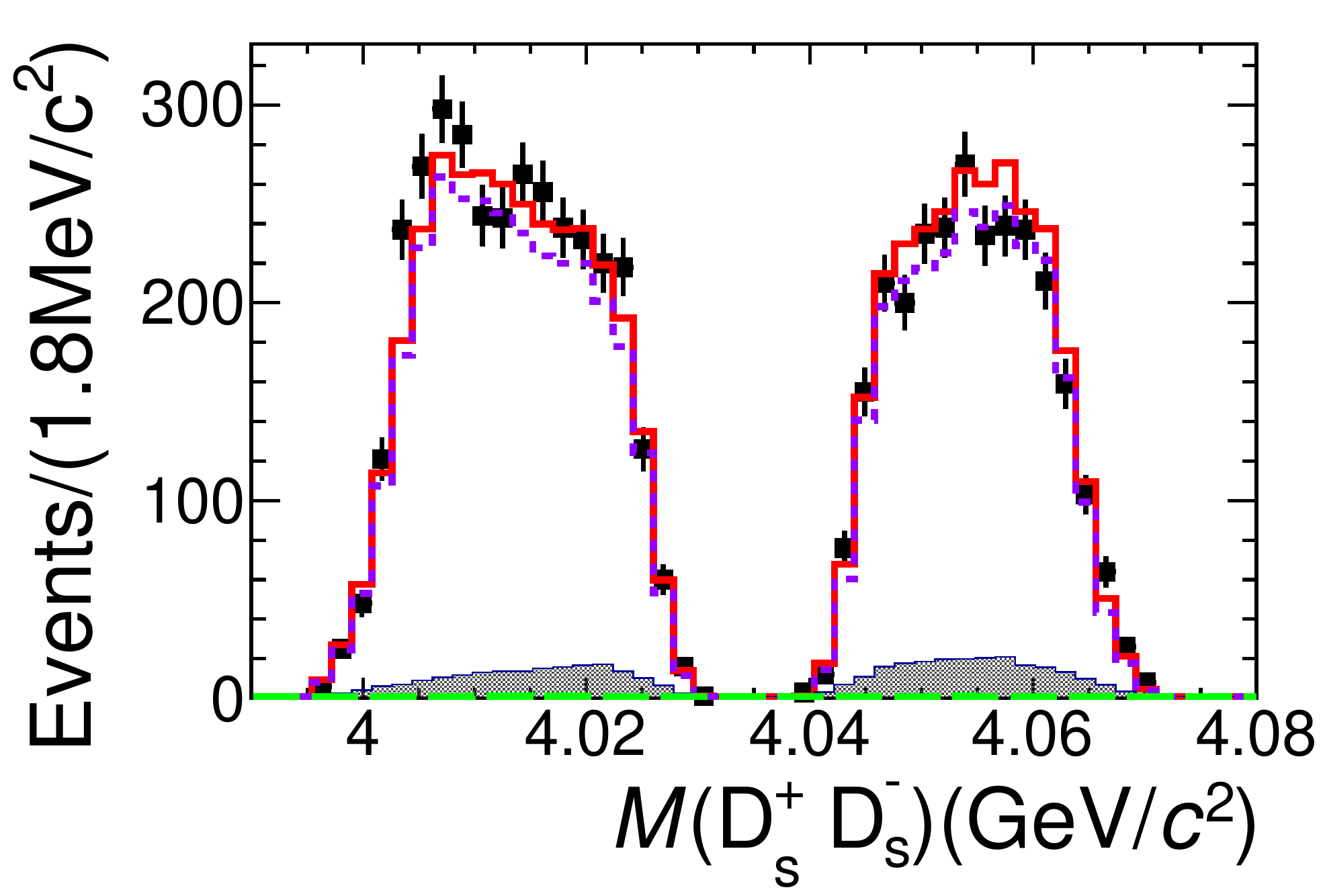}\\
\includegraphics[width=0.68\columnwidth]{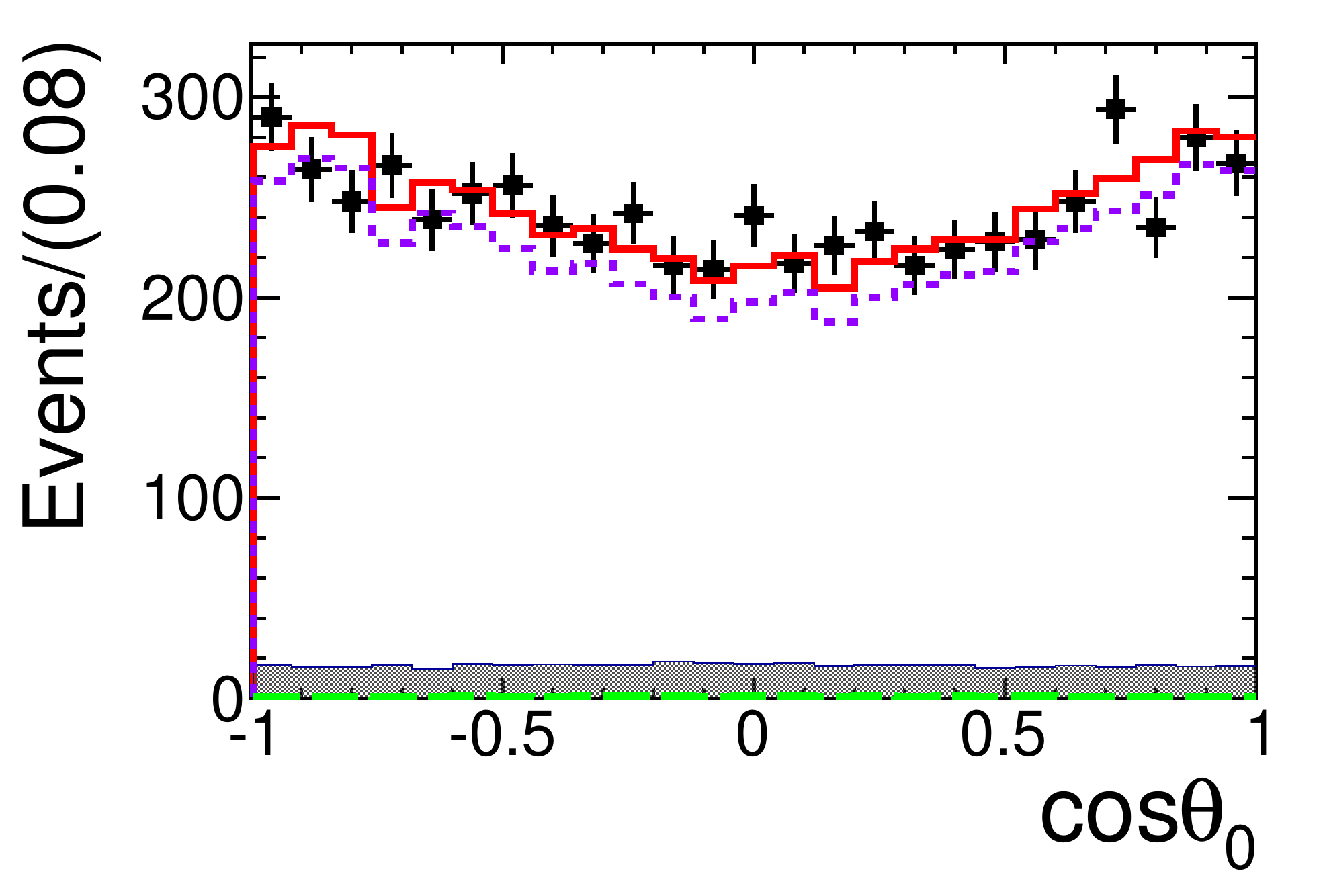}
\includegraphics[width=0.68\columnwidth]{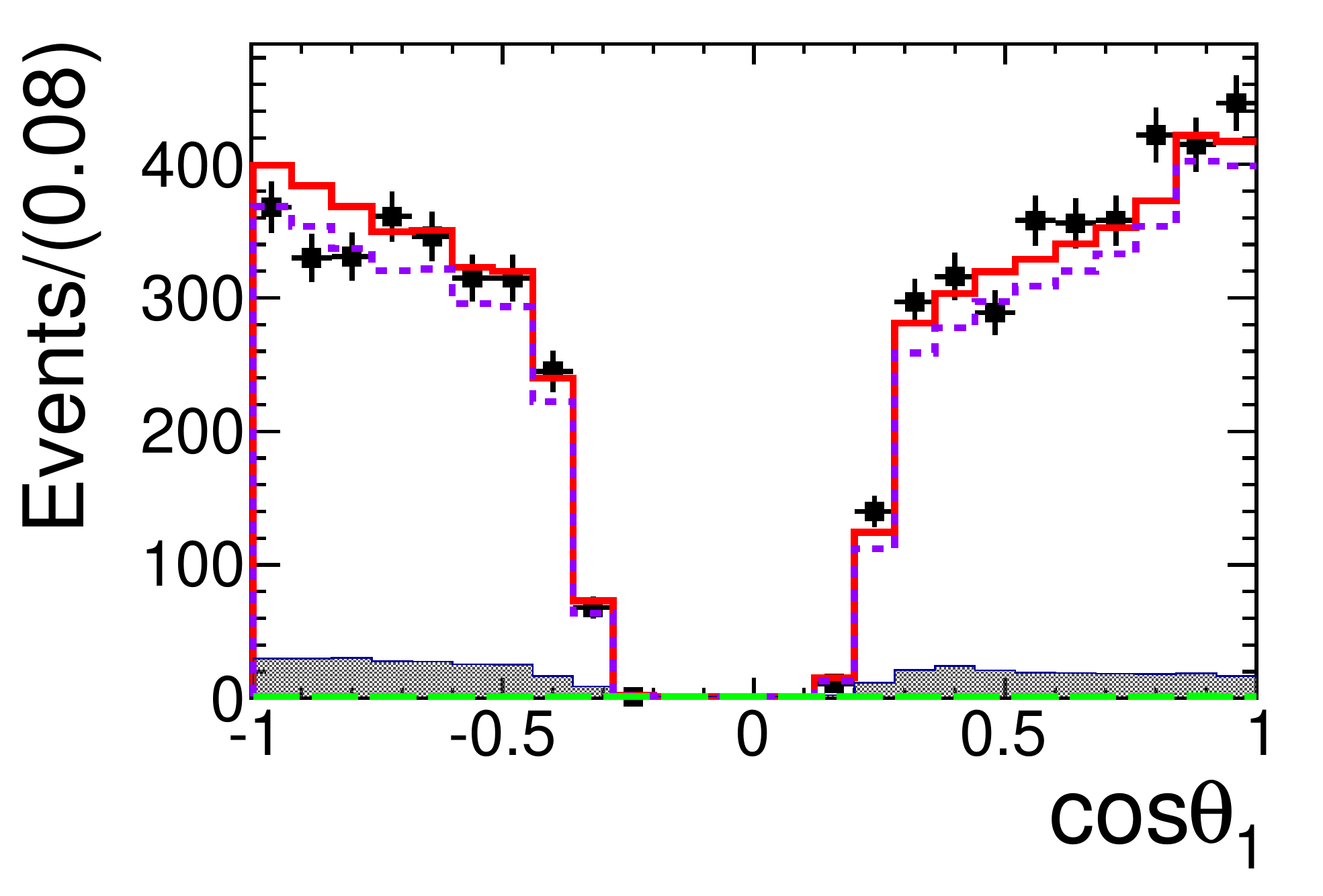}
\includegraphics[width=0.68\columnwidth]{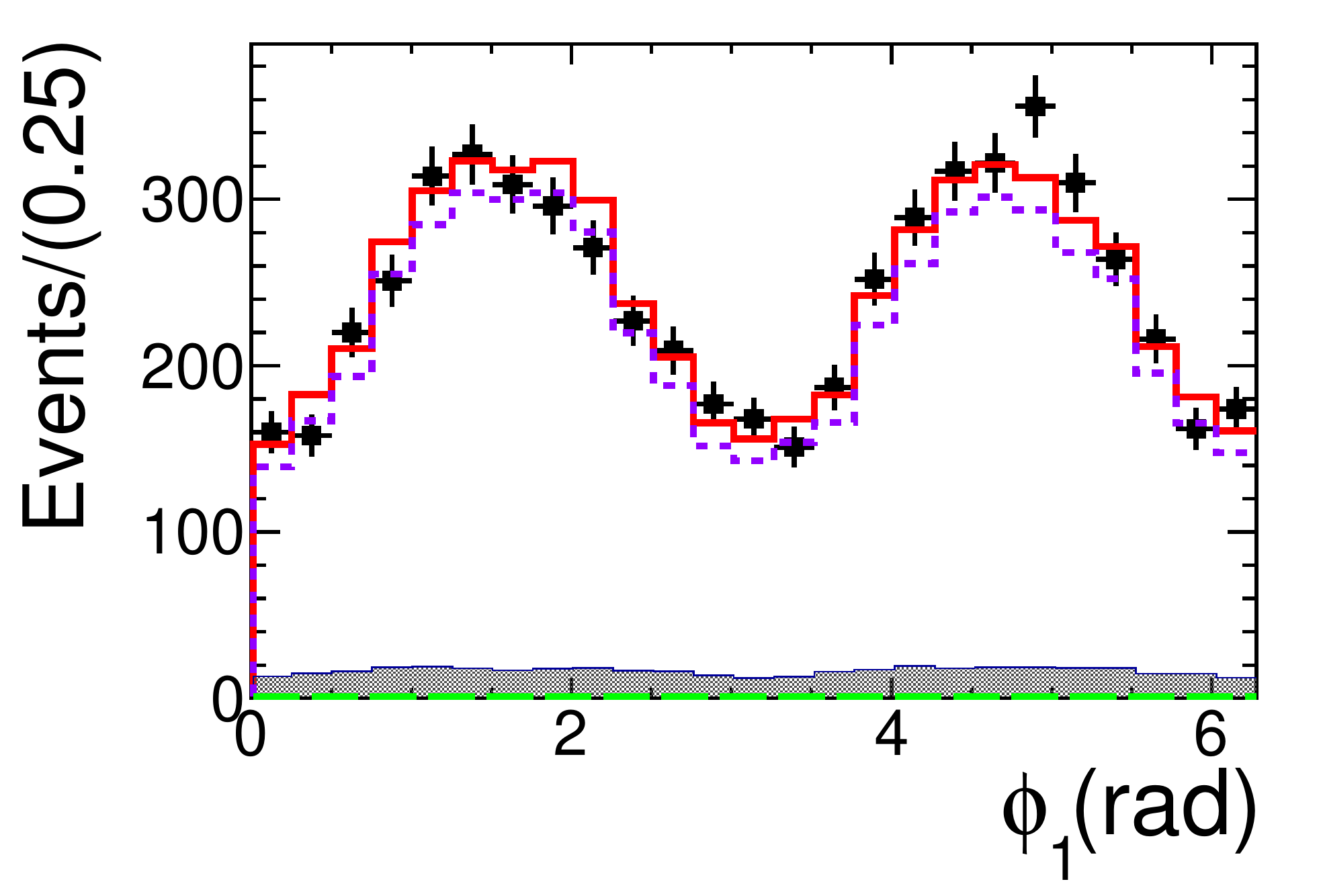}
\caption{Mass and angular distributions of $D^{*+}_s$ tag sample for $J^P=1^-$ hypothesis. Points with error bars are data. The solid red line, dotted purple line, and long dashed green line are the total fit results, $D^{*+}_{s}D^{-}_{s}$, and $\gamma D^{+}_{s}D^{-}_{s}$ processes, respectively. The shaded black histograms indicate the scaled backgrounds derived from the inclusive MC samples.}
\label{fig:fitangular}
\end{figure*}

\begin{figure*}[!hbt]
\centering
\includegraphics[width=0.68\columnwidth]{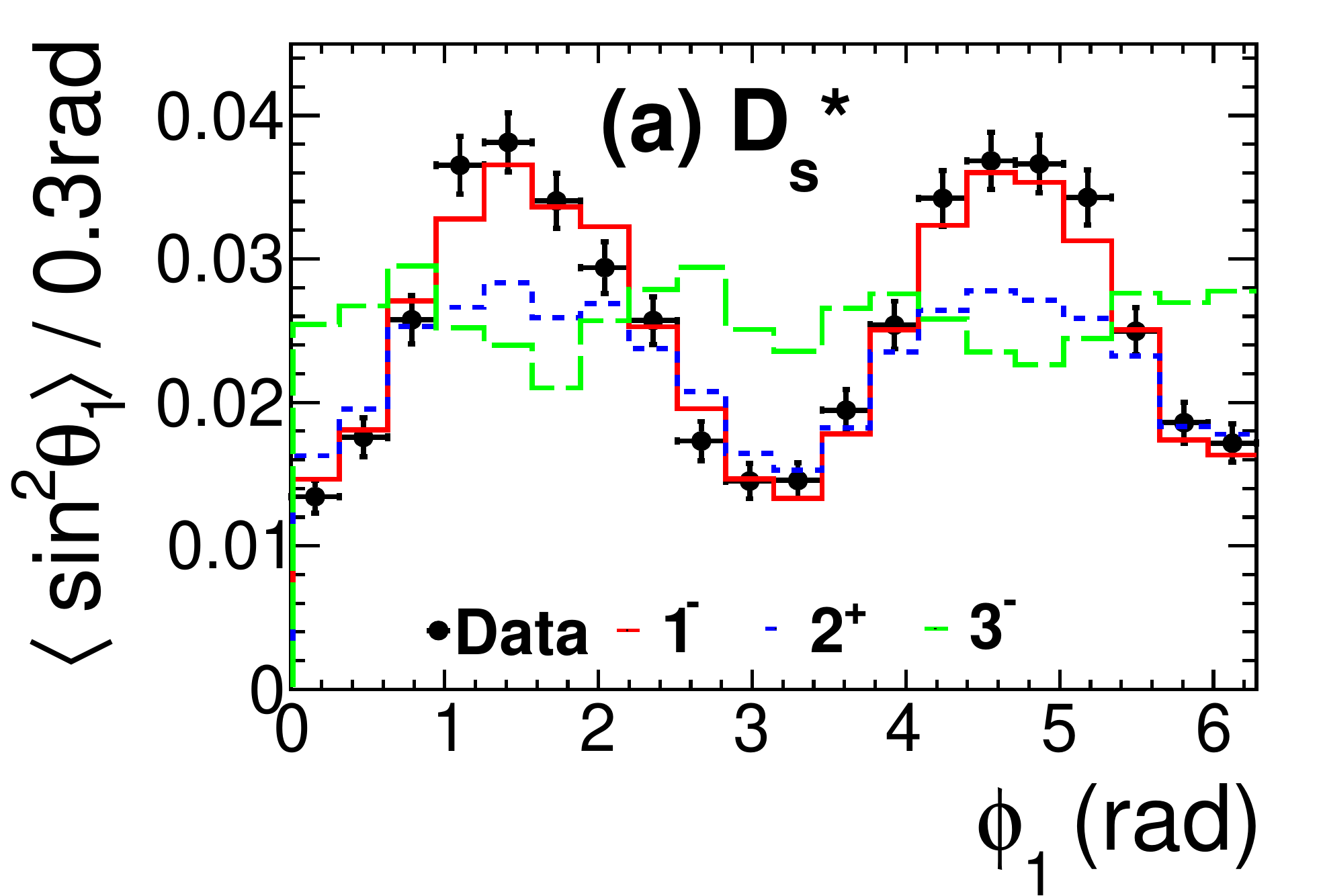}
\includegraphics[width=0.68\columnwidth]{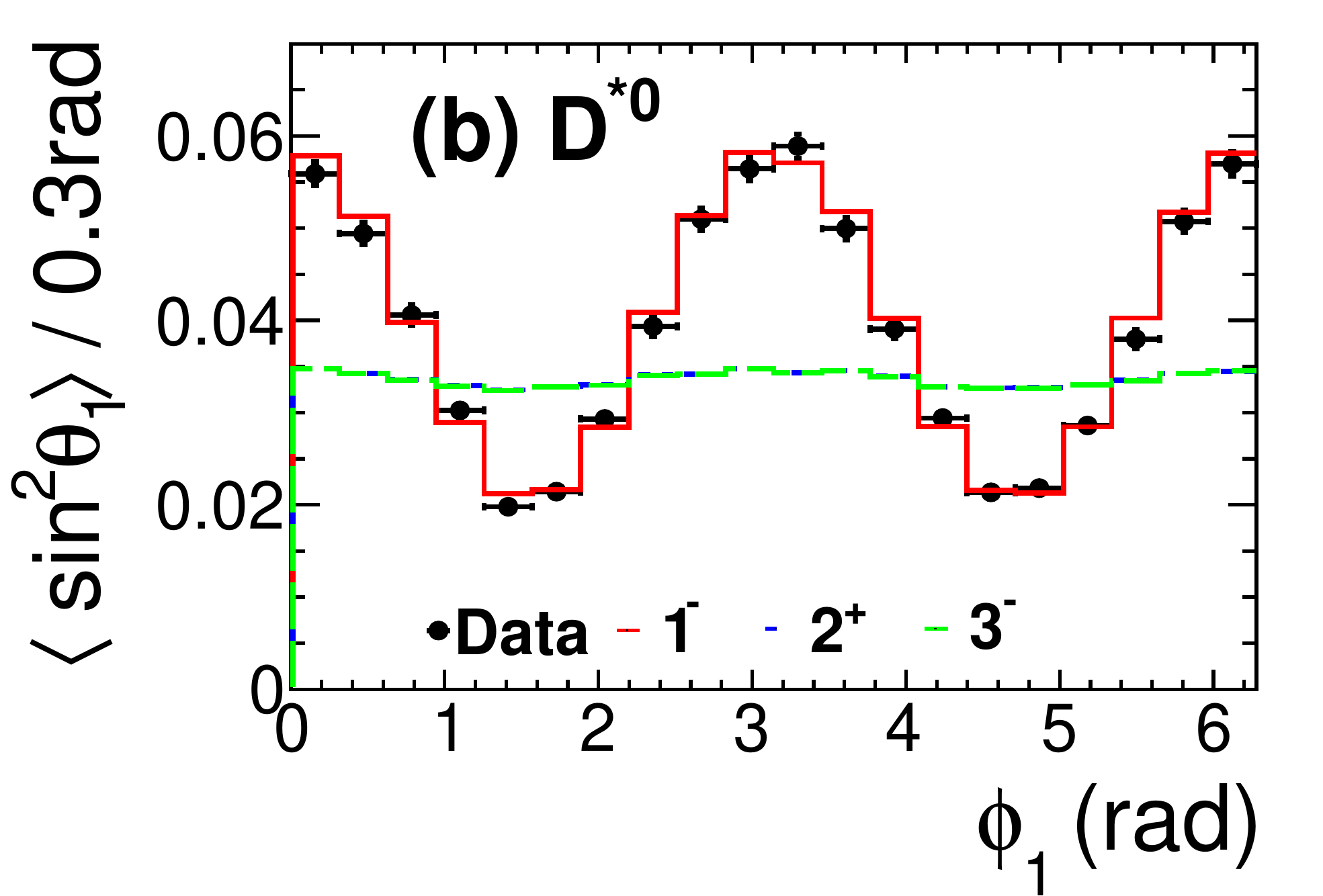}
\includegraphics[width=0.68\columnwidth]{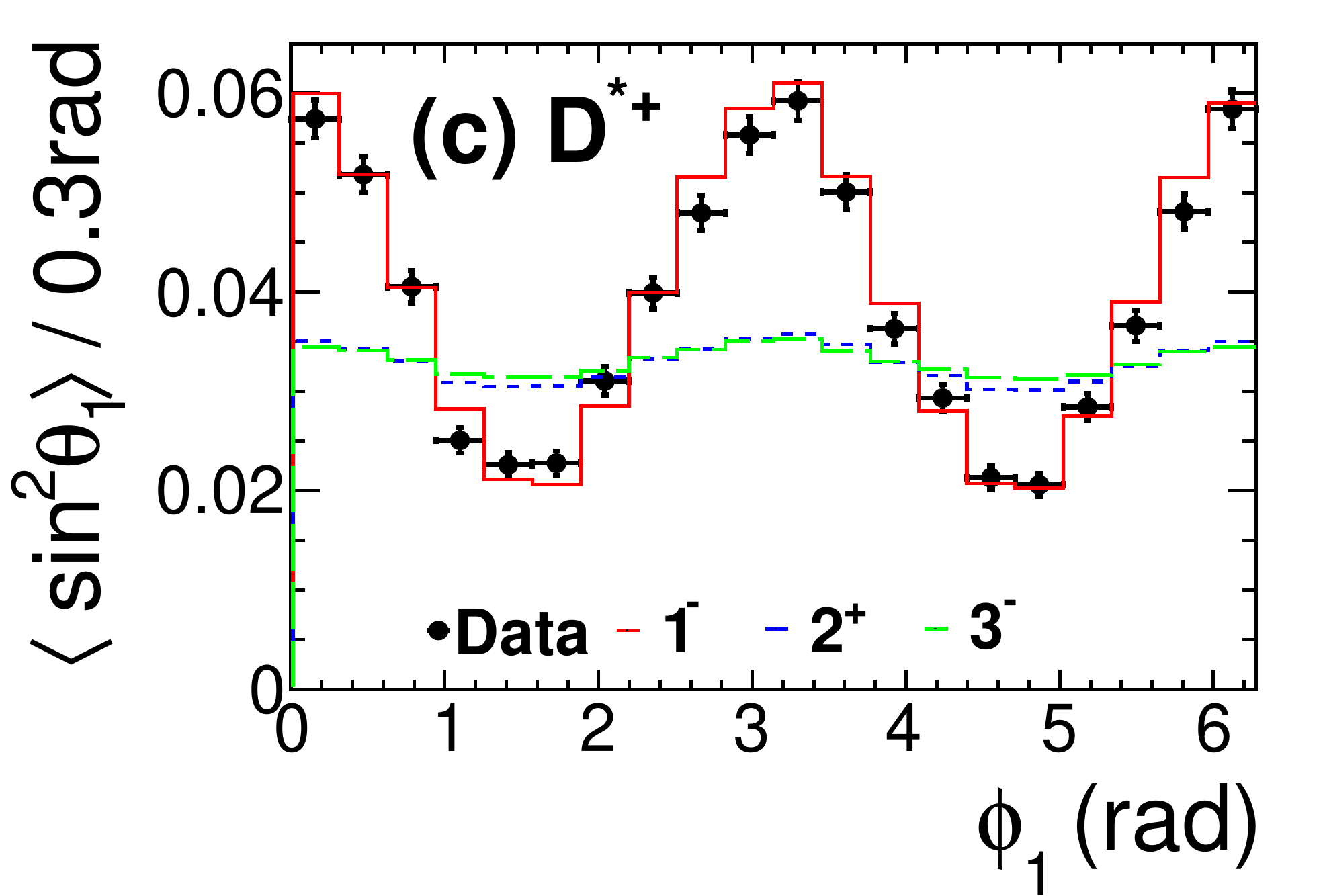}
\caption{
Distributions of the normalized moment $\left\langle \sin^{2}\theta_{1}\right \rangle$ versus $\phi_{1}$ for the processes $e^+e^-\to$  (a) $D_s^{*+}D_s^-$, (b) $D^{*0}\bar{D}^0$ and (c) $D^{*+}D^-$.
Points with error bars are data summing over $D^{*}$-tag sample and $D$-recoil sample with background events subtracted.
The solid red line, dotted blue line, and the long-dashed green line are the fit results for the $J^P=1^{-}$, $2^{+}$ , and  $3^{-}$ hypothesis, respectively.
}
\label{fig:fitdatamoment}
\end{figure*}

\begin{table}[H]
\caption{The significance of $J^P=1^-$ over other quantum number hypotheses. The significance is obtained considering changes in the difference of NDF. In each case that change is $\Delta(\text{NDF}) = 4$, accounting for the mass and width of the $D^{*}$ and the magnitude and phase for each component.}
\scriptsize
   \begin{center}
   \begin{tabular}{c|c|c|c}

\hline \hline
 Process& Hypothesis&$\Delta(-2\ln\mathcal{L})$& Significance\\ \hline
 \multirow{2}{*}{$D_{s}^{*+}D_s^-$}&$1^-$ over $2^+$& 1102  & $>$10$\sigma$ \\
 &$1^-$ over $3^-$& 2104  & $>$10$\sigma$ \\ \hline
 \multirow{2}{*}{$D^{*0}\bar{D}^0$}& $1^-$ over $2^+$& 12134 & $>$10$\sigma$\\
 &$1^-$ over $3^-$& 12096  & $>$10$\sigma$ \\ \hline
 \multirow{2}{*}{$D^{*+}D^-$}& $1^-$ over $2^+$& 11308 & $>$10$\sigma$\\
 &$1^-$ over $3^-$& 11222  & $>$10$\sigma$ \\
	\hline \hline
   \end{tabular}
    \end{center}
 \label{tab:significance}
\end{table}

Fit results for the mass and angular distributions of the $D^{*}_{(s)}$-tag sample  for the hypotheses $J^P$=$1^-$ are shown in Fig.~\ref{fig:fitangular}.
Mass and angular distributions of the  hypotheses $J^P=2^+$ and $3^-$ can be found in the supplemental material~\cite{Supplemental}.
The moment $\langle \sin^{2}\theta_{1}\rangle$, which represents an average observed in each $\phi_1$ bin, is a useful observable illustrating the different behaviors expected for different hypotheses ($J^{P}$ = $1^{-}$, $2^{+}$ and $3^{-}$).
Figure~\ref{fig:fitdatamoment} shows the $\langle  \sin^{2}\theta_{1}\rangle$ distributions for the three spin-parity hypotheses, which combine the $D^{*}_{(s)}$-tag sample and the $D_{(s)}$-recoil sample.

The significances to accept the $1^-$ hypothesis for $D_s^{*+}$, $D^{*0}$, and $D^{*+}$ are determined by the continuous test following Refs.~\cite{Narsky:1999kt,Zhu:2006pfm}, where the null hypothesis ($H_0$) represents the $J^P$ of $D^{*}_{(s)}$ taken as $2^{+}$ or $3^{-}$ and the alternative hypothesis ($H_1$) represents $J^P$ of $D^{*}_{(s)}$ taken as a linear combination of either [$2^{+},1^{-}$] or [$3^{-},1^{-}$].
The angular fits based on the two hypotheses are performed to data and the likelihood function values are denoted as $\ln(\mathcal{L}_{H_0})$ and $\ln(\mathcal{L}_{H_1})$.

The significance of the $J^P$=$1^-$ hypothesis is found to be larger than $10\sigma$, based on the change in 
 likelihood ($-2\ln\mathcal{L}$) and number of degrees of freedom (NDF) listed in Table~\ref{tab:significance}.


\section{Systematic uncertainties}
The results of the helicity amplitude analysis can be 
affected by detector effects and analysis procedures. In order to 
quantify the impact of such systematic effects on this work, dedicated 
studies on control samples are performed and corrections for eventual 
data/MC differences are evaluated.

 The sources of systematic uncertainties include the tracking and PID efficiencies for $K^{\pm}$ and $\pi^{\pm}$, which are studied with control samples of $e^+e^-\rightarrow K^+K^-\pi^+\pi^-$, $K^+K^-K^+K^-$, $K^+K^-\pi^+\pi^-\pi^0$, $\pi^+\pi^-\pi^+\pi^-$, and $\pi^+\pi^-\pi^+\pi^-\pi^0$ events~\cite{BESIII:2019kfh}.	
The photon reconstruction efficiency is studied using $J/\psi \rightarrow \rho^0\pi^0$ events~\cite{BESIII:2010ank}.						
The $K_S^0$ reconstruction efficiency is studied with control samples of $J/\psi \rightarrow K^*(892)^{\pm}K^{\mp},K^*(892)^{\pm}\rightarrow K^0_S\pi^{\pm}$ and $J/\psi \rightarrow \phi K^0_SK^{\mp}\pi^{\pm}$~\cite{BESIII:2015jmz}. 
The $\pi^0$ reconstruction efficiency is studied by the double-tag $D\bar{D}$ hadronic decays $D^0\rightarrow K^-\pi^+,K^-\pi^+\pi^+\pi^-$ versus $\bar{D}^0\rightarrow K^+\pi^-\pi^0,K_S^0\pi^0$~\cite{BESIII:2016gbw,BESIII:2016hko}.					According to the efficiency differences determined above, an overall weighting of MC events is performed to match the data events in the fit.

Additionally, inconsistencies between data and MC in the description of the track helix parameters may result in systematic effects. 
The helix parameters in MC simulations are, therefore, corrected following the procedure described in Ref.~\cite{BESIII:2012mpj}.

To estimate potential bias due to differences between the data and MC simulation in the selected regions, the distributions of the kinematic variables $\Delta E$ and invariant mass ($M(D_{(s)})$, $M(D_{(s)}\pi^0(\gamma))$, and $RM(D_{(s)})$) in MC simulations are smeared with Gaussian functions to match the corresponding distributions in the data.

The systematic uncertainty due to the background weight factor ($\omega_{\rm bkg}$)  is estimated by setting it to 0.
For the non-perfect resolution description, the values of the $BW$ parameters are varied by one standard deviation ($\pm 1 \sigma$) from the nominal fit result and the result with the lowest significance is assigned as the systematic uncertainty.

Taking into account all of the systematic uncertainties, the $J^P = 1^-$ hypothesis is always unambiguous and its significance is always greater than 10$\sigma$, confirming the results of the main analysis.



\section{Summary}

The spin and parity of $D_{s}^{*+}$, $D^{*0}$, and $D^{*+}$ are determined in the processes $\ee\to D^{*+}_{s}D^{-}_{s}$, $D^{*0}\bar{D}^{0}$, and $D^{*+}D^{-}$ , based on 3.19~fb$^{-1}$ of $e^+e^-$ collision data accumulated at $\sqrt s=4.178$~GeV with the \mbox
{BESIII} detector.
The application of a helicity amplitude analysis 
results in a preference of the quantum number $J^{P}=1^{-}$ over the hypotheses $2^{+}$ and $3^{-}$ with a significance of more than $10\sigma$, 
thus confirming the quark model predictions.
This is the first experimental determination of the spin and parity of the $D^{*}_{(s)}$ mesons, which are the cornerstone for the exploration of the properties of heavier charm and beauty mesons~\cite{Asner:2008nq}.

\section{Acknowledgments}
The BESIII Collaboration thanks the staff of BEPCII and the IHEP computing center for their strong support. This work is supported in part by National Key R\&D Program of China under Contracts Nos. 2020YFA0406400, 2020YFA0406300; National Natural Science Foundation of China (NSFC) under Contracts Nos. 11635010, 11735014, 11835012, 11875262, 11935015, 11935016, 11935018, 11961141012, 12022510, 12025502, 12035009, 12035013, 12061131003, 12175244, 12192260, 12192261, 12192262, 12192263, 12192264, 12192265, 12221005; the Chinese Academy of Sciences (CAS) Large-Scale Scientific Facility Program; the CAS Center for Excellence in Particle Physics (CCEPP); Joint Large-Scale Scientific Facility Funds of the NSFC and CAS under Contract No. U1832207; CAS Key Research Program of Frontier Sciences under Contracts Nos. QYZDJ-SSW-SLH003, QYZDJ-SSW-SLH040; 100 Talents Program of CAS; Fundamental Research Funds for the Central Universities, Lanzhou University, University of Chinese Academy of Sciences; The Institute of Nuclear and Particle Physics (INPAC) and Shanghai Key Laboratory for Particle Physics and Cosmology; ERC under Contract No. 758462; European Union's Horizon 2020 research and innovation programme under Marie Sklodowska-Curie grant agreement under Contract No. 894790; German Research Foundation DFG under Contracts Nos. 443159800, 455635585, Collaborative Research Center CRC 1044, FOR5327, GRK 2149; Istituto Nazionale di Fisica Nucleare, Italy; Ministry of Development of Turkey under Contract No. DPT2006K-120470; National Research Foundation of Korea under Contract No. NRF-2022R1A2C1092335; National Science and Technology fund; National Science Research and Innovation Fund (NSRF) via the Program Management Unit for Human Resources \& Institutional Development, Research and Innovation under Contract No. B16F640076; Polish National Science Centre under Contract No. 2019/35/O/ST2/02907; Suranaree University of Technology (SUT), Thailand Science Research and Innovation (TSRI), and National Science Research and Innovation Fund (NSRF) under Contract No. 160355; The Royal Society, UK under Contract No. DH160214; The Swedish Research Council; U. S. Department of Energy under Contract No. DE-FG02-05ER41374.


\input{bibitem.tex}
\end{multicols}
\end{document}


\title{\bf\boldmath Supplemental Material for “Determination of spin and parity of $D^{*}_{(s)}$ mesons"}

\date{\it \small \bf \today}

 \author{\input{authors.tex}}
\maketitle
\newpage 
\section{Helicity Amplitude}
$B_L$ is the Blatt-Weisskopf factor, which depends on the angular momenta $L$ reaching from 0 up to 4 and can be written as
\begin{eqnarray}
B_0(r)/B_0(r_0)&=&1,\\
B_1(r)/B_1(r_0)&=&\frac{\sqrt{1+(qr_0)^{2}}}{\sqrt{1+(qr)^{2}}},\\
B_2(r)/B_2(r_0)&=&\frac{\sqrt{9+3(qr_0)^{2}+(qr_0)^{4}}}{\sqrt{9+3(qr)^{2}+(qr)^{4}}},\\
B_3(r)/B_3(r_0)&=&\frac{\sqrt{225+45(qr_0)^{2}+6(qr_0)^{4}+(qr_0)^{6}}}{\sqrt{225+45(qr)^{2}+6(qr)^{4}+(qr)^{6}}},\\
B_4(r)/B_4(r_0)&=&\frac{\sqrt{11025+1575(qr_0)^{2}+135(qr_0)^{4}+10(qr_0)^{6}+(qr_0)^{8}}}{\sqrt{11025+1575(qr)^{2}+135(qr)^{4}+10(qr)^{6}+(qr)^{8}}},
\end{eqnarray}
where $r$ is the magnitude of the momentum difference
 between the two final state particles; $r_0$ corresponds to choice by
 setting the resonant invariant
mass equal to its nominal mass; $q$ is a constant fixed to 3 GeV$^{-1}$.

\section{Projected distributions of the fitting results}

\begin{figure}[!hbt]
\centering
\includegraphics[width=0.3\columnwidth]{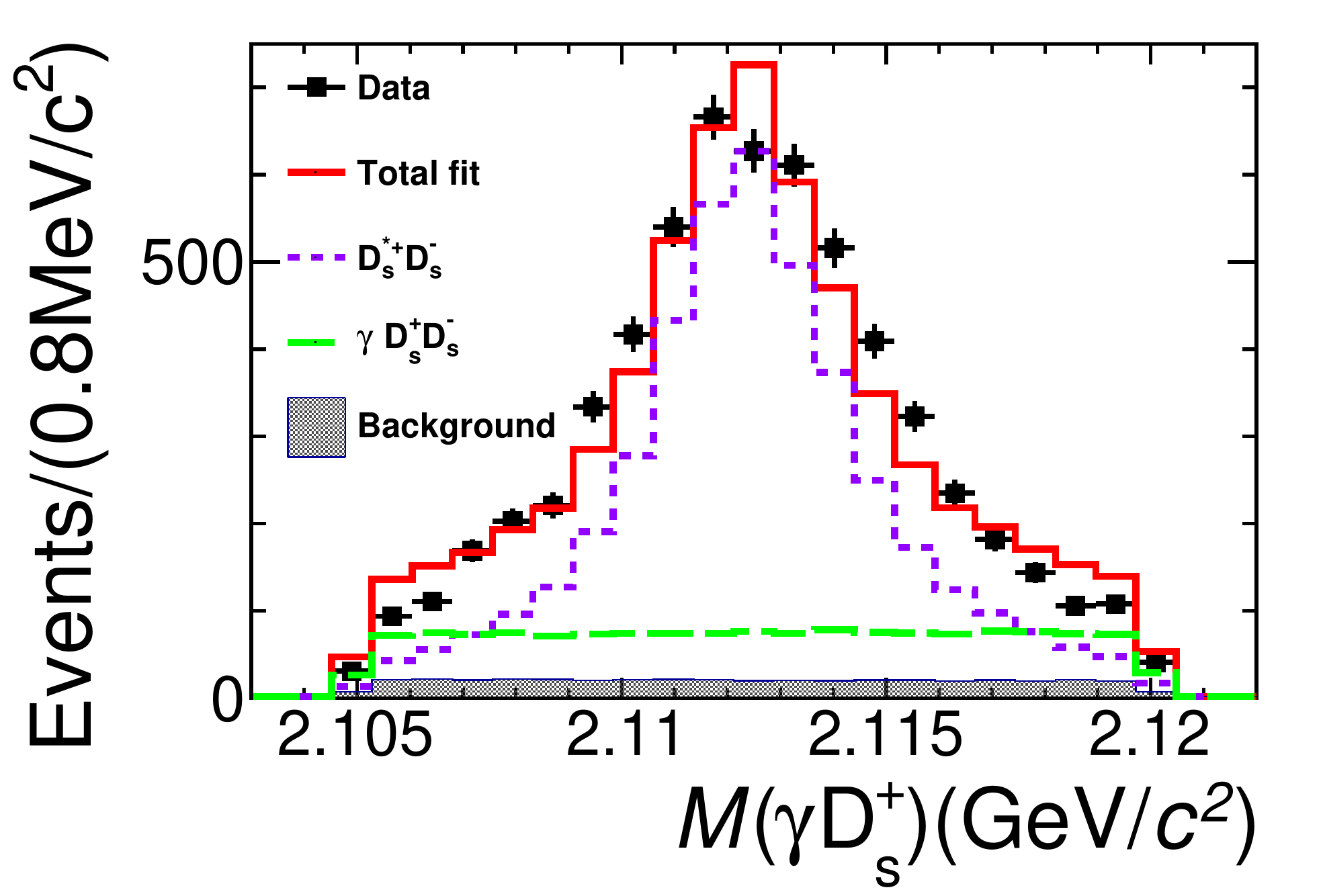}
\includegraphics[width=0.3\columnwidth]{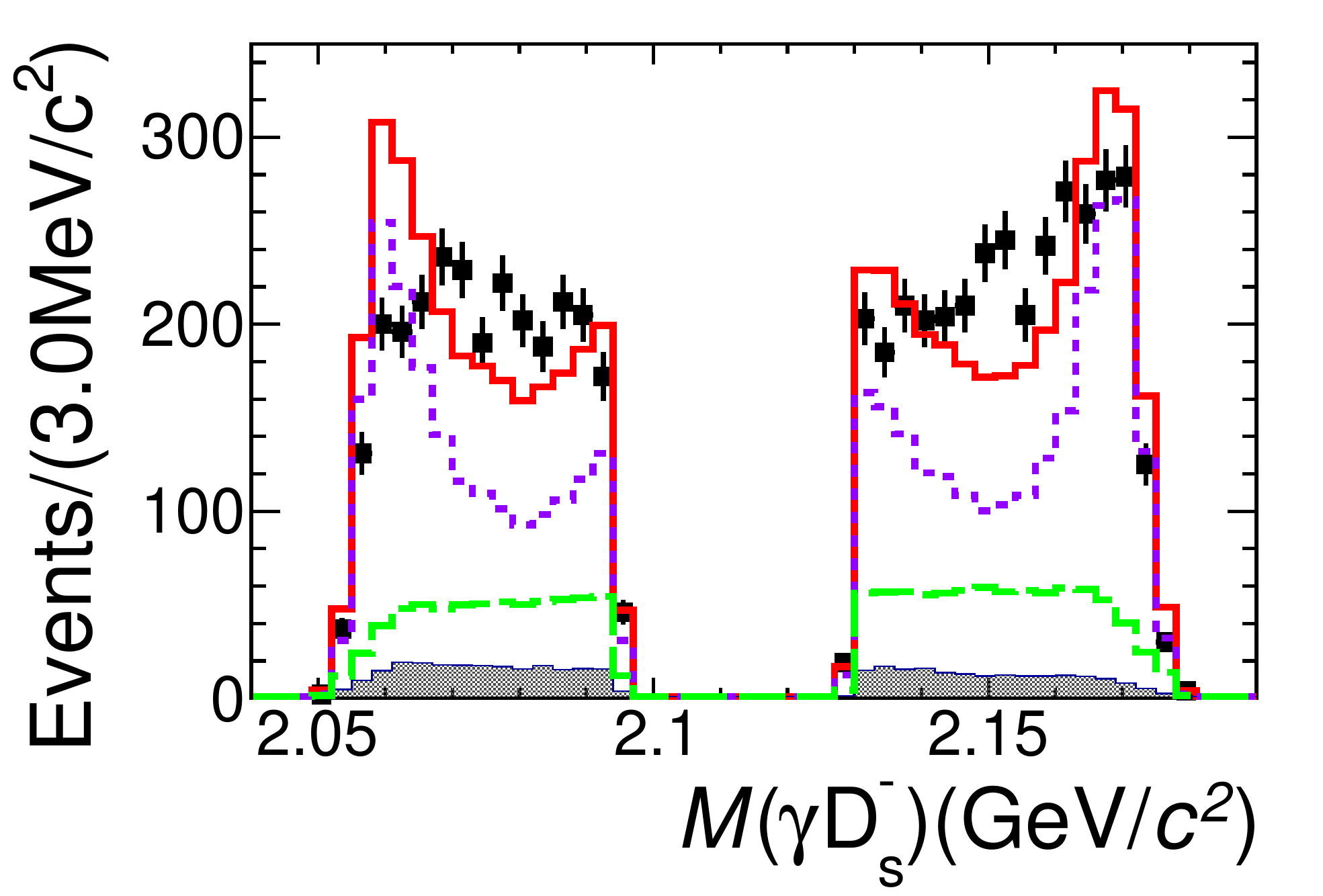}
\includegraphics[width=0.3\columnwidth]{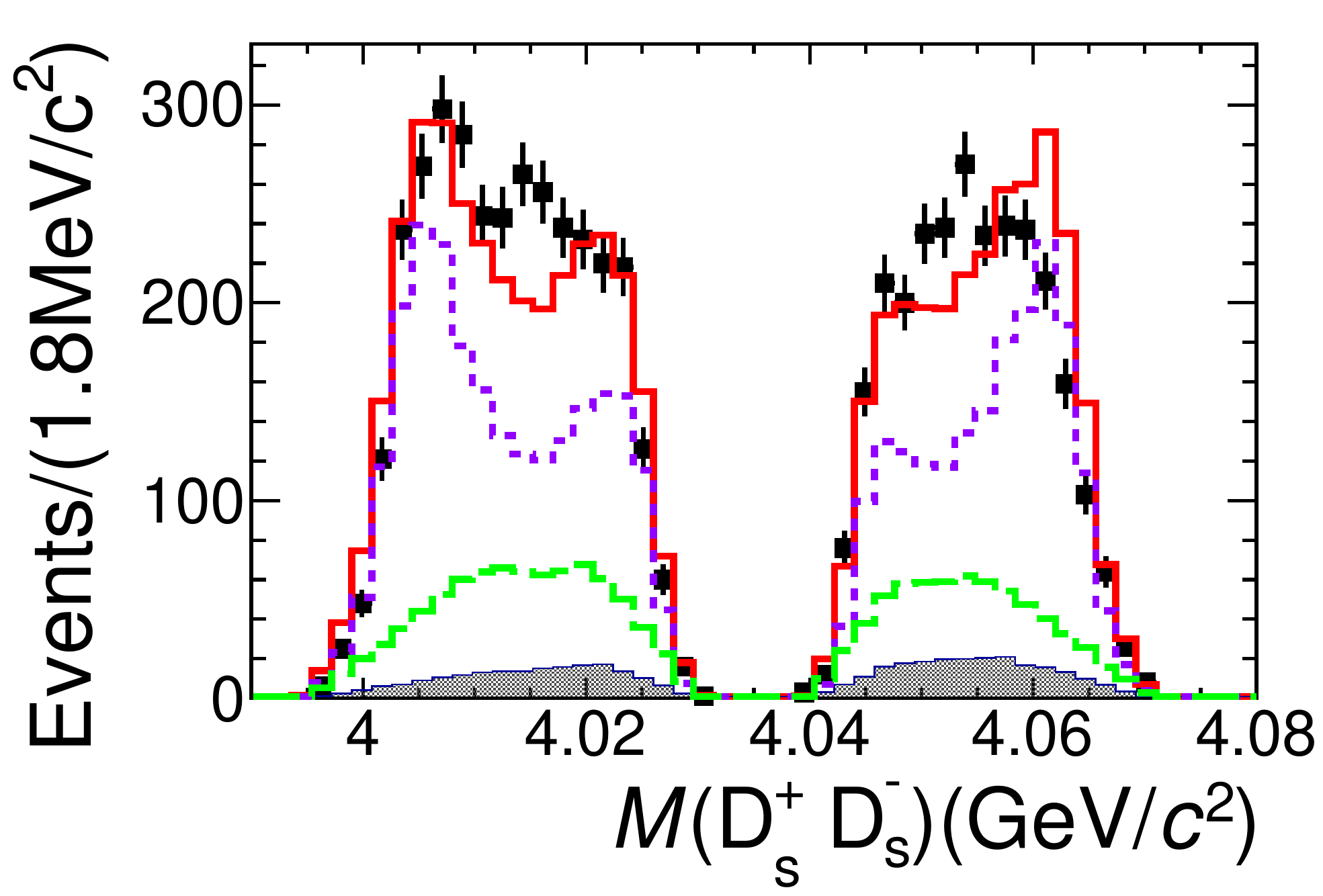}
\includegraphics[width=0.3\columnwidth]{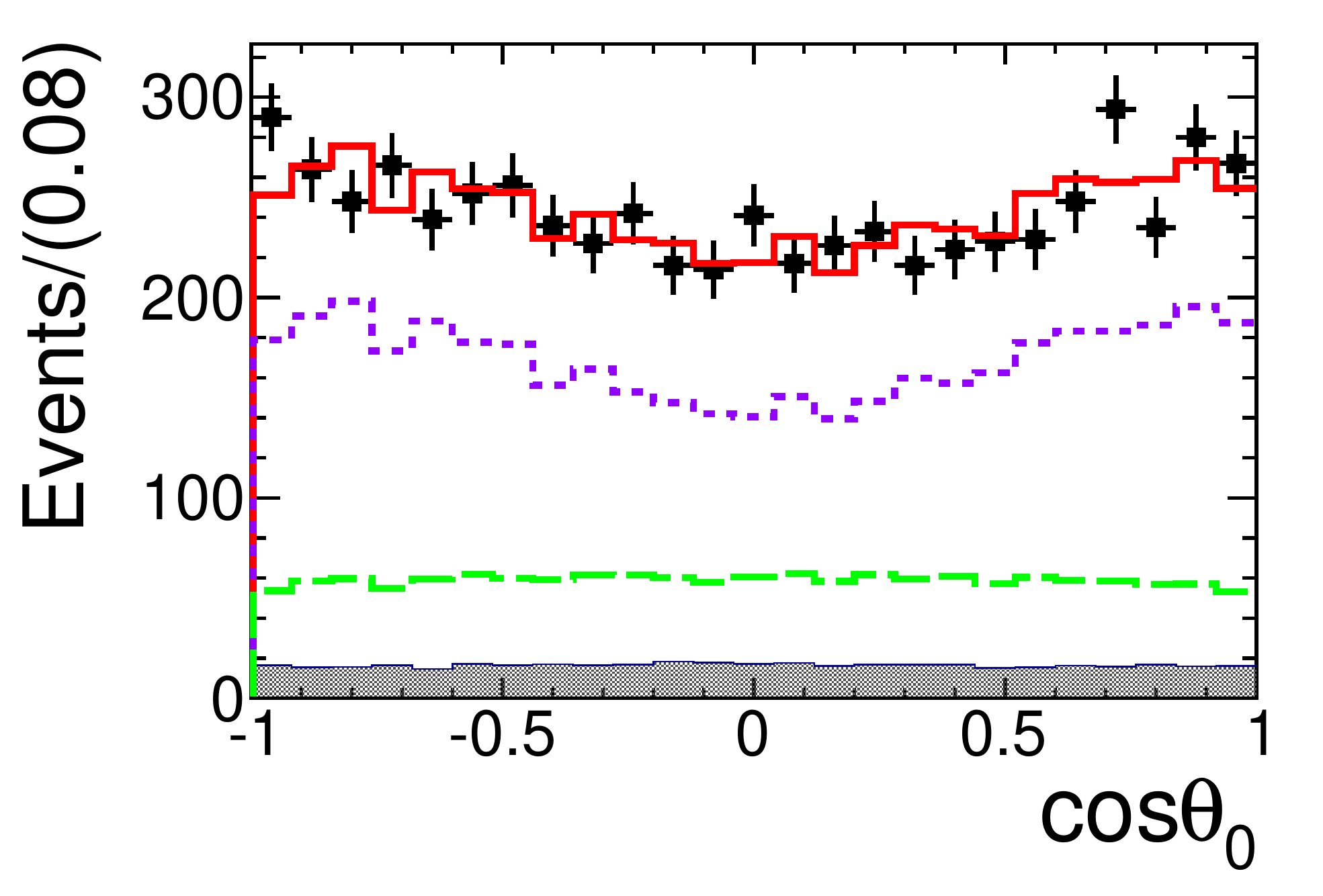}
\includegraphics[width=0.3\columnwidth]{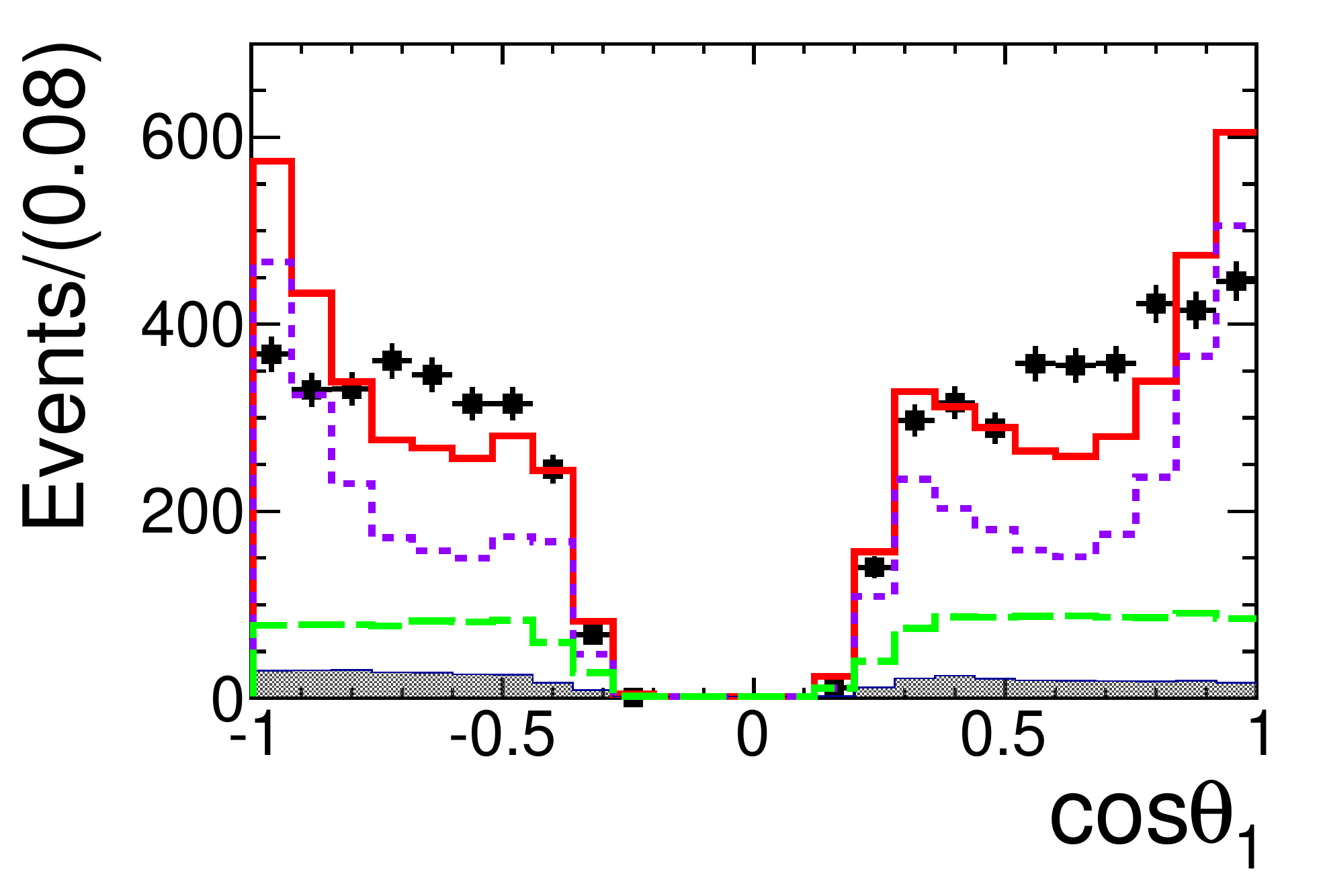}
\includegraphics[width=0.3\columnwidth]{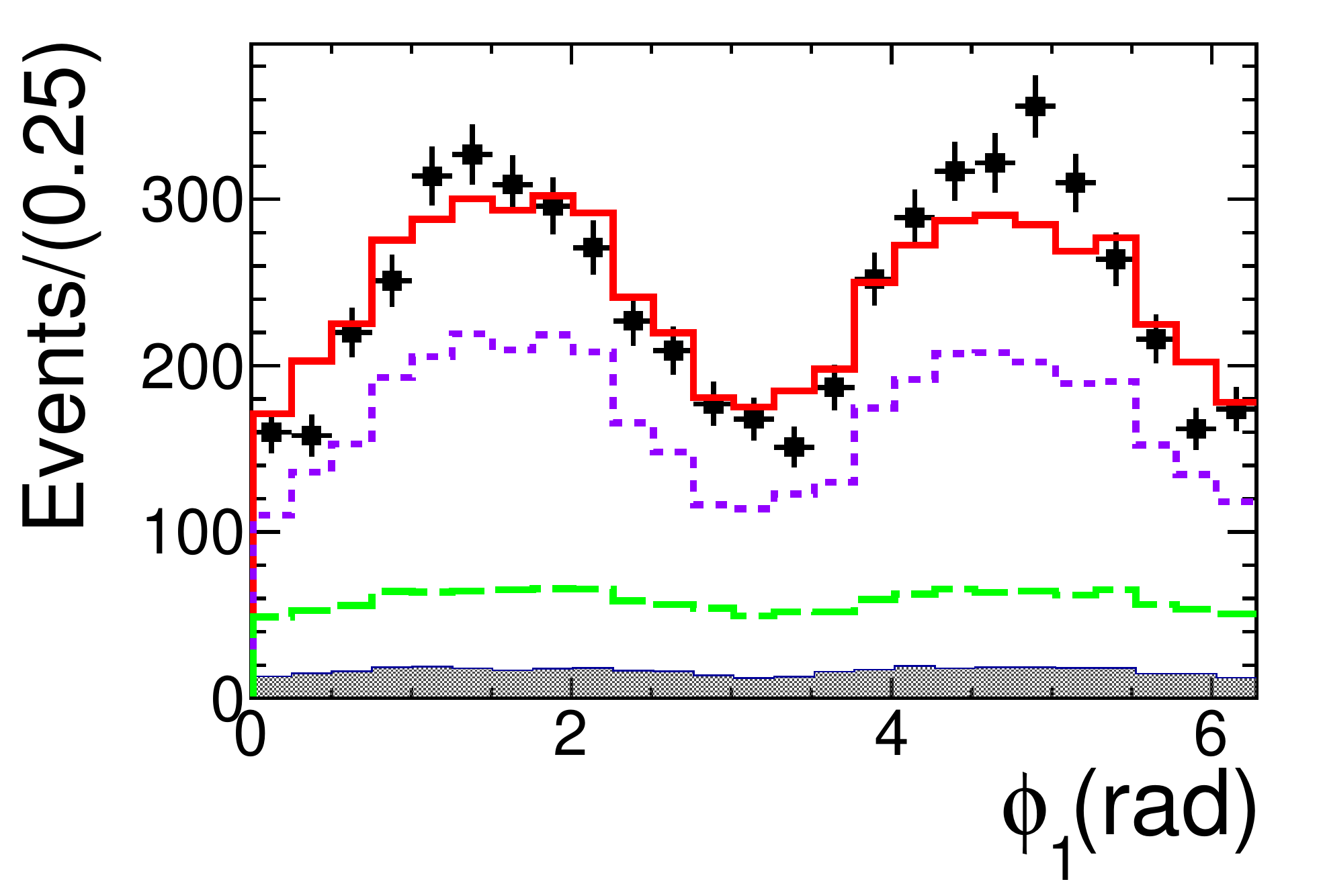}
\caption{
Mass and angular distributions of $D^{*+}_s$-tag sample for $J^P=2^+$ hypothesis. 
}
\label{fig:angular_Dsp_2}

\end{figure}

\begin{figure}[!hbt]
\centering
\includegraphics[width=0.3\columnwidth]{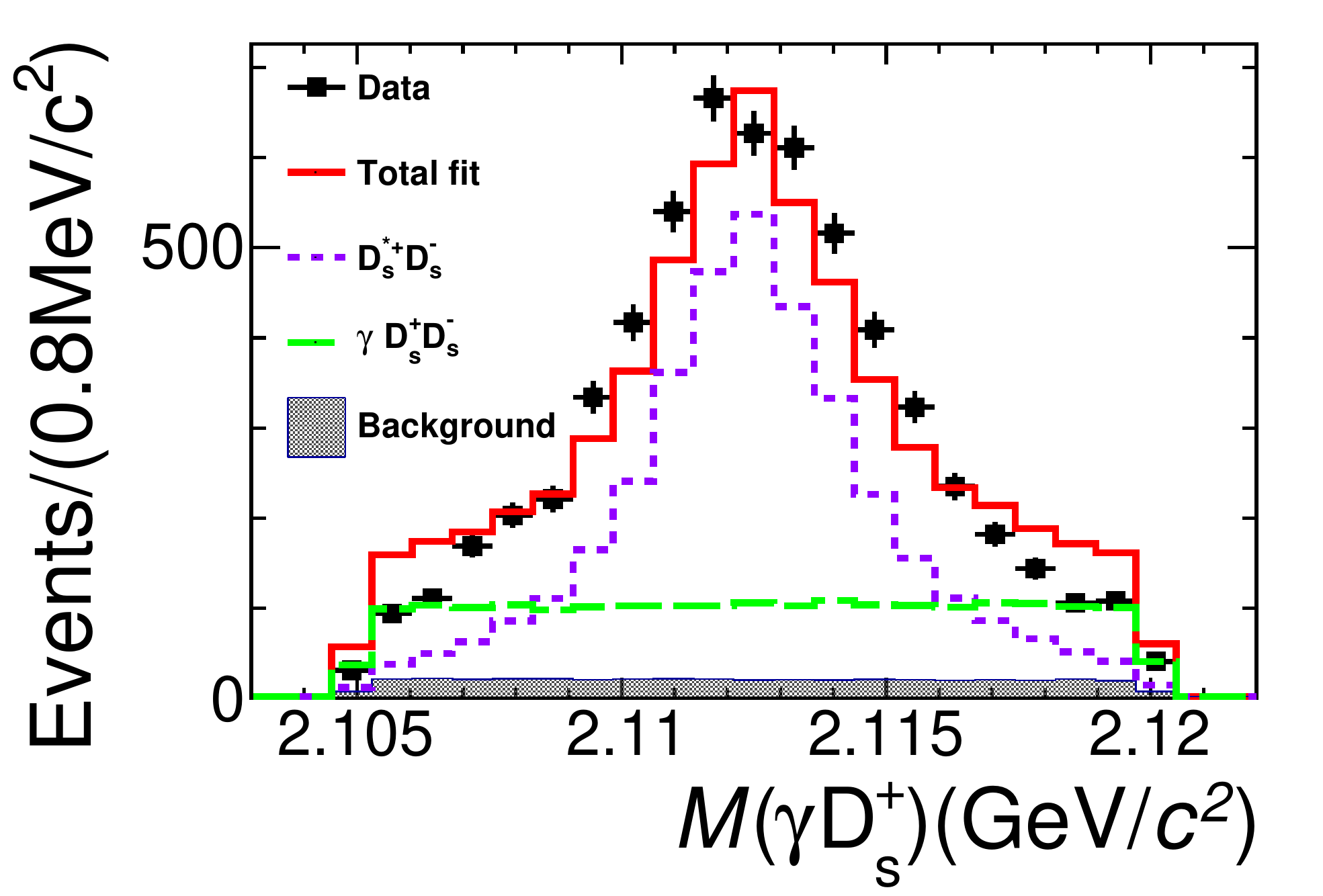}
\includegraphics[width=0.3\columnwidth]{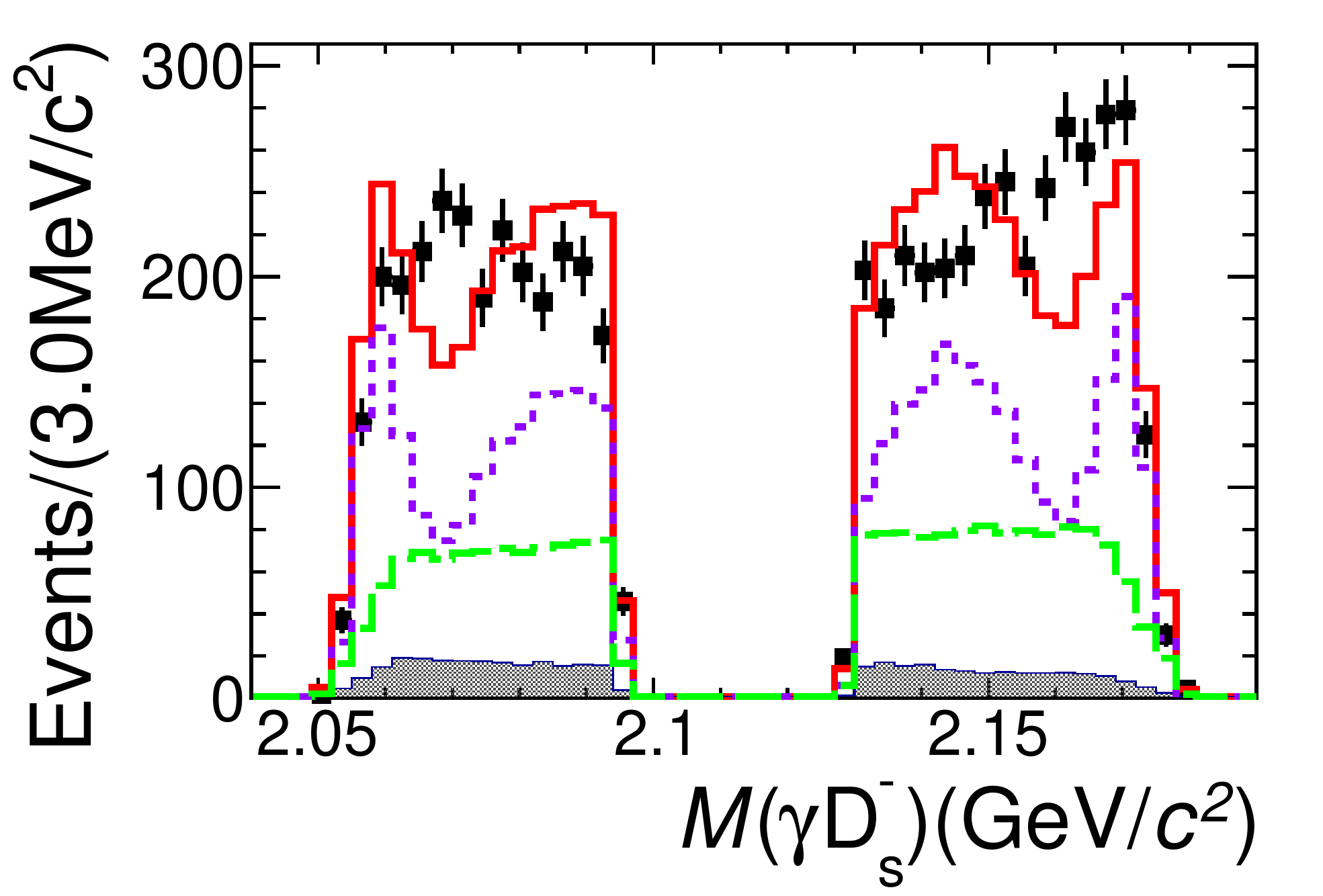}
\includegraphics[width=0.3\columnwidth]{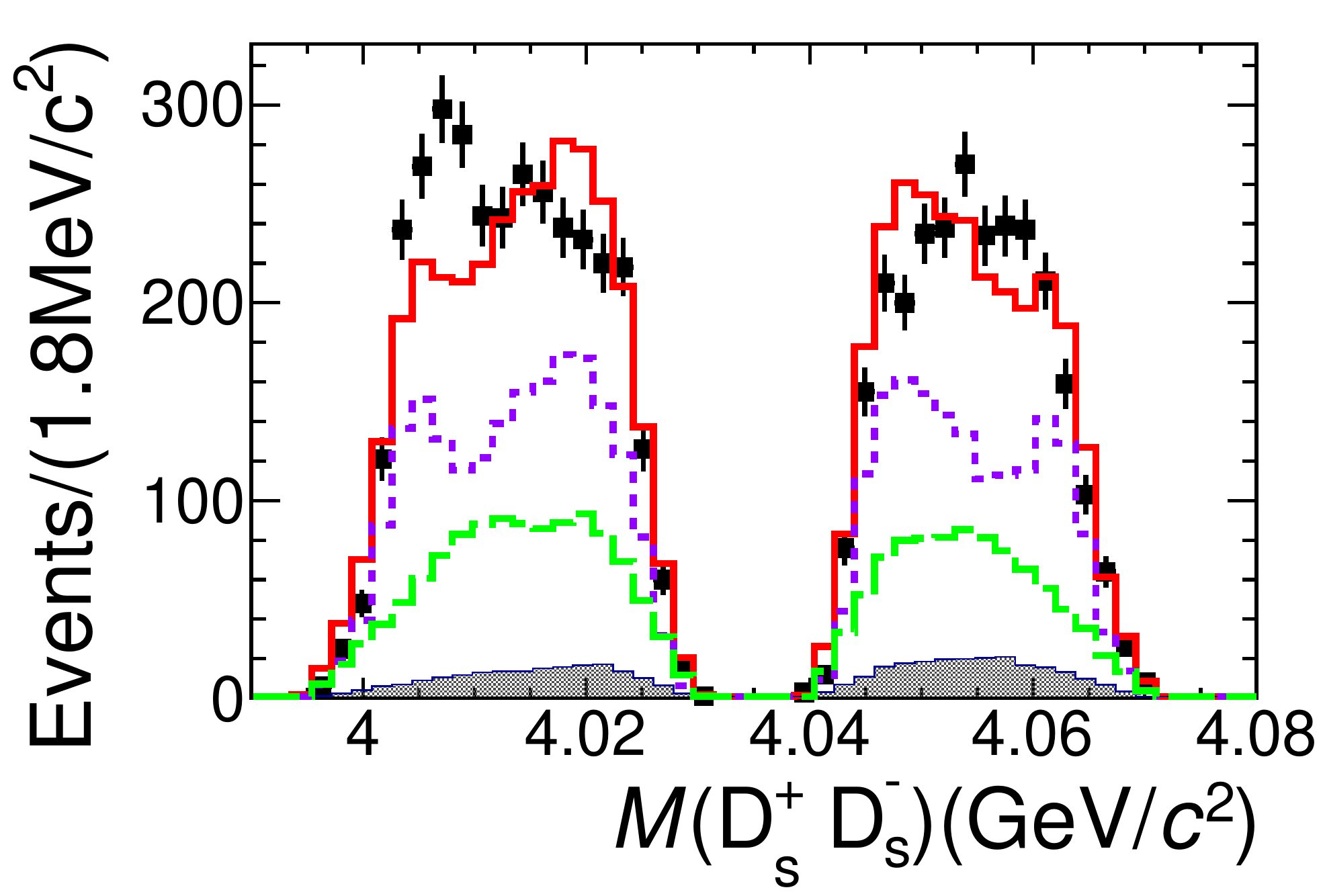}
\includegraphics[width=0.3\columnwidth]{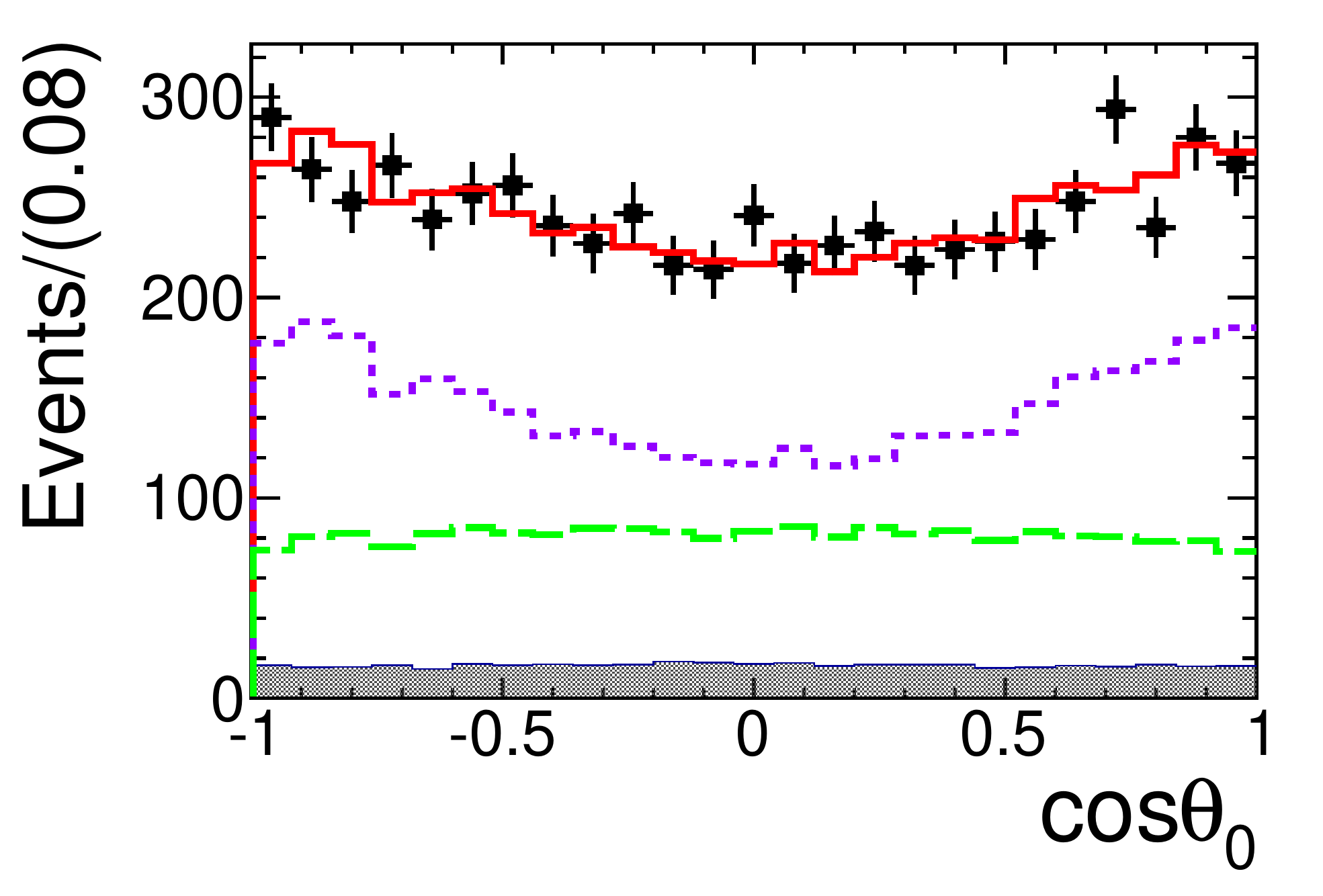}
\includegraphics[width=0.3\columnwidth]{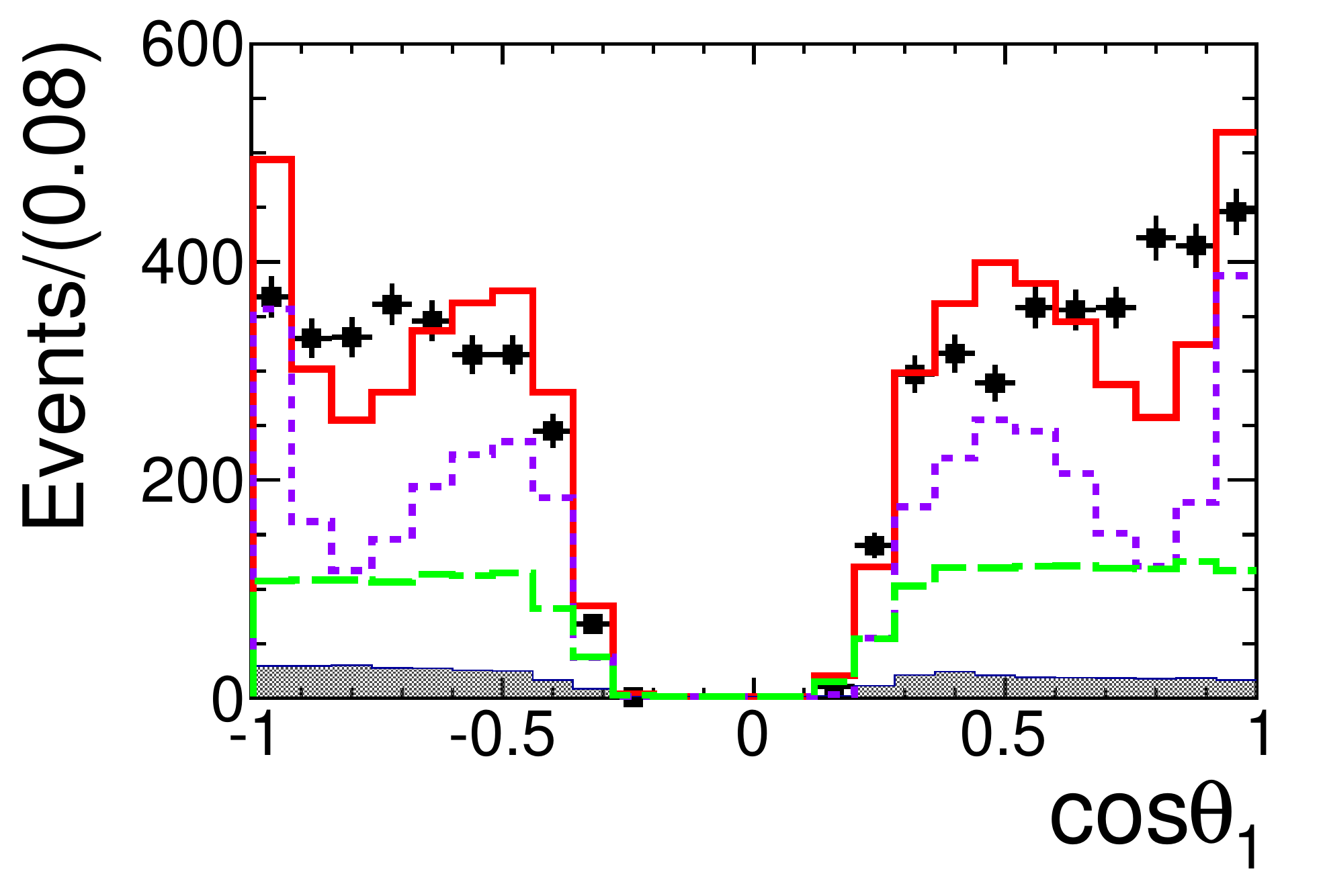}
\includegraphics[width=0.3\columnwidth]{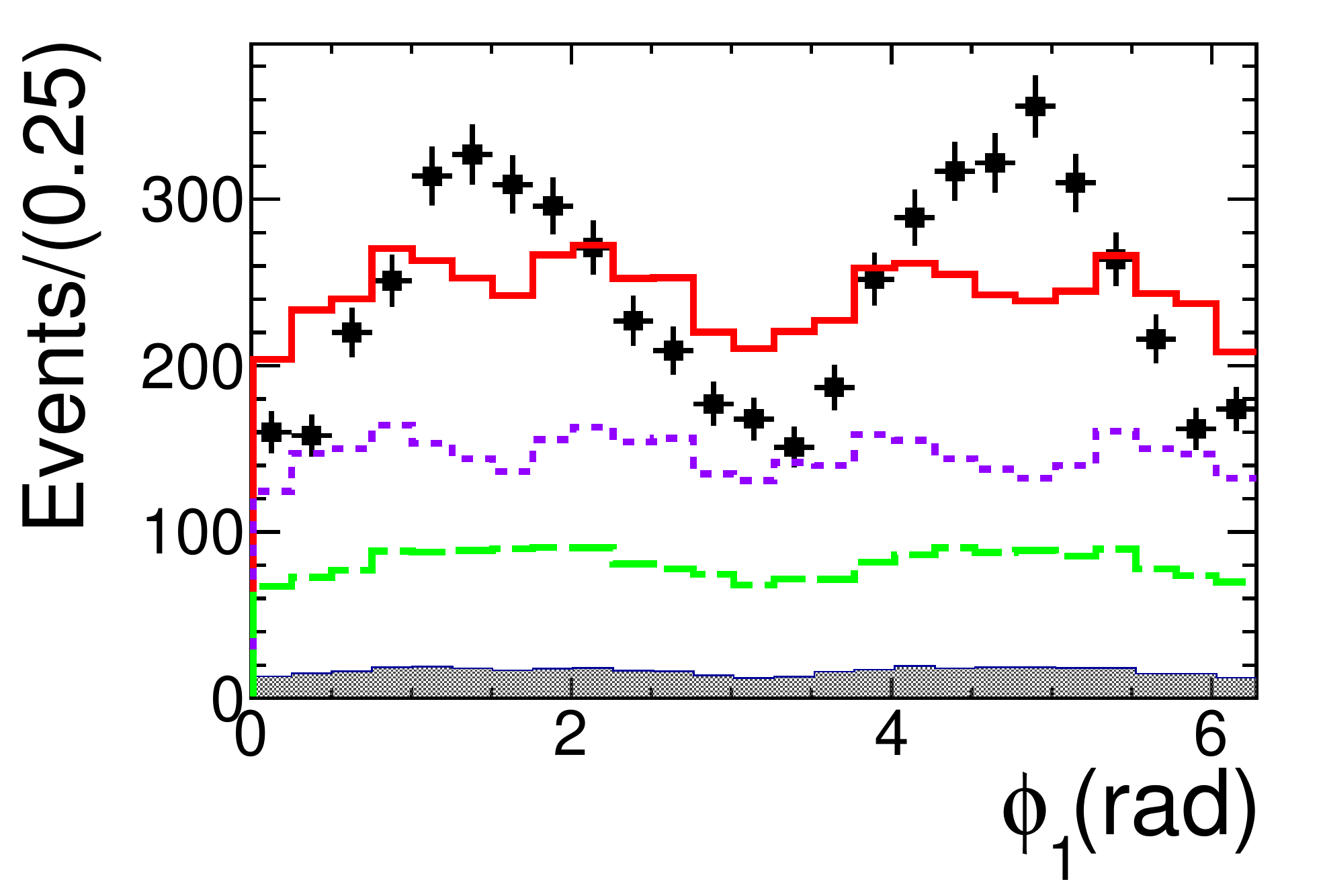}
\caption{
Mass and angular distributions of $D^{*+}_s$-tag sample for $J^P=3^-$ hypothesis.
}
\label{fig:angular_Dsp_3}
\end{figure}

\begin{figure}[!hbt]
\centering
\includegraphics[width=0.3\columnwidth]{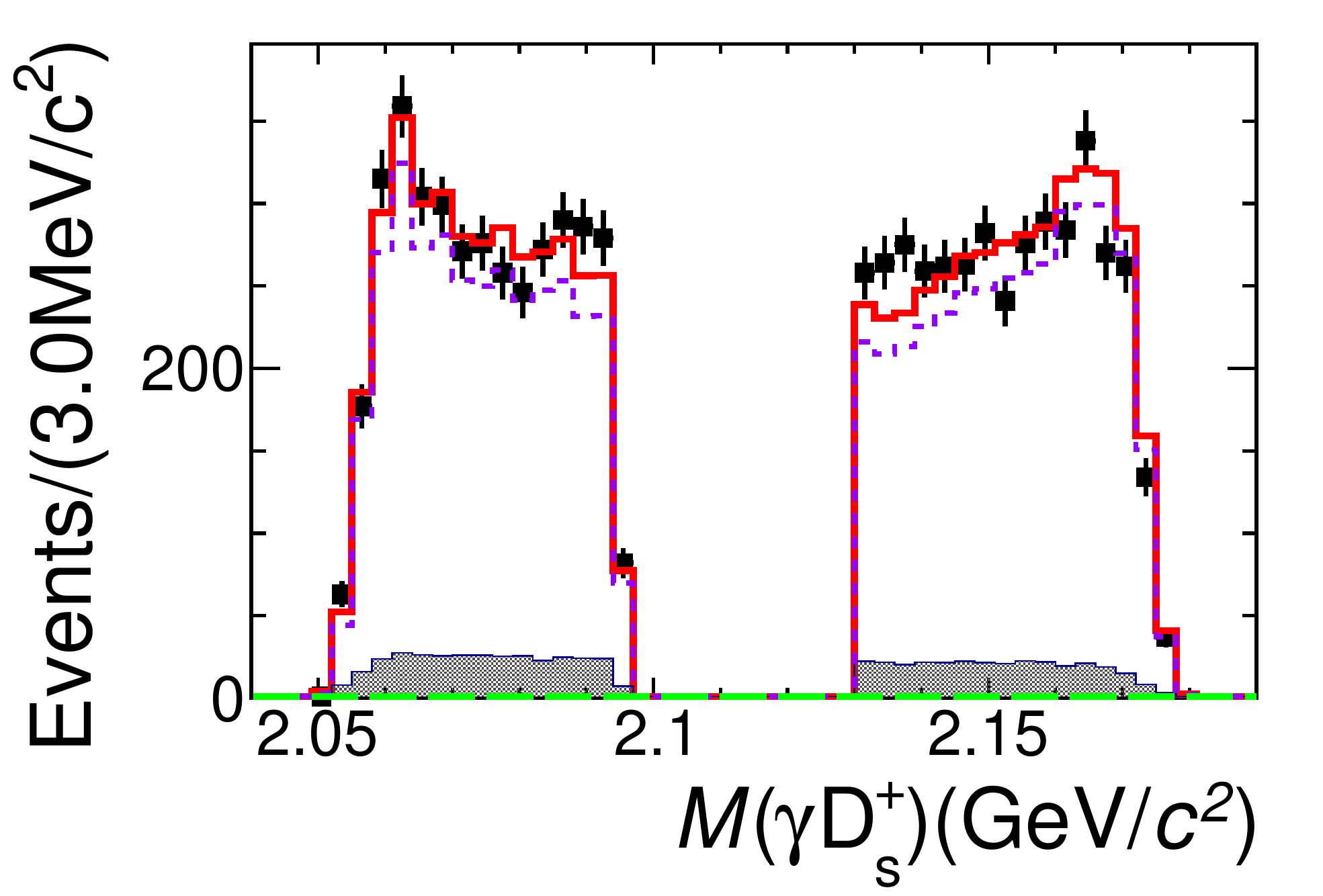}
\includegraphics[width=0.3\columnwidth]{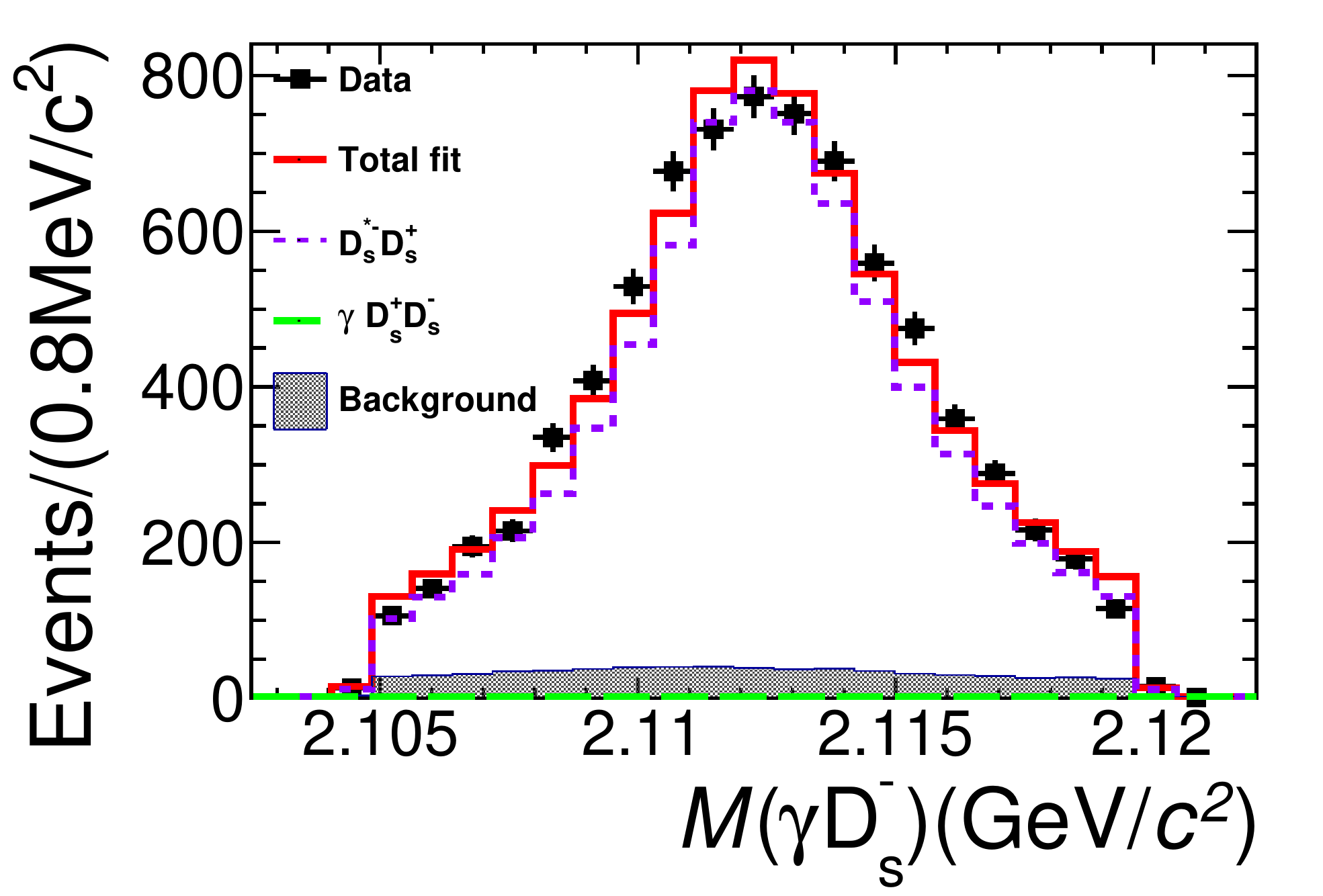}
\includegraphics[width=0.3\columnwidth]{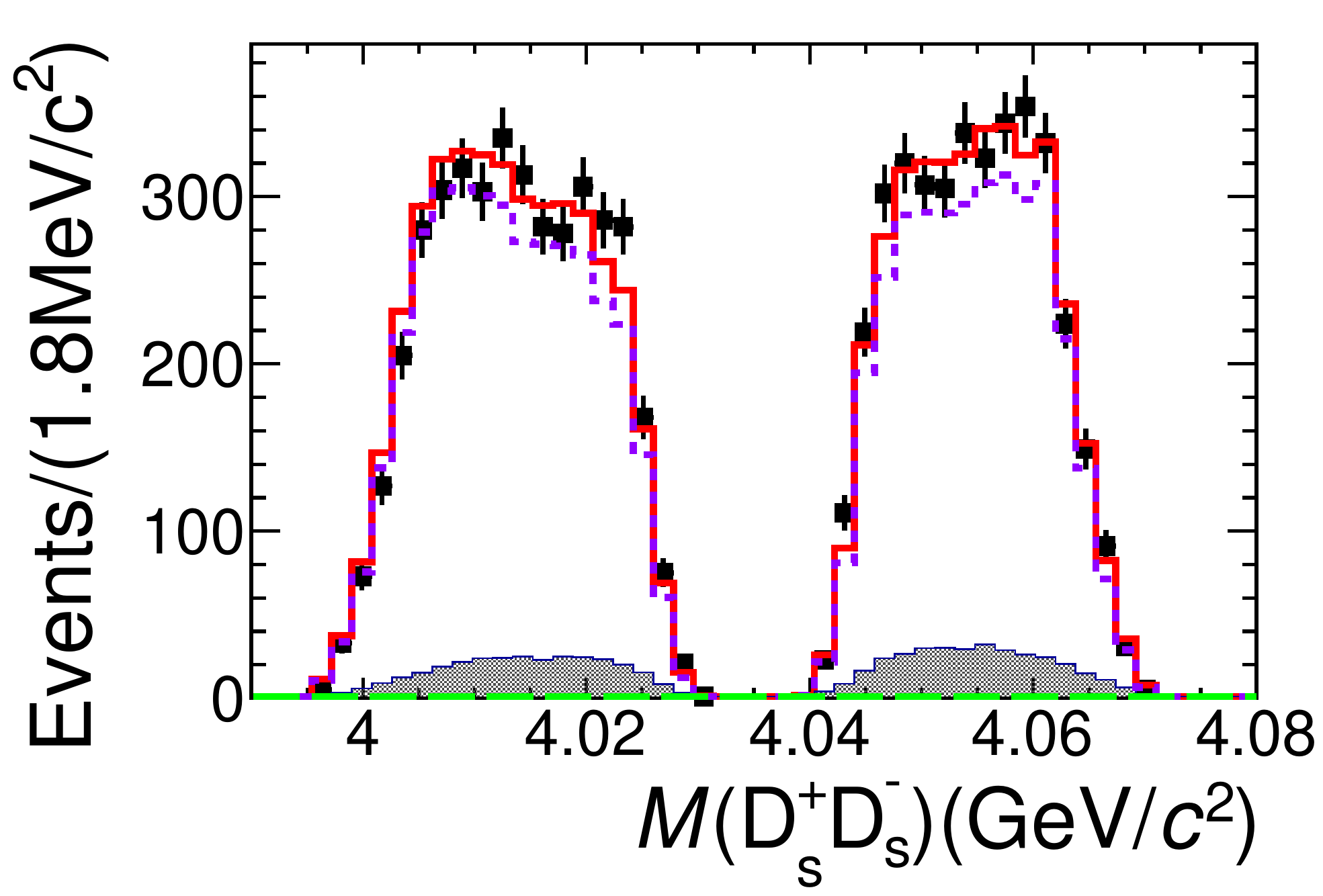}
\includegraphics[width=0.3\columnwidth]{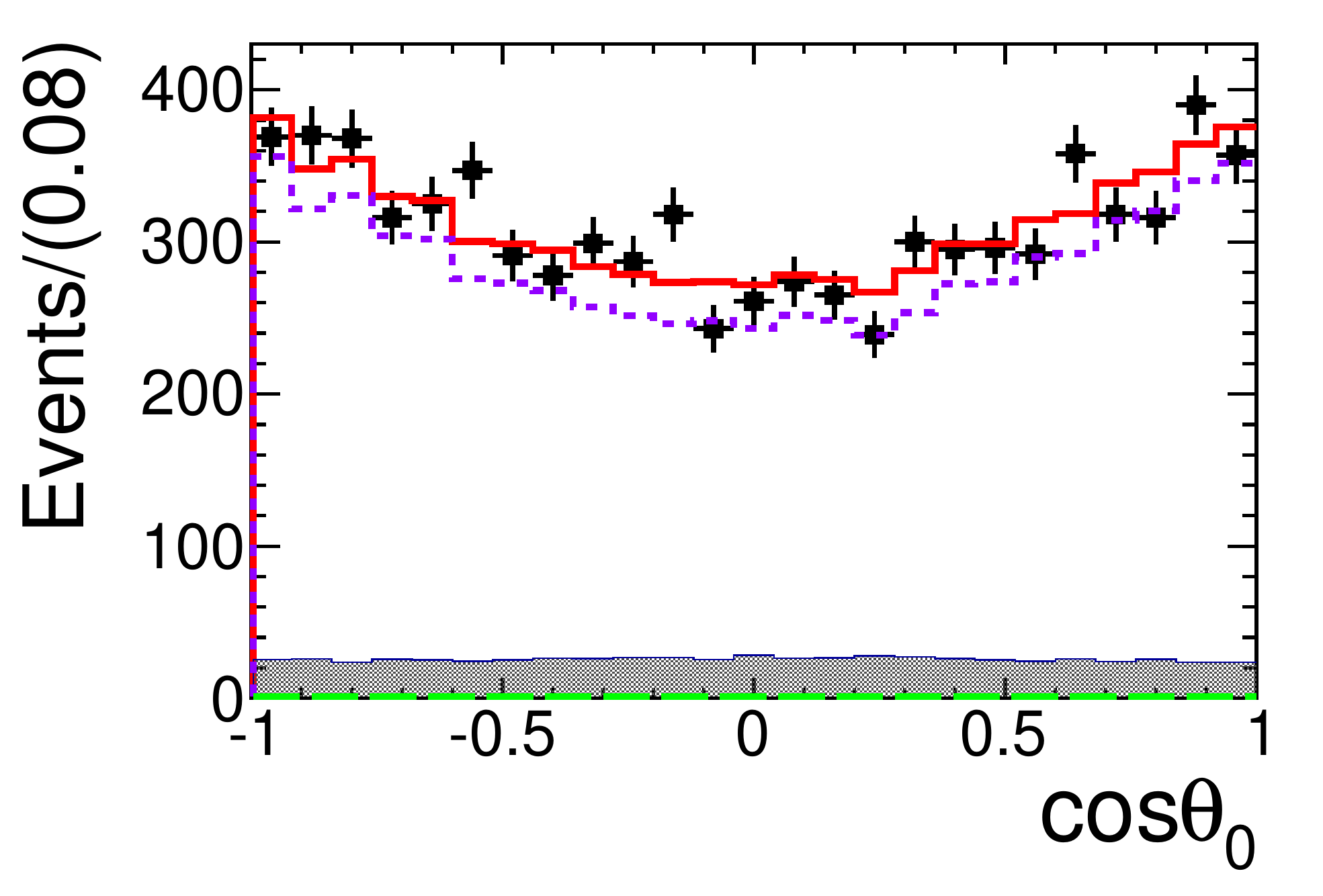}
\includegraphics[width=0.3\columnwidth]{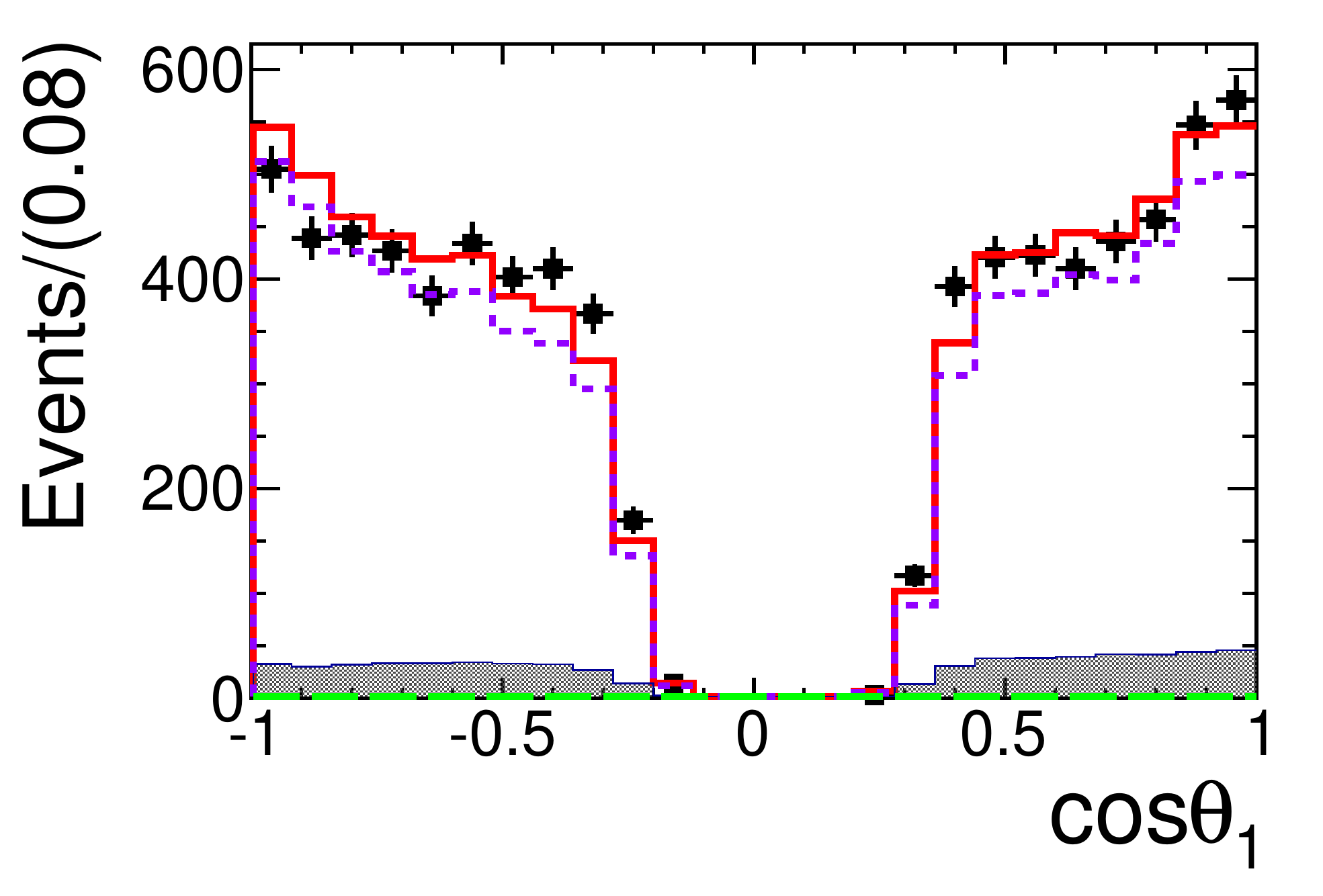}
\includegraphics[width=0.3\columnwidth]{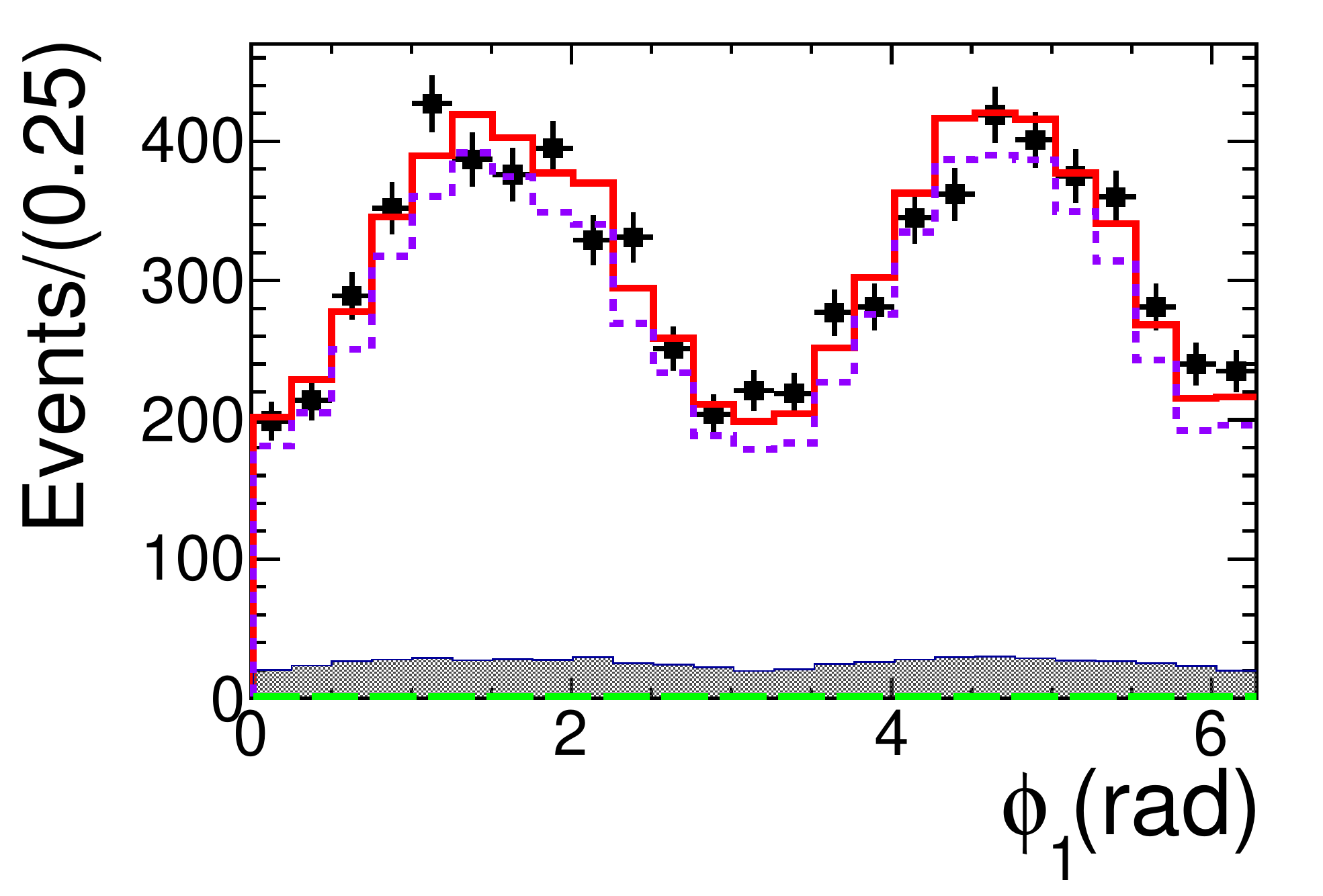}
\caption{
Mass and angular distributions of $D_s^+$-recoil sample for $J^P=1^-$ hypothesis.
}
\label{fig:angular_Dsm_1}
\end{figure}

\begin{figure}[!hbt]
\centering
\includegraphics[width=0.3\columnwidth]{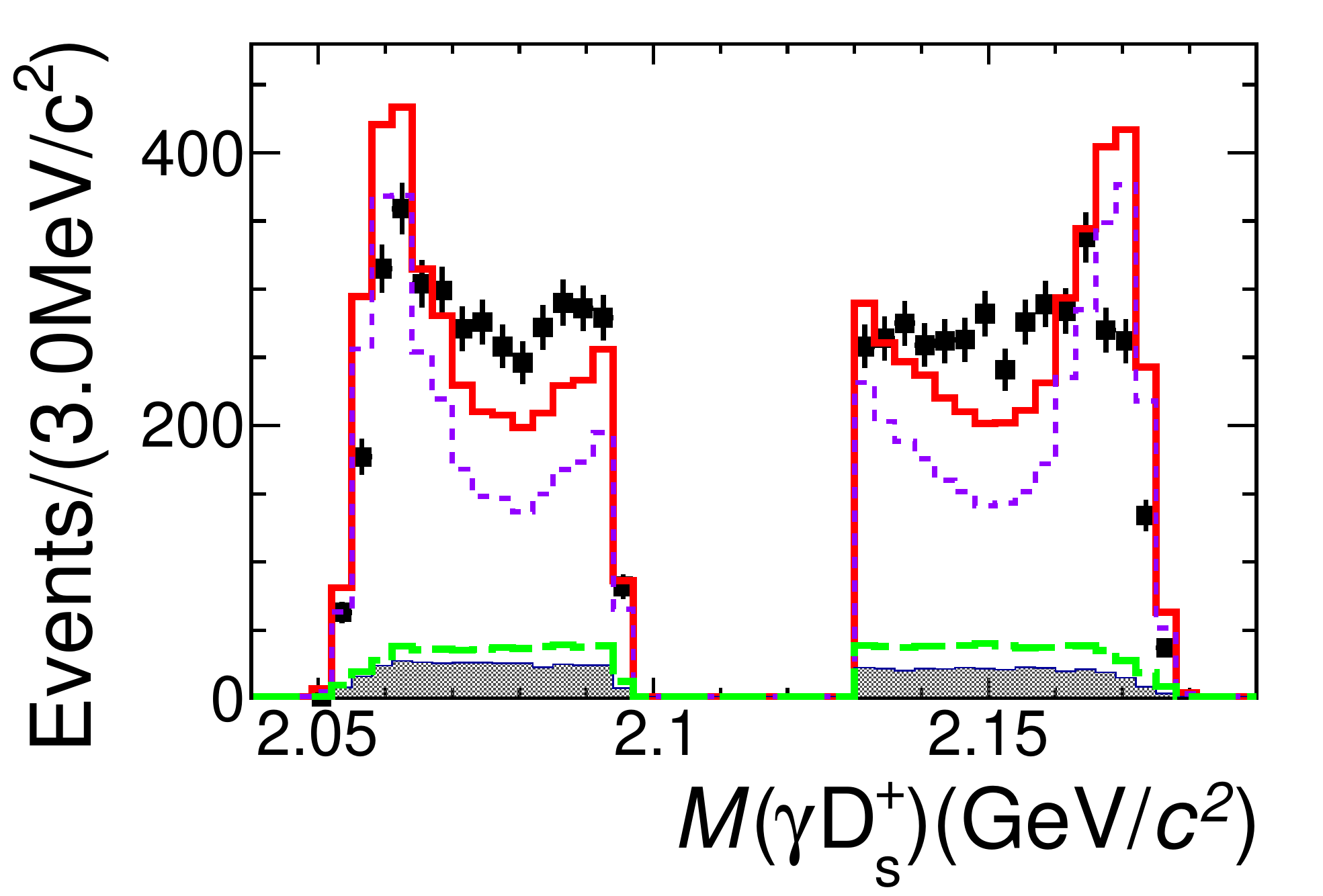}
\includegraphics[width=0.3\columnwidth]{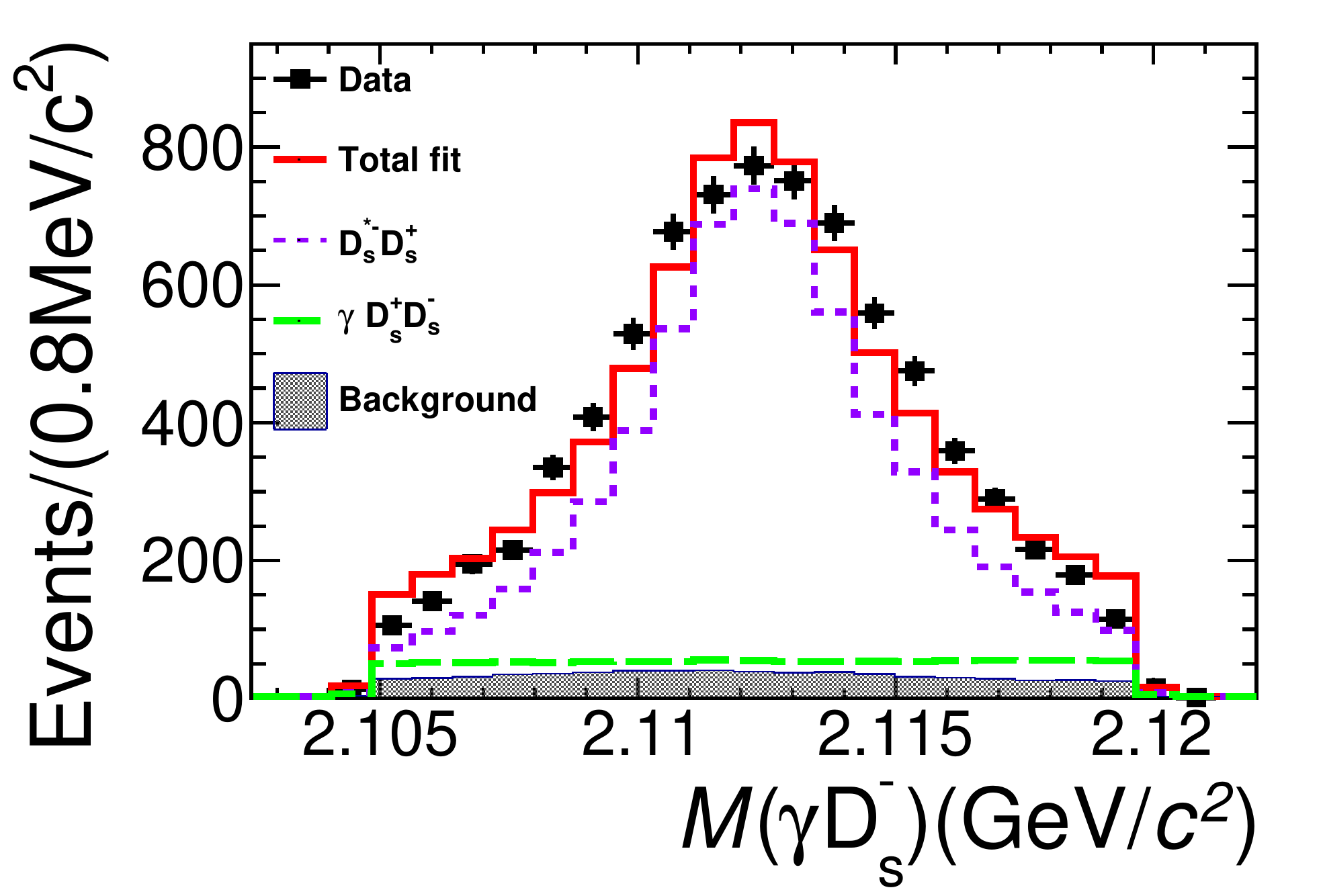}
\includegraphics[width=0.3\columnwidth]{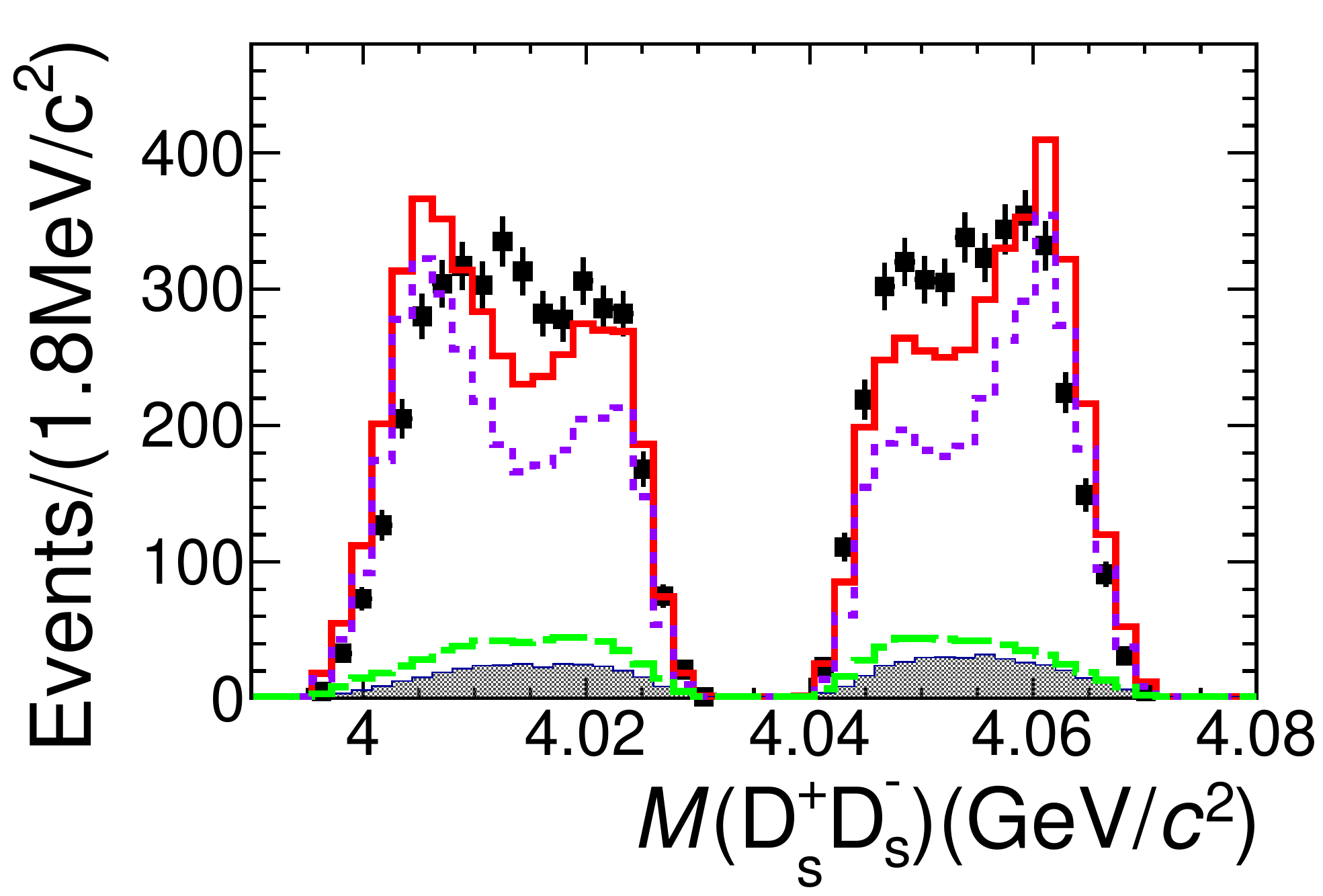}
\includegraphics[width=0.3\columnwidth]{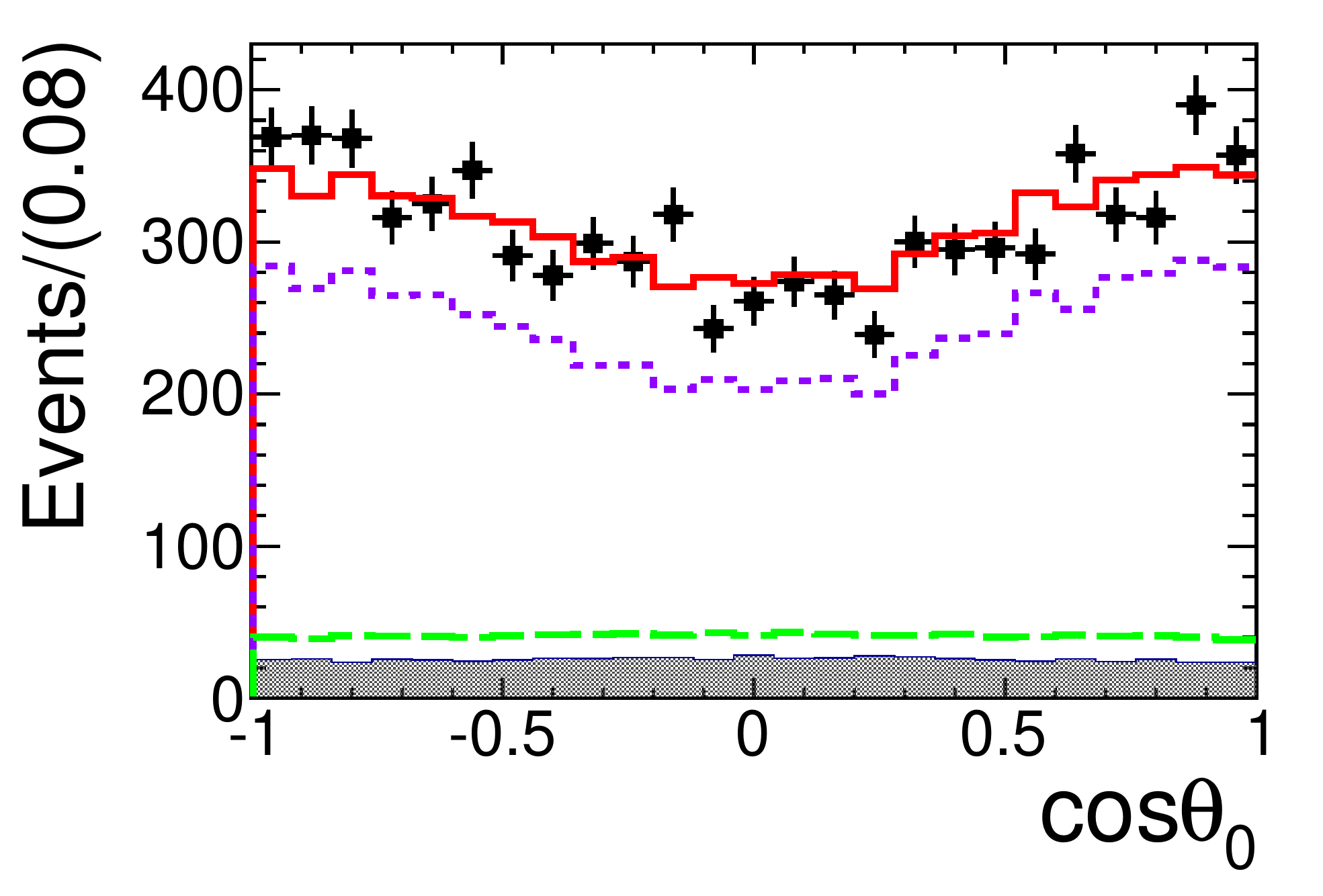}
\includegraphics[width=0.3\columnwidth]{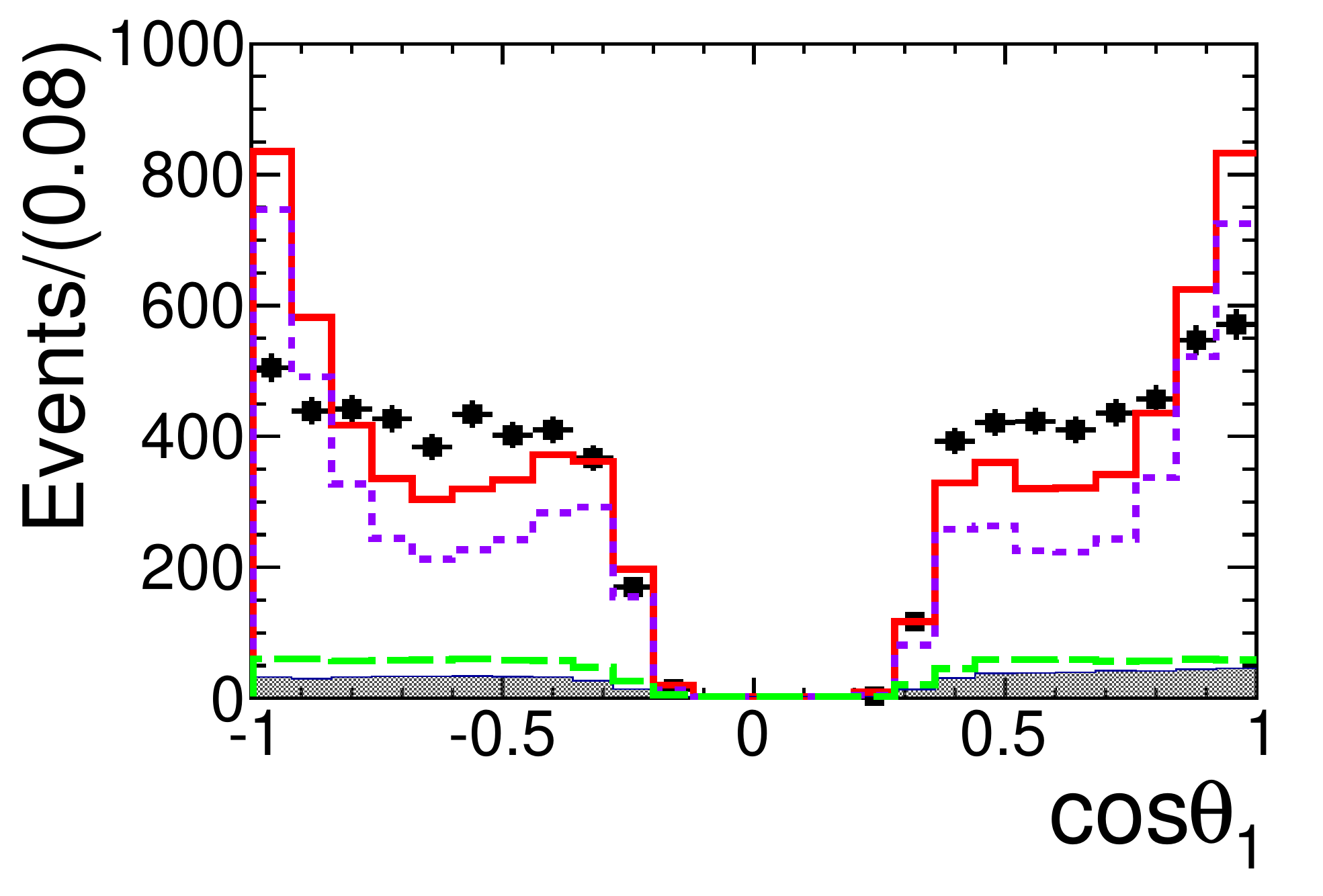}
\includegraphics[width=0.3\columnwidth]{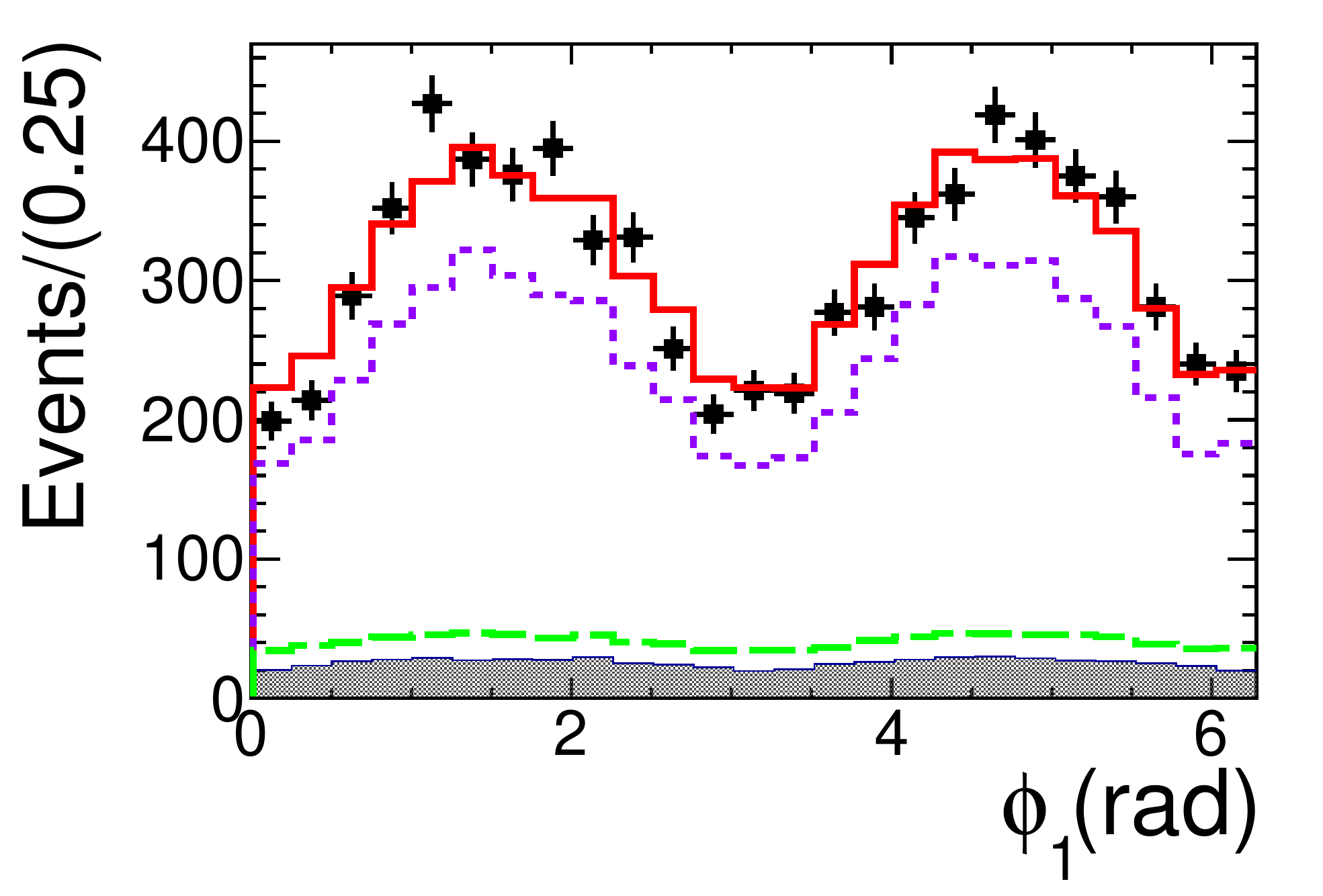}
\caption{
Mass and angular distributions of $D_s^+$-recoil sample for $J^P=2^+$ hypothesis.
}
\label{fig:angular_Dsm_2}
\end{figure}

\begin{figure}[!hbt]
\centering
\includegraphics[width=0.3\columnwidth]{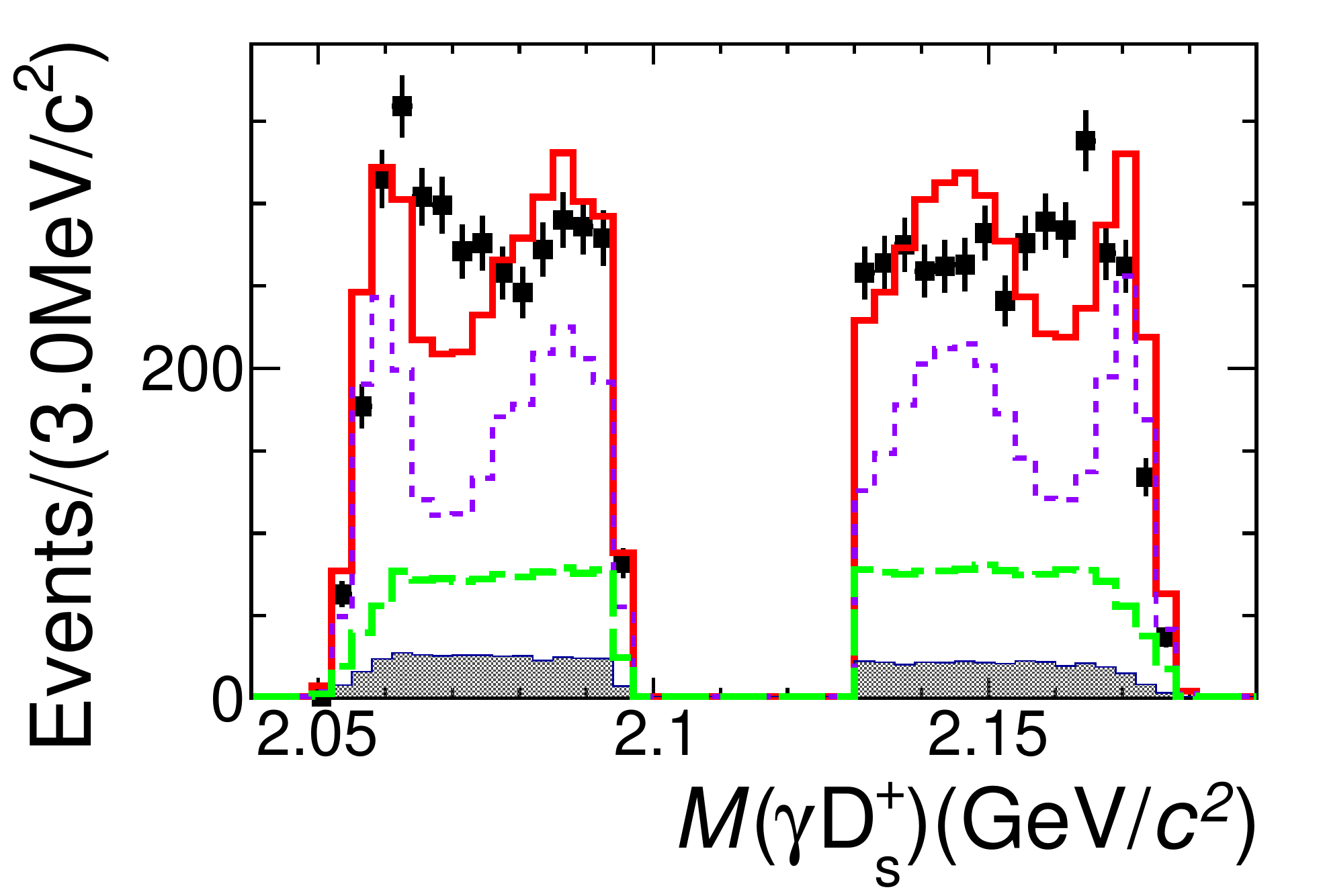}
\includegraphics[width=0.3\columnwidth]{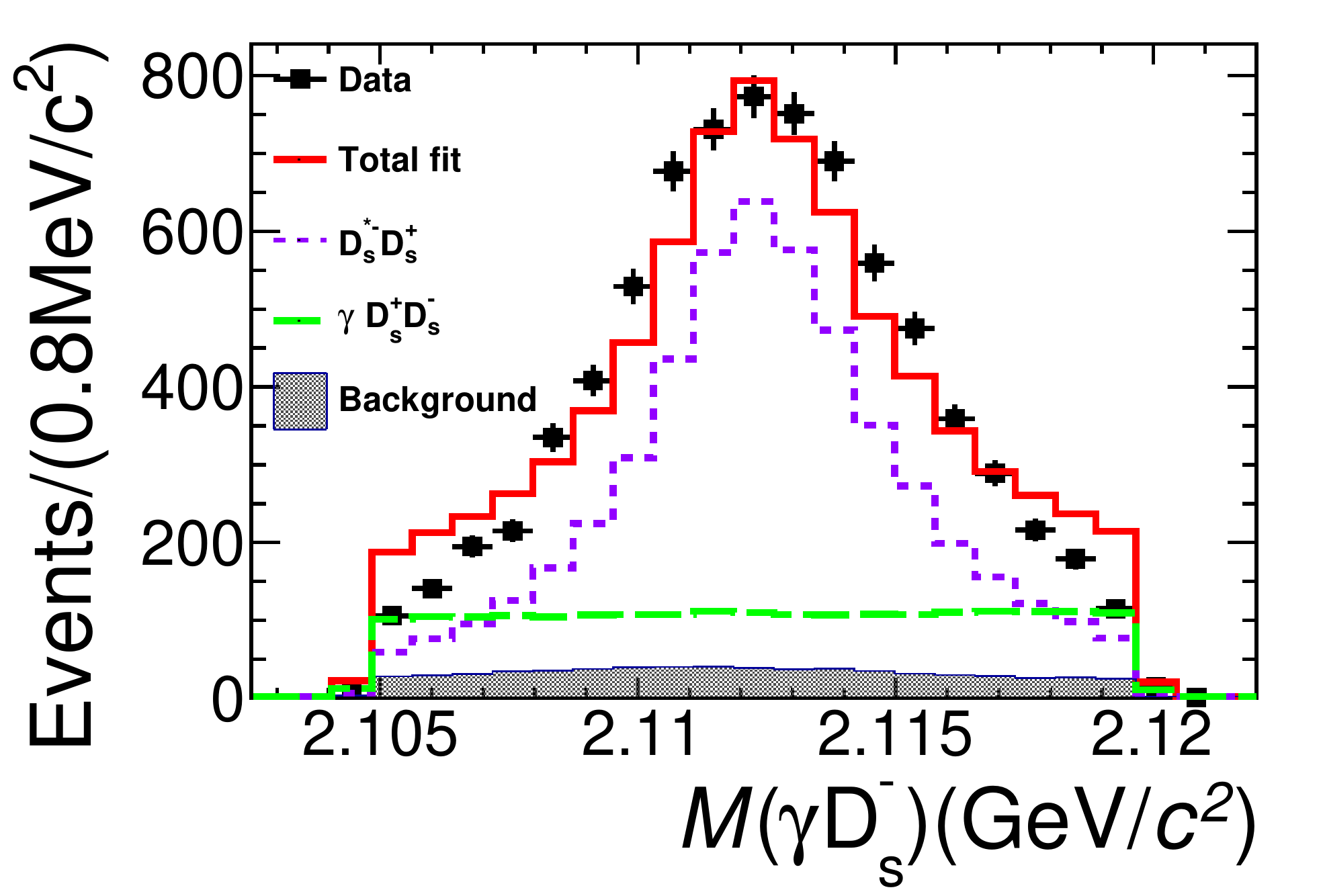}
\includegraphics[width=0.3\columnwidth]{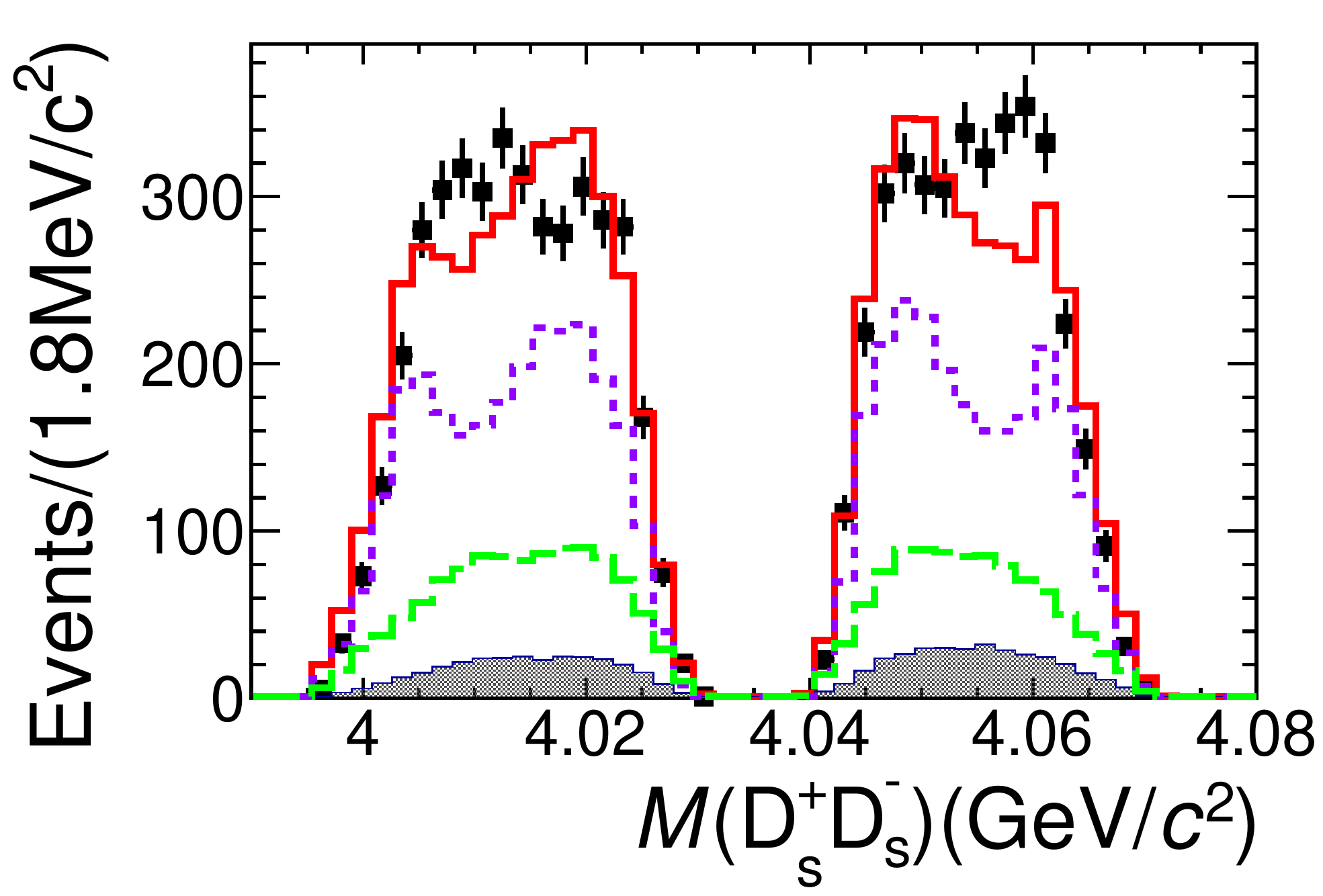}
\includegraphics[width=0.3\columnwidth]{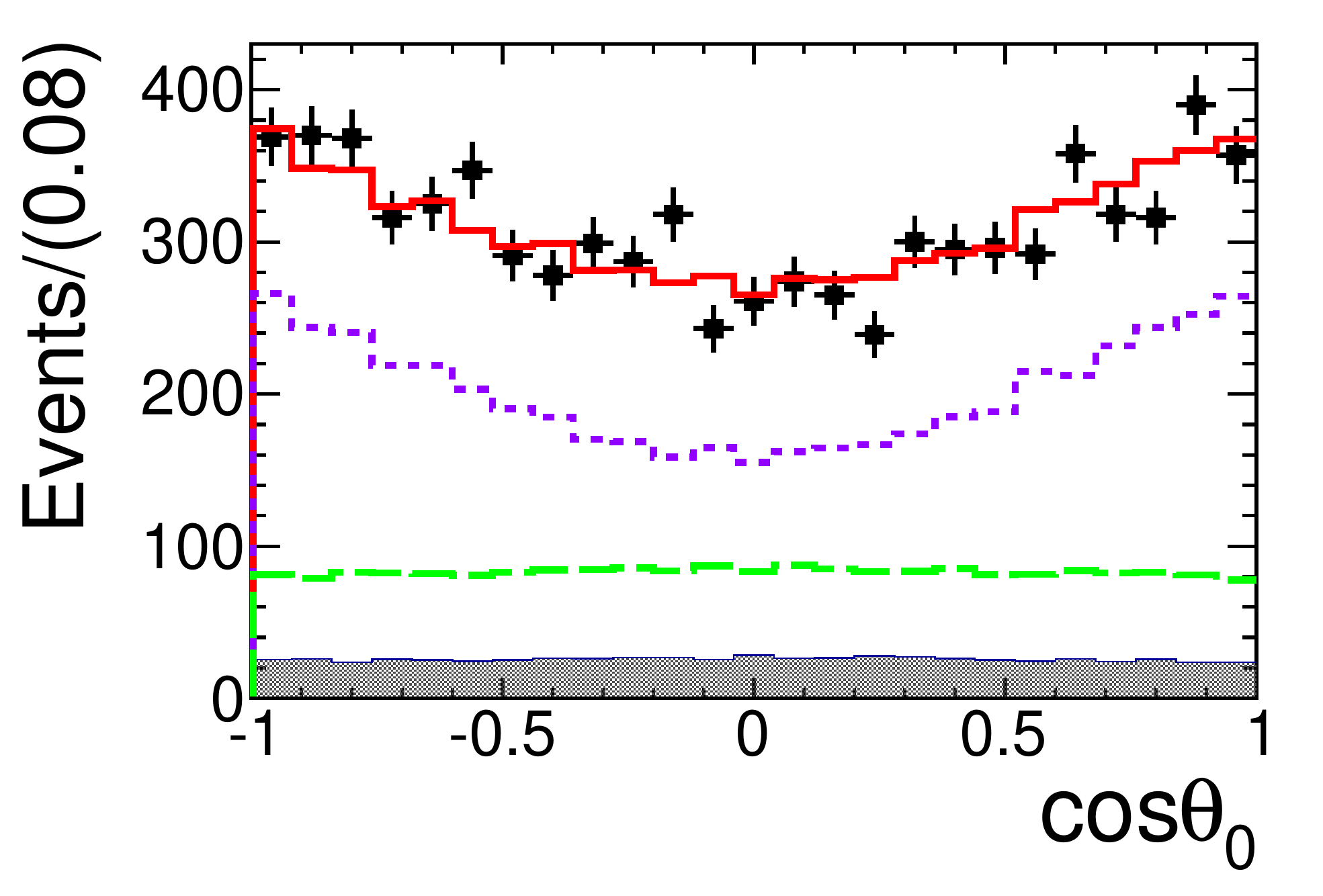}
\includegraphics[width=0.3\columnwidth]{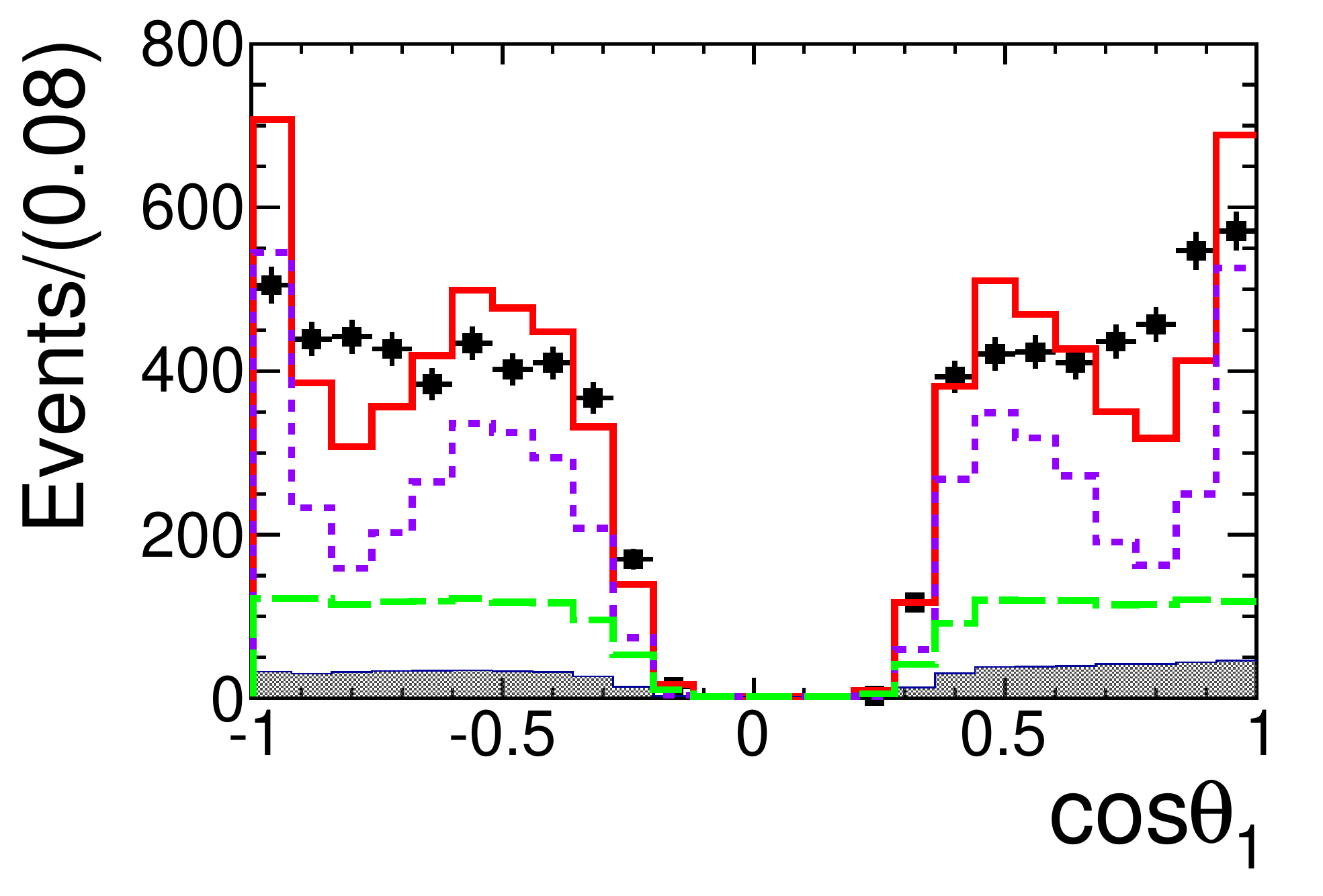}
\includegraphics[width=0.3\columnwidth]{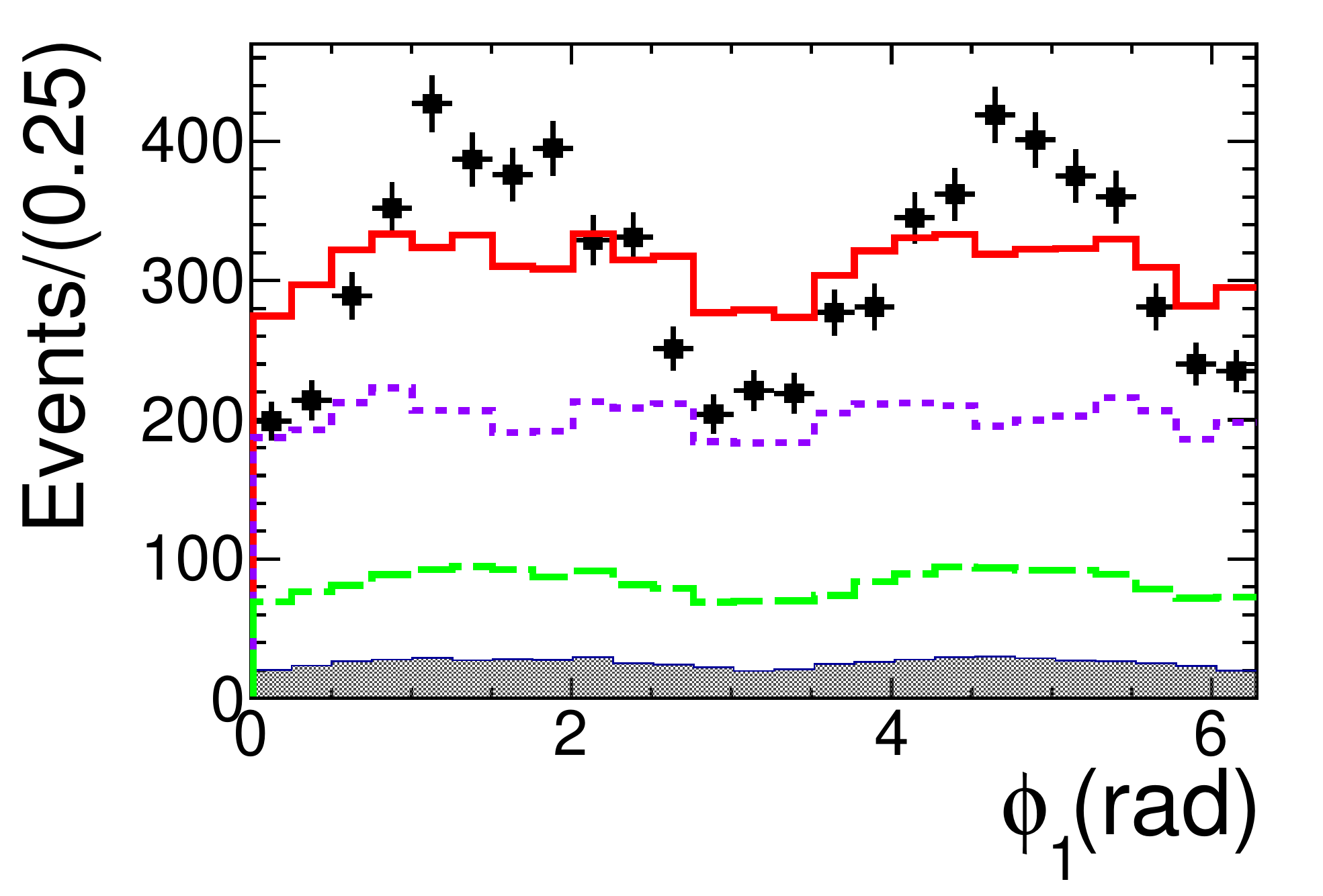}
\caption{
Mass and angular distributions of $D_s^+$-recoil sample for $J^P=3^-$ hypothesis.
}
\label{fig:angular_Dsm_3}
\end{figure}

\begin{figure}[!hbt]
\centering
\includegraphics[width=0.3\columnwidth]{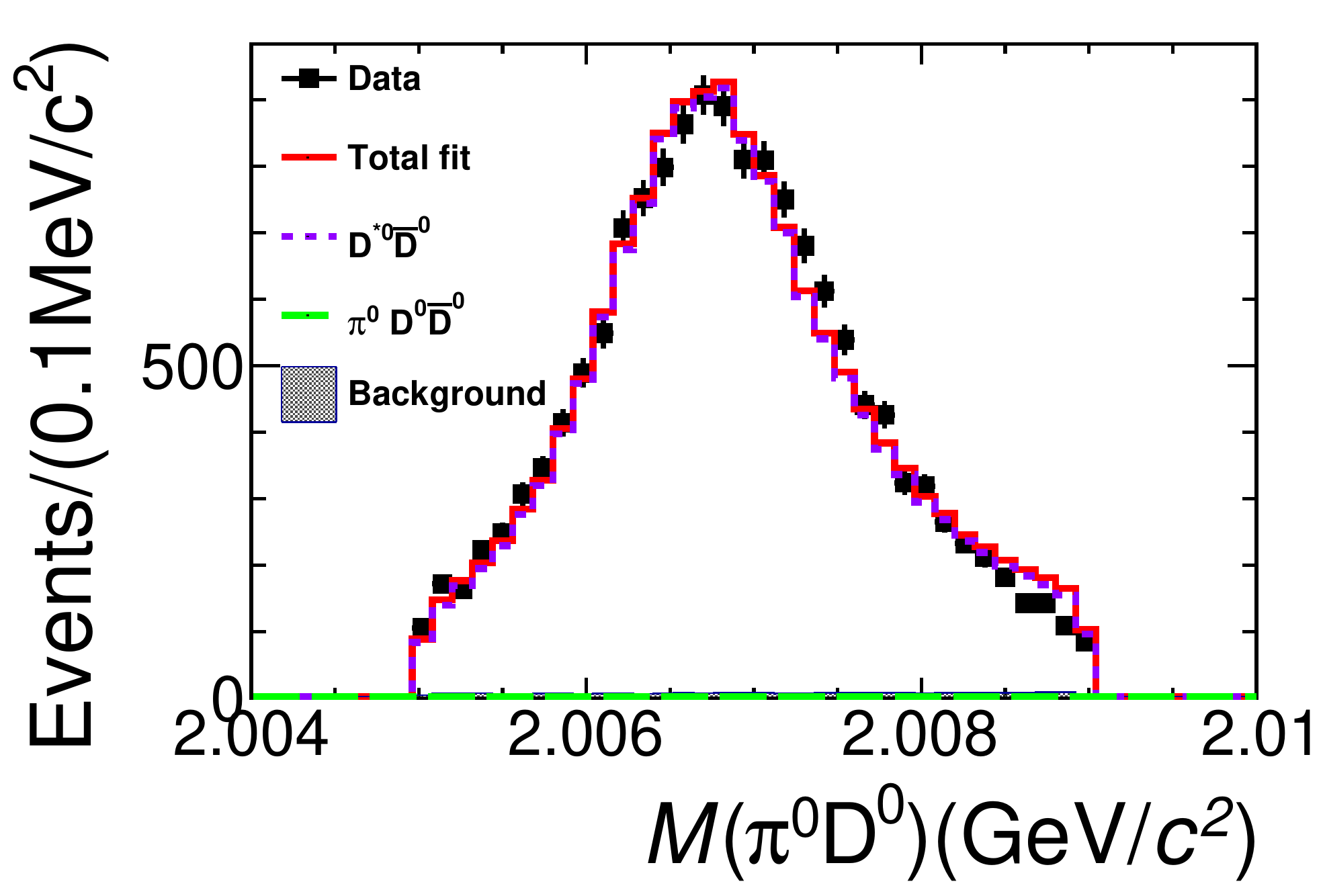}
\includegraphics[width=0.3\columnwidth]{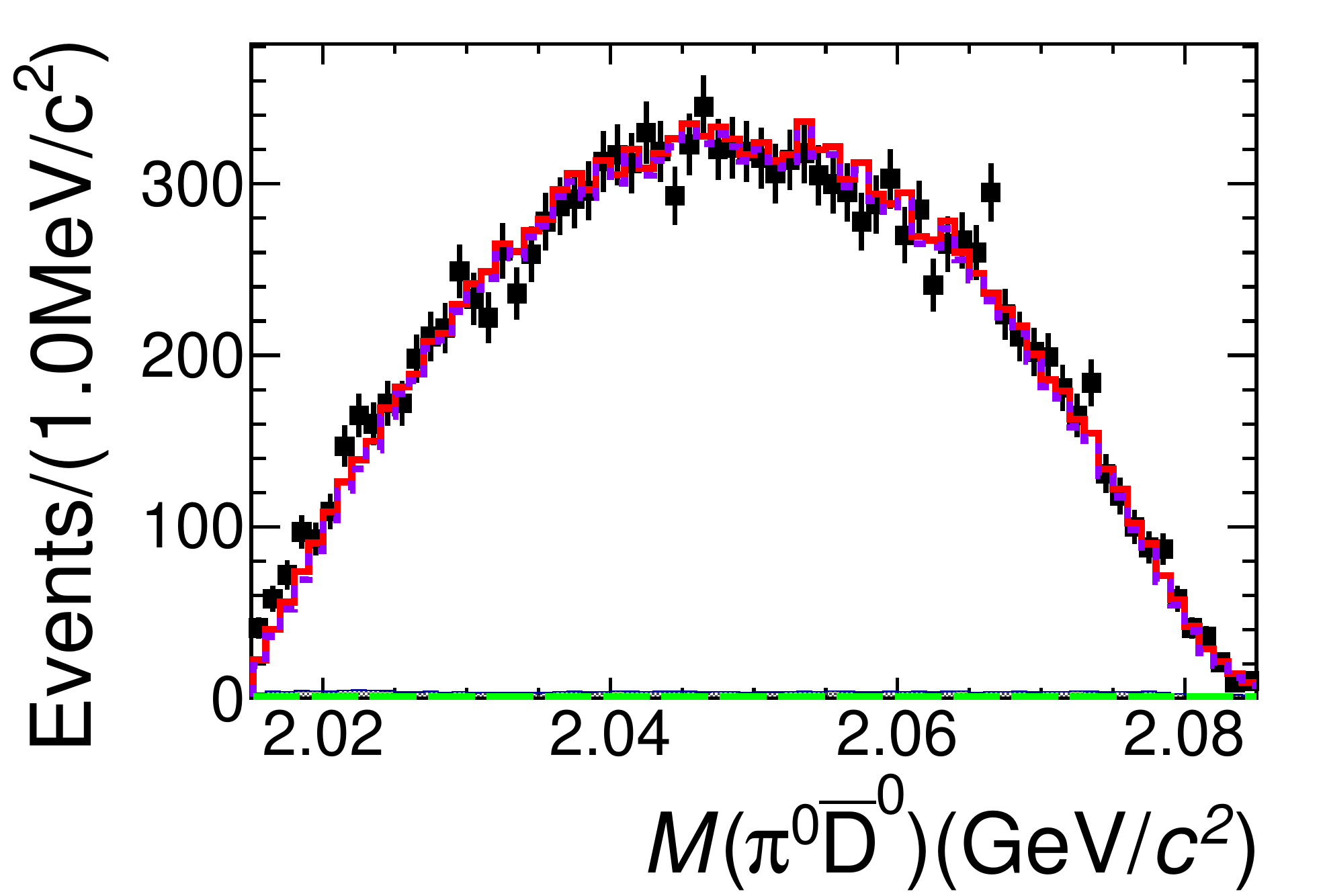}
\includegraphics[width=0.3\columnwidth]{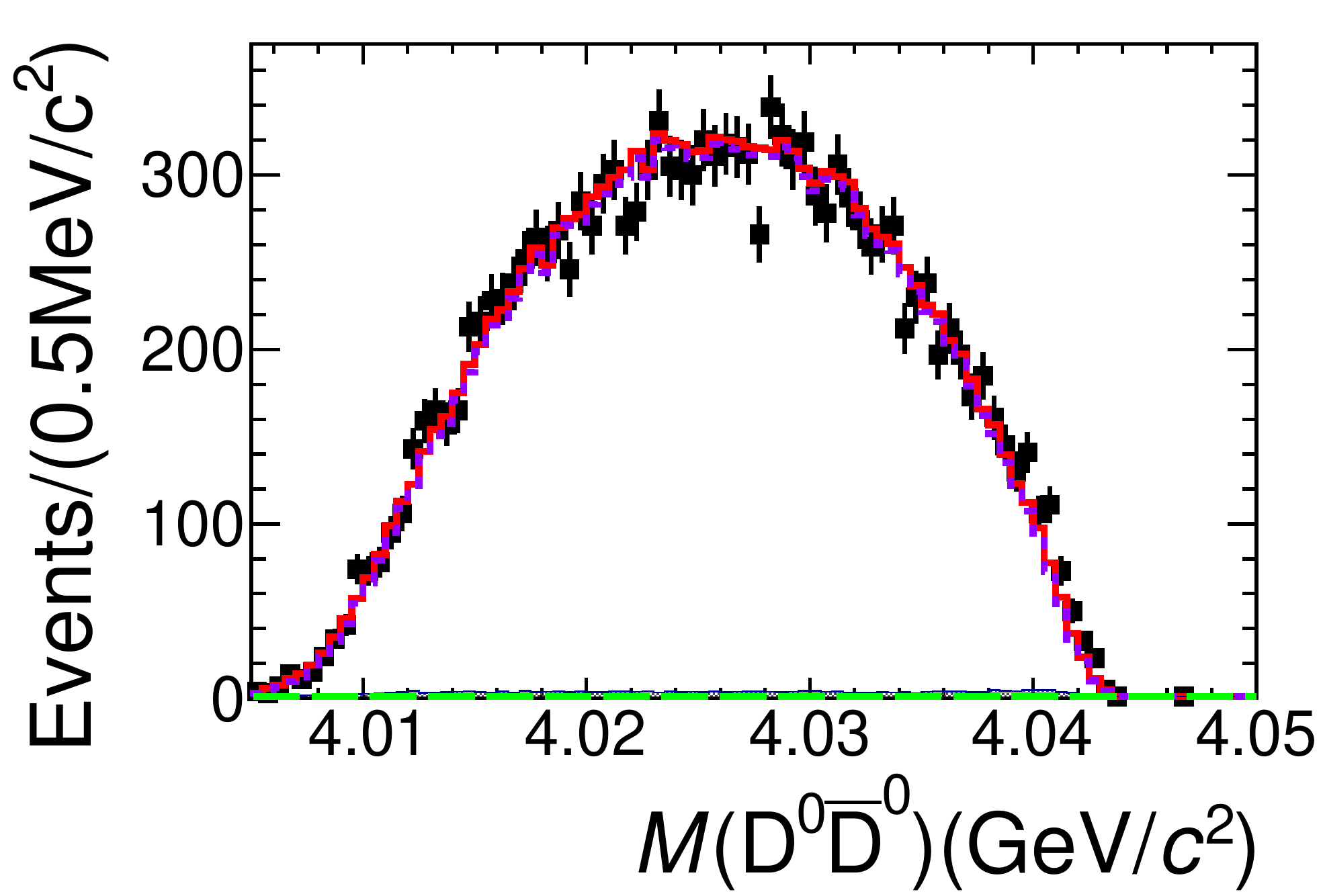}
\includegraphics[width=0.3\columnwidth]{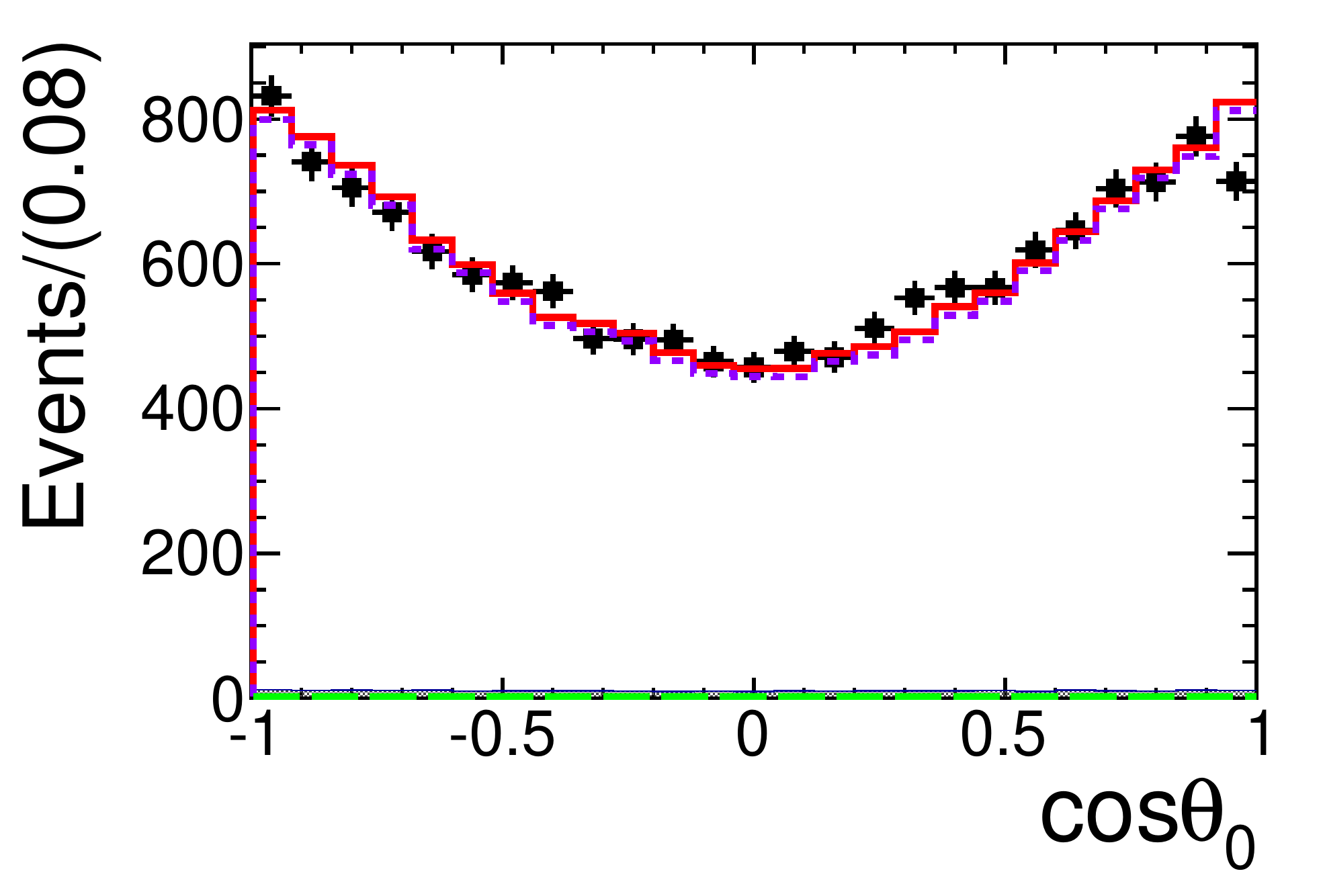}
\includegraphics[width=0.3\columnwidth]{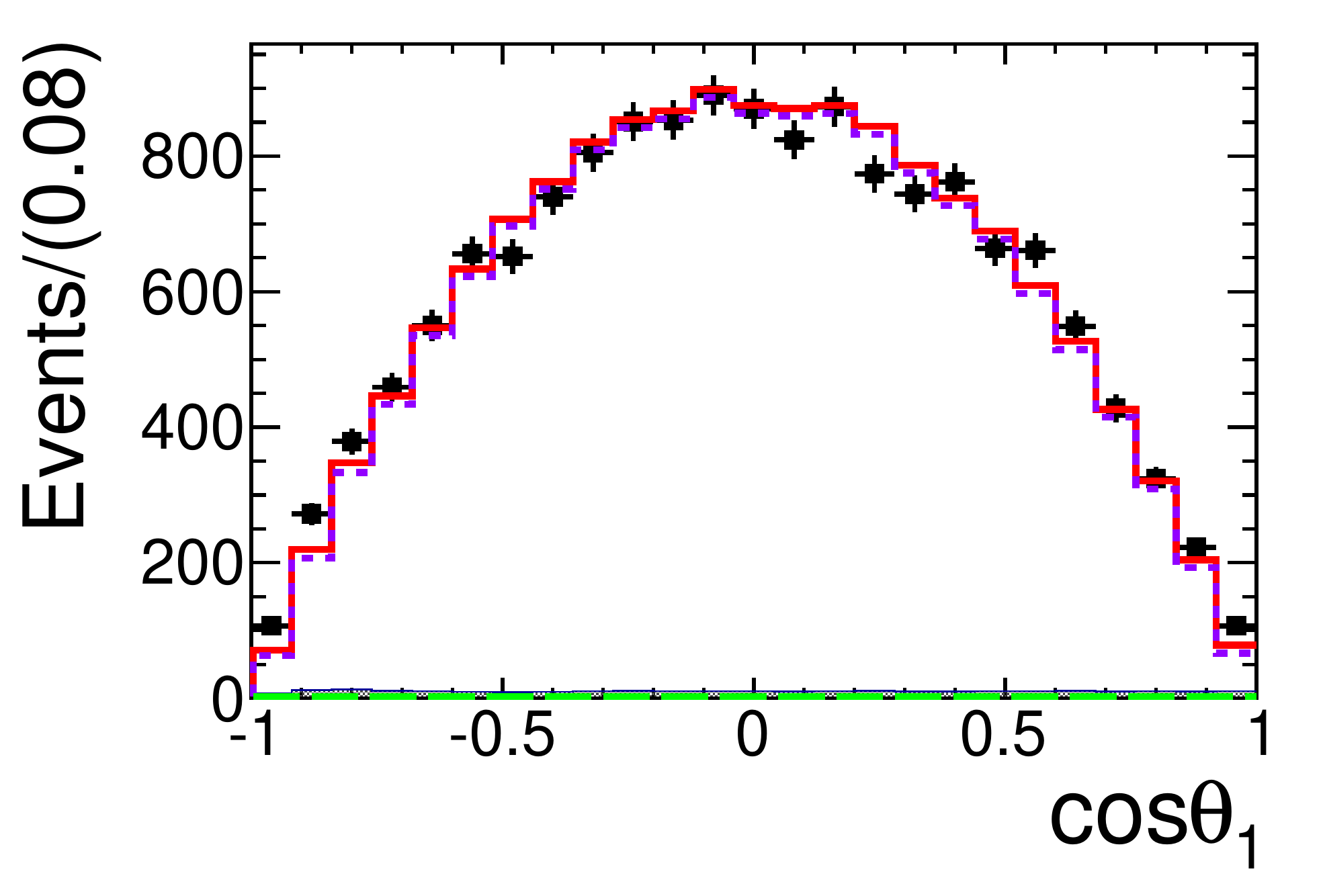}
\includegraphics[width=0.3\columnwidth]{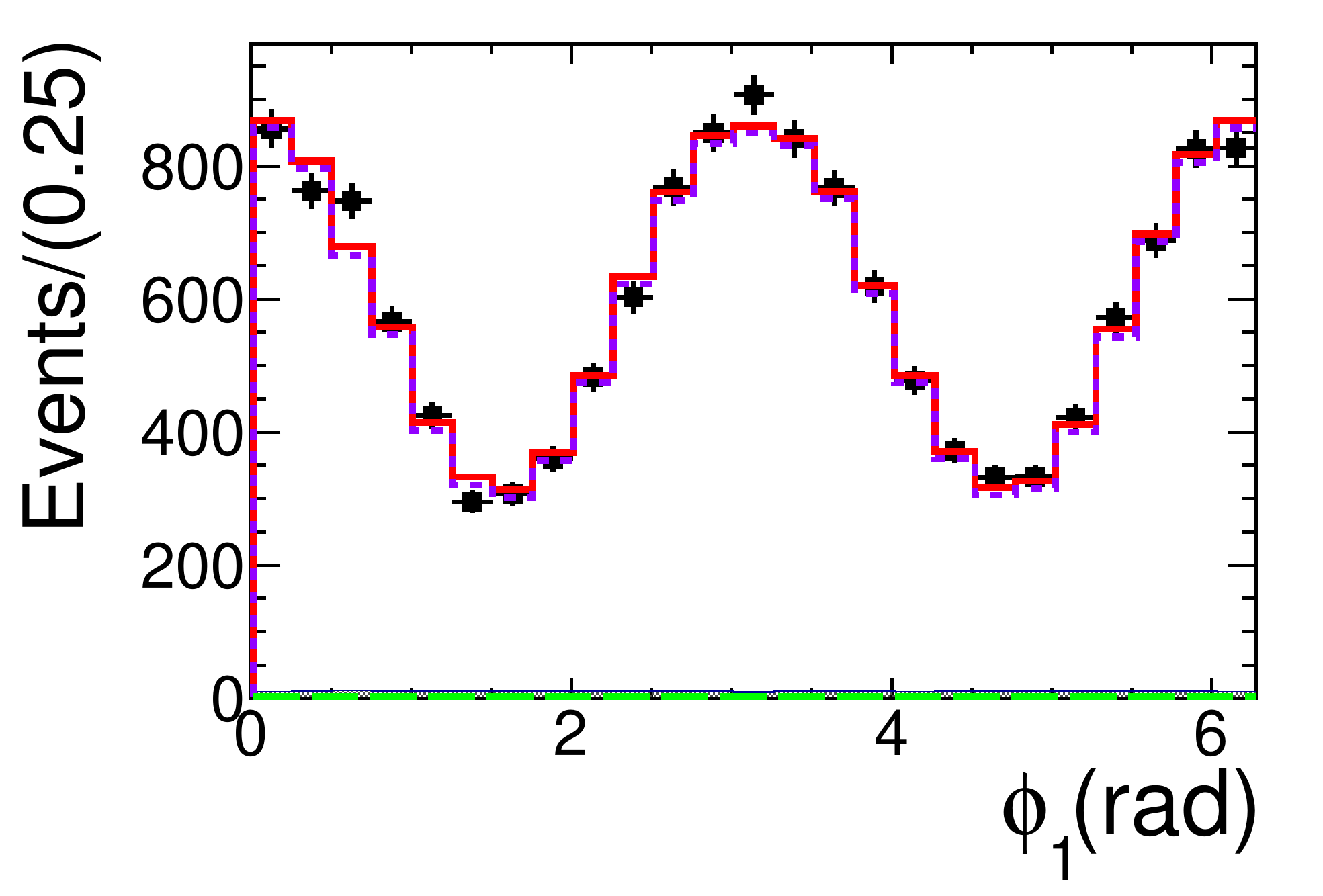}
\caption{
Mass and angular  distributions of $D^{*0}$-tag sample for $J^P=1^-$ hypothesis. 
}
\label{fig:angular_D0pi0_1}
\end{figure}

\begin{figure}[!hbt]
\centering
\includegraphics[width=0.3\columnwidth]{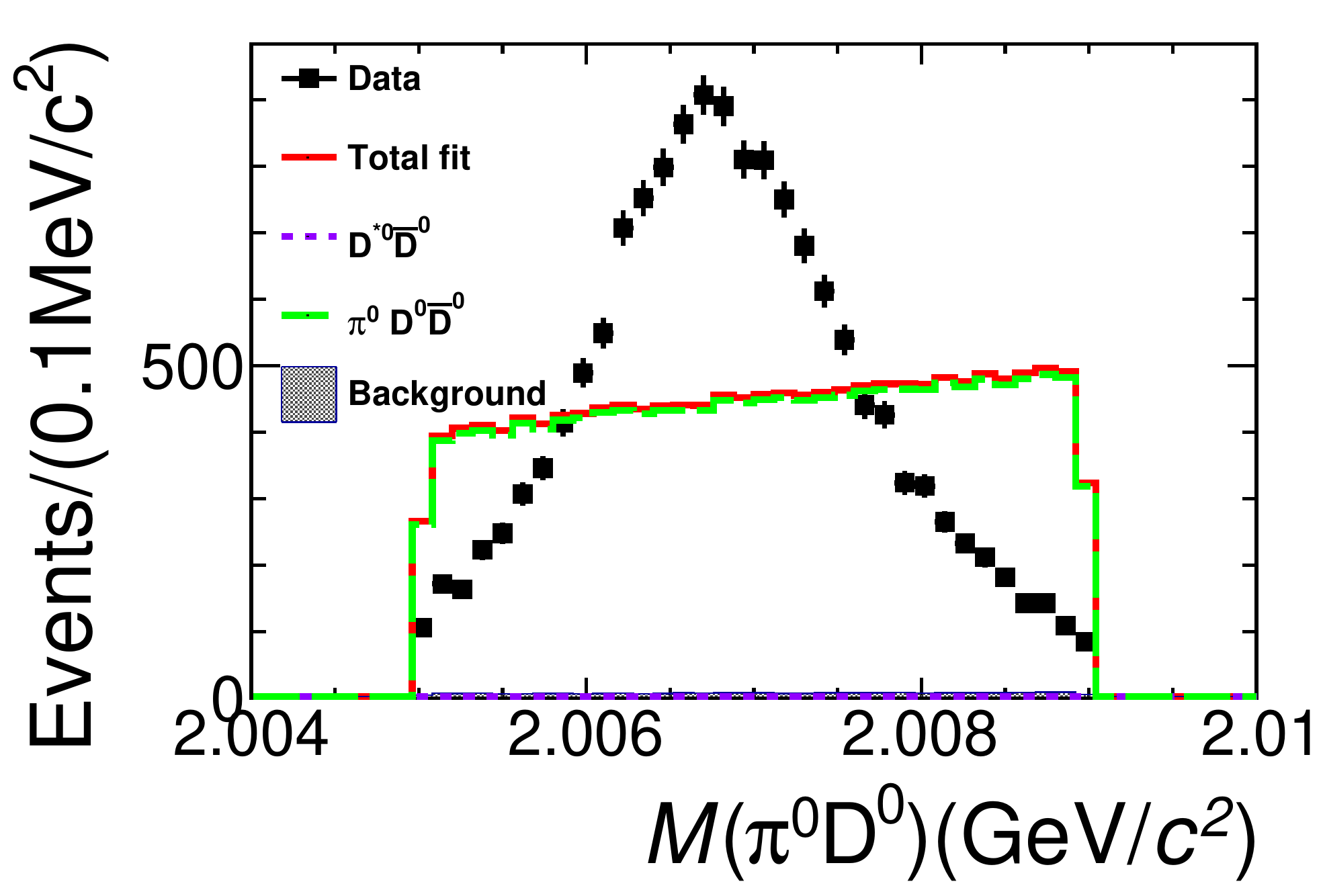}
\includegraphics[width=0.3\columnwidth]{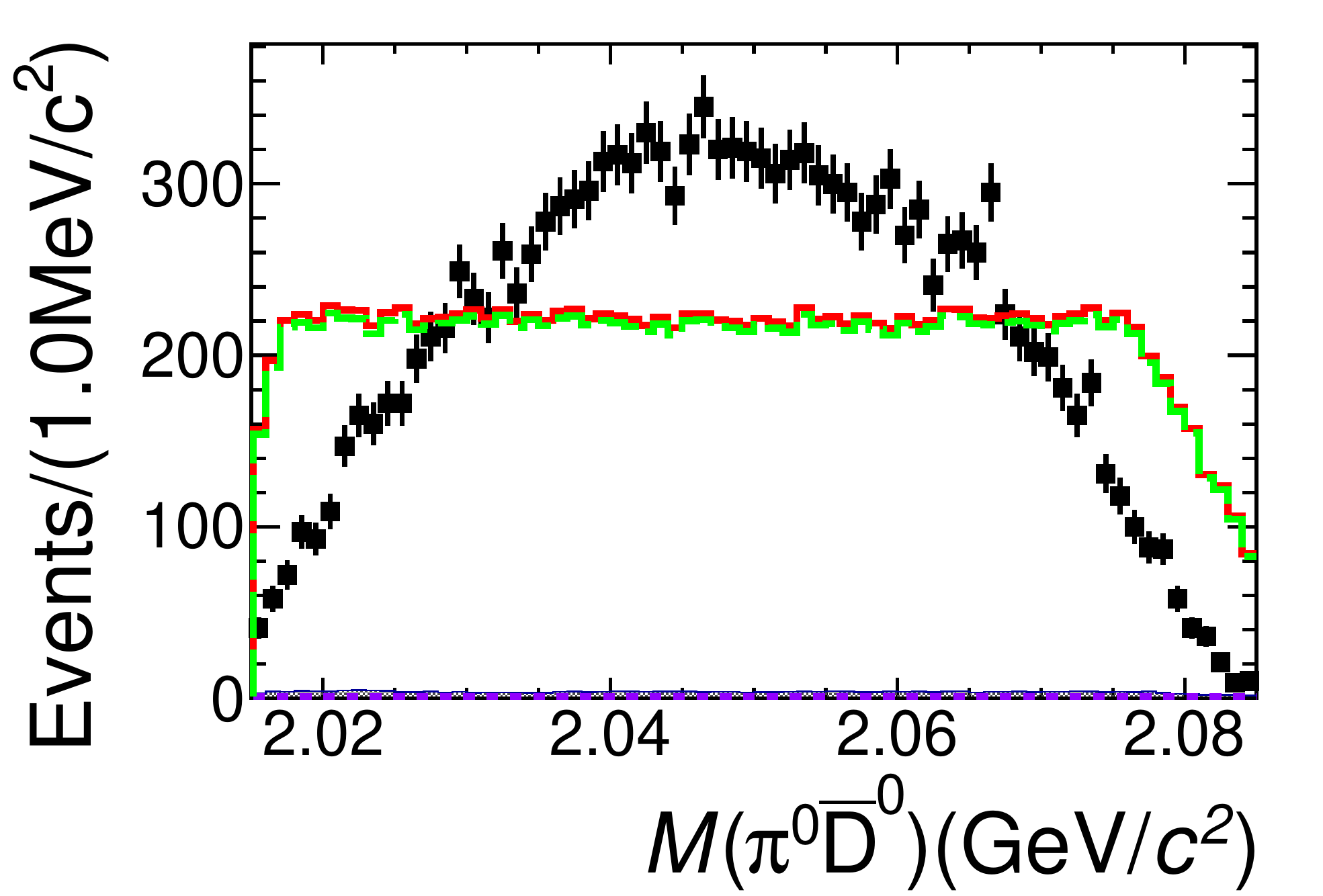}
\includegraphics[width=0.3\columnwidth]{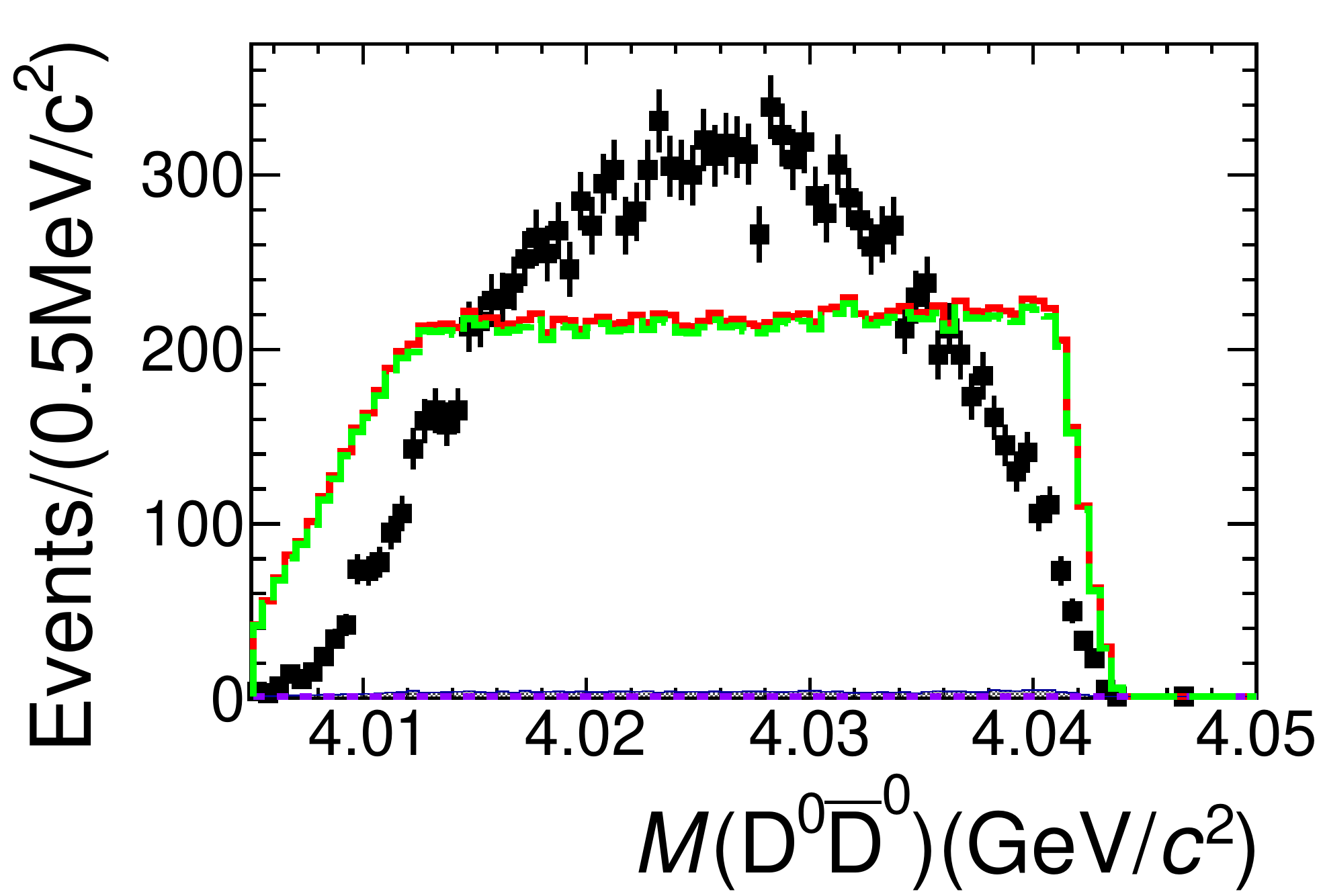}
\includegraphics[width=0.3\columnwidth]{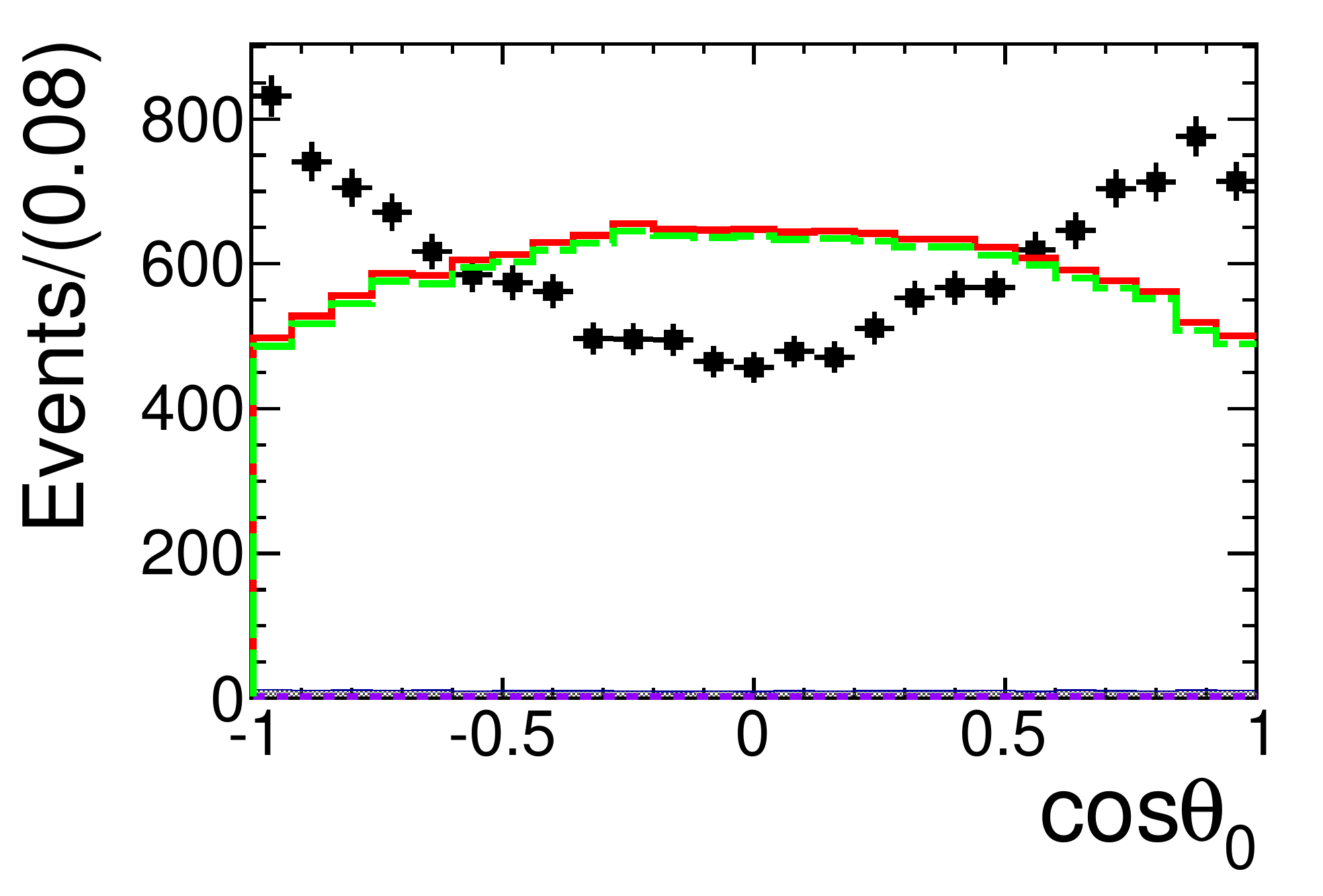}
\includegraphics[width=0.3\columnwidth]{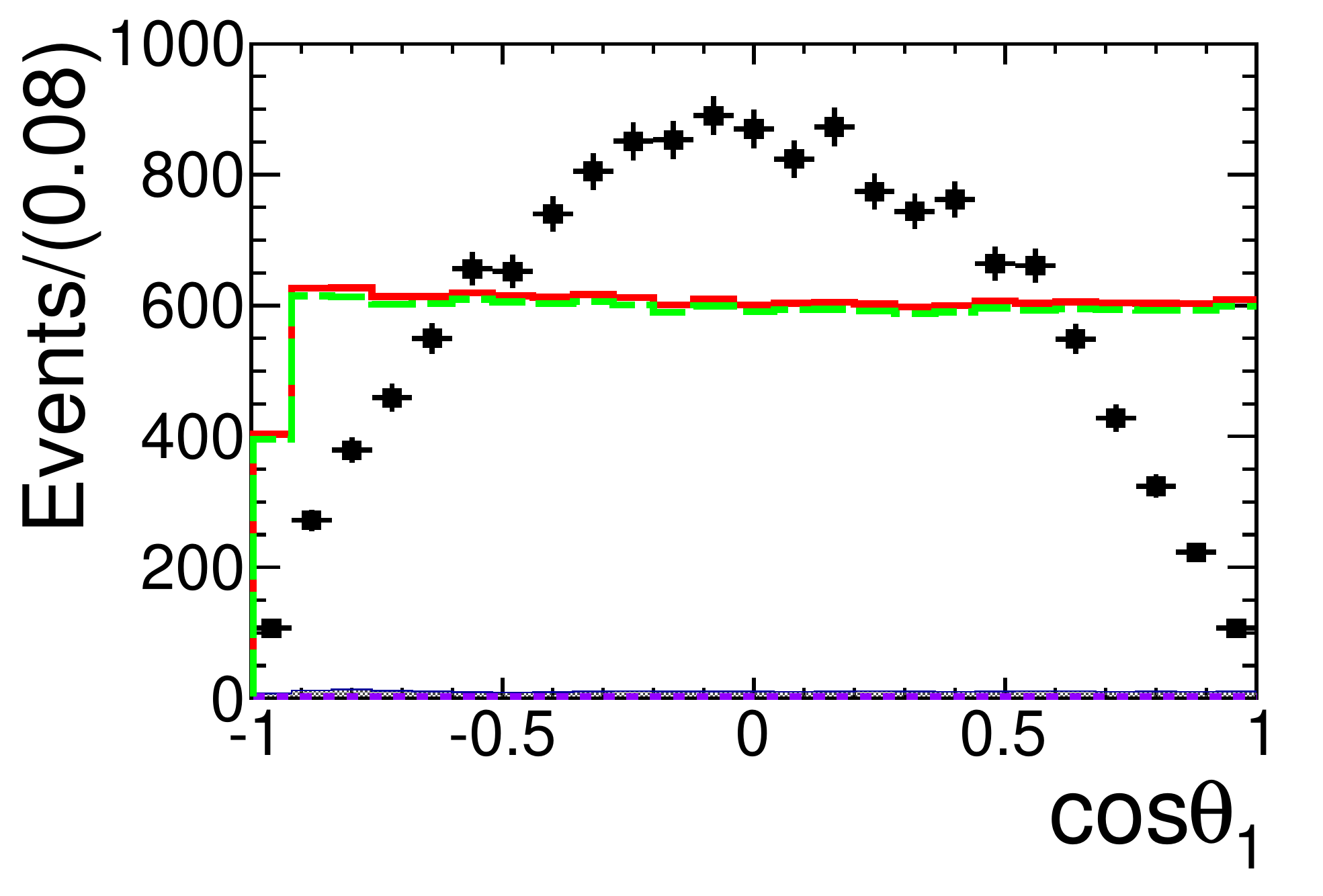}
\includegraphics[width=0.3\columnwidth]{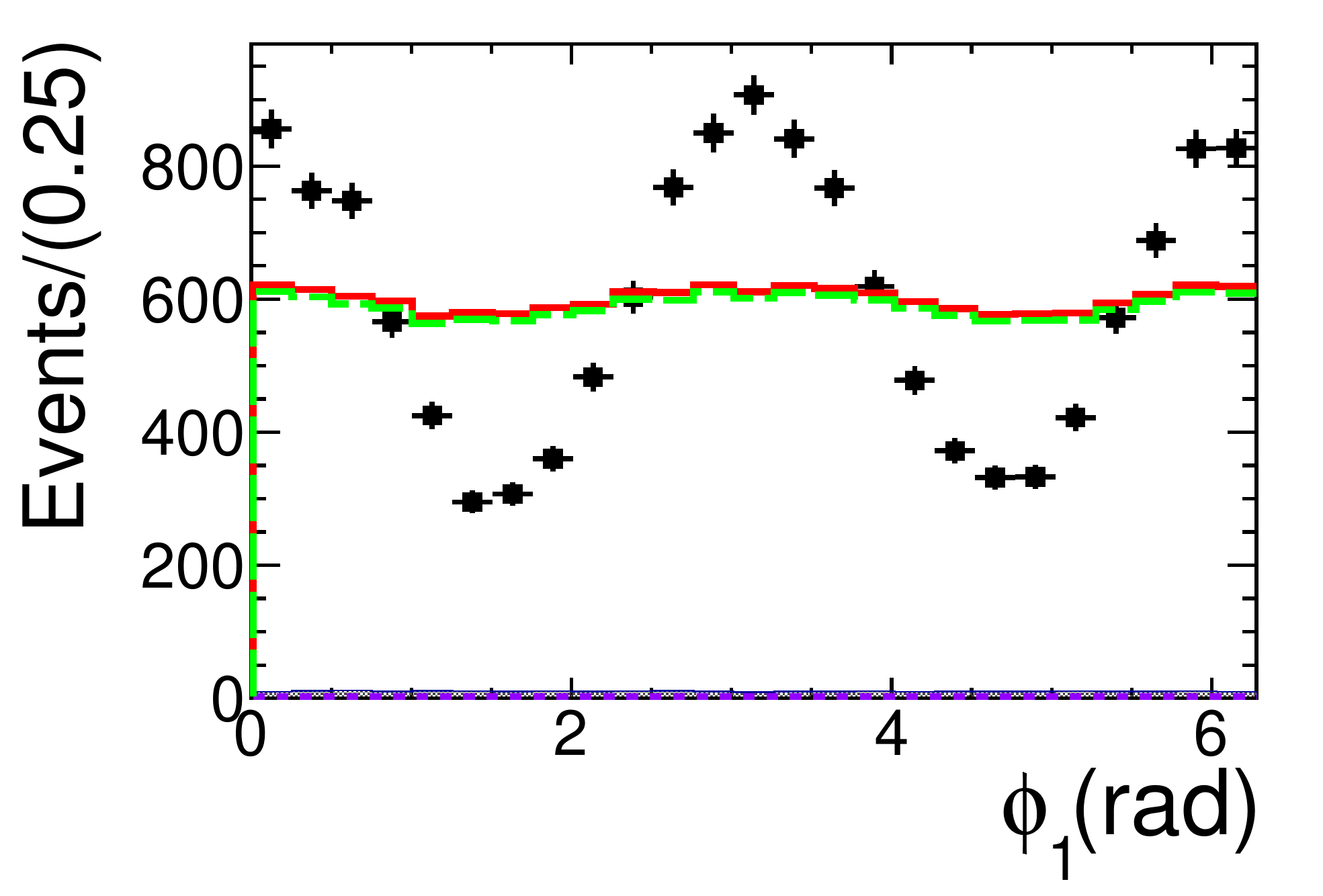}
\caption{
Mass and angular distributions of $D^{*0}$-tag sample for $J^P=2^+$ hypothesis.
}
\label{fig:angular_D0pi0_2}

\end{figure}

\begin{figure}[!hbt]
\centering
\includegraphics[width=0.3\columnwidth]{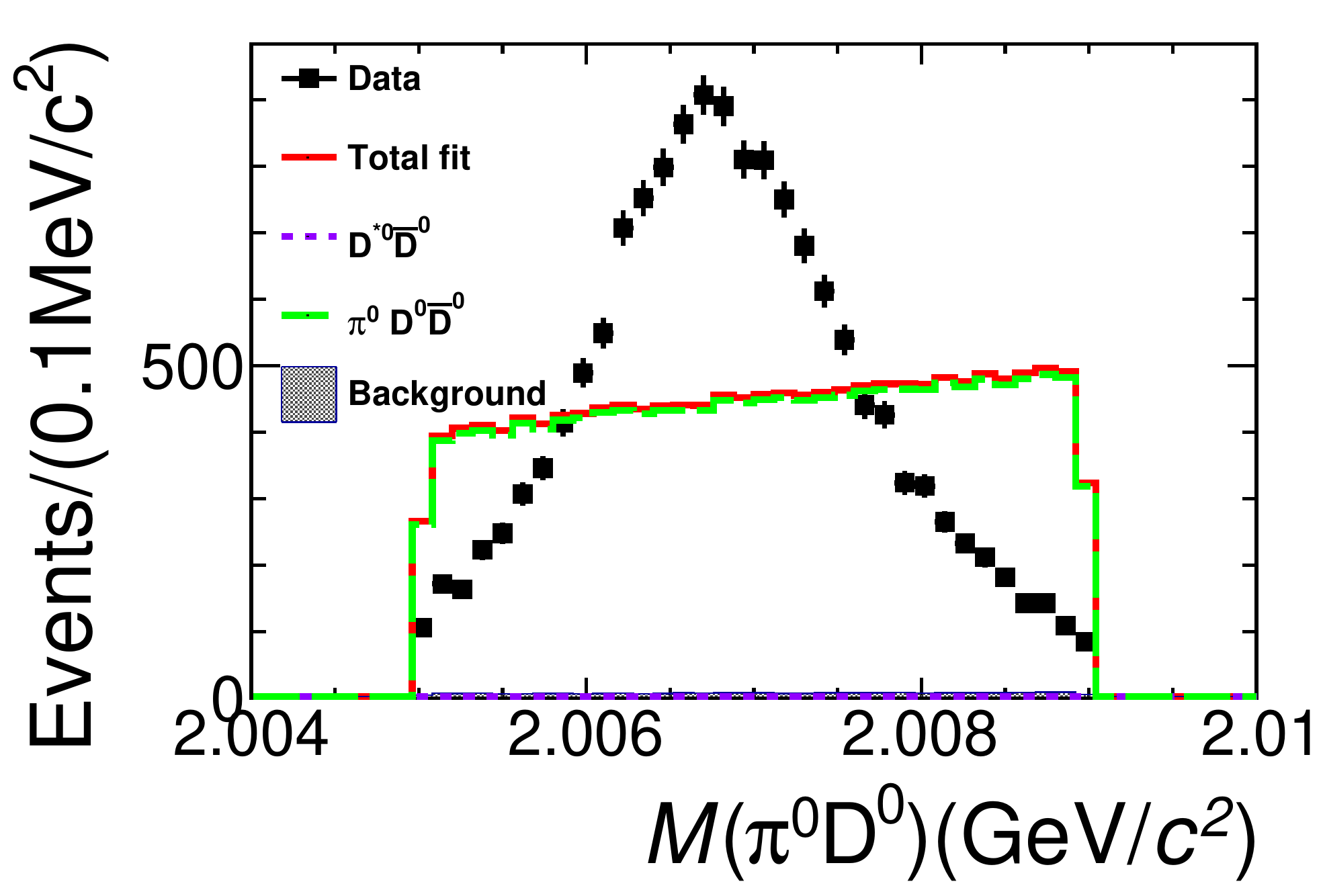}
\includegraphics[width=0.3\columnwidth]{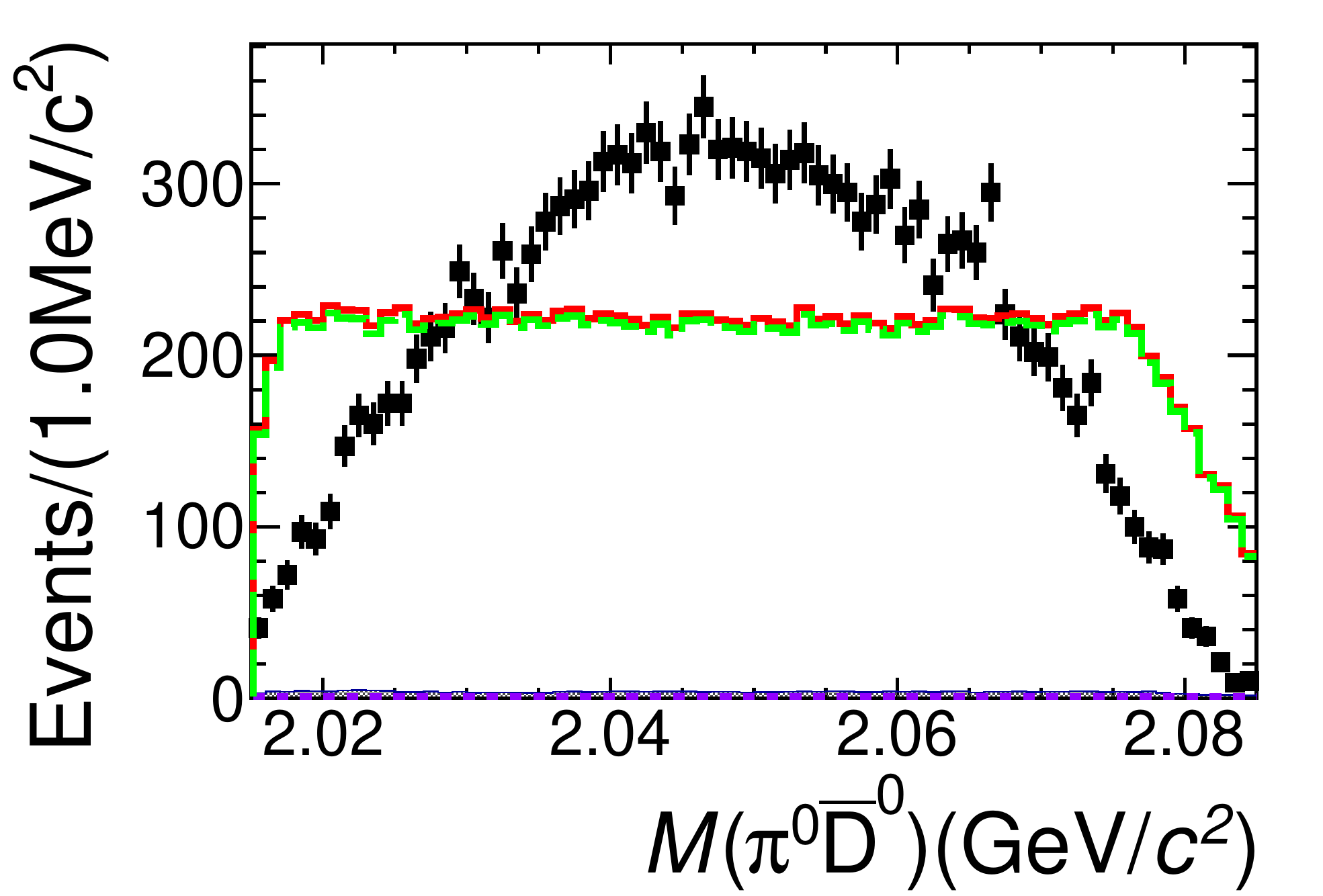}
\includegraphics[width=0.3\columnwidth]{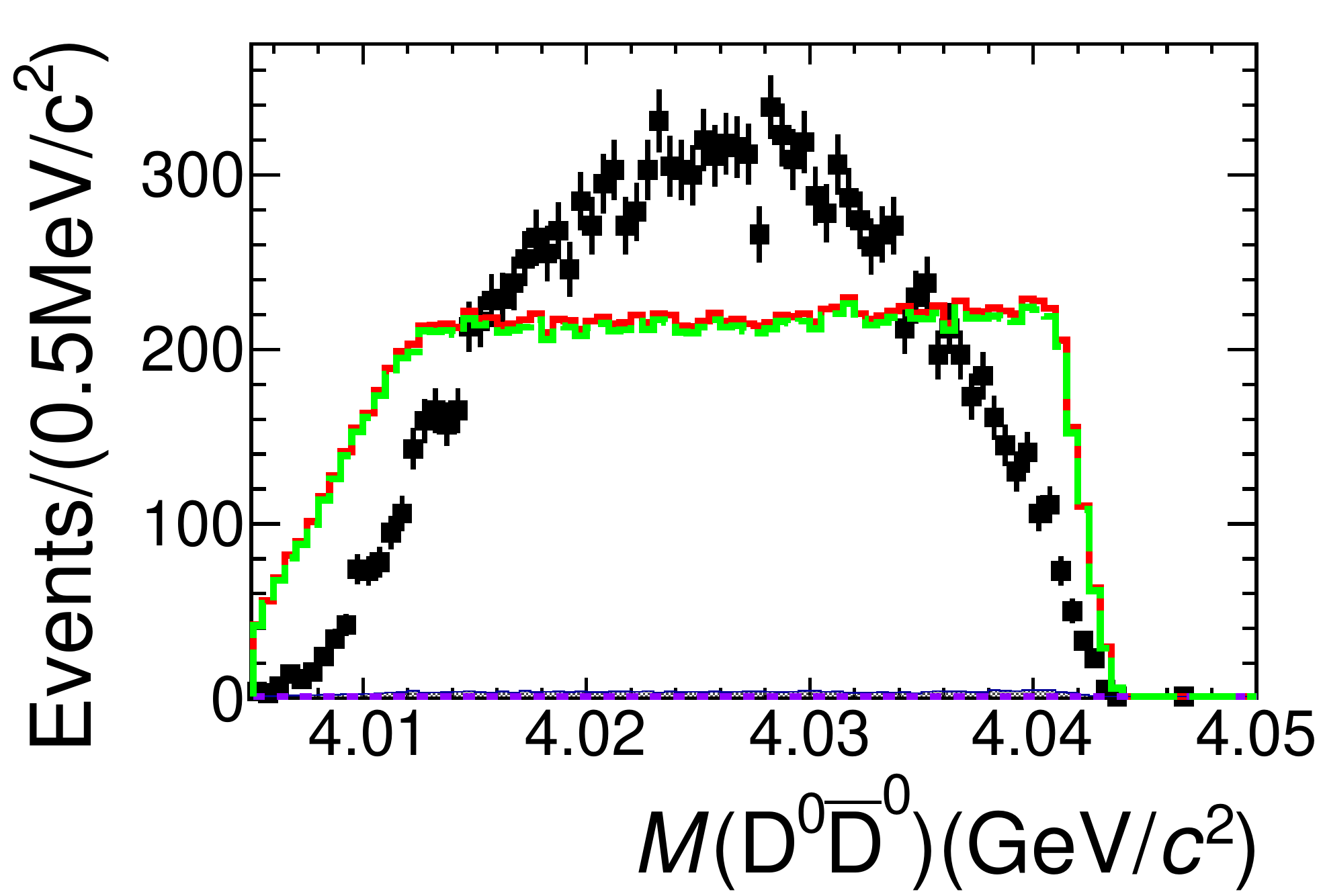}
\includegraphics[width=0.3\columnwidth]{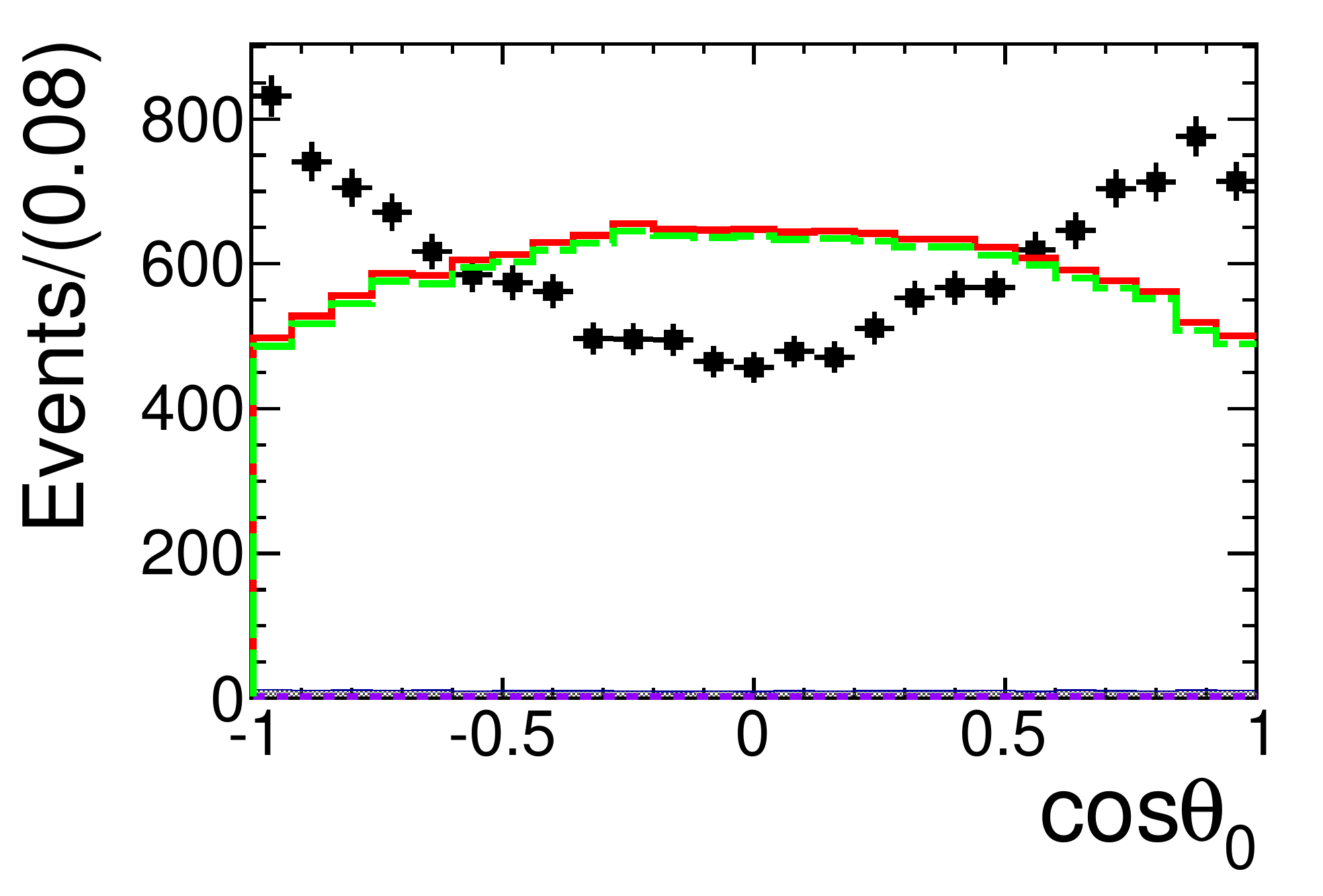}
\includegraphics[width=0.3\columnwidth]{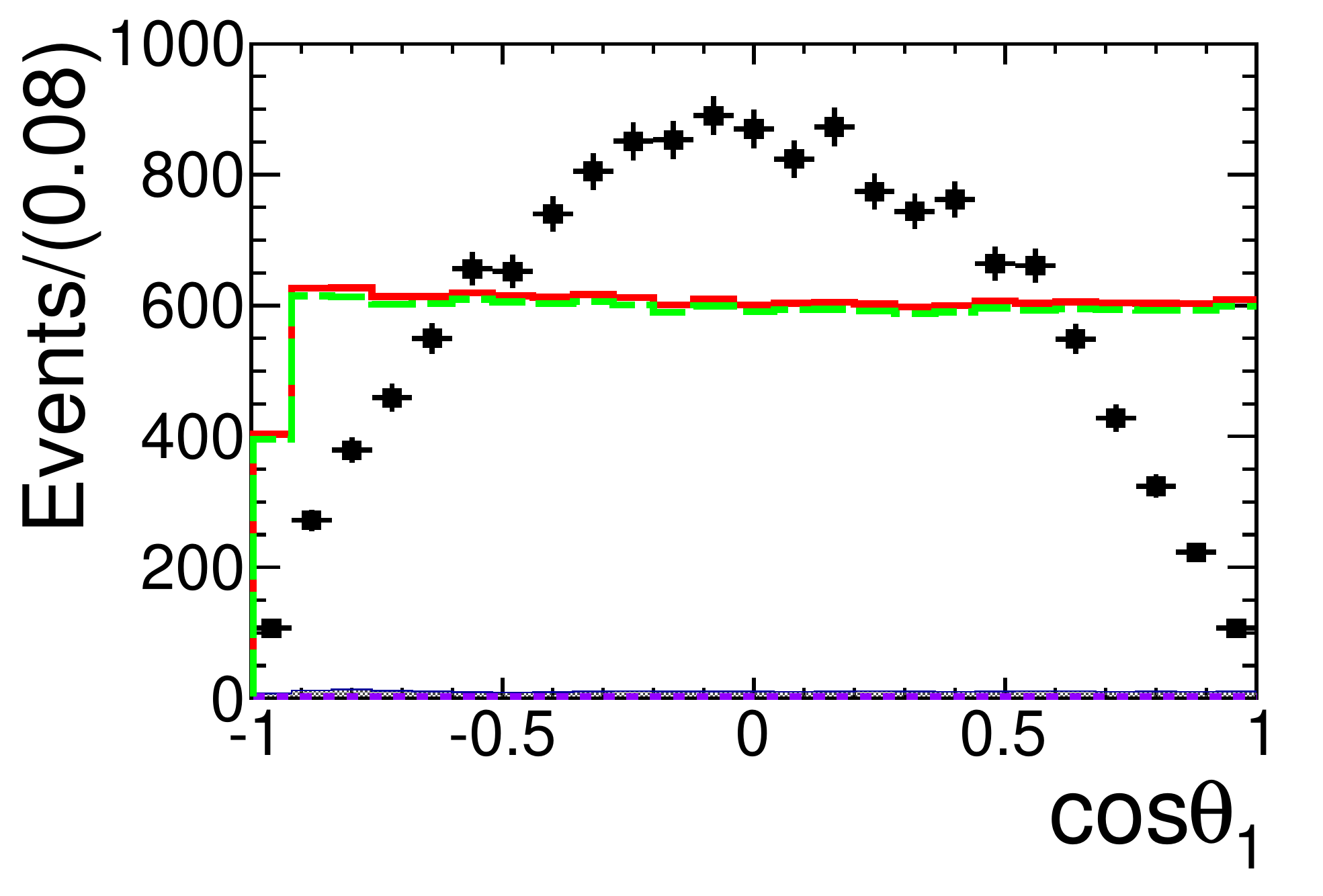}
\includegraphics[width=0.3\columnwidth]{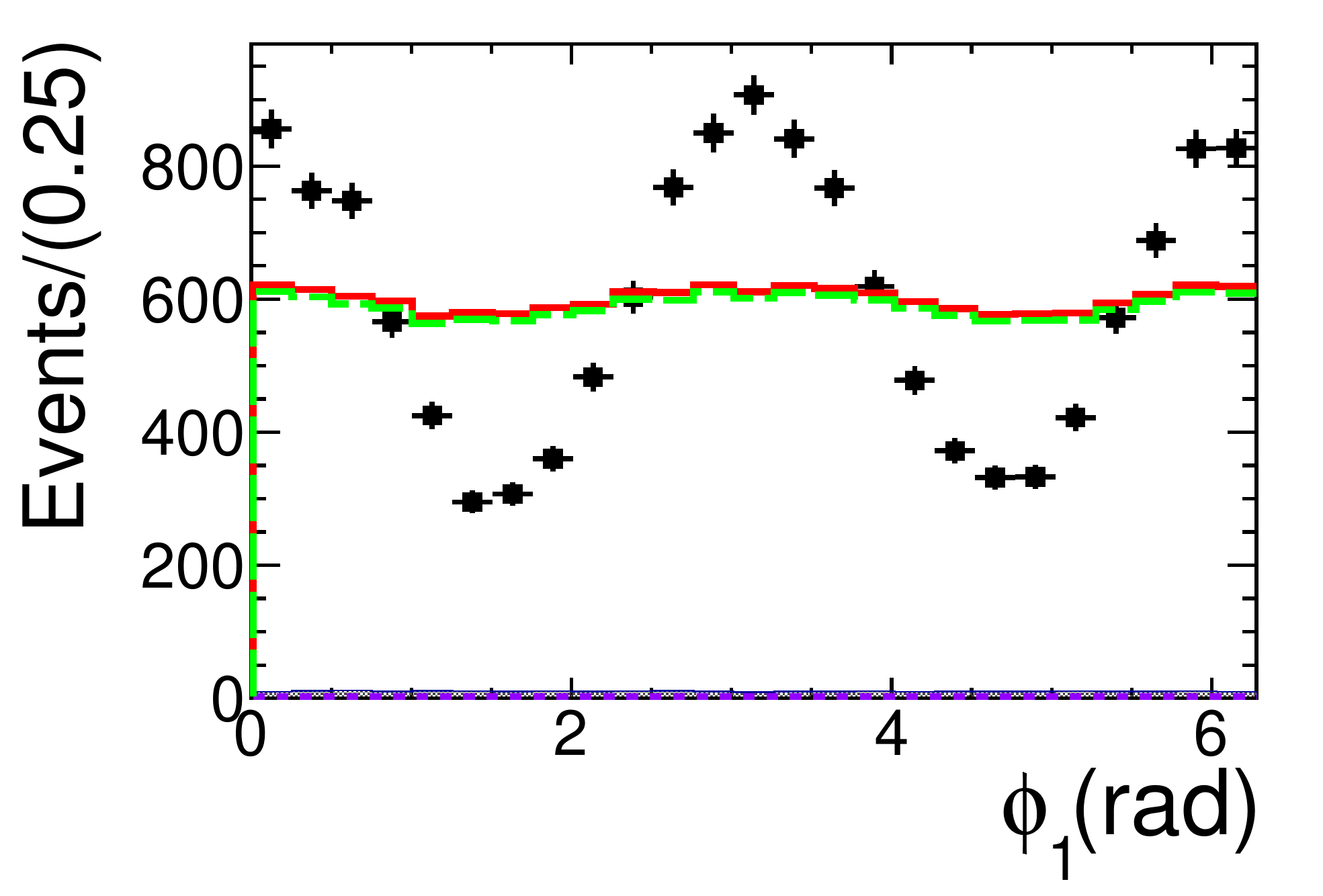}
\caption{
Mass and angular distributions of $D^{*0}$-tag sample for $J^P=3^-$ hypothesis.
}
\label{fig:angular_D0pi0_3}
\end{figure}

\begin{figure}[!hbt]
\centering
\includegraphics[width=0.3\columnwidth]{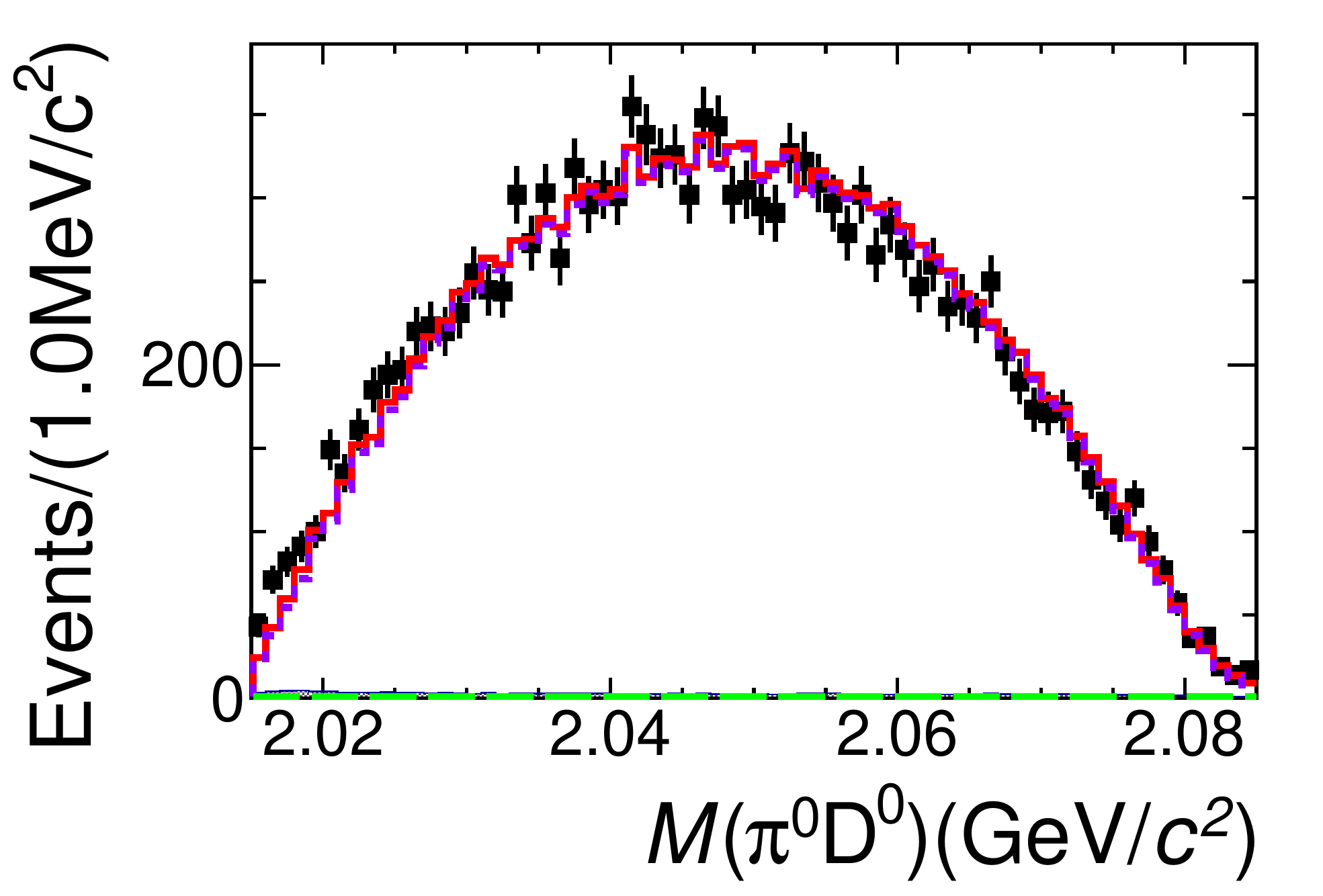}
\includegraphics[width=0.3\columnwidth]{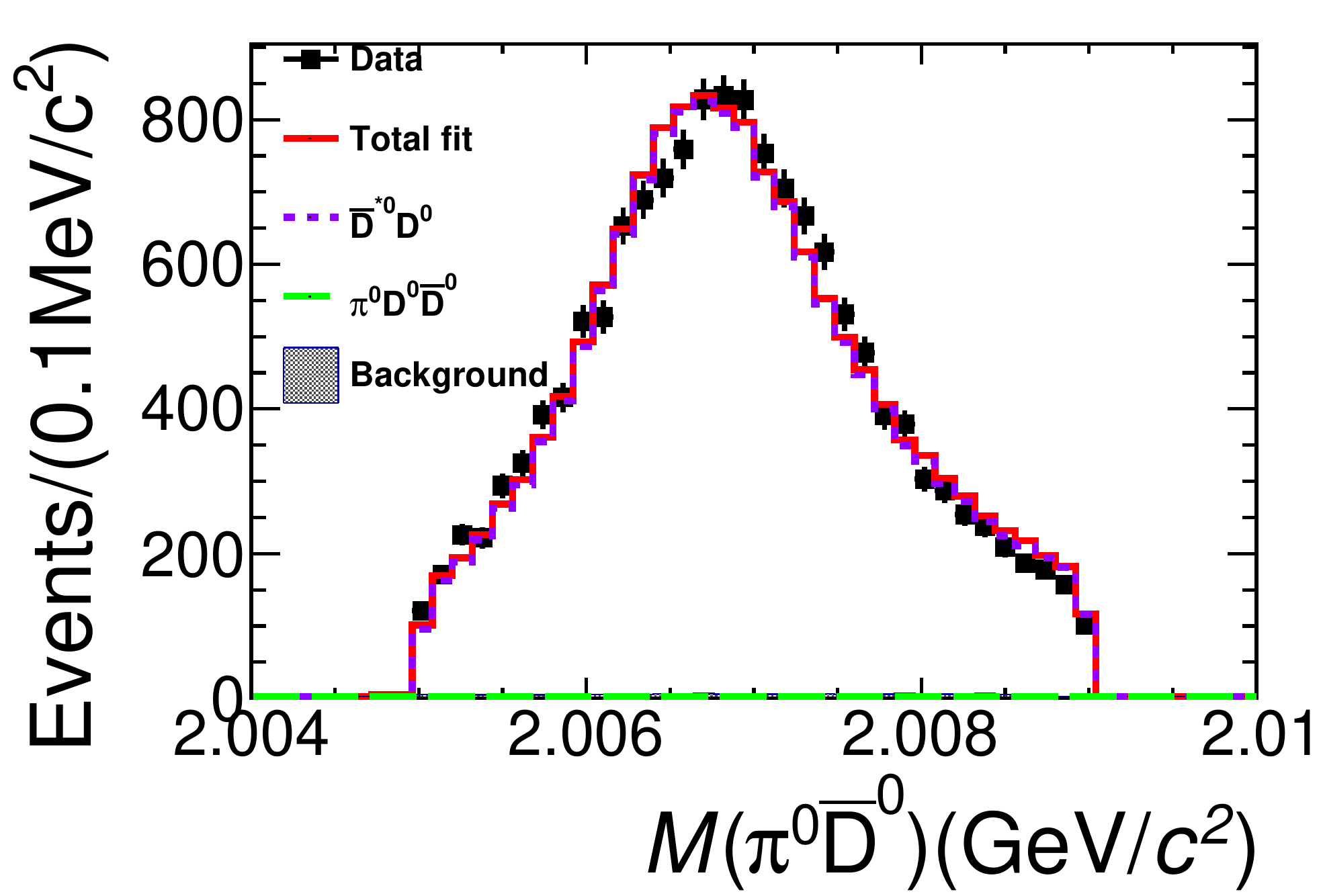}
\includegraphics[width=0.3\columnwidth]{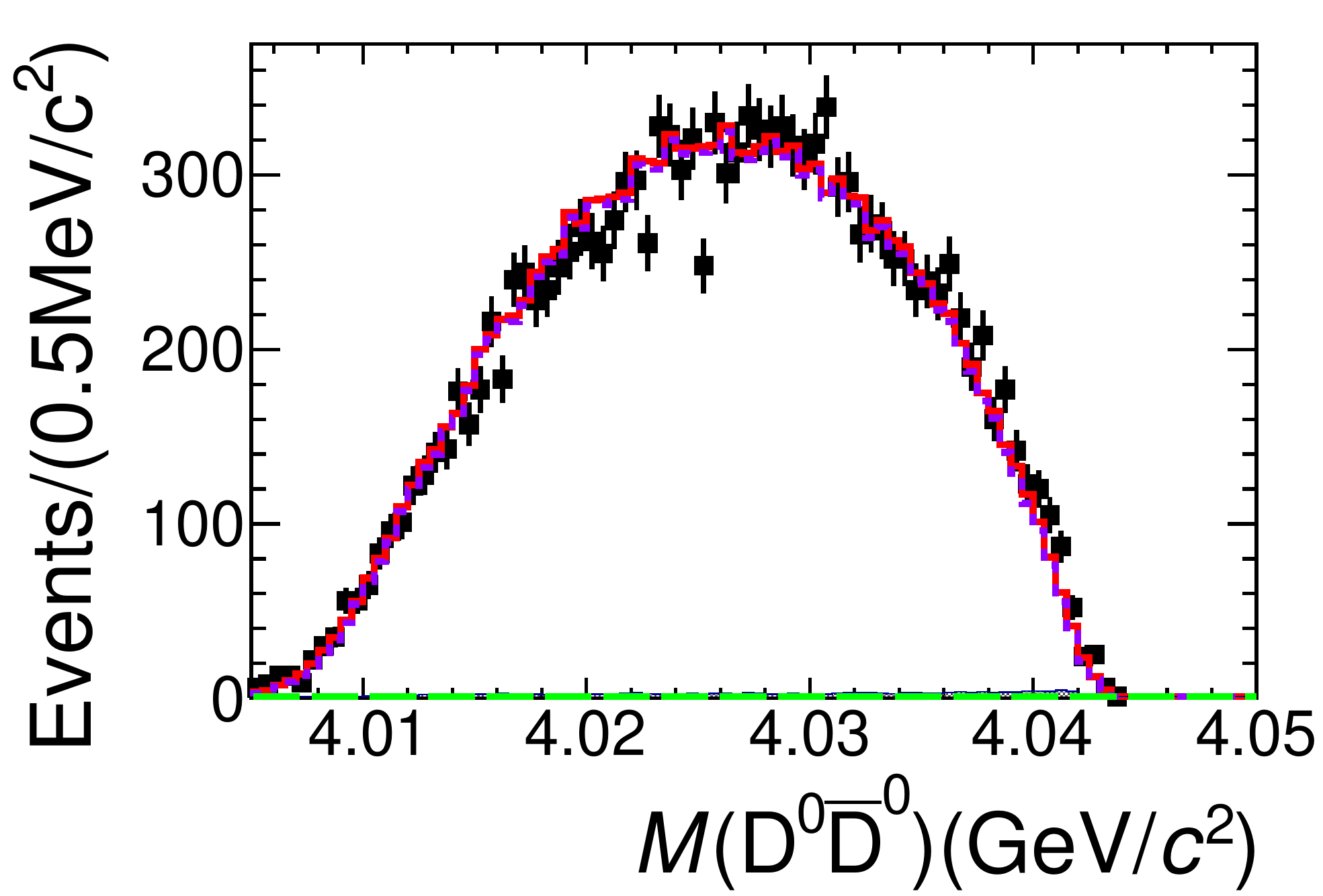}
\includegraphics[width=0.3\columnwidth]{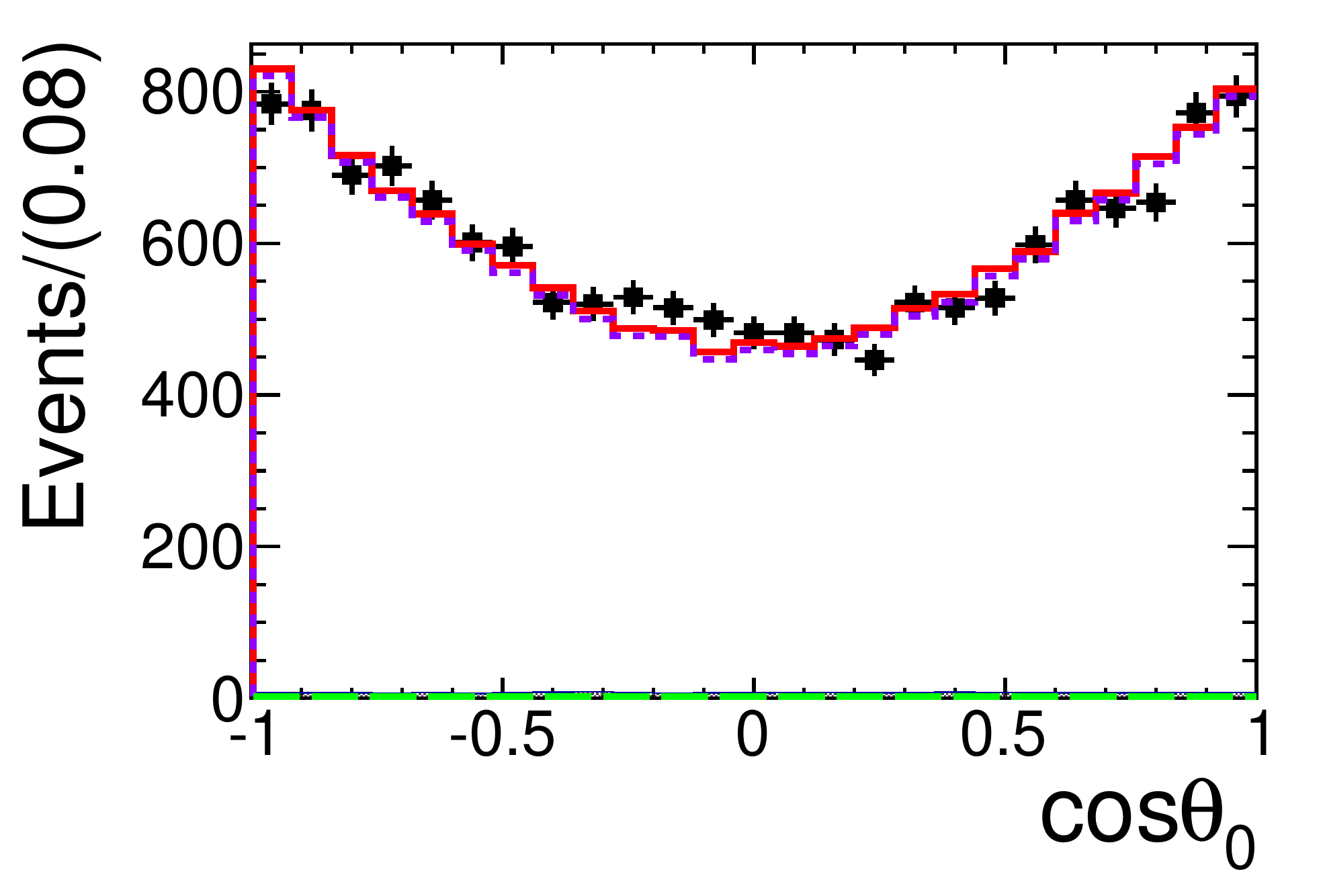}
\includegraphics[width=0.3\columnwidth]{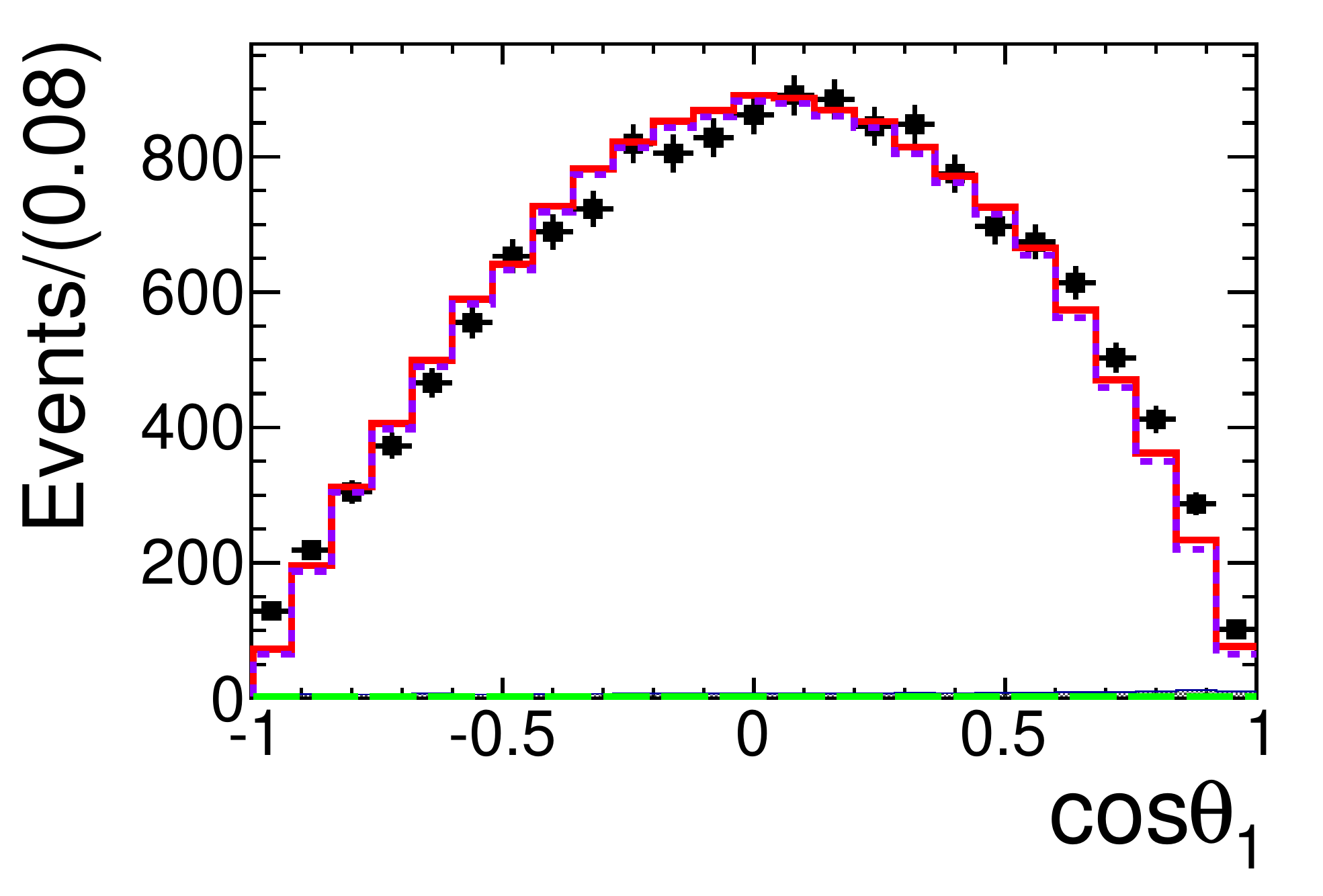}
\includegraphics[width=0.3\columnwidth]{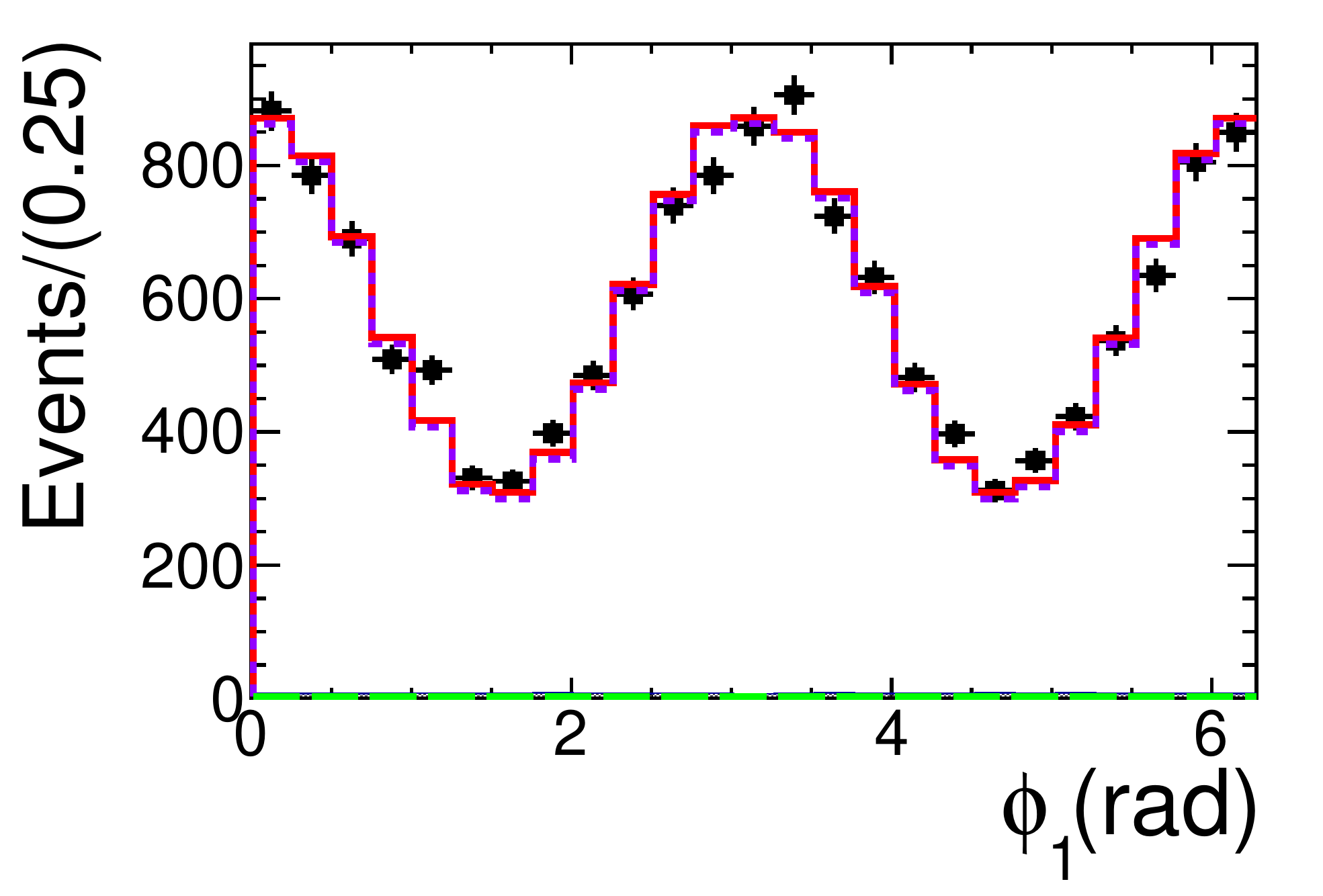}
\caption{
Mass and angular distributions of $D^0$-recoil sample for $J^P=1^-$ hypothesis.
}
\label{fig:angular_rmD0_1}
\end{figure}

\begin{figure}[!hbt]
\centering
\includegraphics[width=0.3\columnwidth]{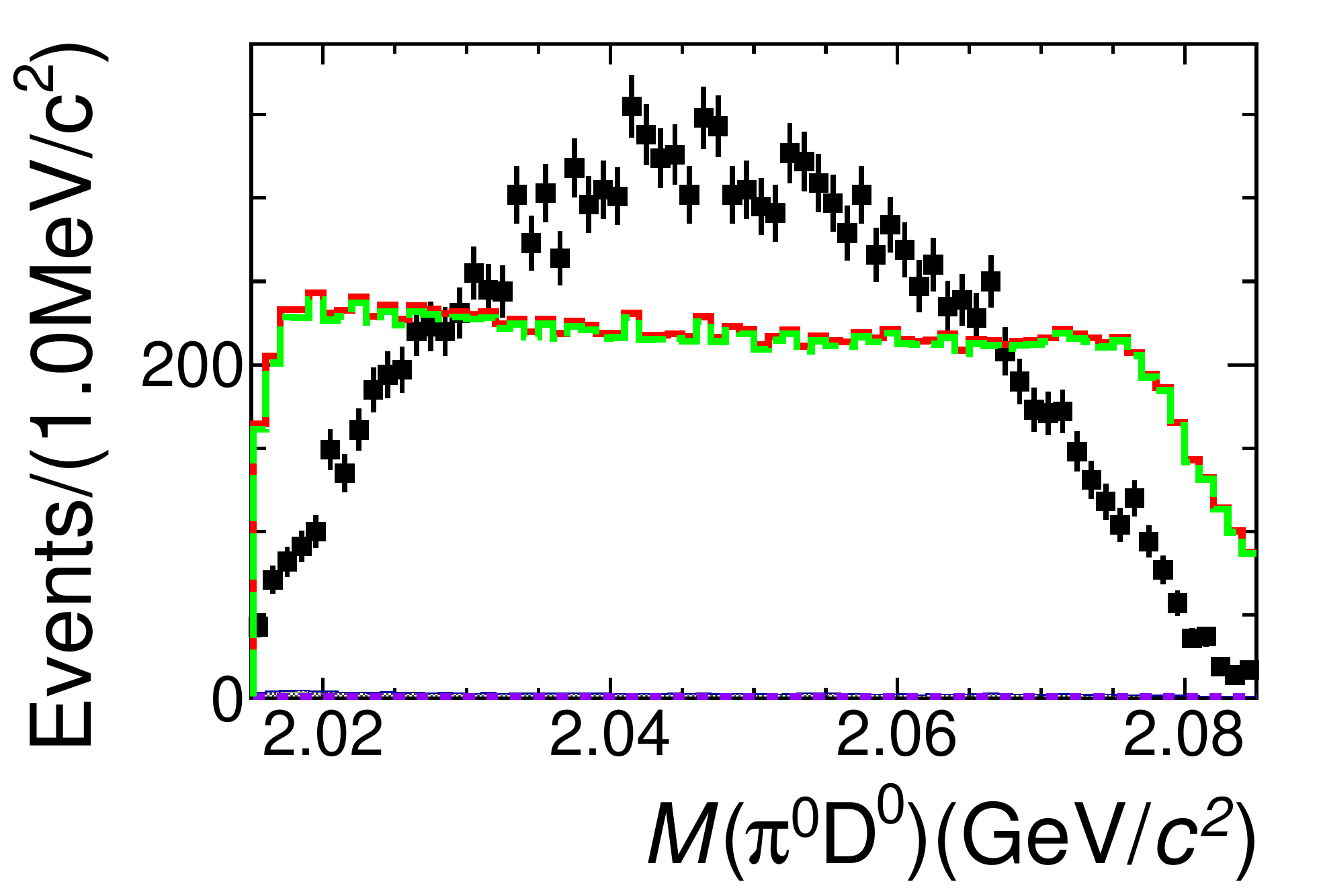}
\includegraphics[width=0.3\columnwidth]{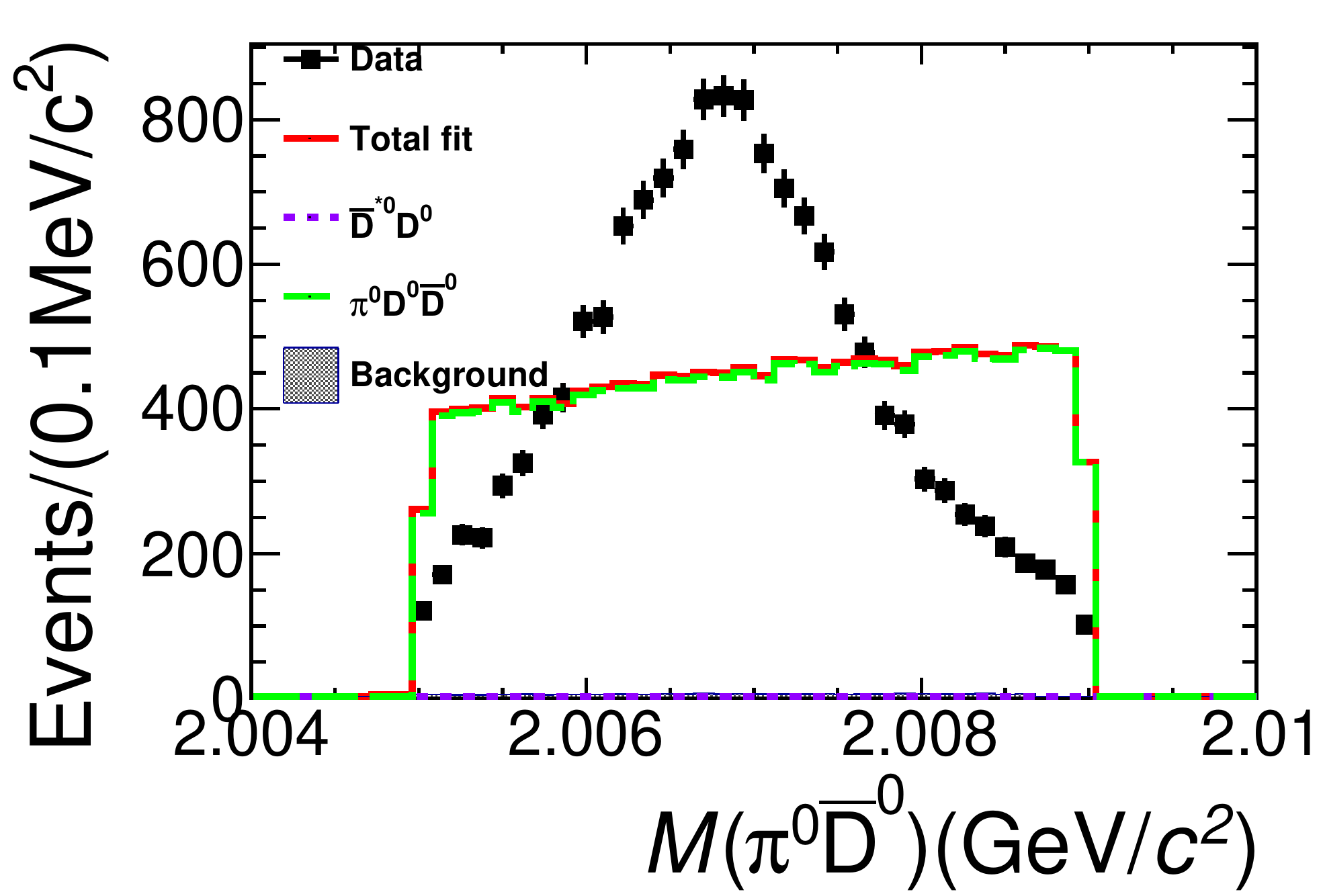}
\includegraphics[width=0.3\columnwidth]{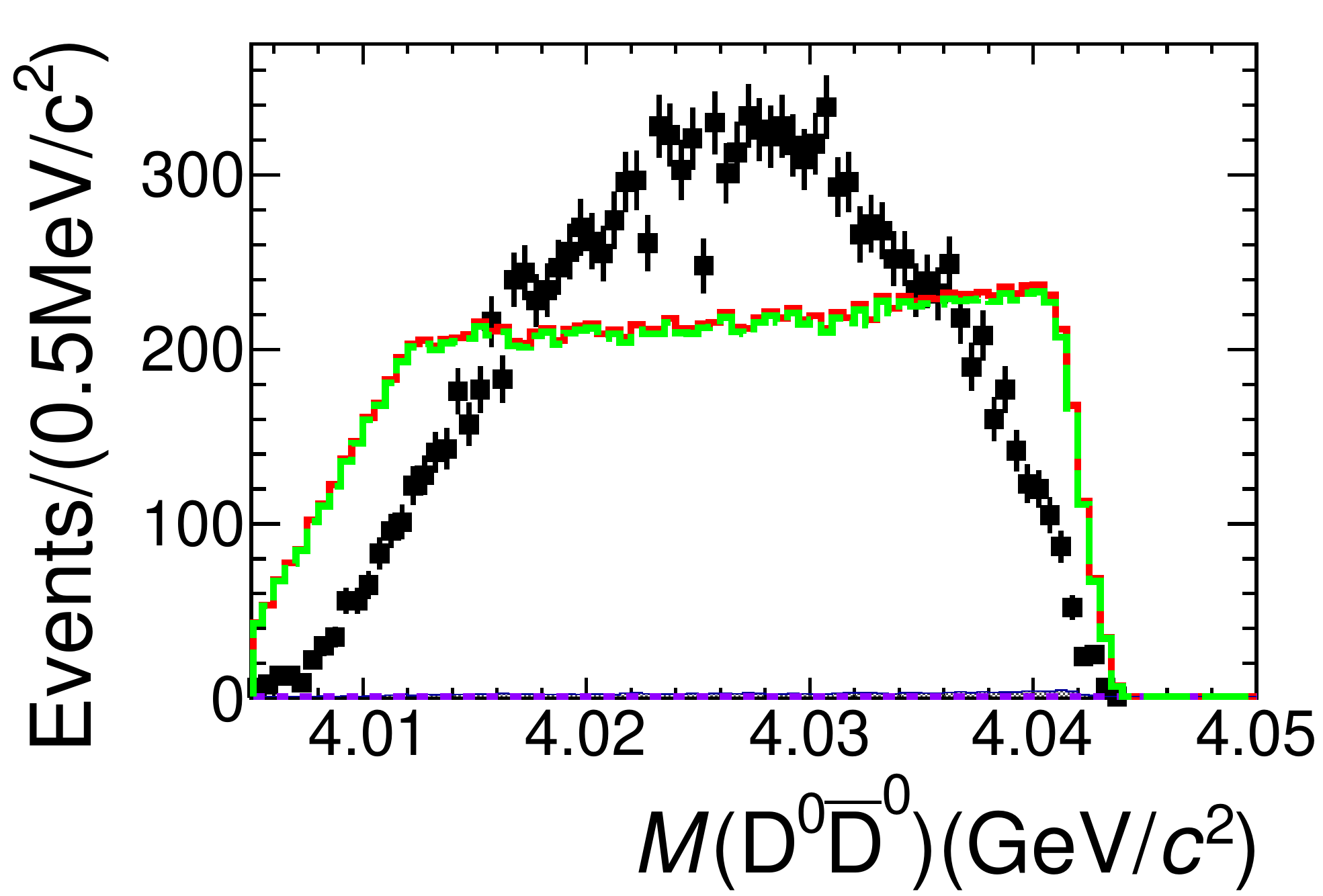}
\includegraphics[width=0.3\columnwidth]{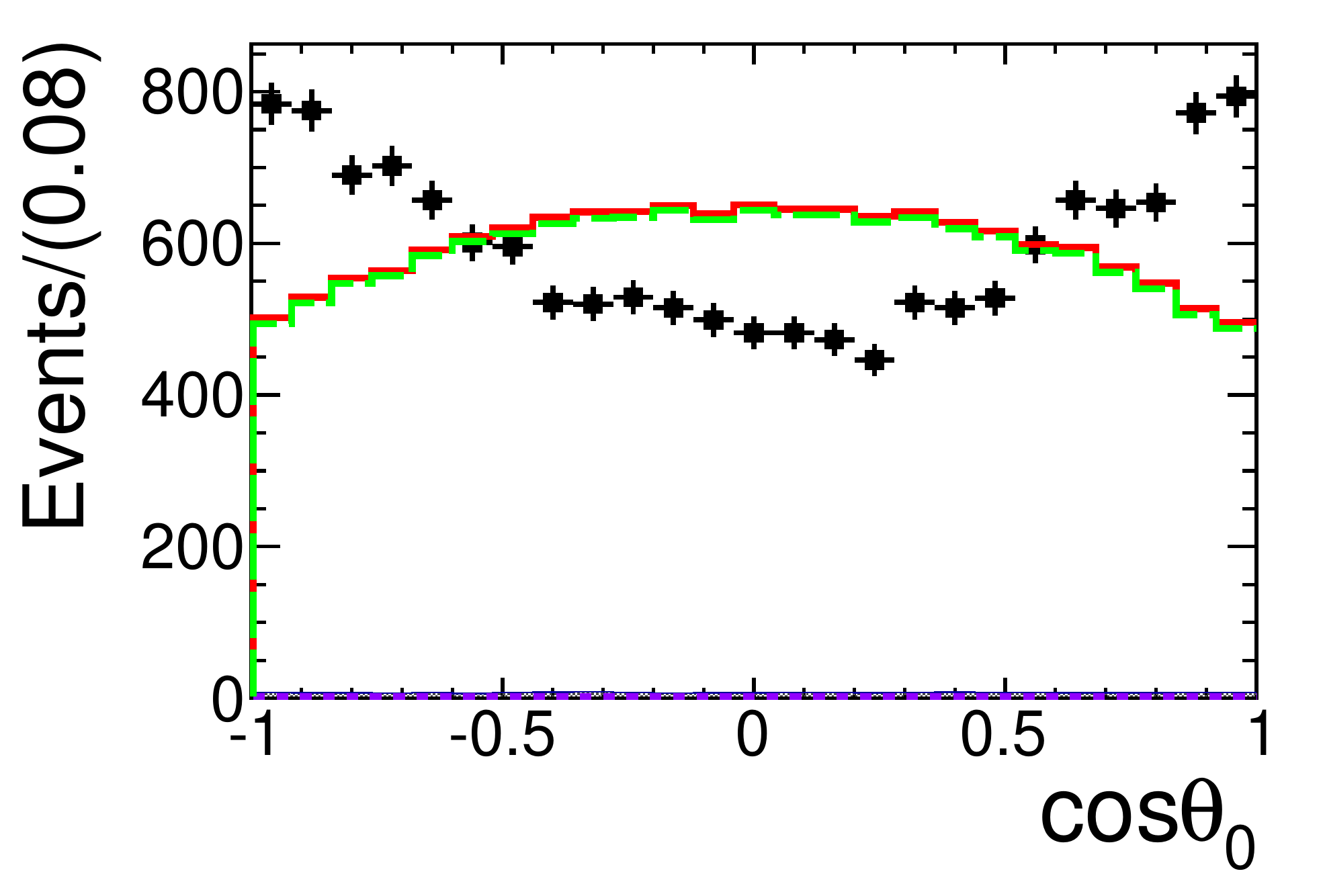}
\includegraphics[width=0.3\columnwidth]{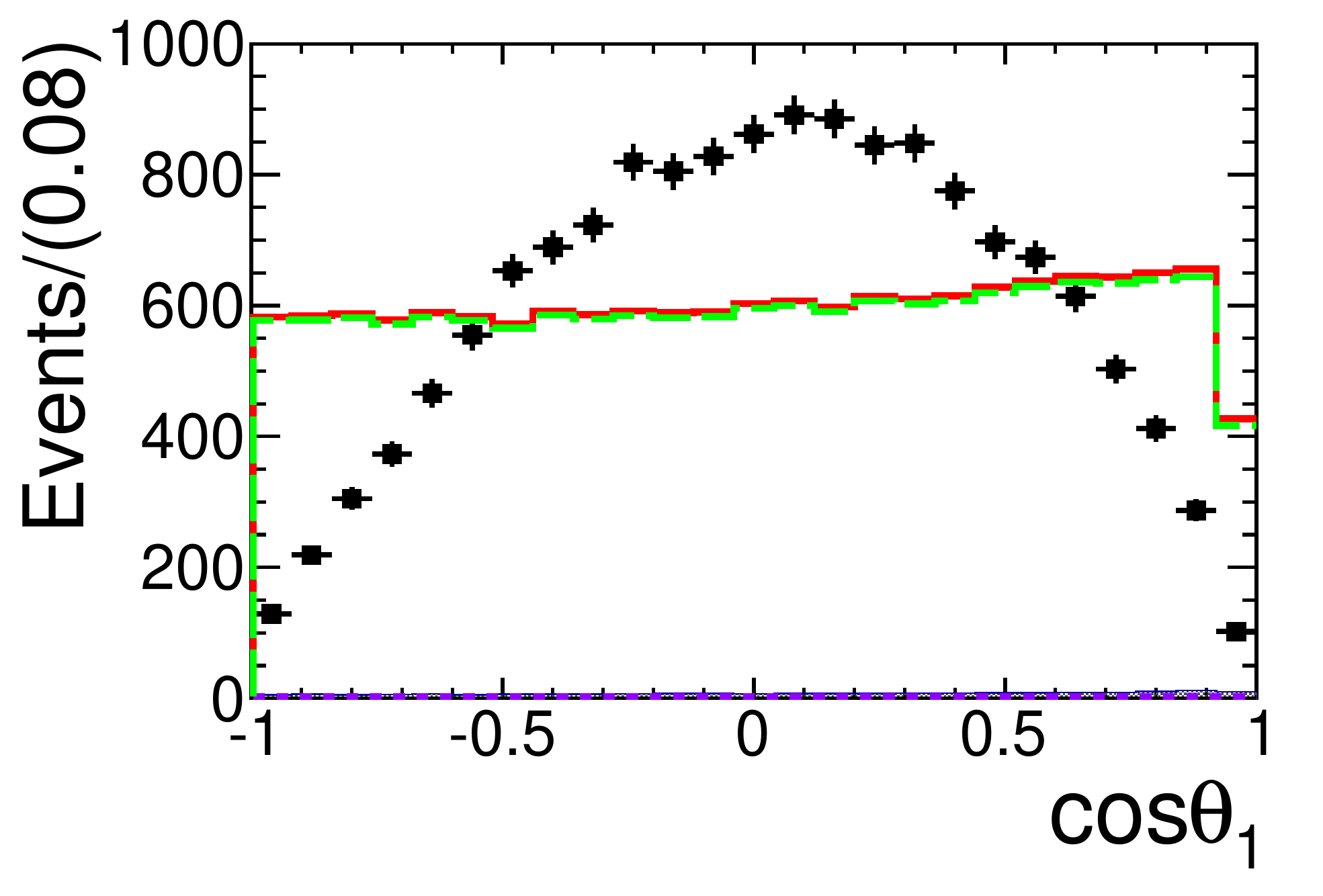}
\includegraphics[width=0.3\columnwidth]{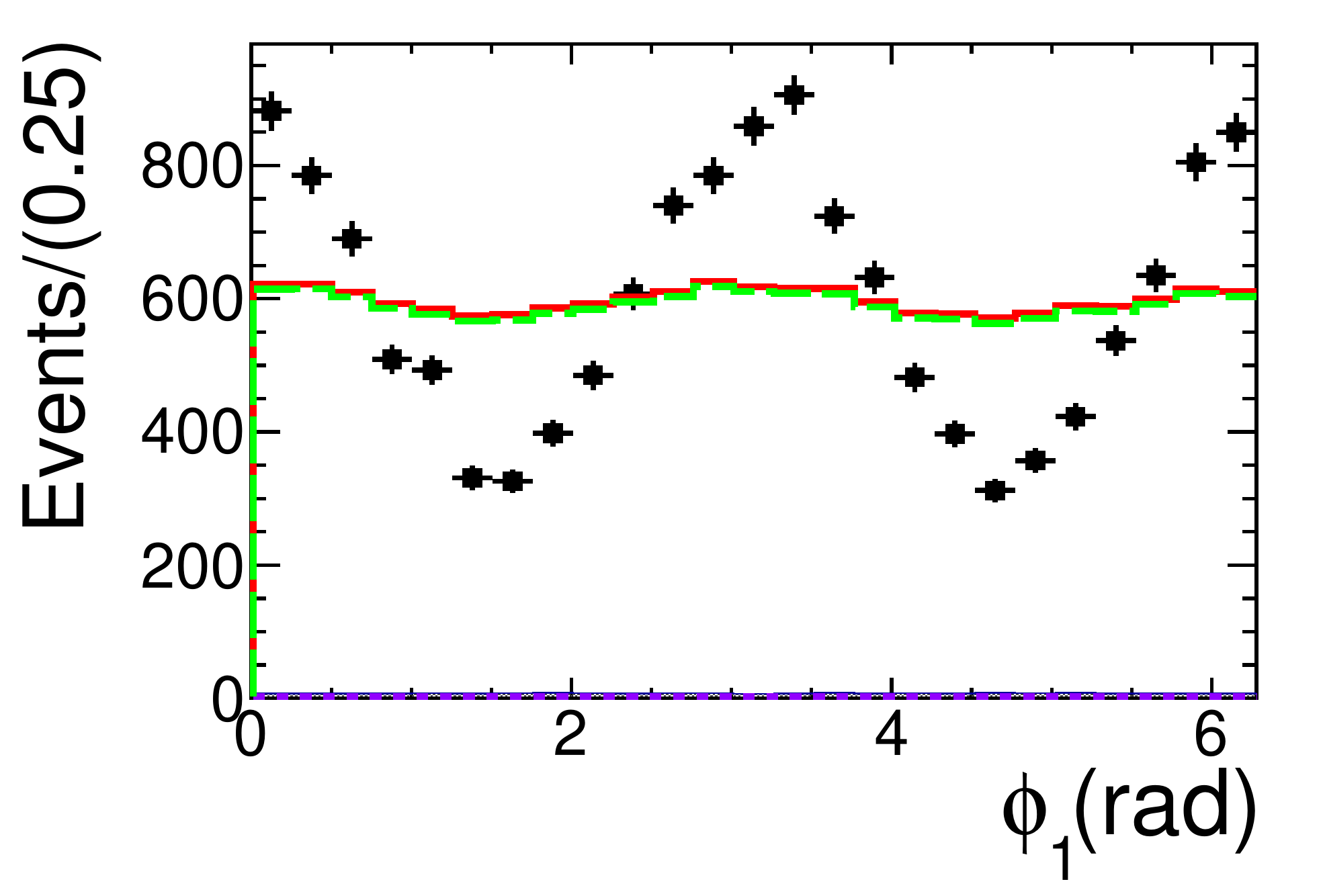}
\caption{
Mass and angular distributions of $D^0$-recoil sample for $J^P=2^+$ hypothesis.
}
\label{fig:angular_rmD0_2}
\end{figure}

\begin{figure}[!hbt]
\centering
\includegraphics[width=0.3\columnwidth]{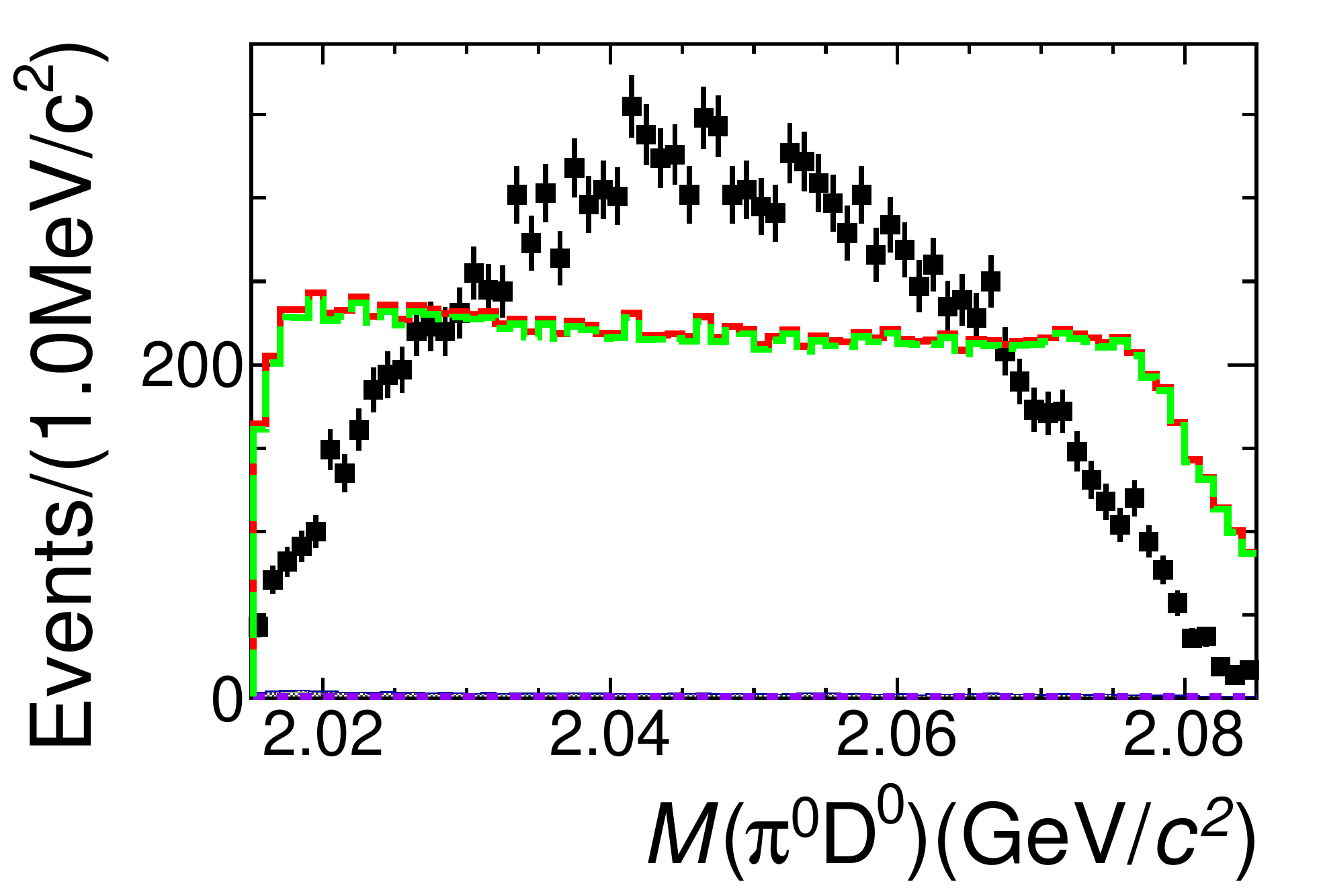}
\includegraphics[width=0.3\columnwidth]{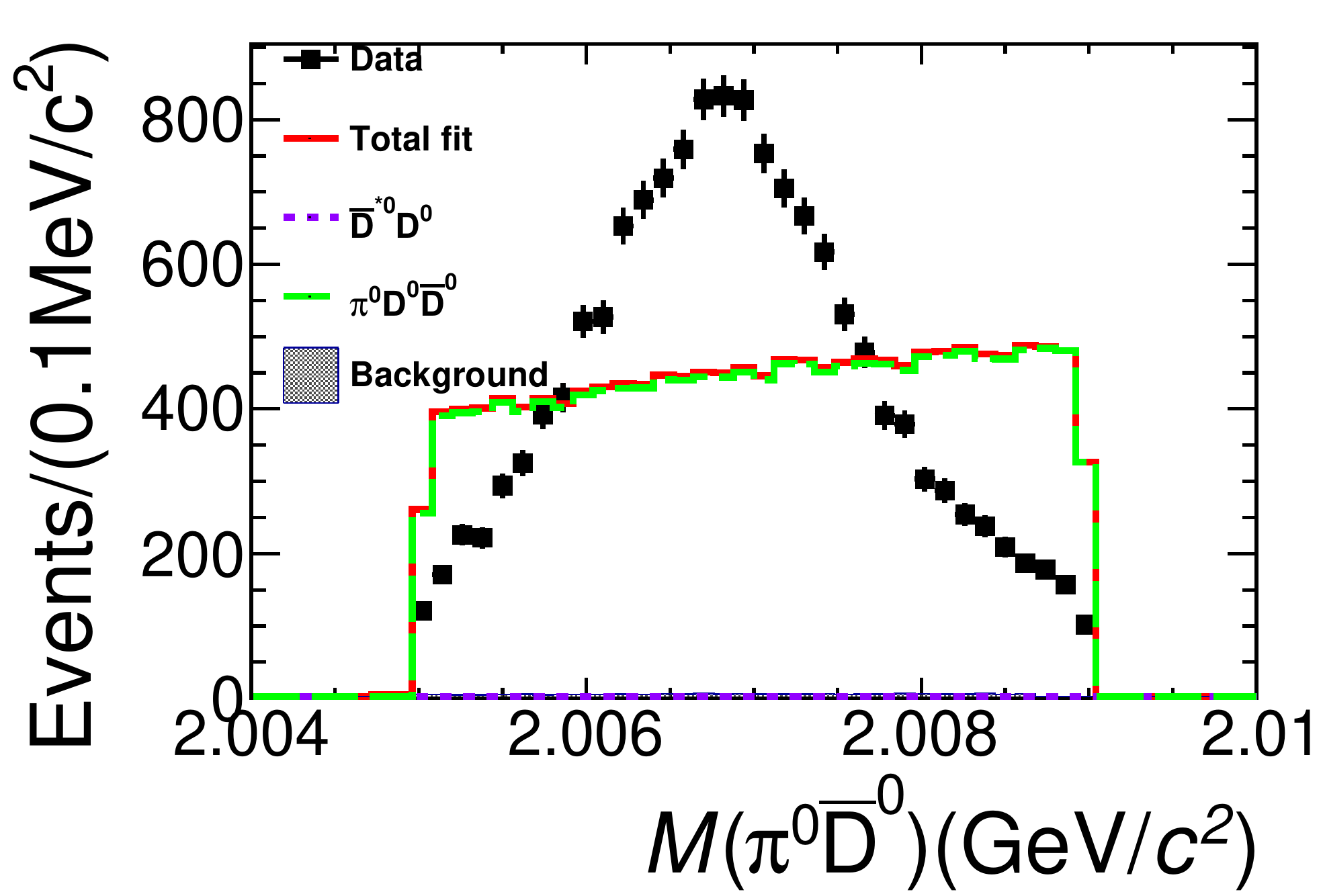}
\includegraphics[width=0.3\columnwidth]{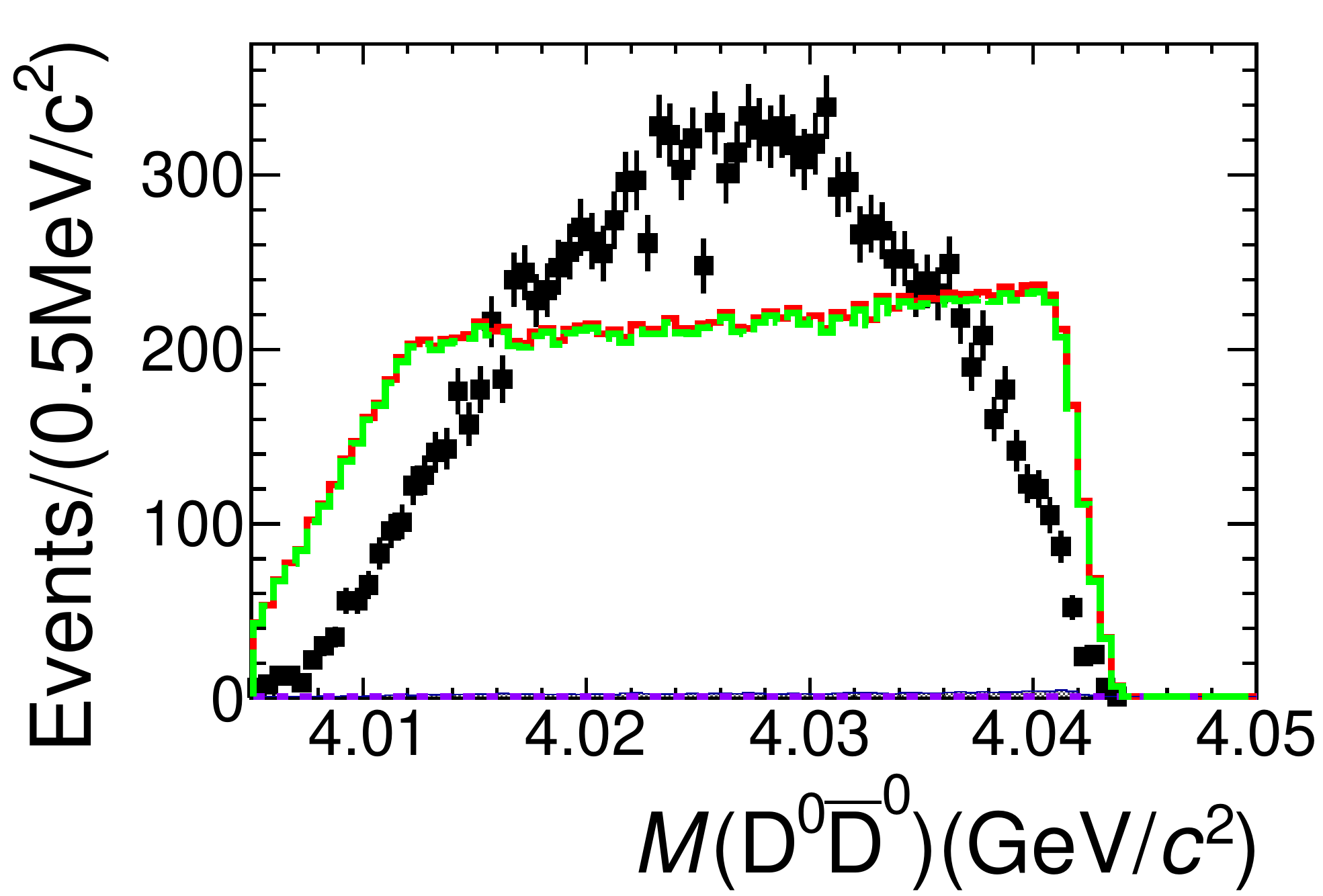}
\includegraphics[width=0.3\columnwidth]{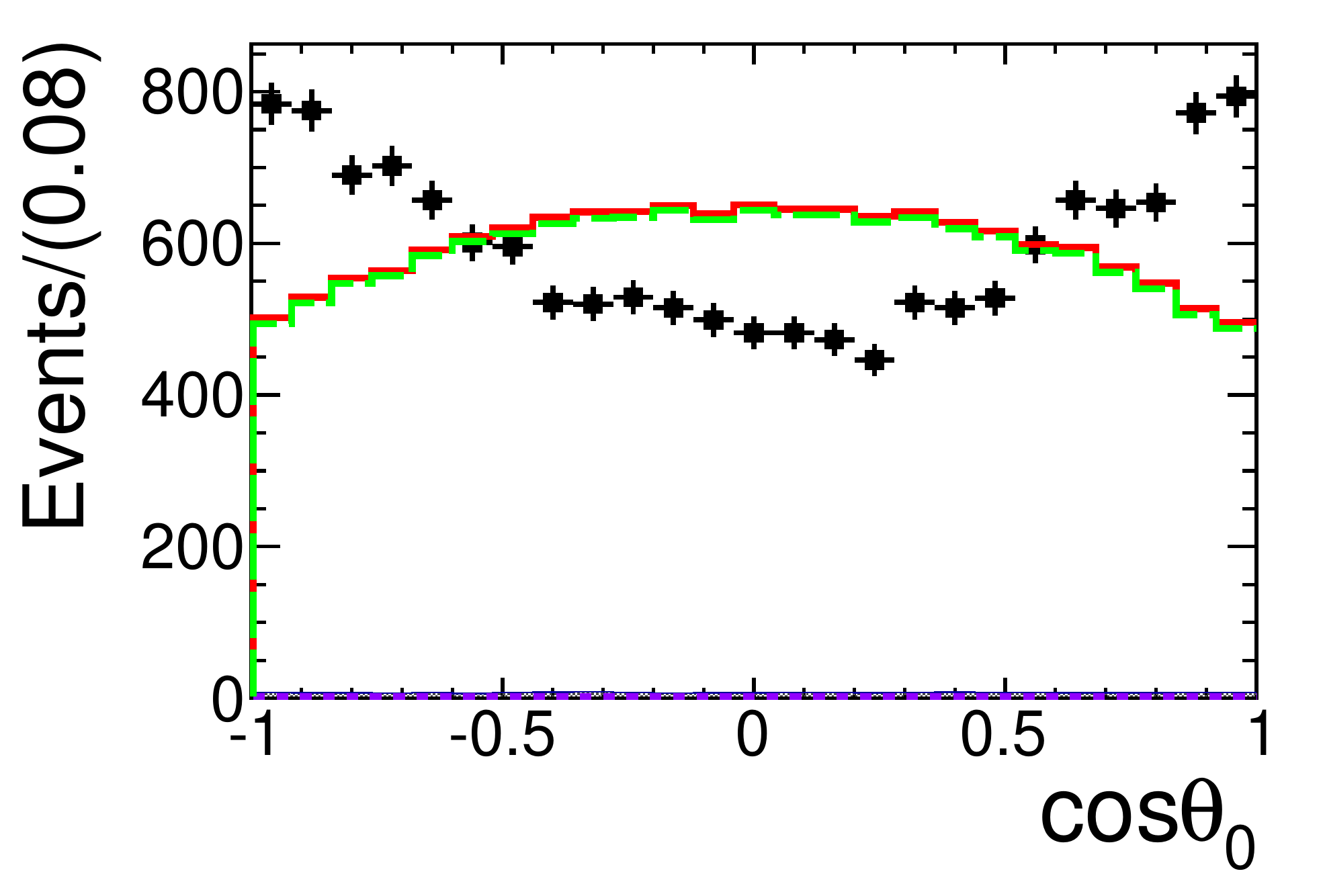}
\includegraphics[width=0.3\columnwidth]{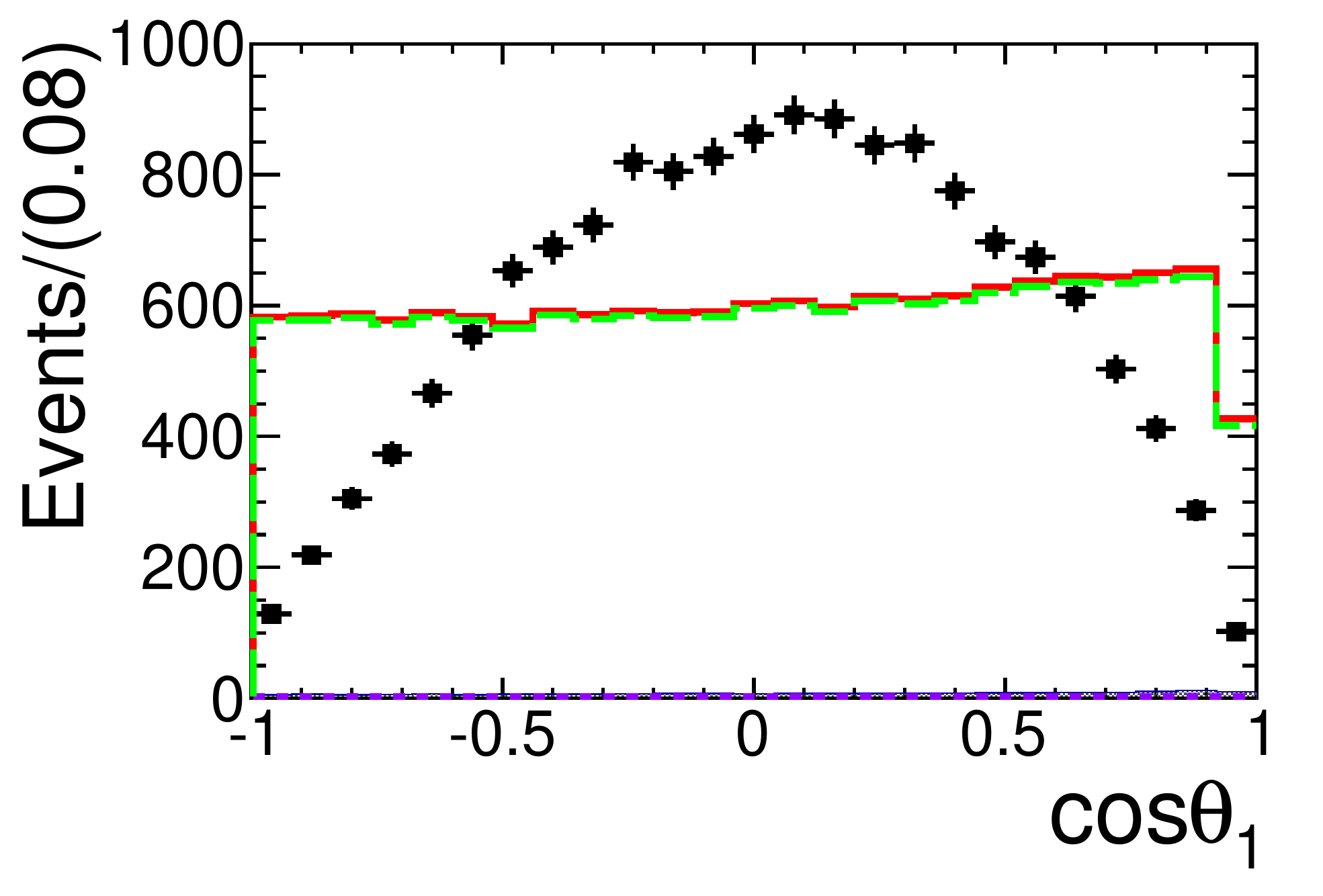}
\includegraphics[width=0.3\columnwidth]{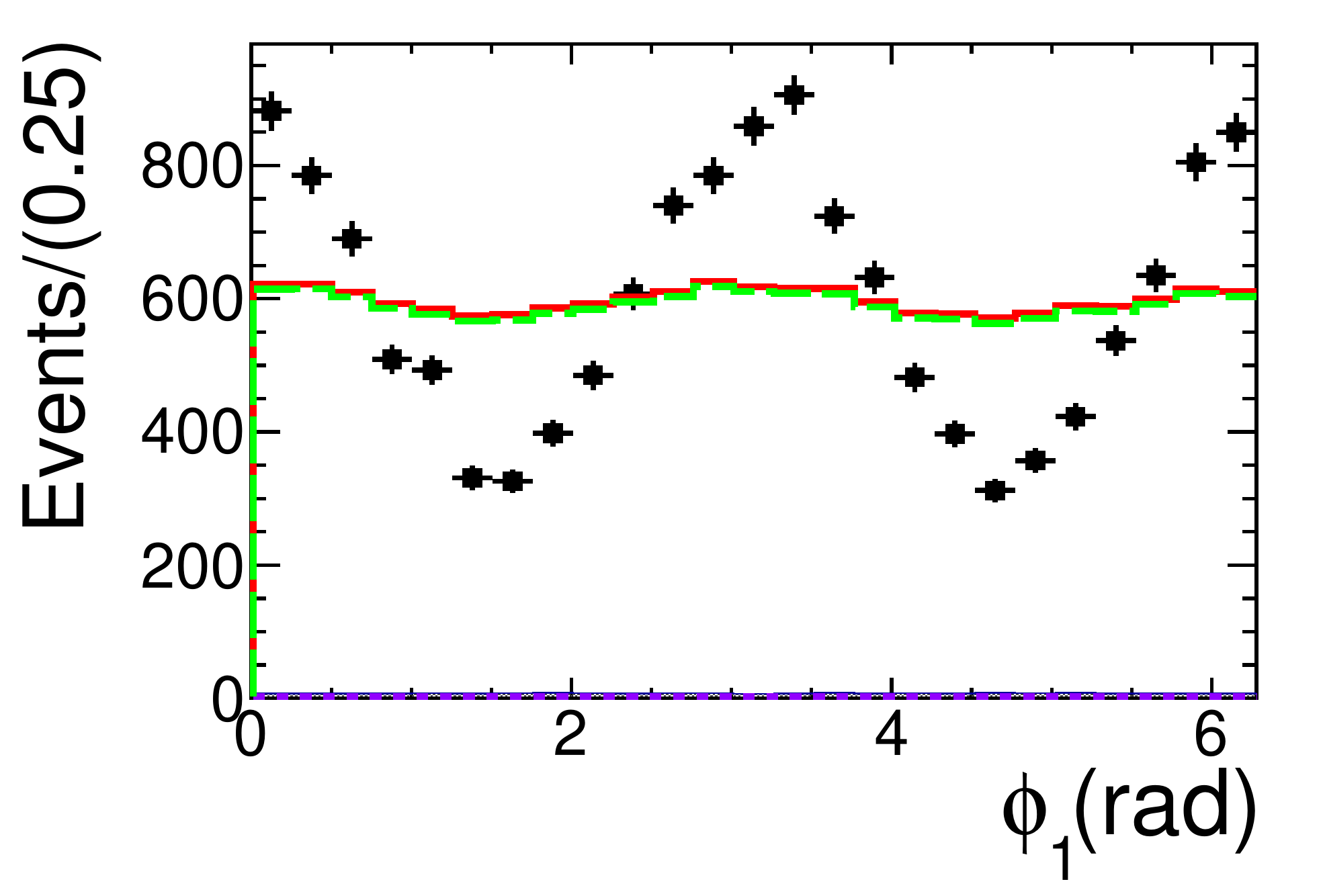}
\caption{
Mass and angular distributions of $D^0$-recoil sample for $J^P=3^-$ hypothesis.
}
\label{fig:angular_rmD0_3}
\end{figure}

\begin{figure}[!hbt]
\centering
\includegraphics[width=0.3\columnwidth]{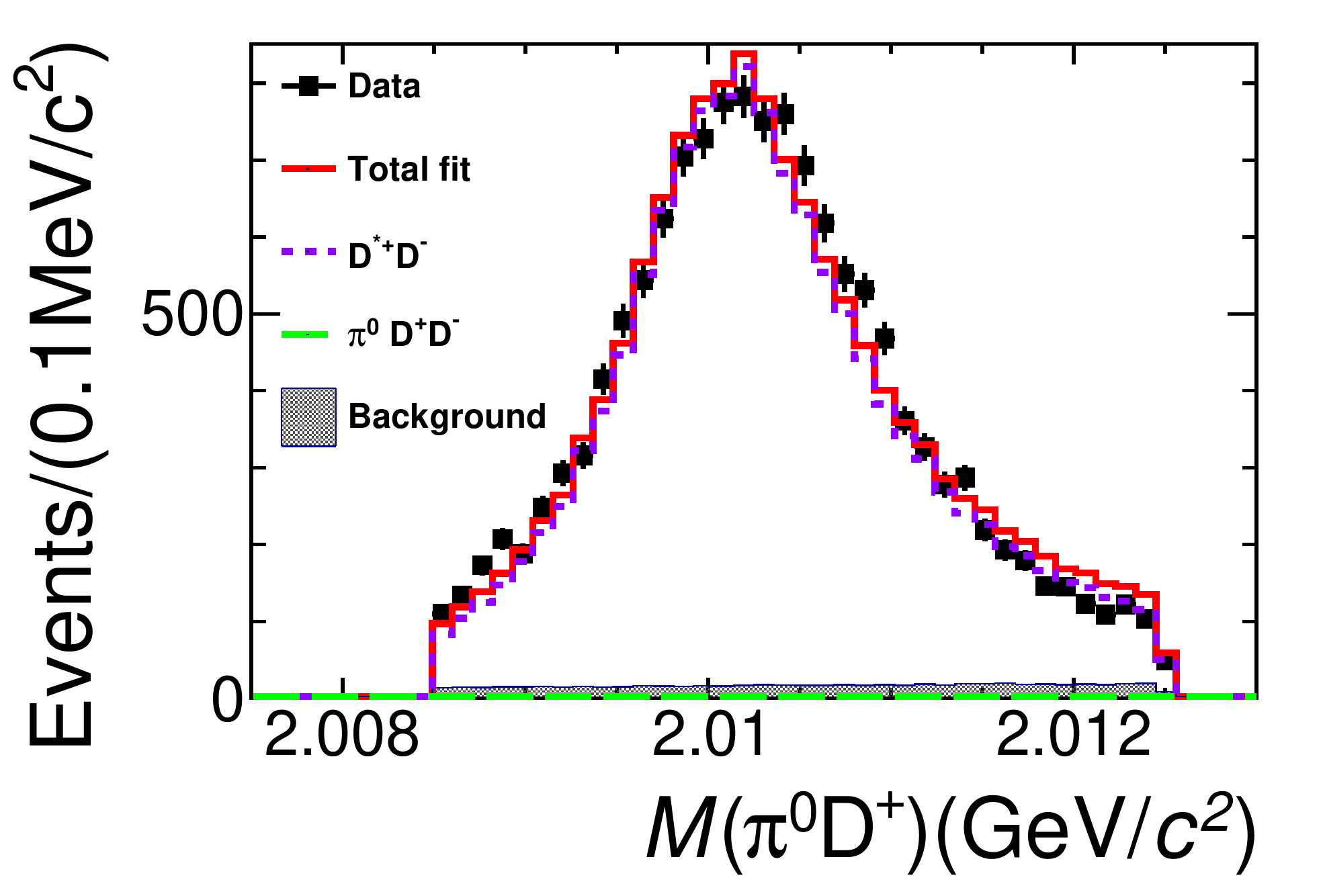}
\includegraphics[width=0.3\columnwidth]{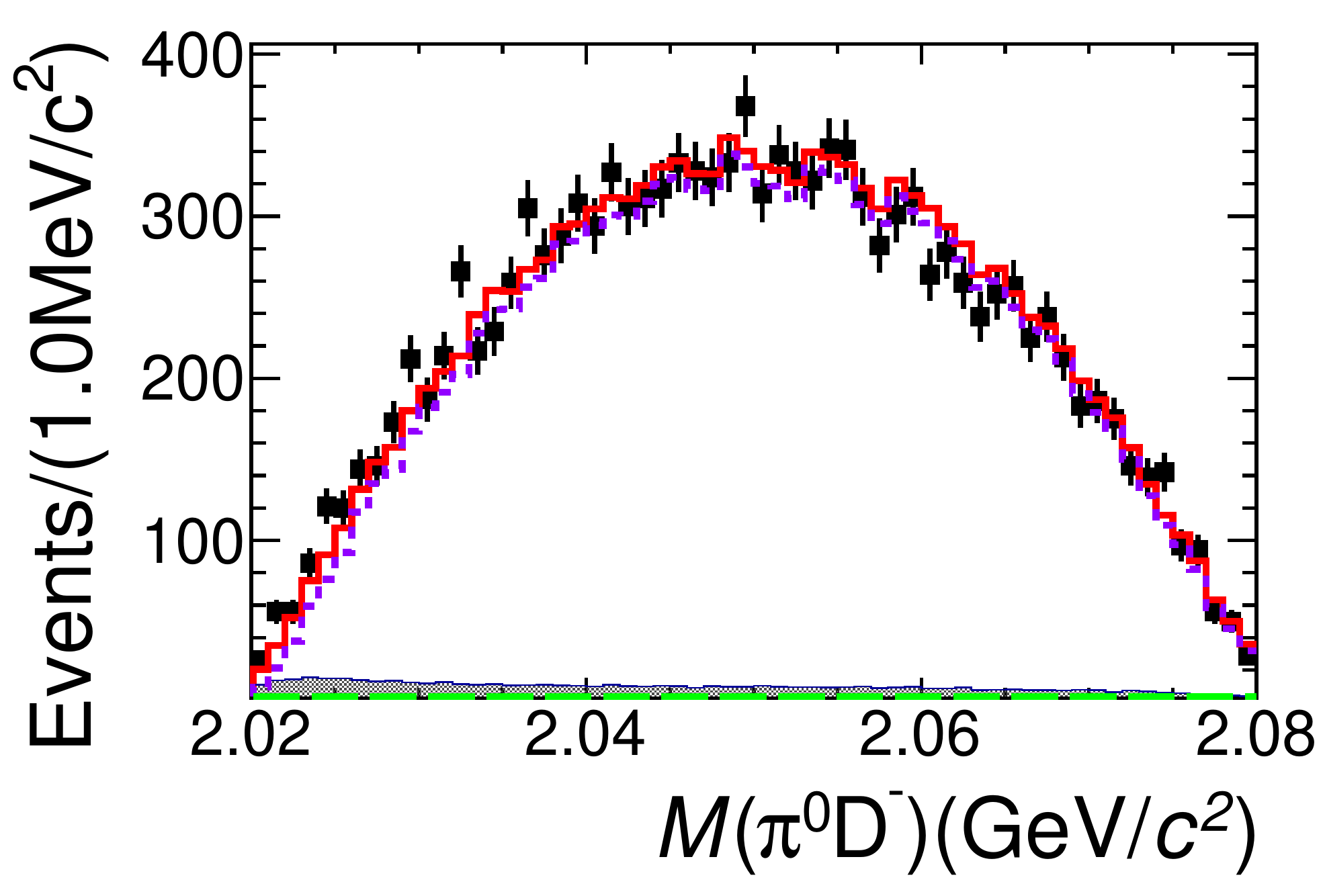}
\includegraphics[width=0.3\columnwidth]{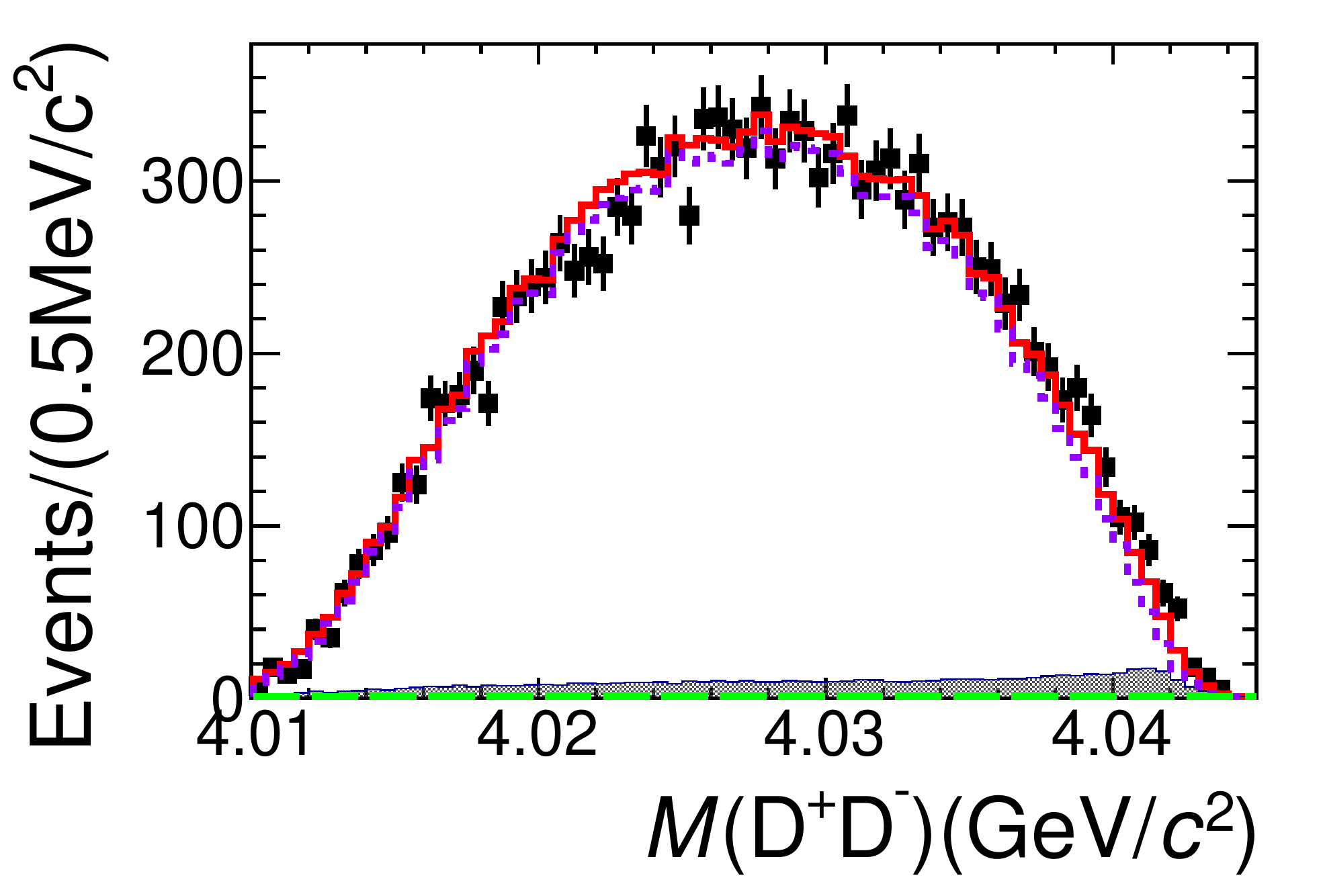}
\includegraphics[width=0.3\columnwidth]{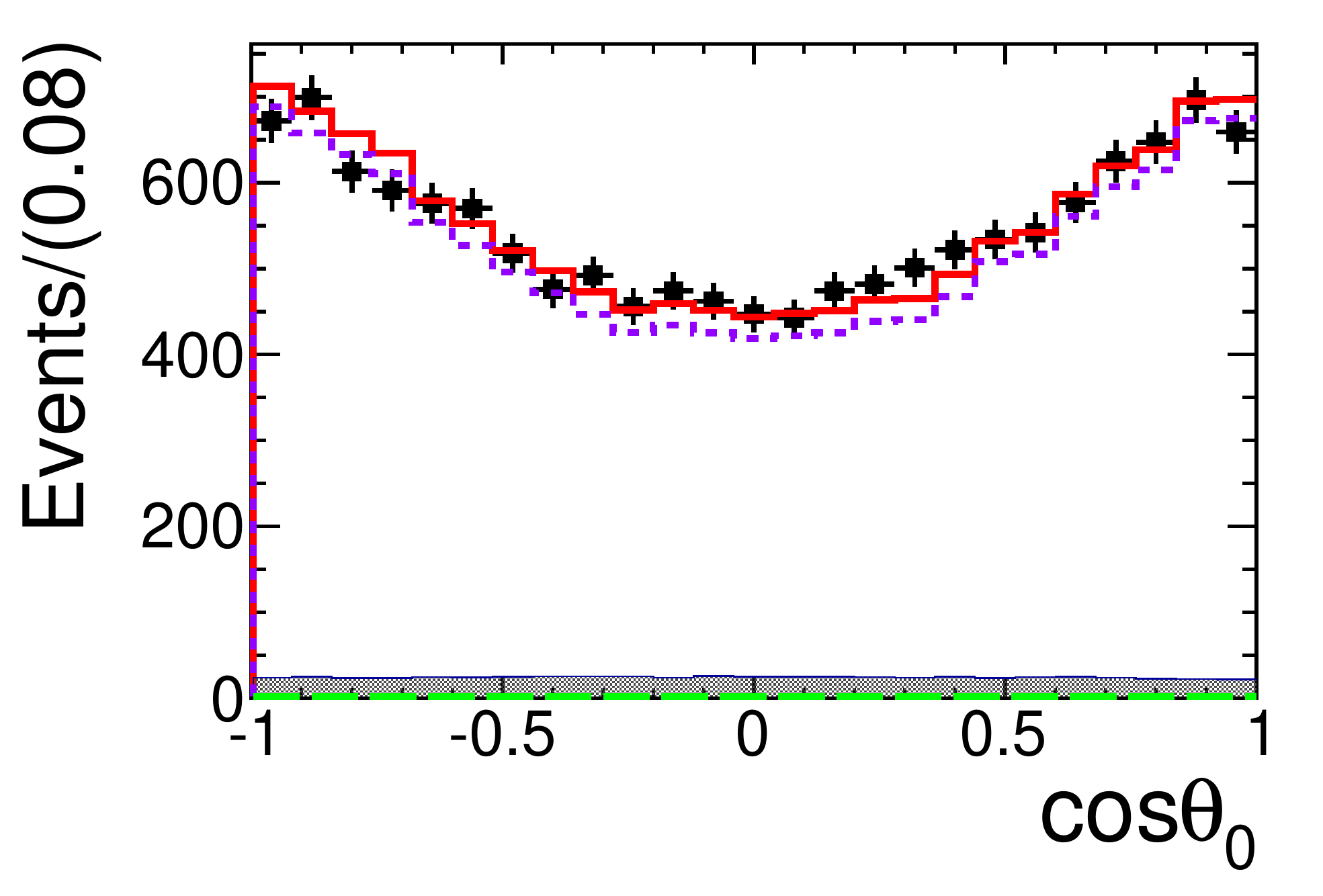}
\includegraphics[width=0.3\columnwidth]{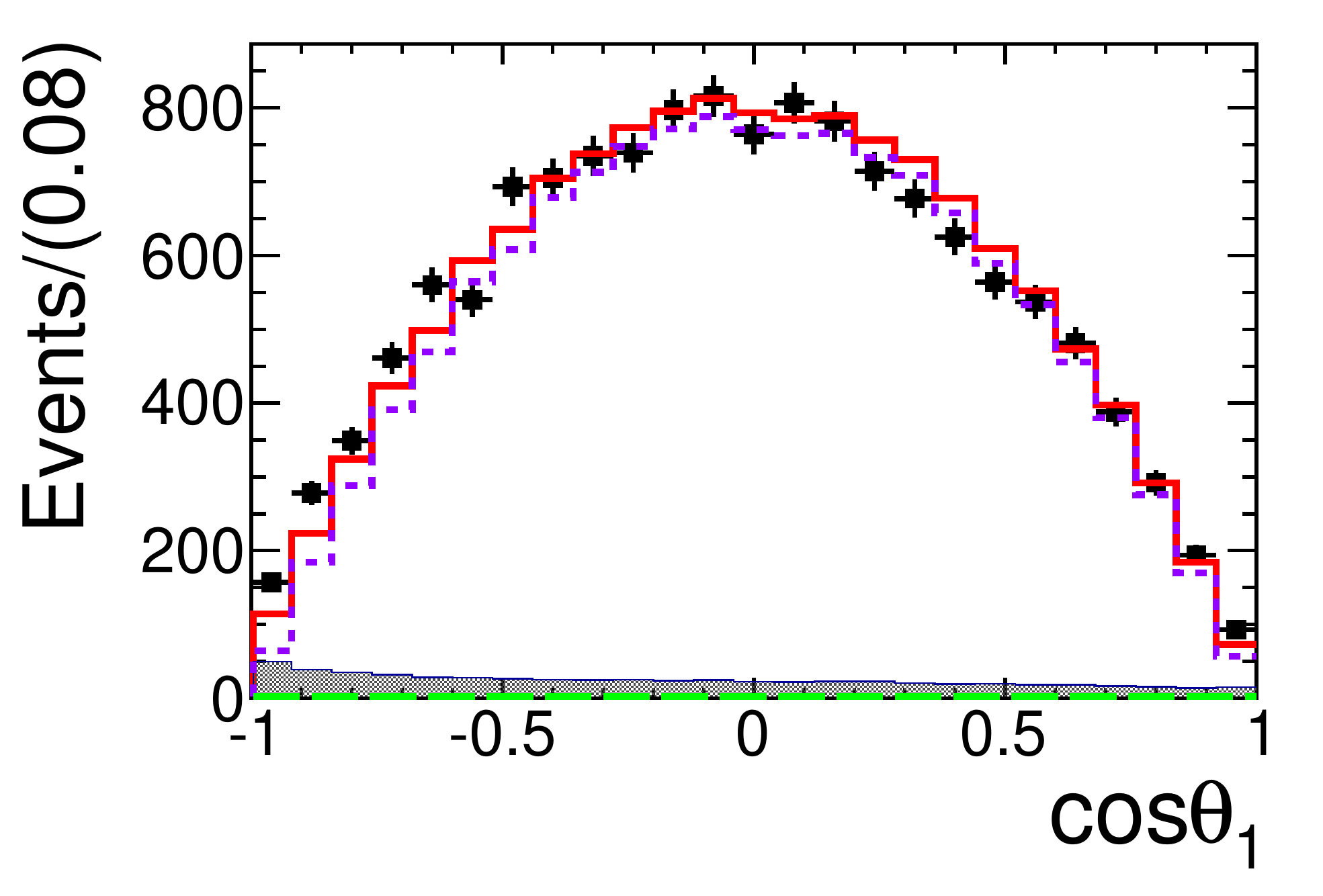}
\includegraphics[width=0.3\columnwidth]{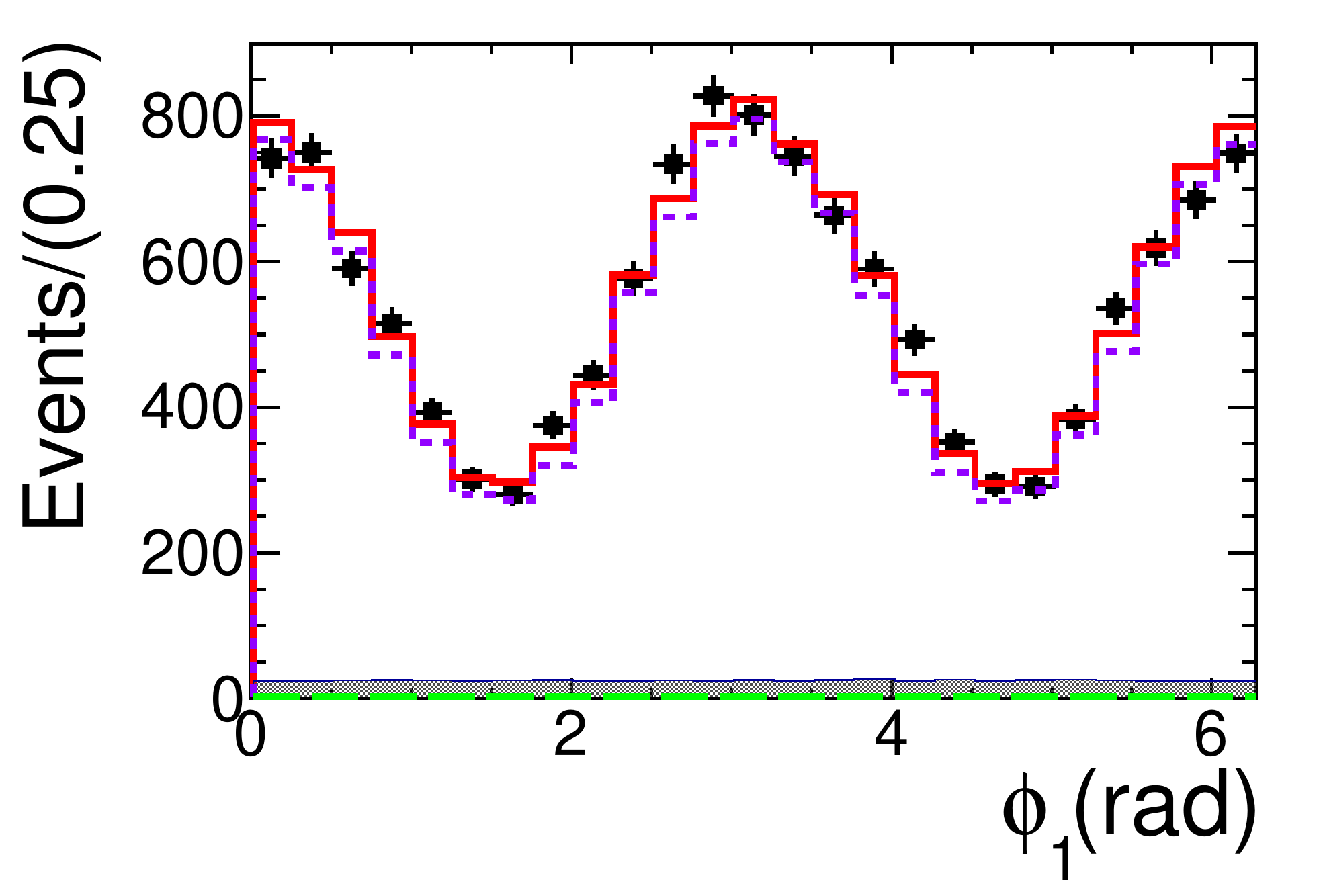}
\caption{
Mass and angular distributions of $D^{*+}$-tag sample for $J^P=1^-$ hypothesis. }
\label{fig:angular_Dppi0_1}
\end{figure}

\begin{figure}[!hbt]
\centering
\includegraphics[width=0.3\columnwidth]{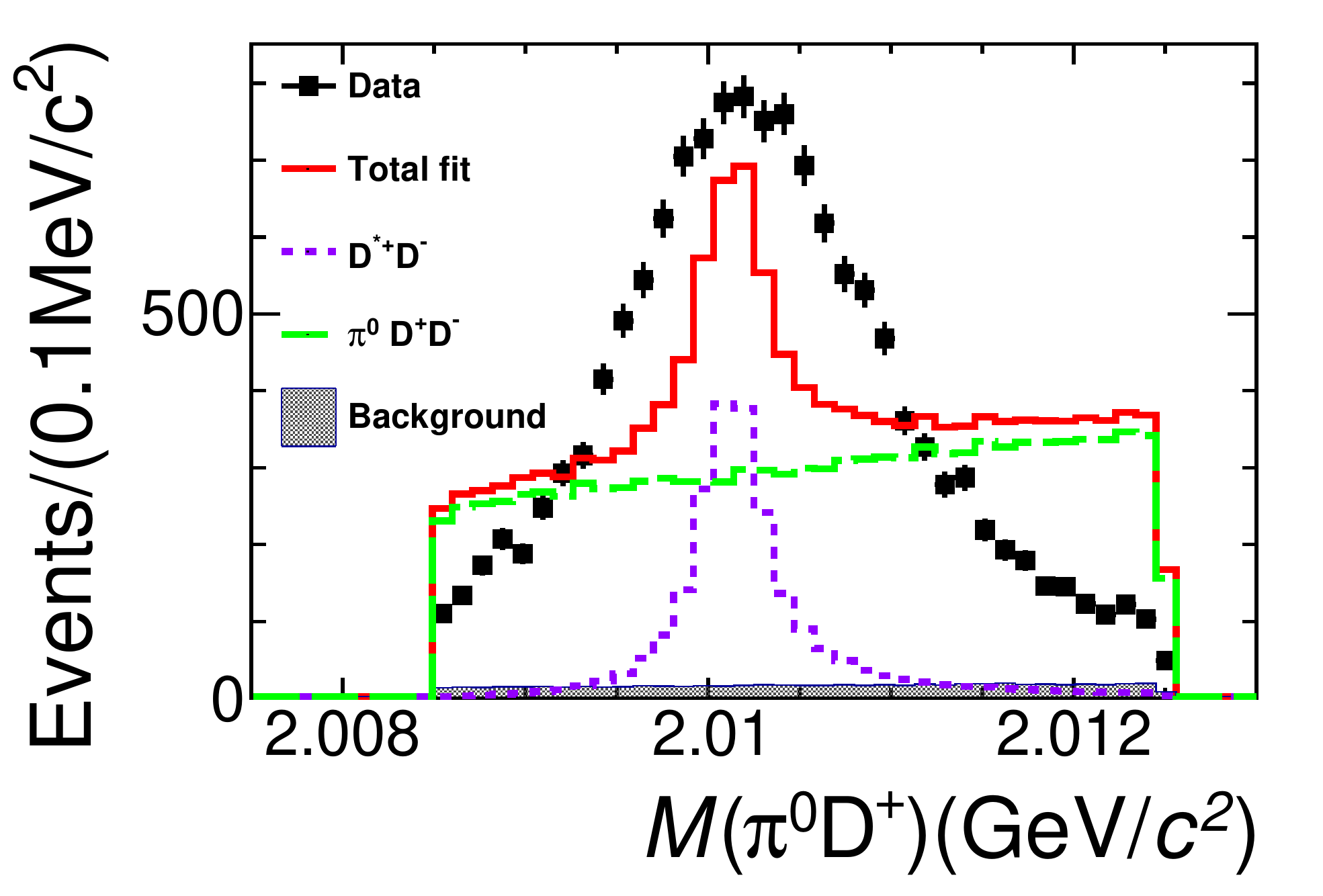}
\includegraphics[width=0.3\columnwidth]{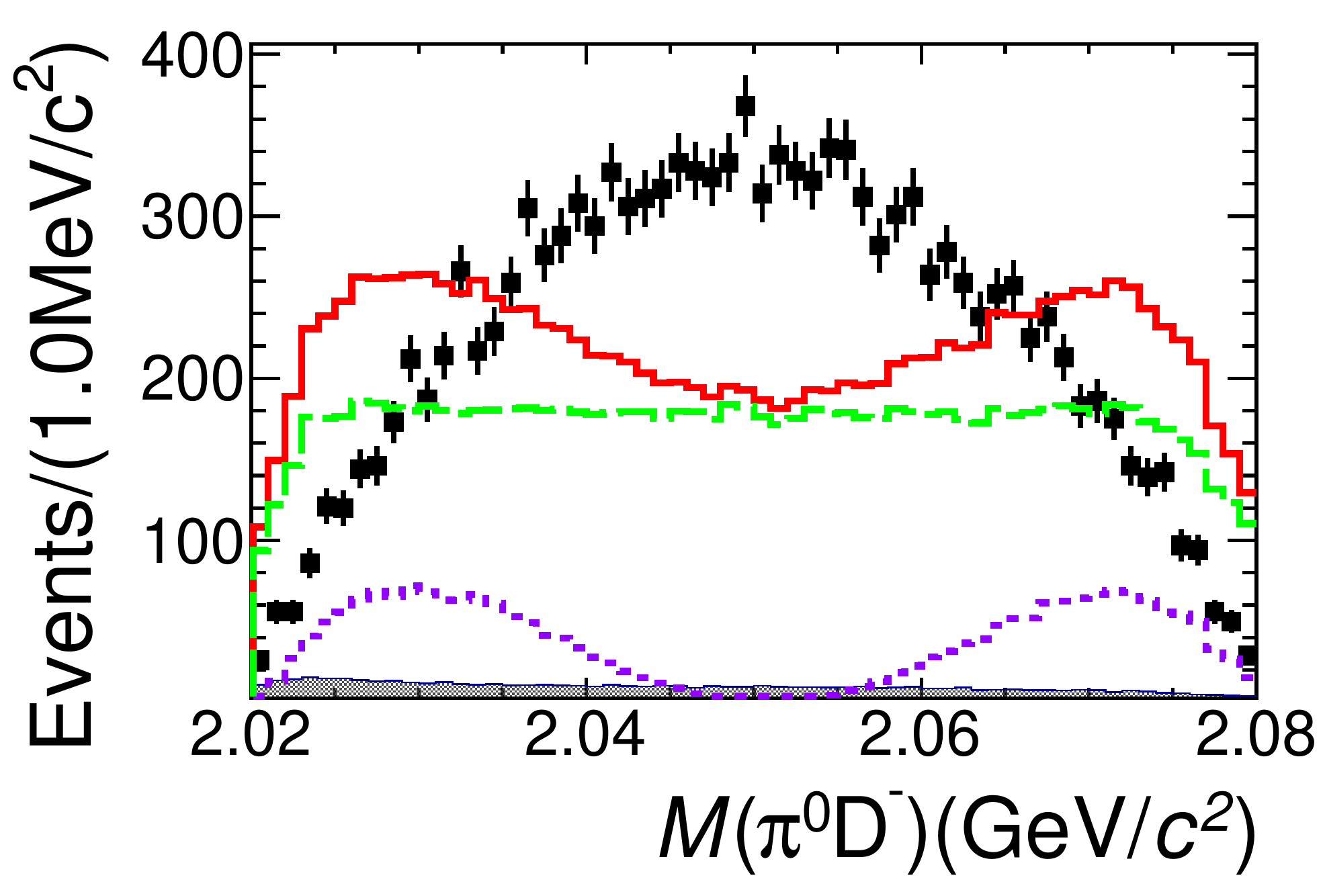}
\includegraphics[width=0.3\columnwidth]{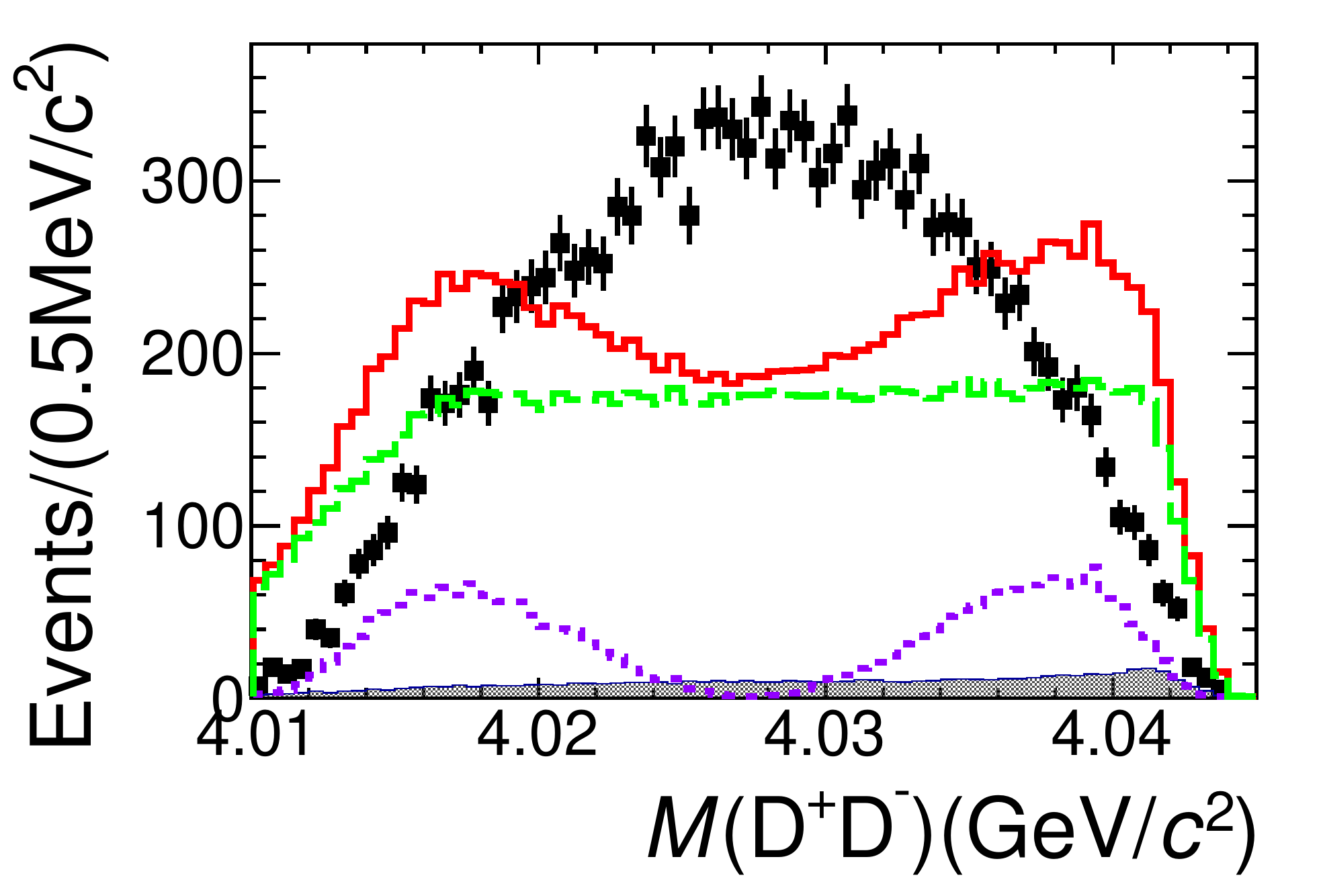}
\includegraphics[width=0.3\columnwidth]{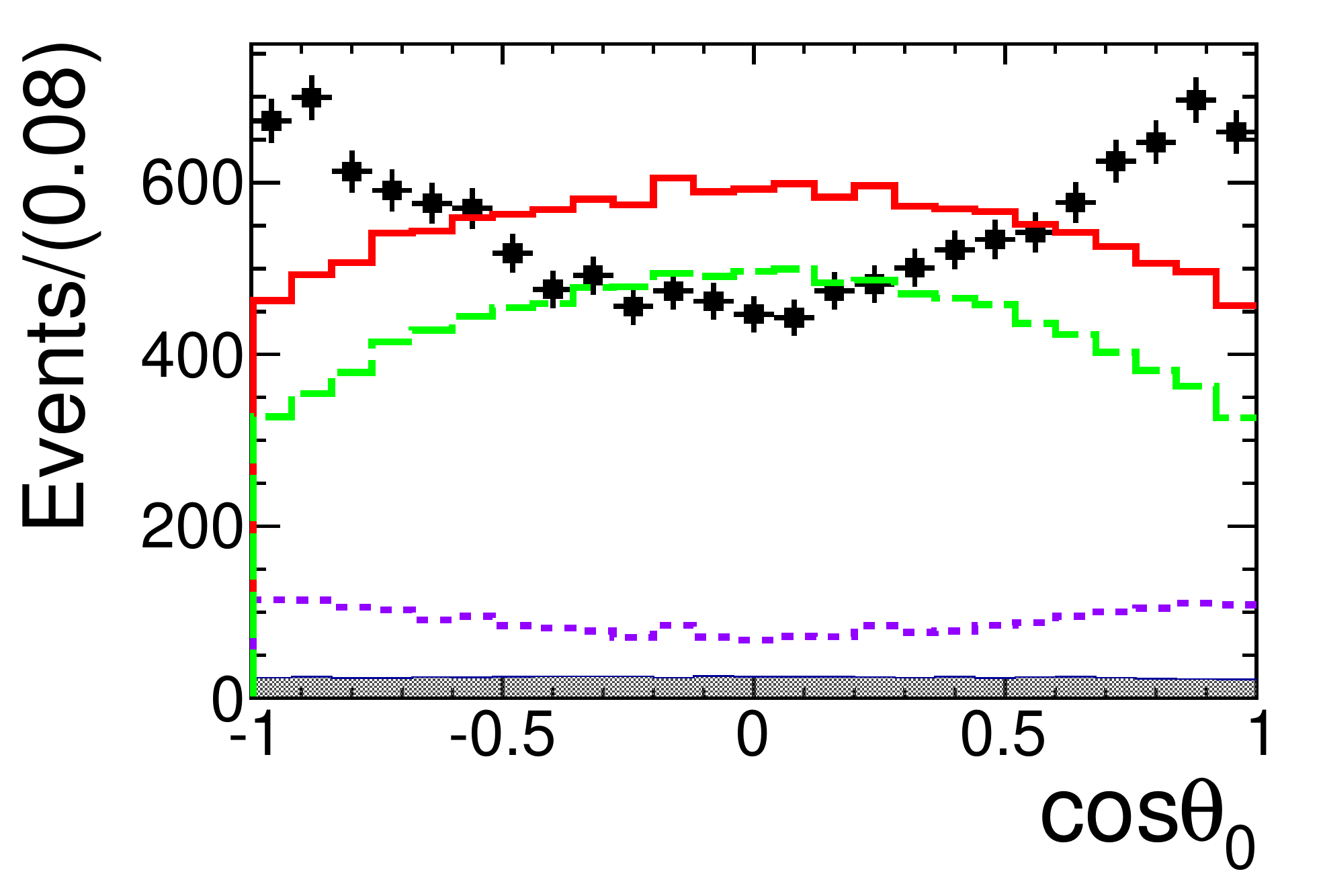}
\includegraphics[width=0.3\columnwidth]{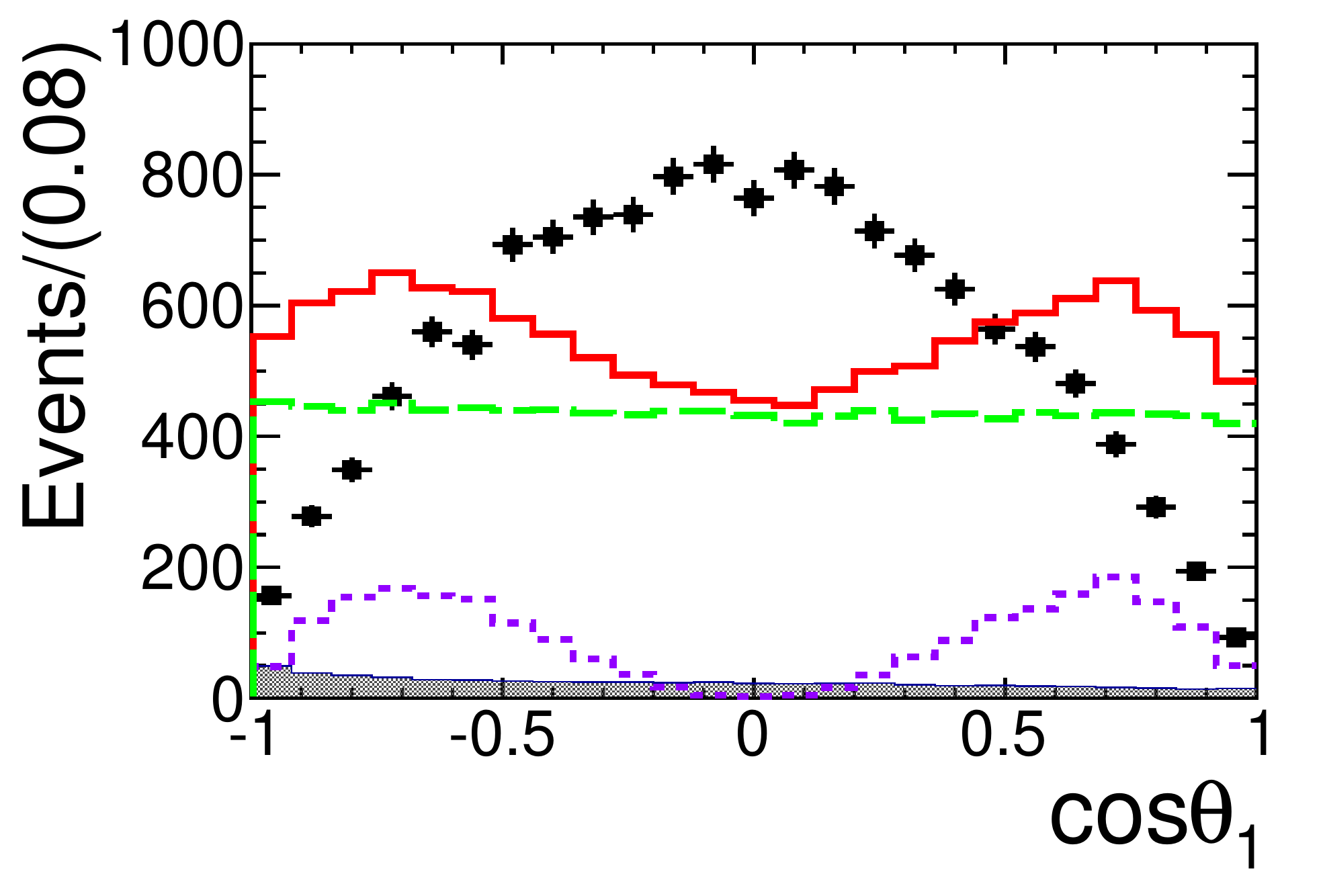}
\includegraphics[width=0.3\columnwidth]{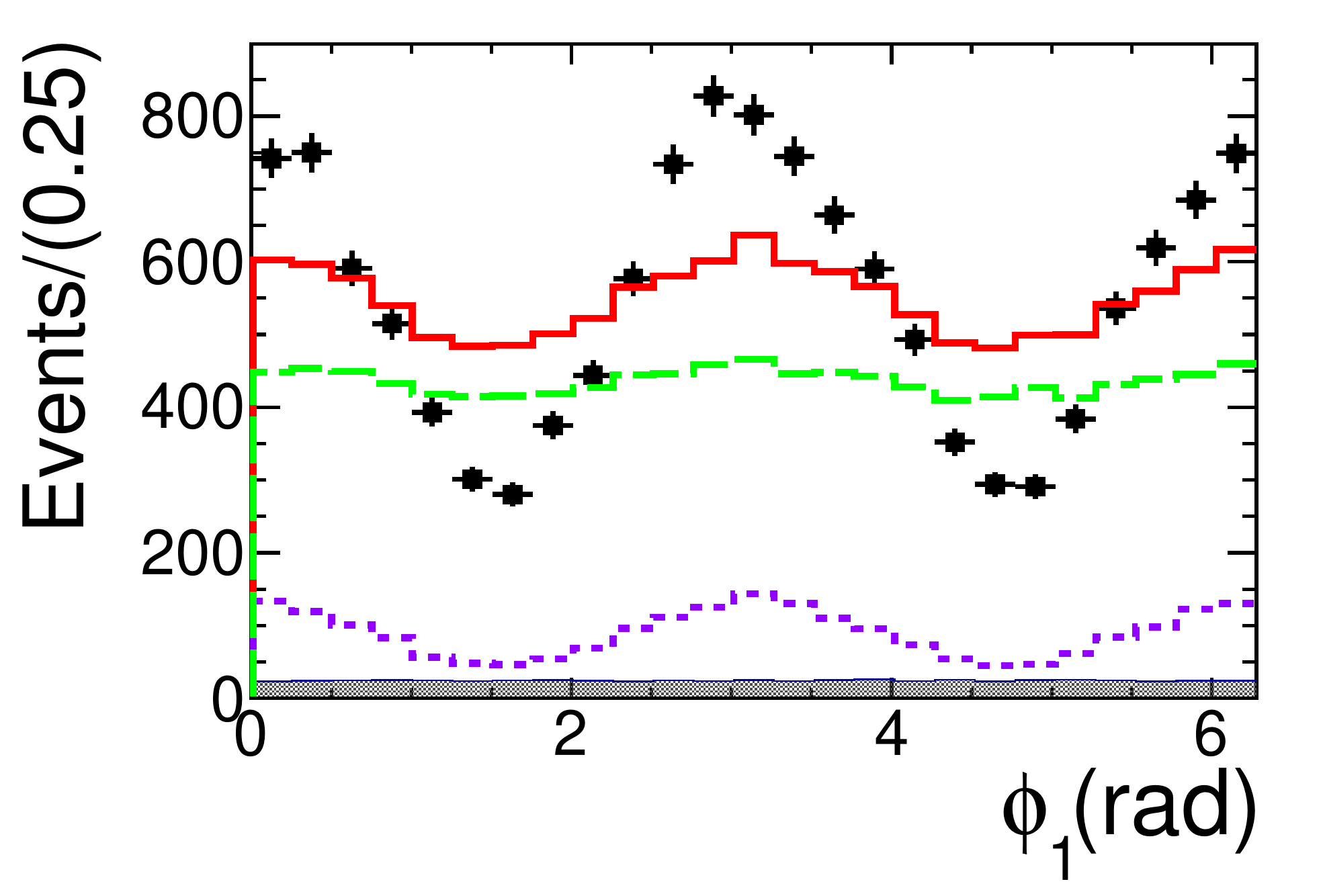}
\caption{
Mass and angular distributions of $D^{*+}$-tag sample for $J^P=2^+$ hypothesis. 
}
\label{fig:angular_Dppi0_2}

\end{figure}

\begin{figure}[!hbt]
\centering
\includegraphics[width=0.3\columnwidth]{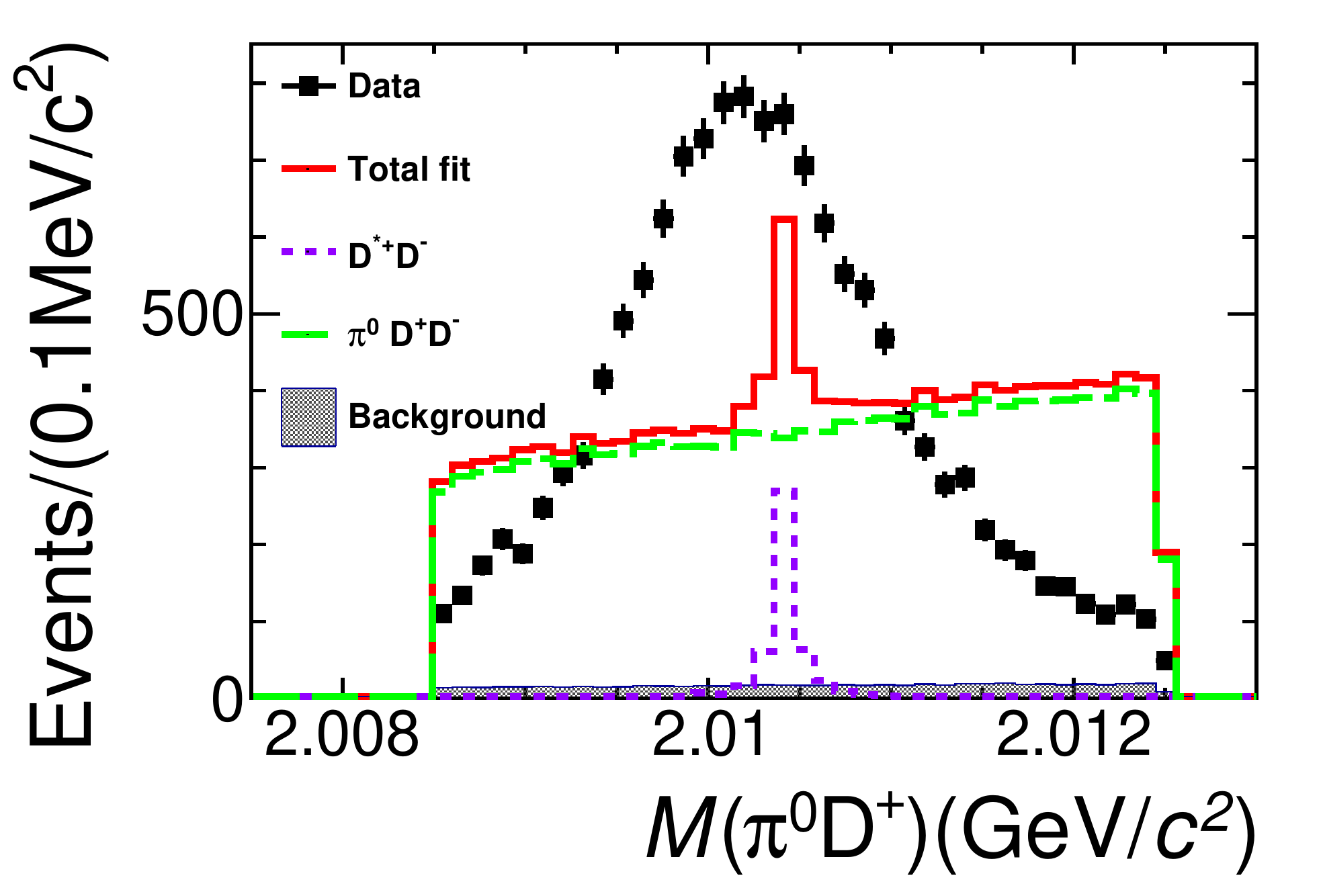}
\includegraphics[width=0.3\columnwidth]{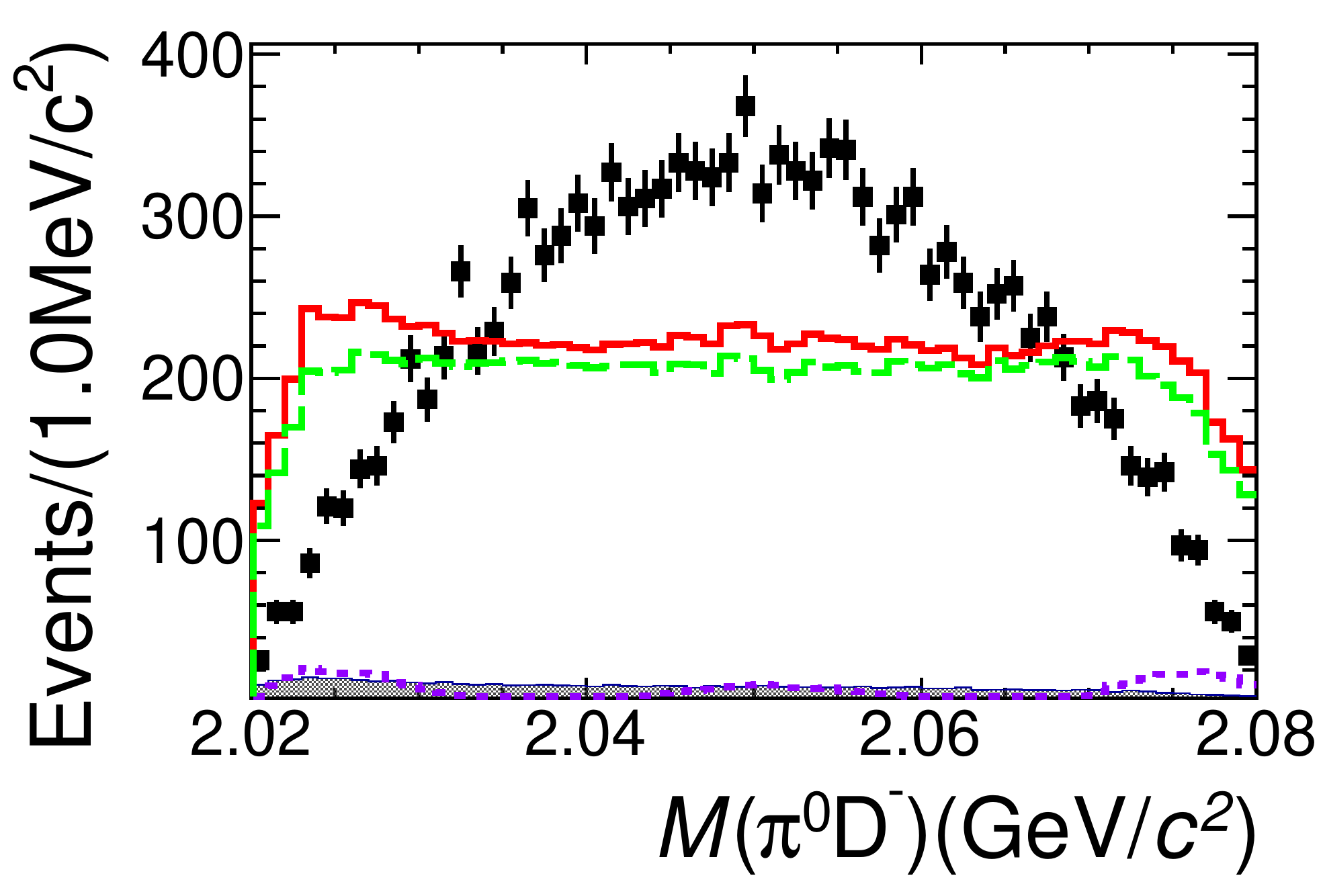}
\includegraphics[width=0.3\columnwidth]{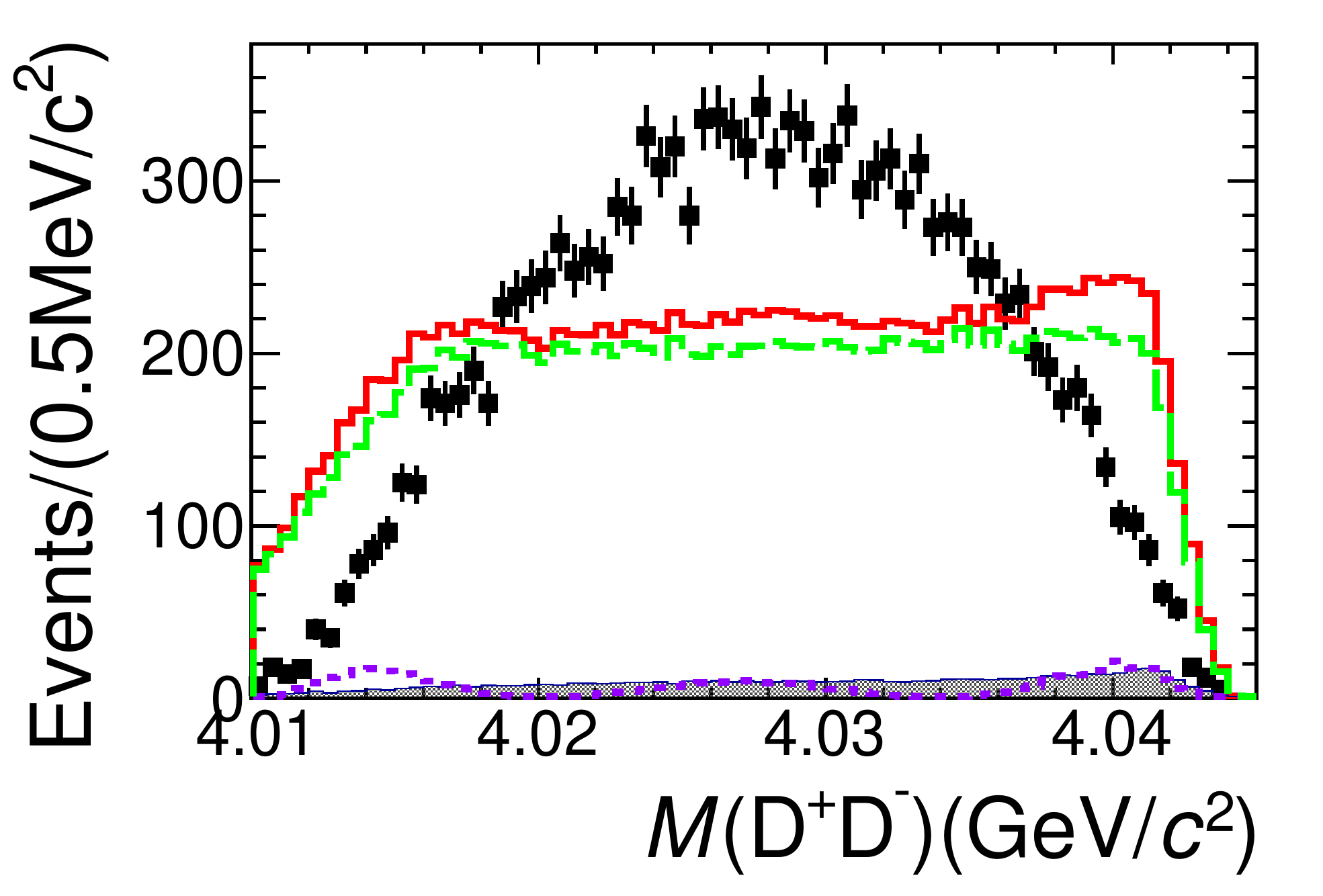}
\includegraphics[width=0.3\columnwidth]{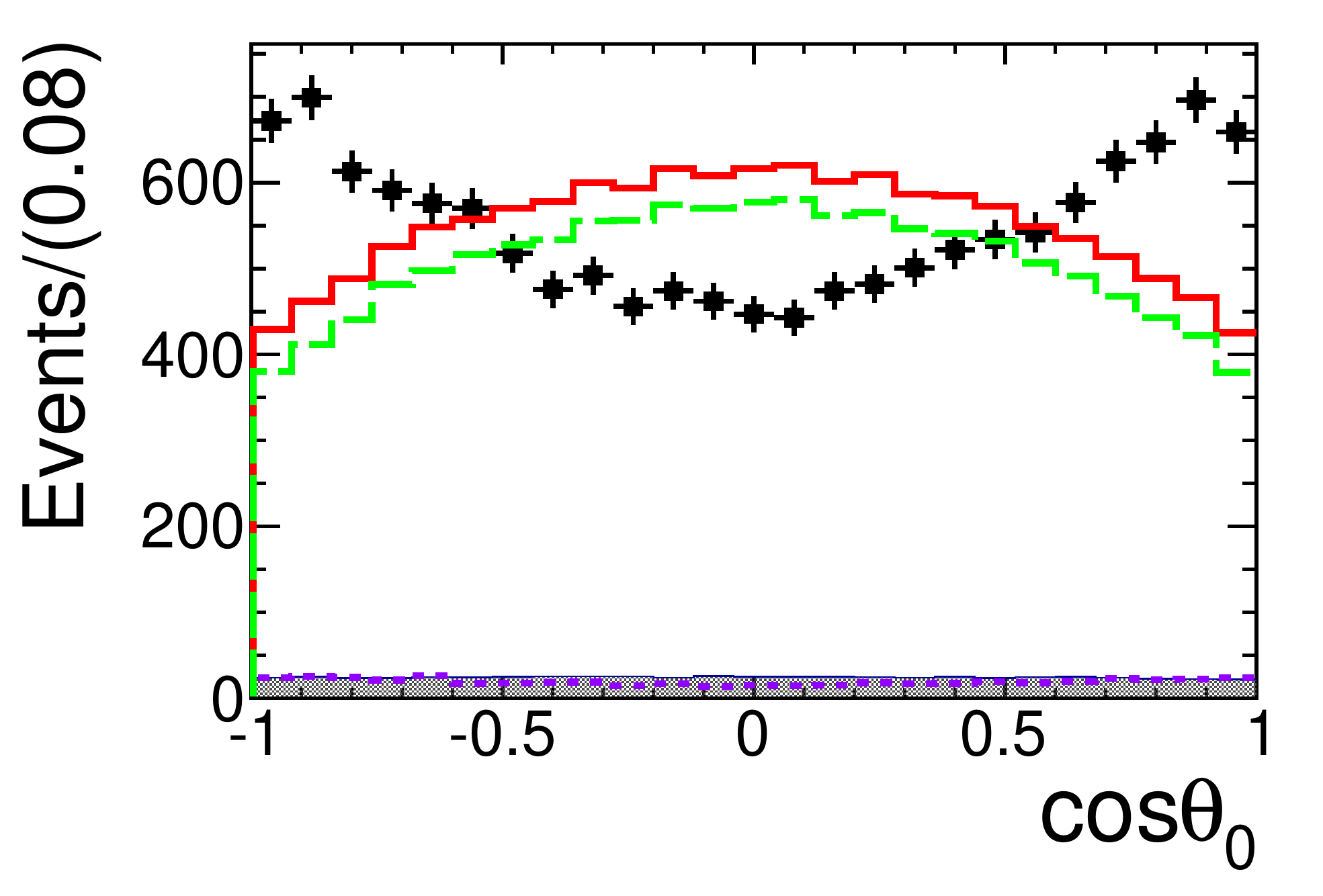}
\includegraphics[width=0.3\columnwidth]{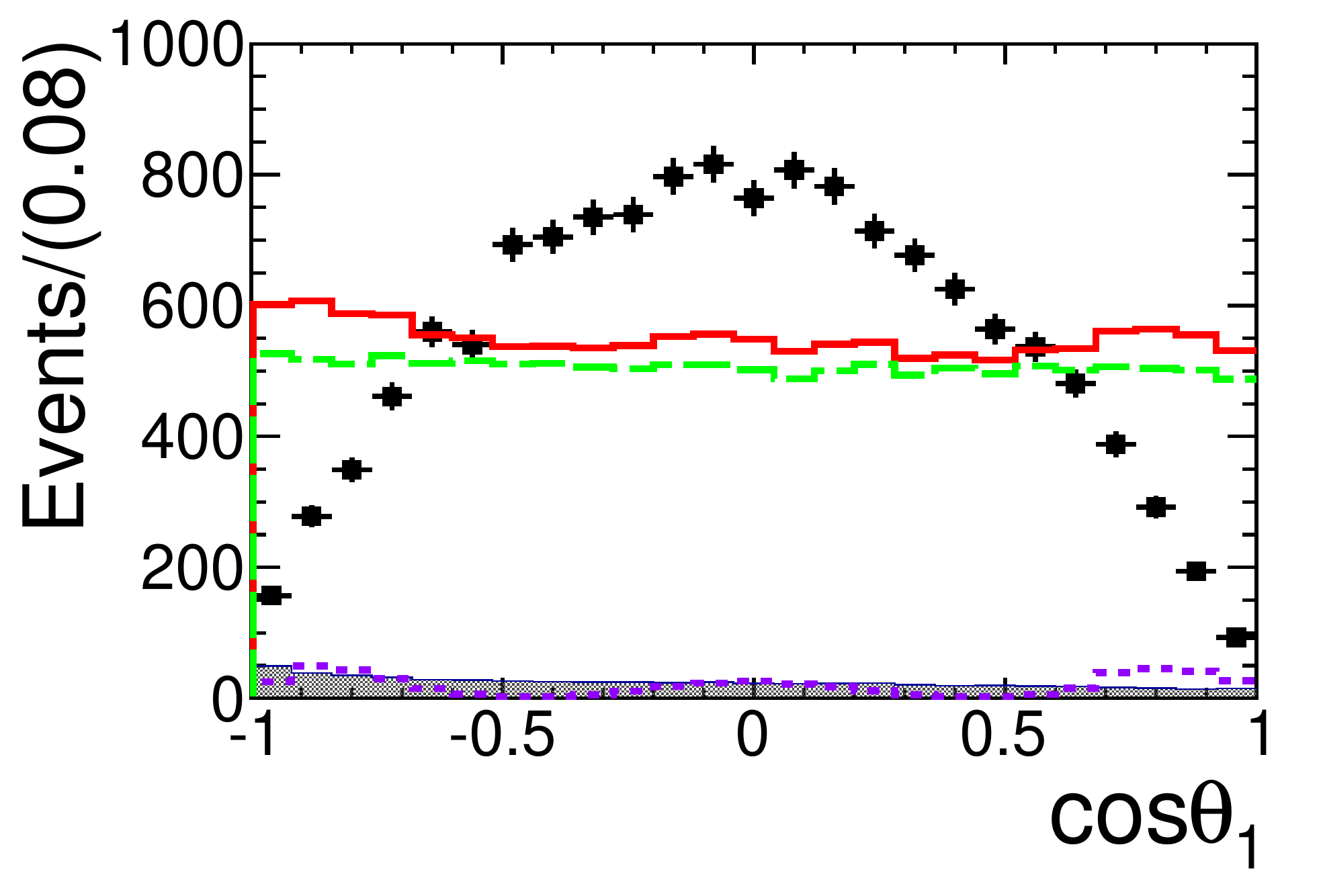}
\includegraphics[width=0.3\columnwidth]{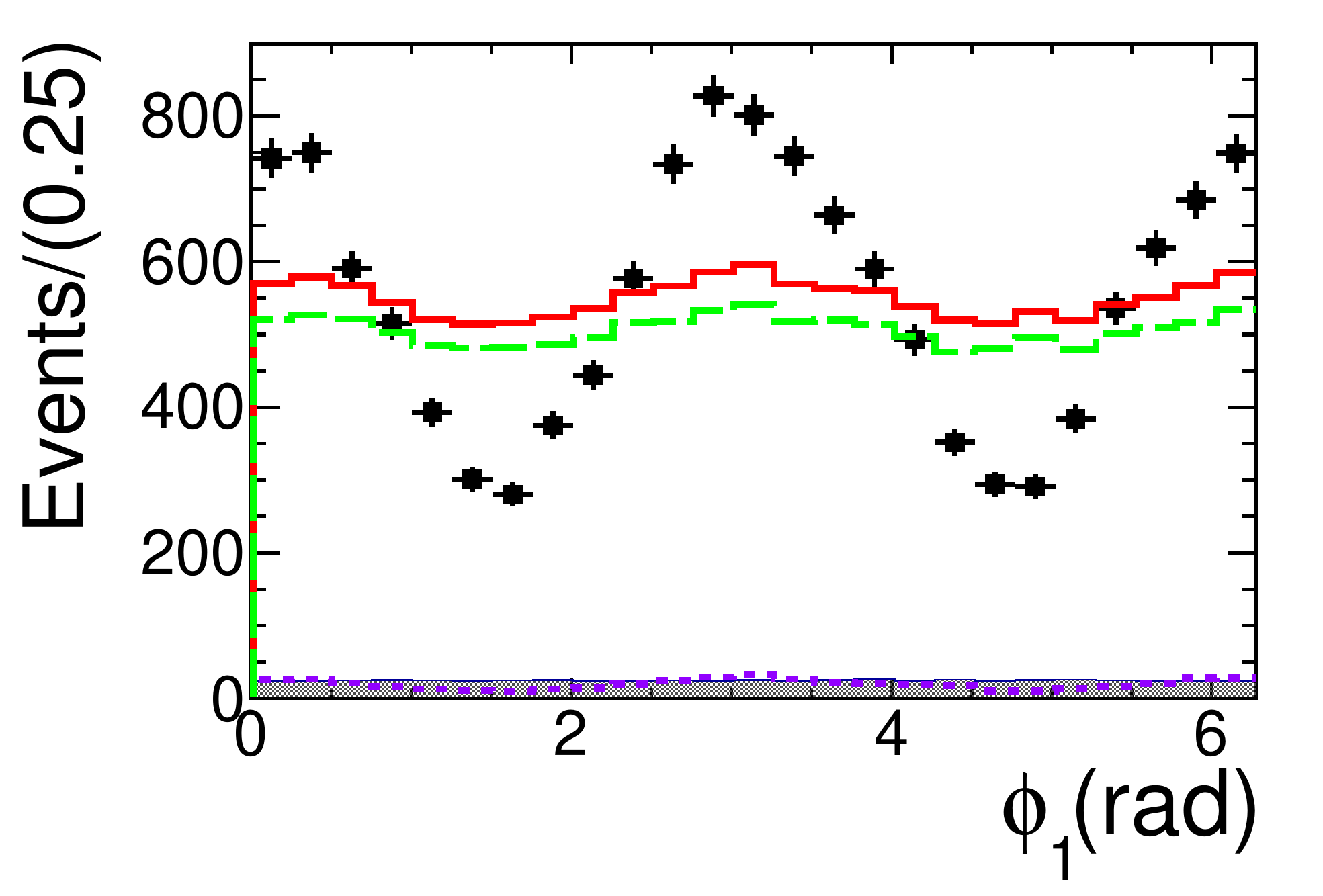}
\caption{
Mass and angular distributions of $D^{*+}$-tag sample for $J^P=3^-$ hypothesis.
}
\label{fig:angular_Dppi0_3}
\end{figure}

\begin{figure}[!hbt]
\centering
\includegraphics[width=0.3\columnwidth]{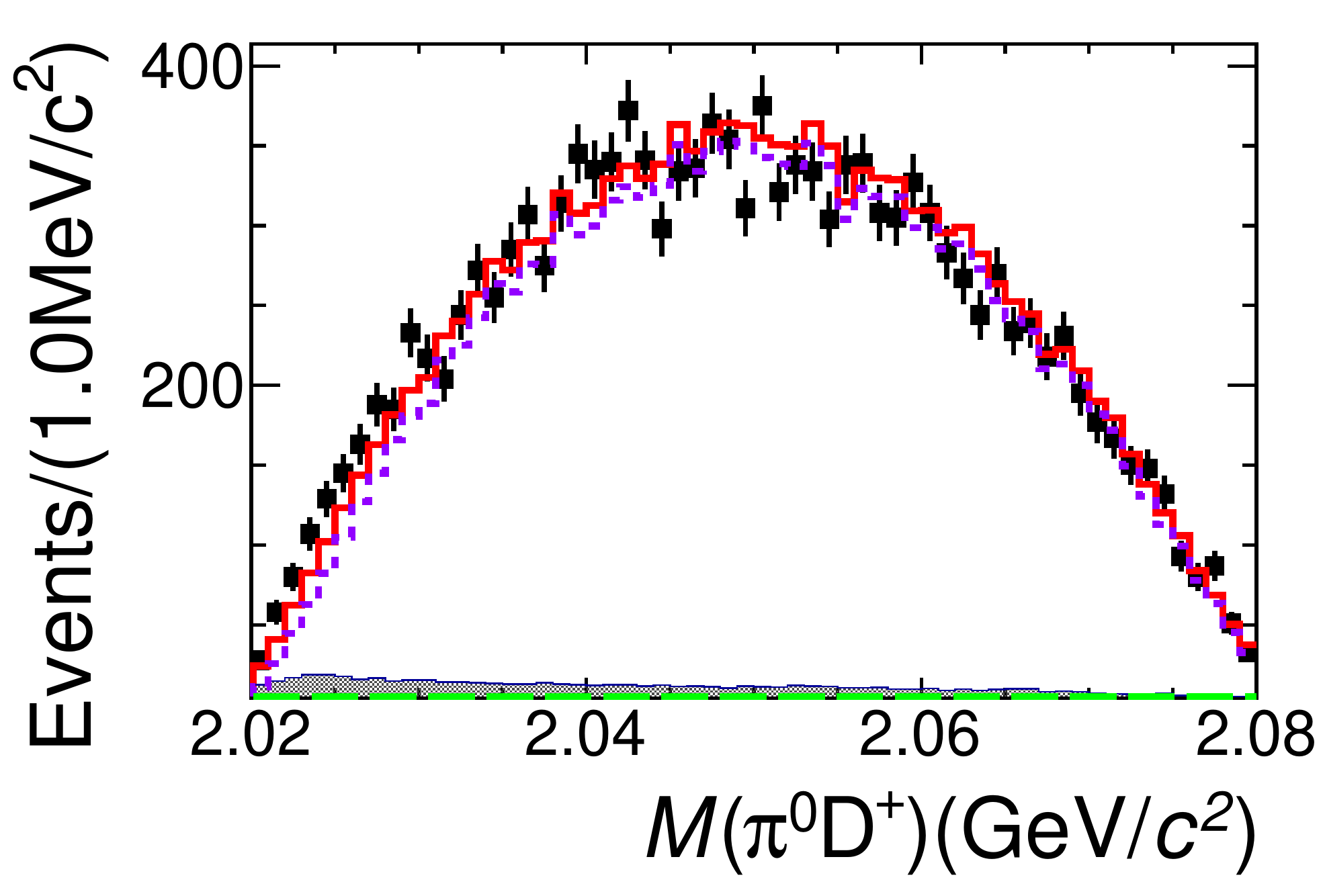}
\includegraphics[width=0.3\columnwidth]{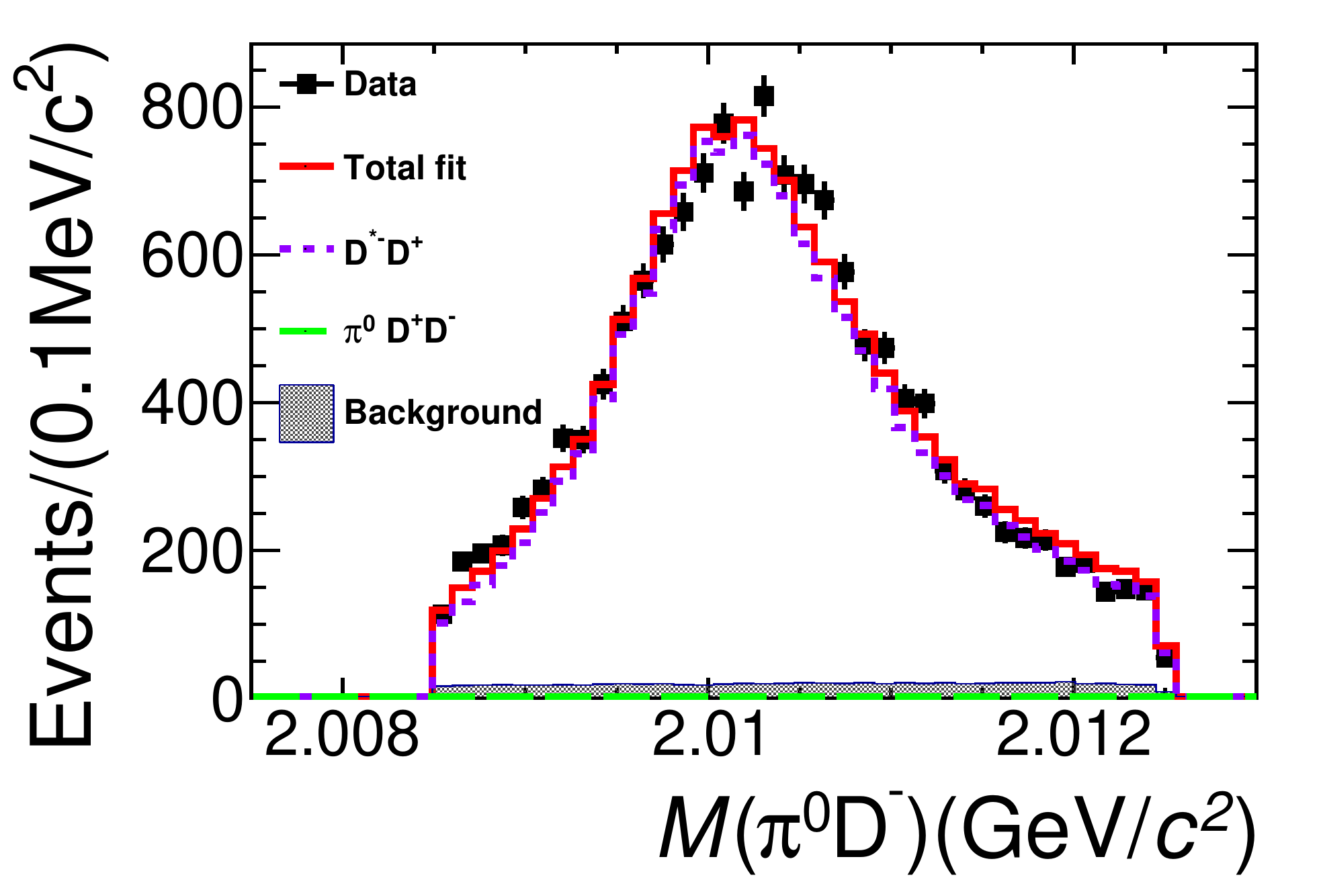}
\includegraphics[width=0.3\columnwidth]{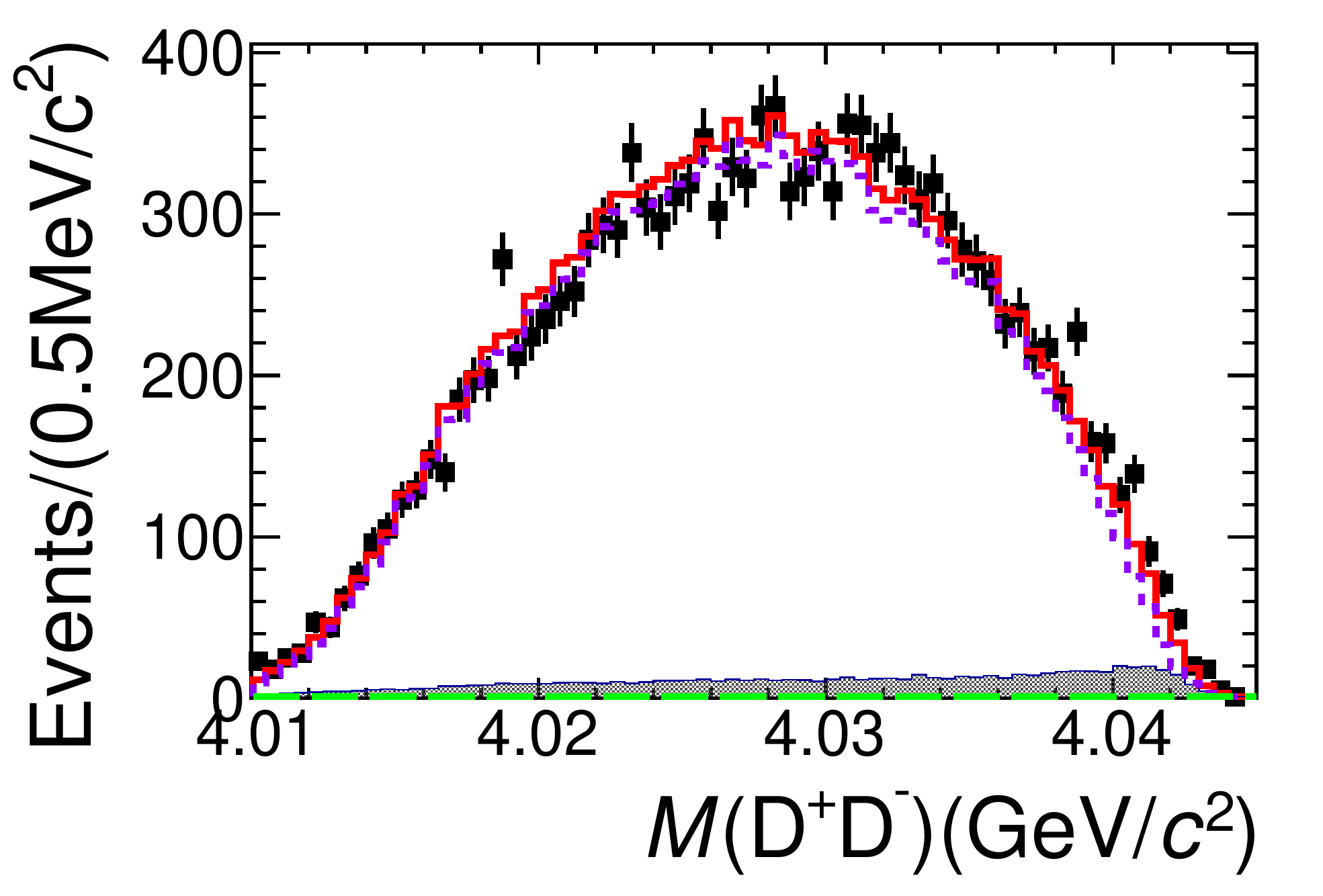}
\includegraphics[width=0.3\columnwidth]{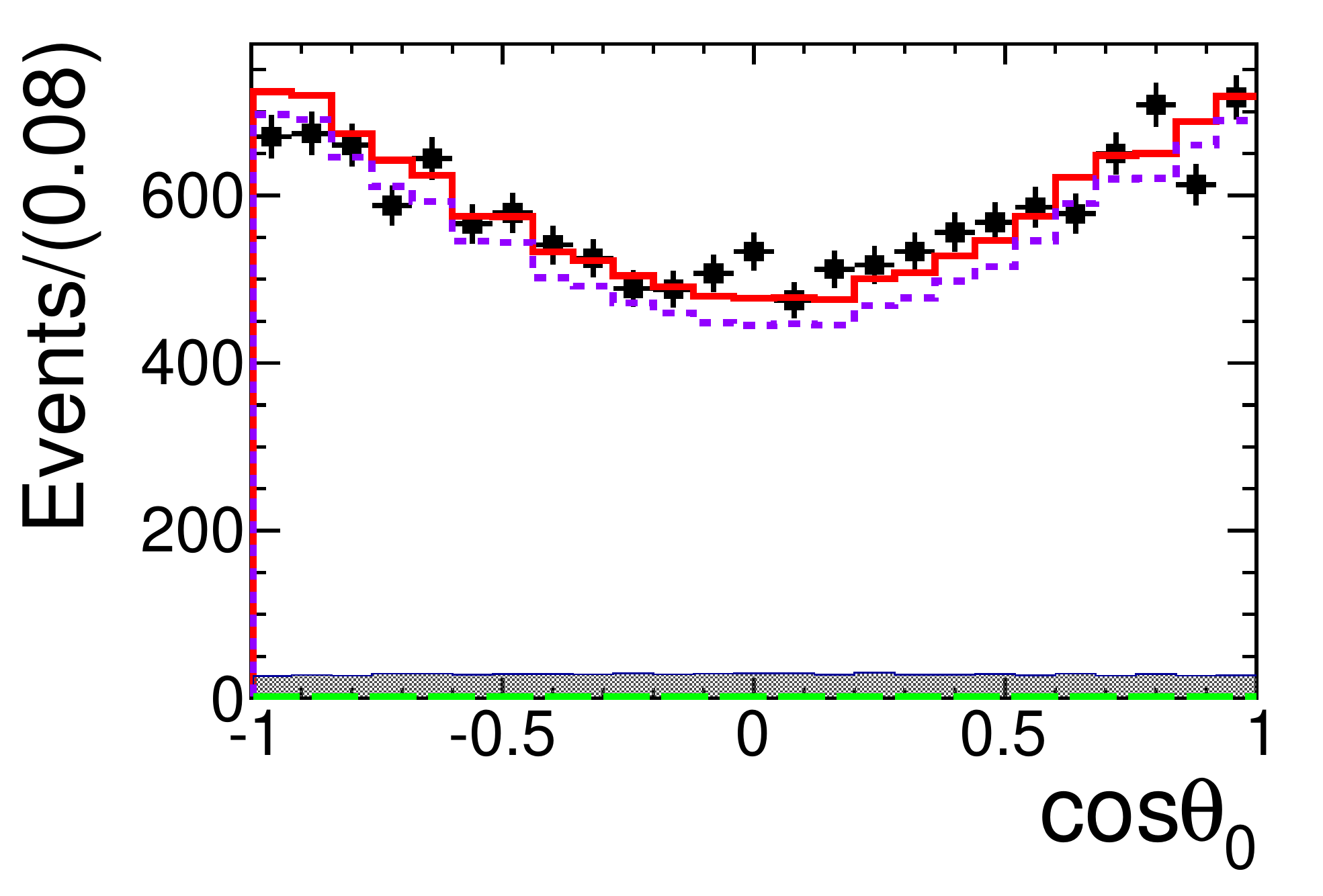}
\includegraphics[width=0.3\columnwidth]{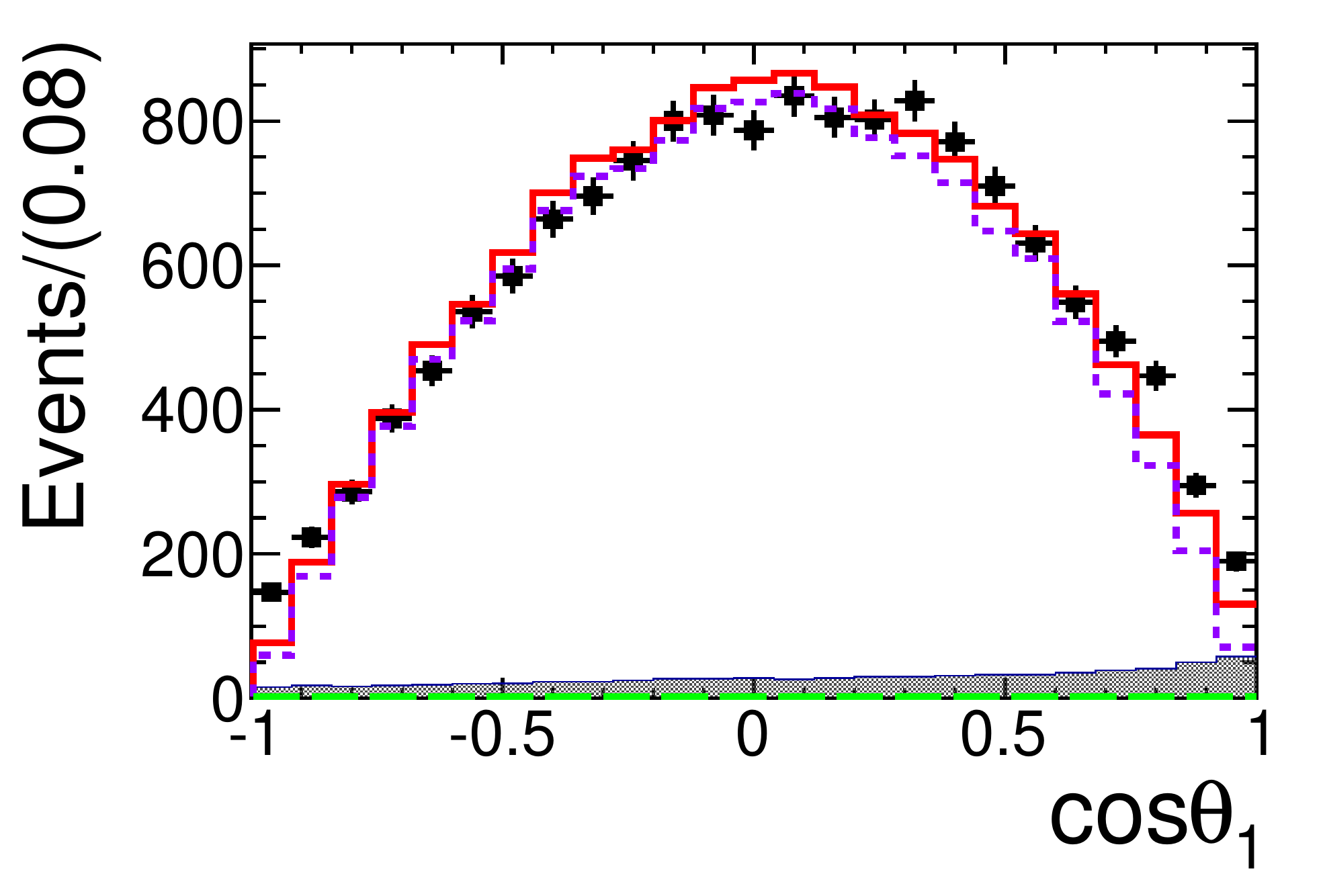}
\includegraphics[width=0.3\columnwidth]{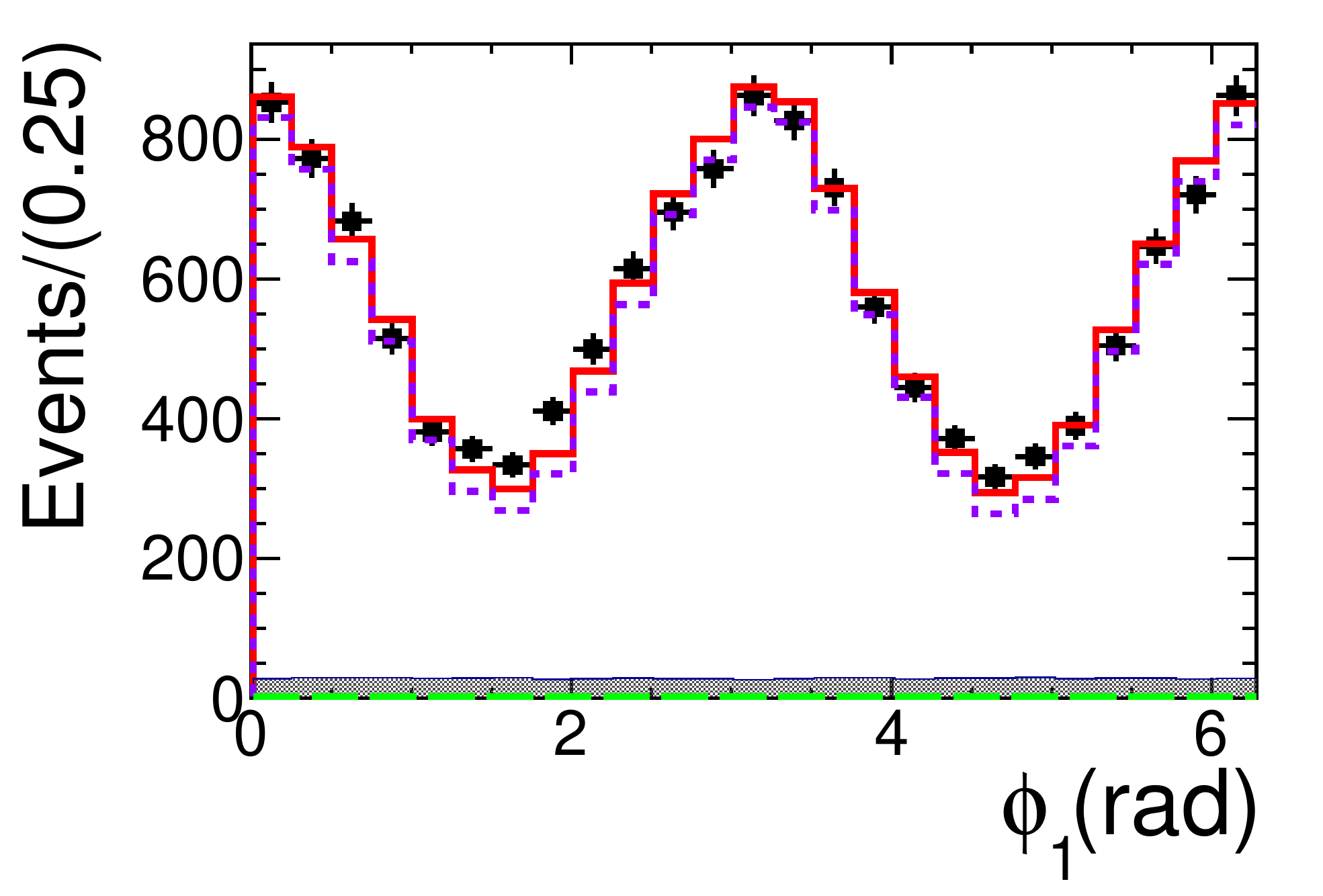}

\caption{
Mass and angular distributions of $D^+$-recoil sample for$J^P=1^-$ hypothesis.
}
\label{fig:angular_rmDp_1}
\end{figure}

\begin{figure}[!hbt]
\centering
\includegraphics[width=0.3\columnwidth]{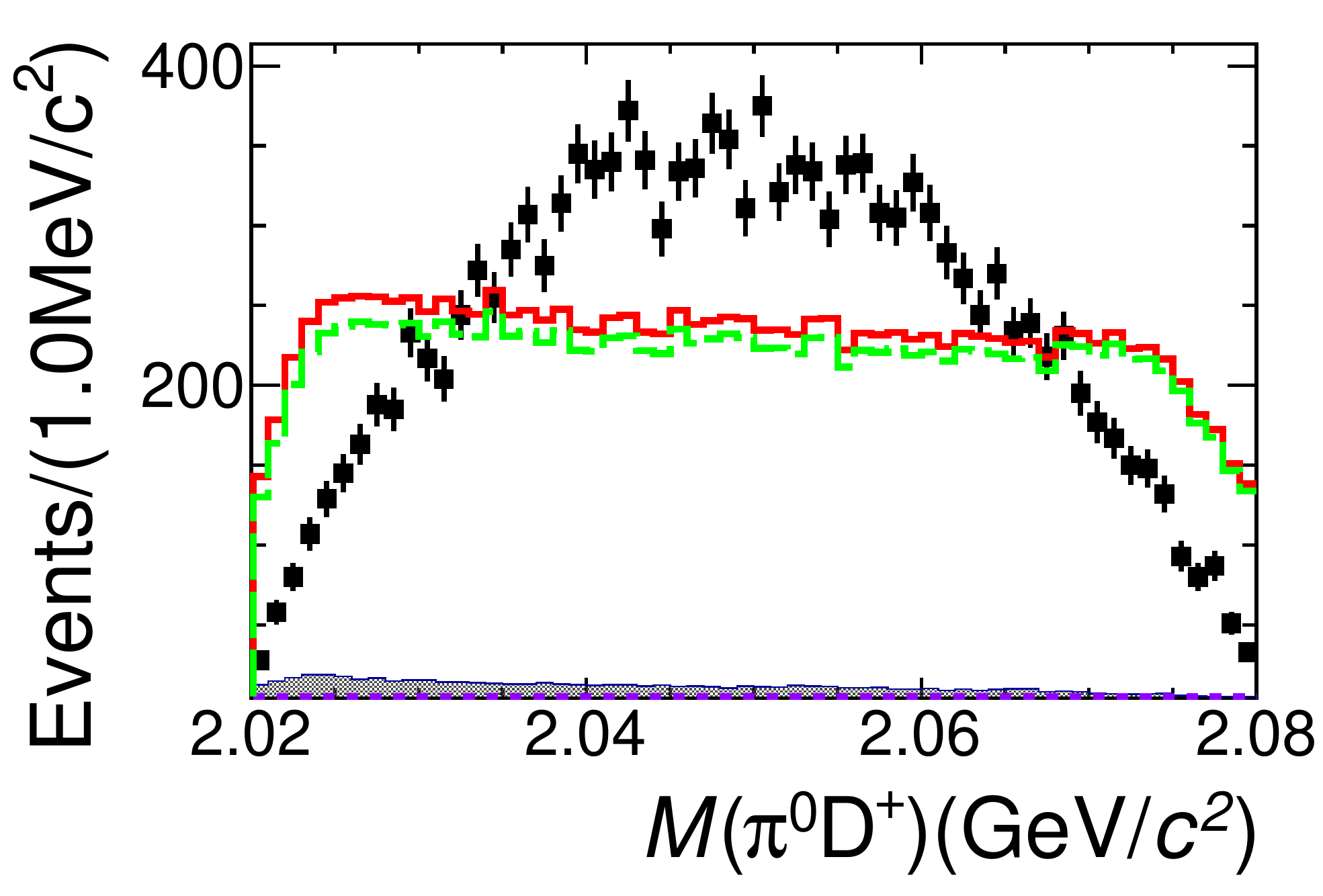}
\includegraphics[width=0.3\columnwidth]{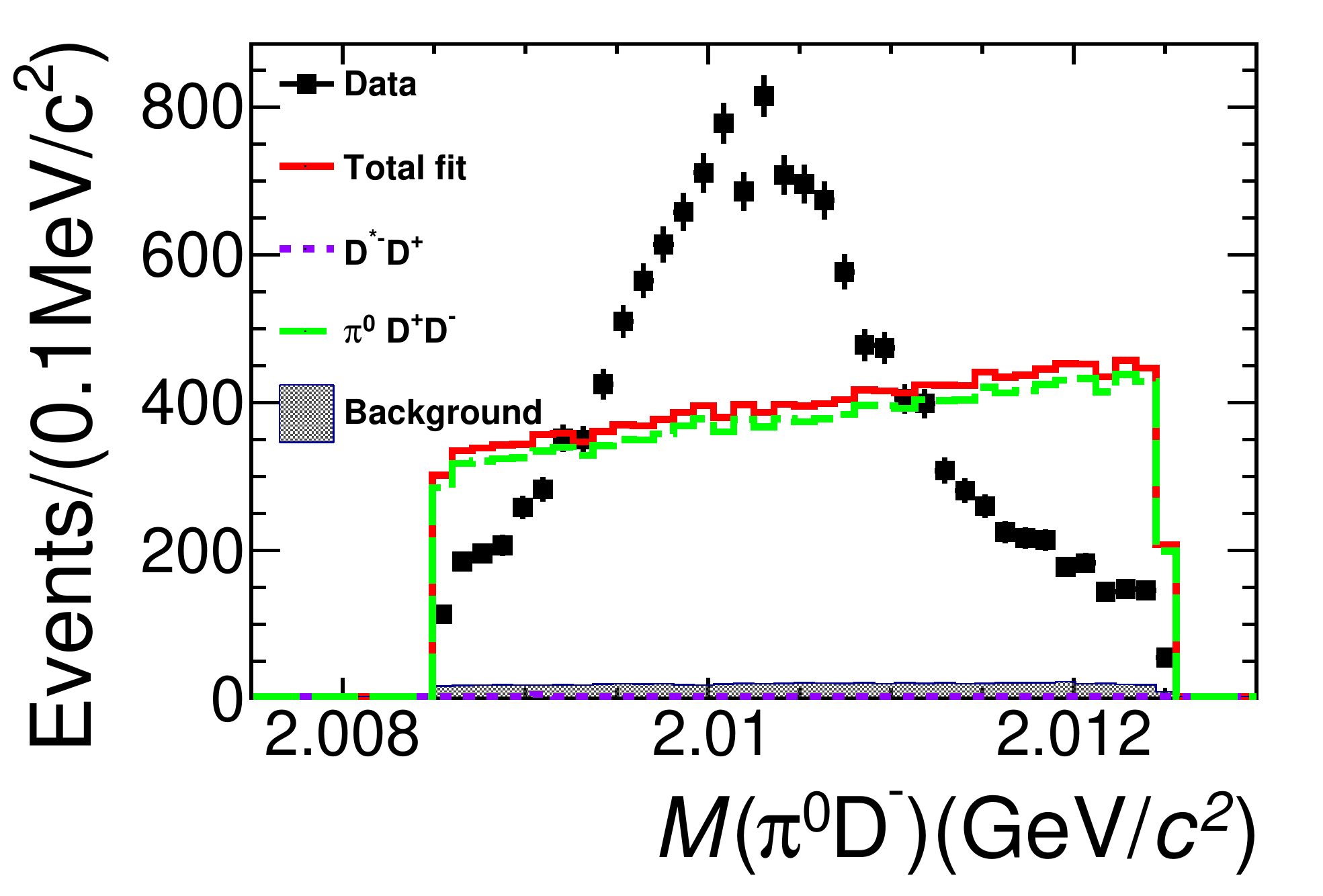}
\includegraphics[width=0.3\columnwidth]{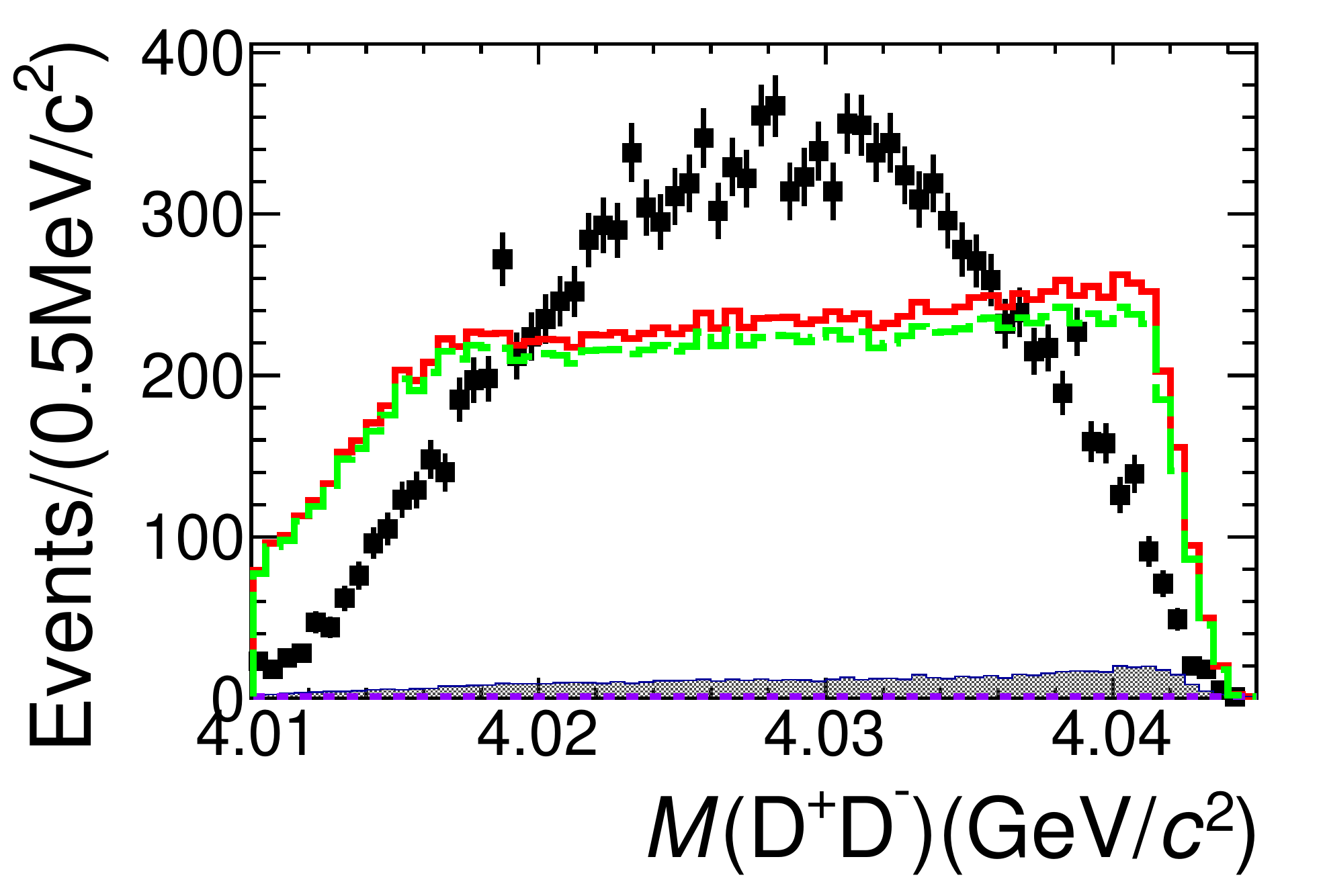}
\includegraphics[width=0.3\columnwidth]{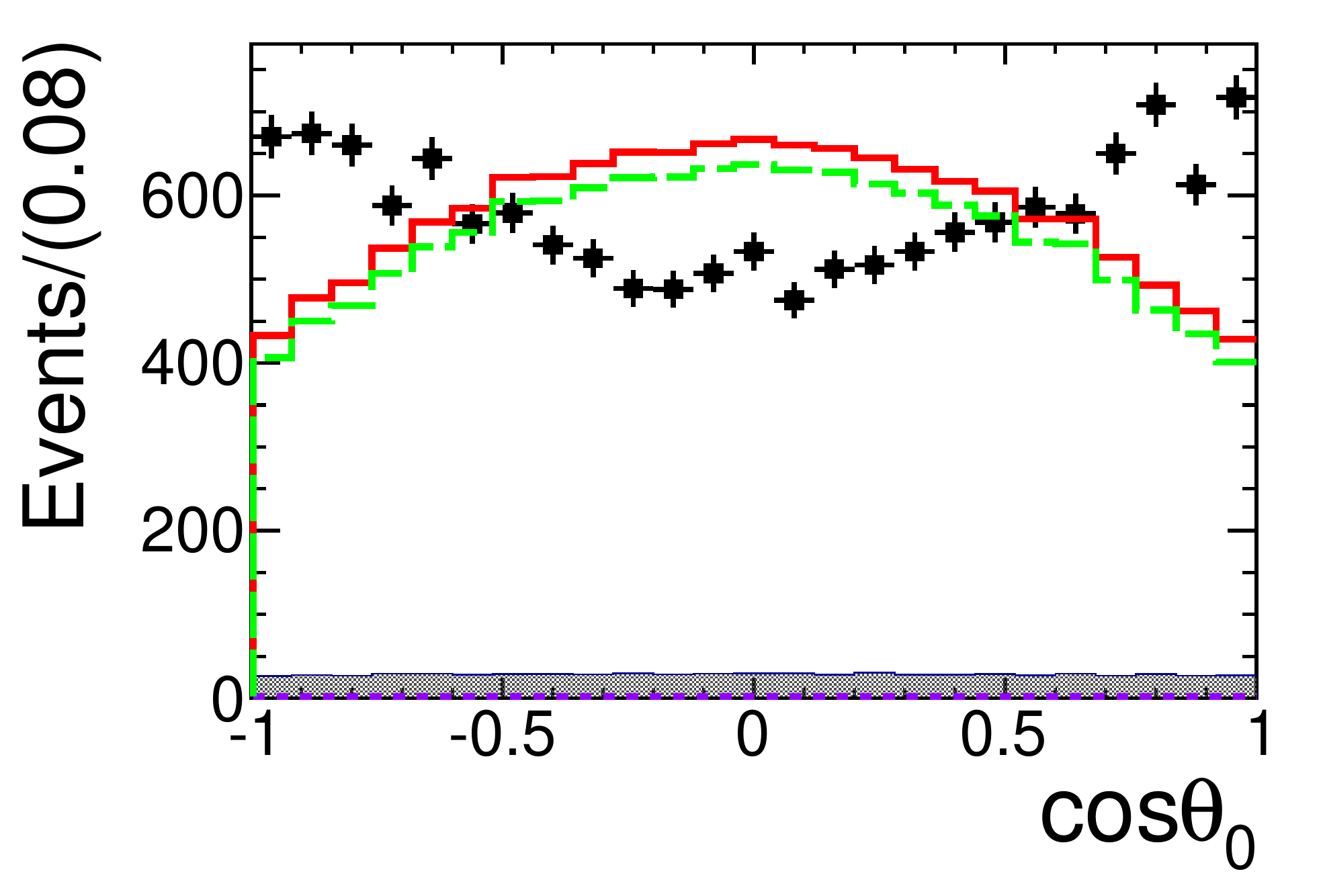}
\includegraphics[width=0.3\columnwidth]{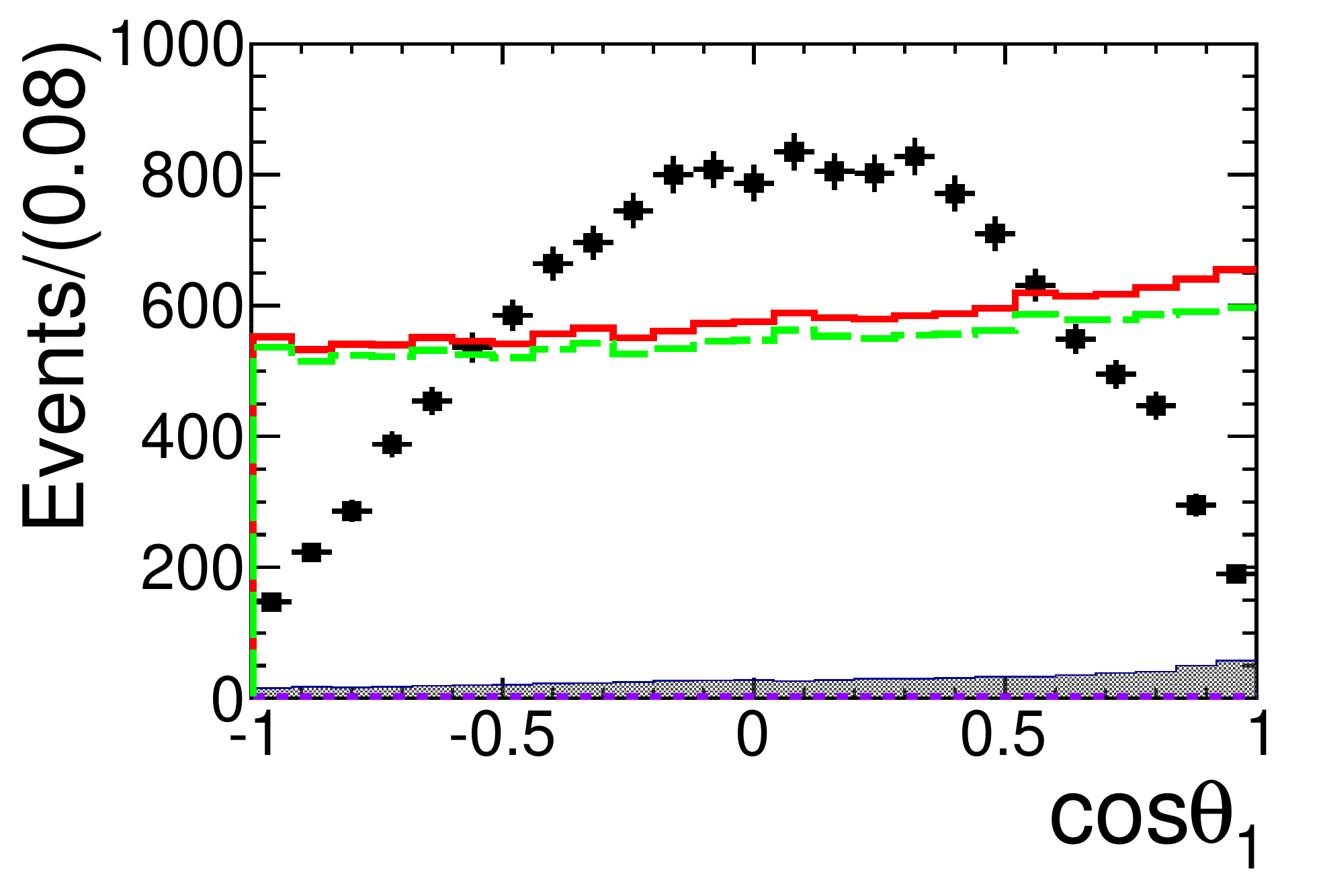}
\includegraphics[width=0.3\columnwidth]{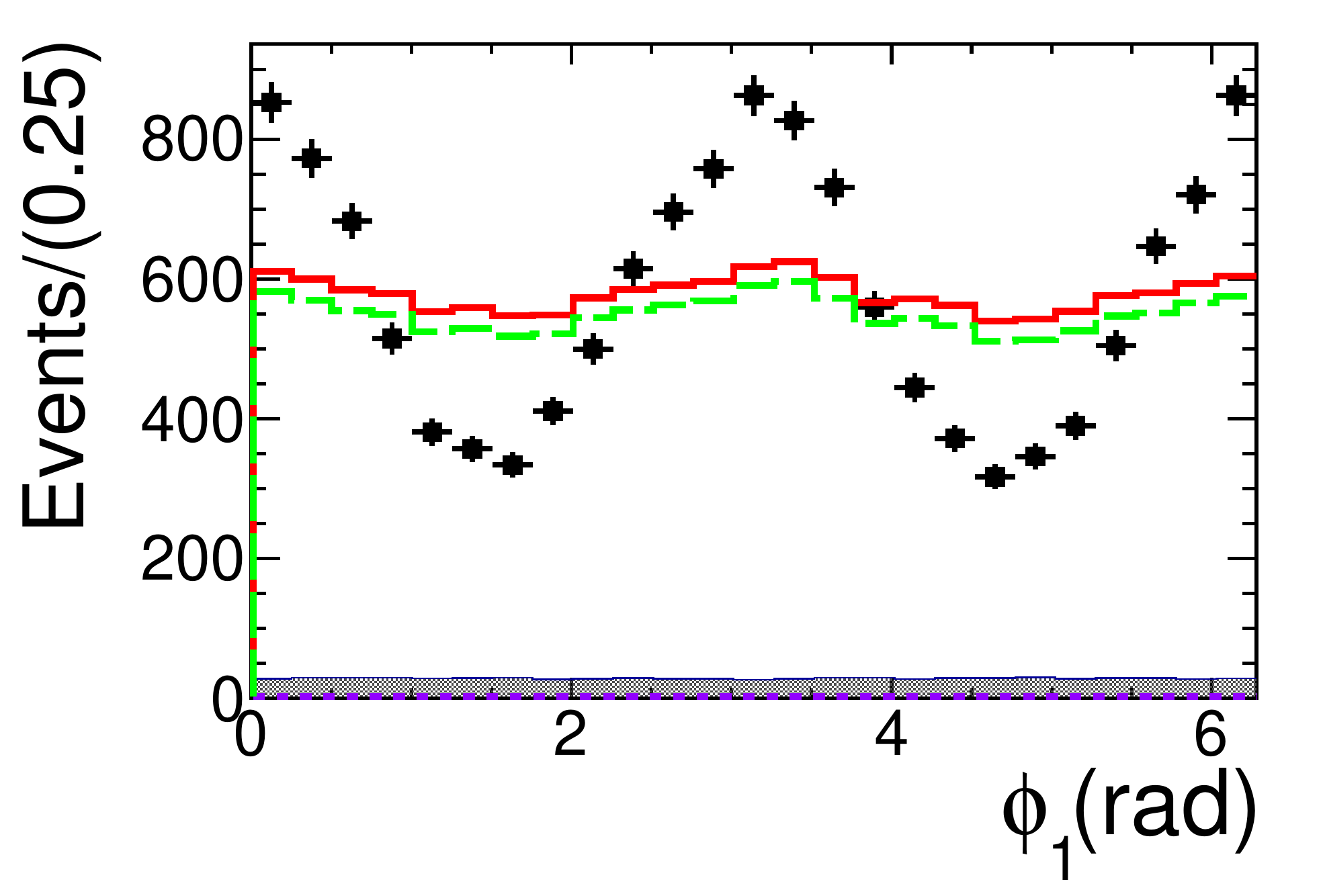}
\caption{
Mass and angular distributions of $D^+$-recoil sample for $J^P=2^+$ hypothesis.
}
\label{fig:angular_rmDp_2}
\end{figure}

\begin{figure}[!hbt]
\centering
\includegraphics[width=0.3\columnwidth]{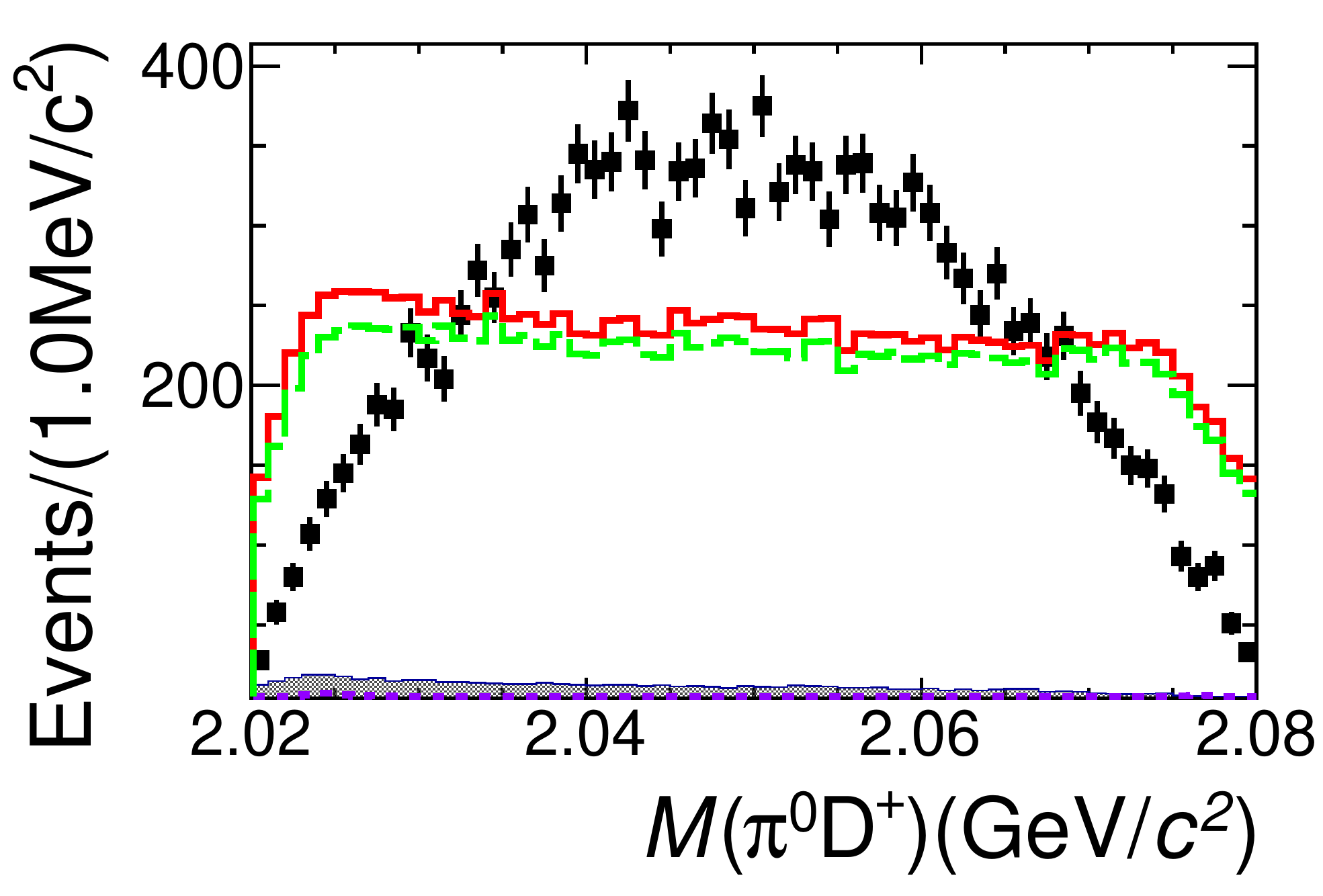}
\includegraphics[width=0.3\columnwidth]{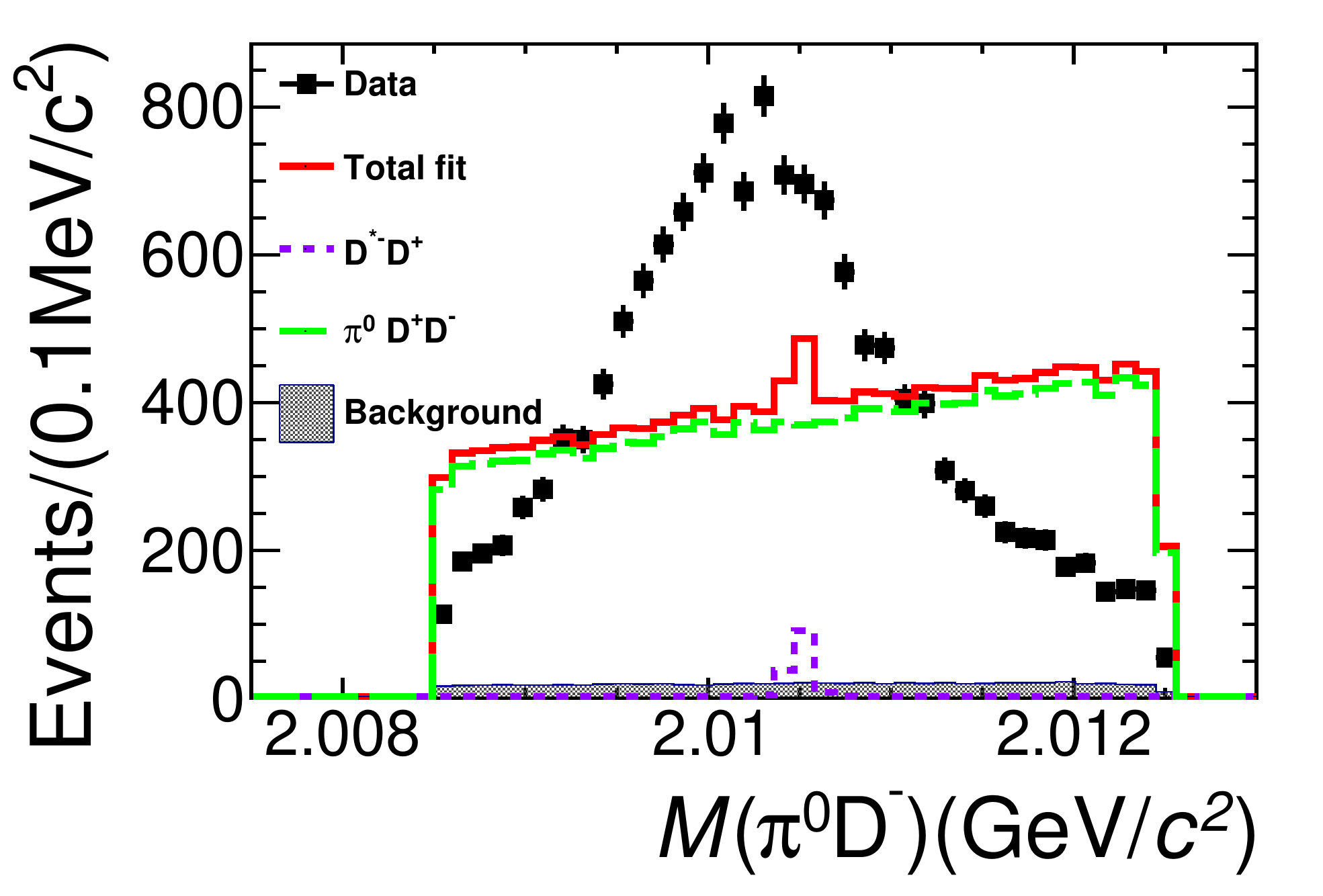}
\includegraphics[width=0.3\columnwidth]{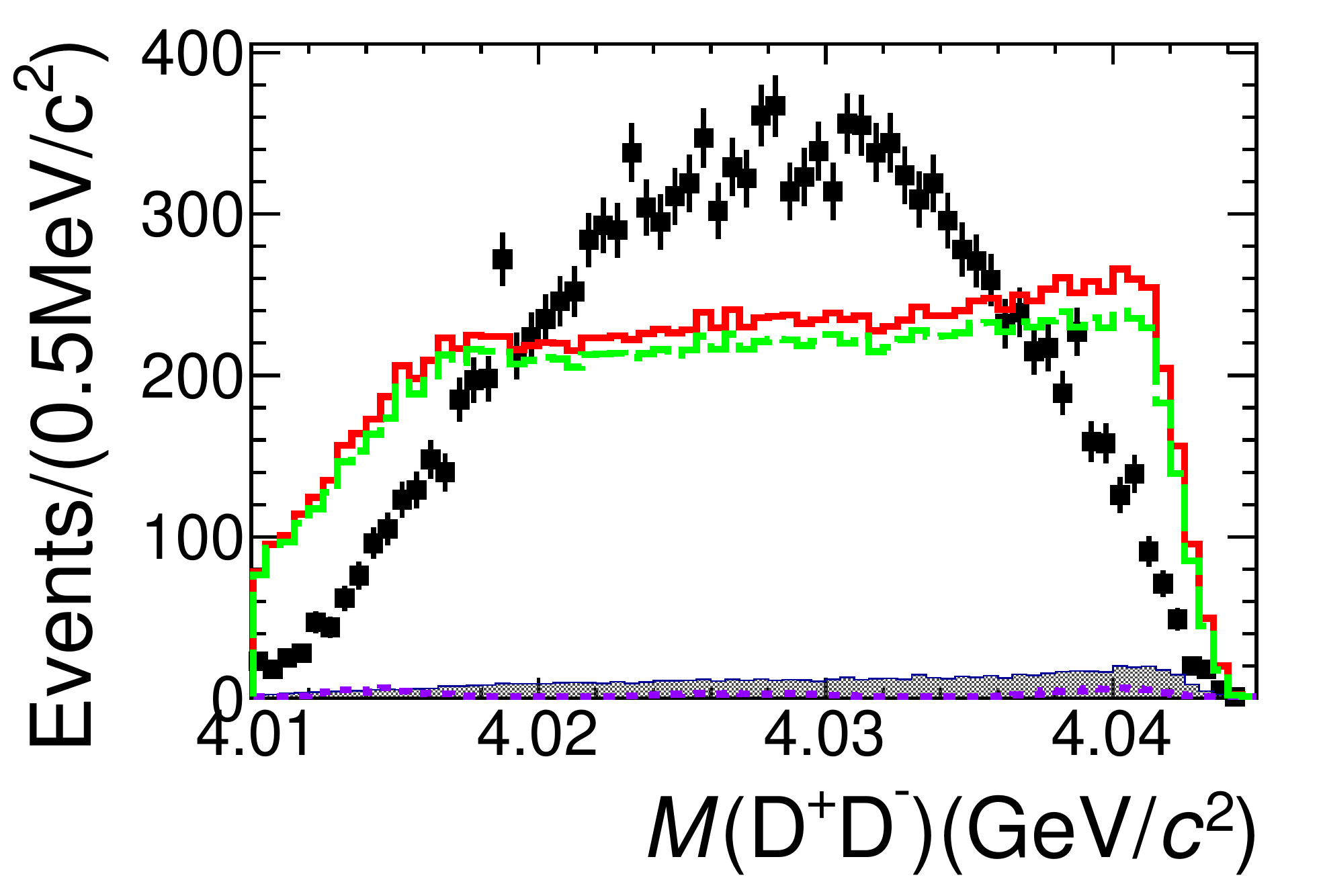}
\includegraphics[width=0.3\columnwidth]{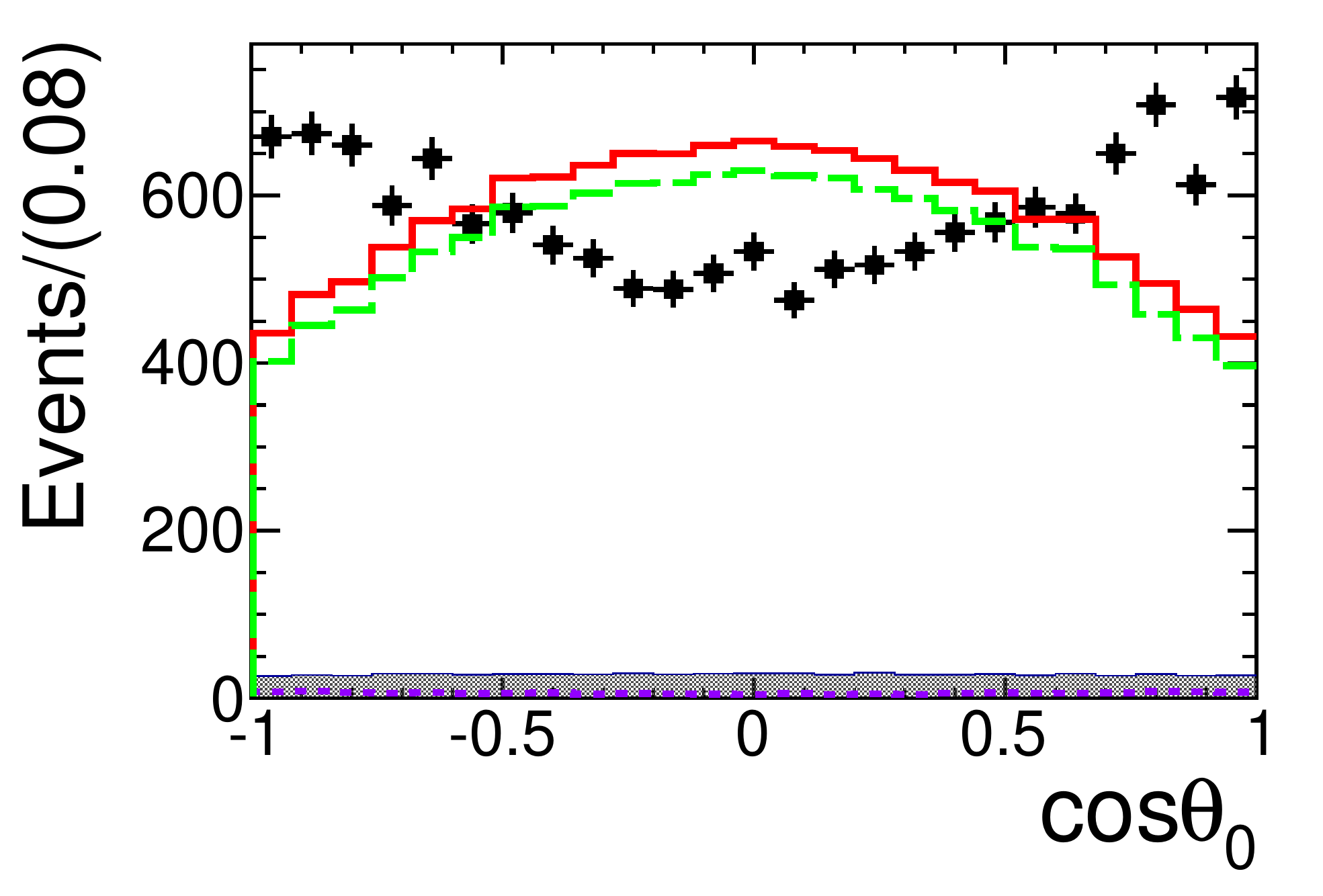}
\includegraphics[width=0.3\columnwidth]{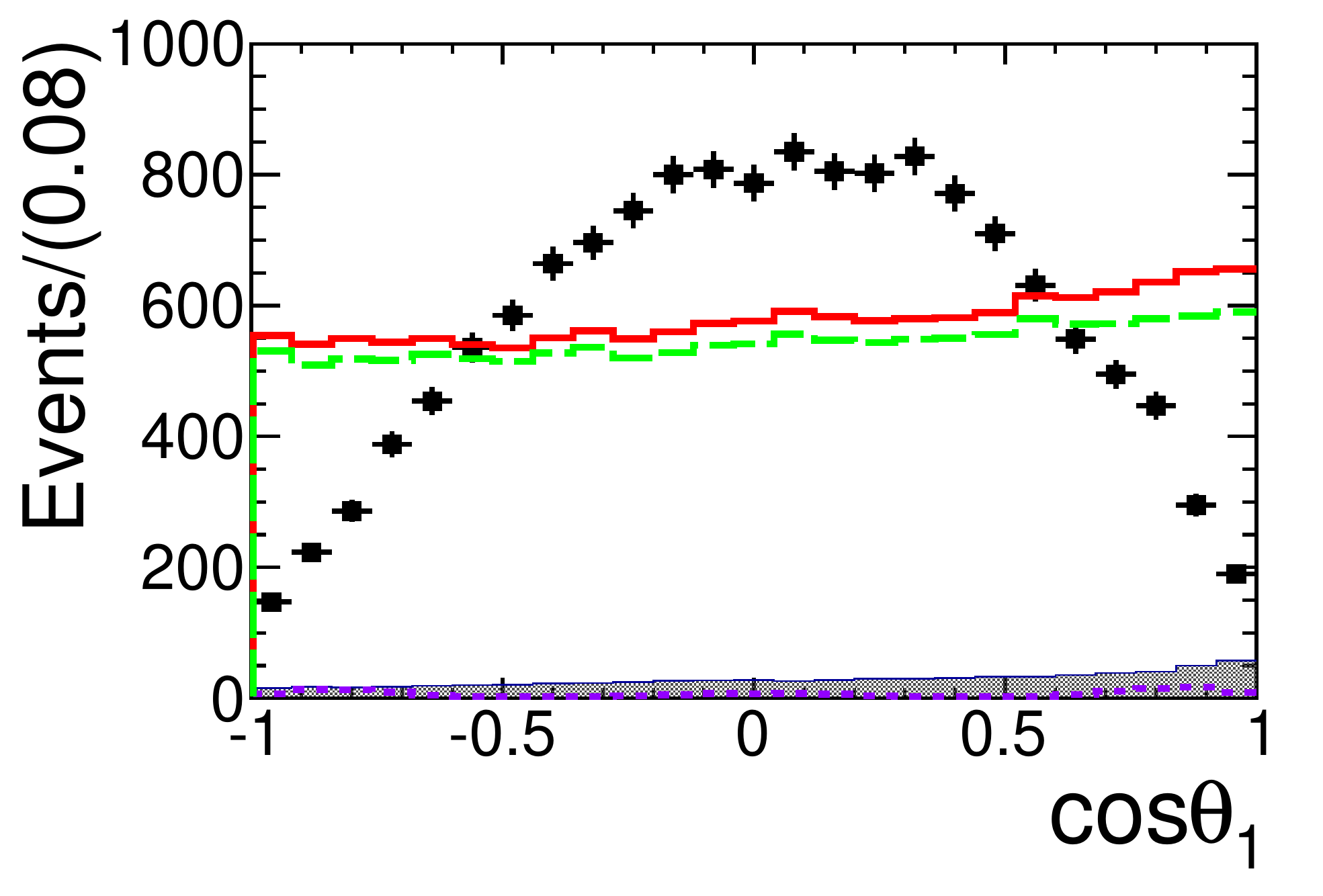}
\includegraphics[width=0.3\columnwidth]{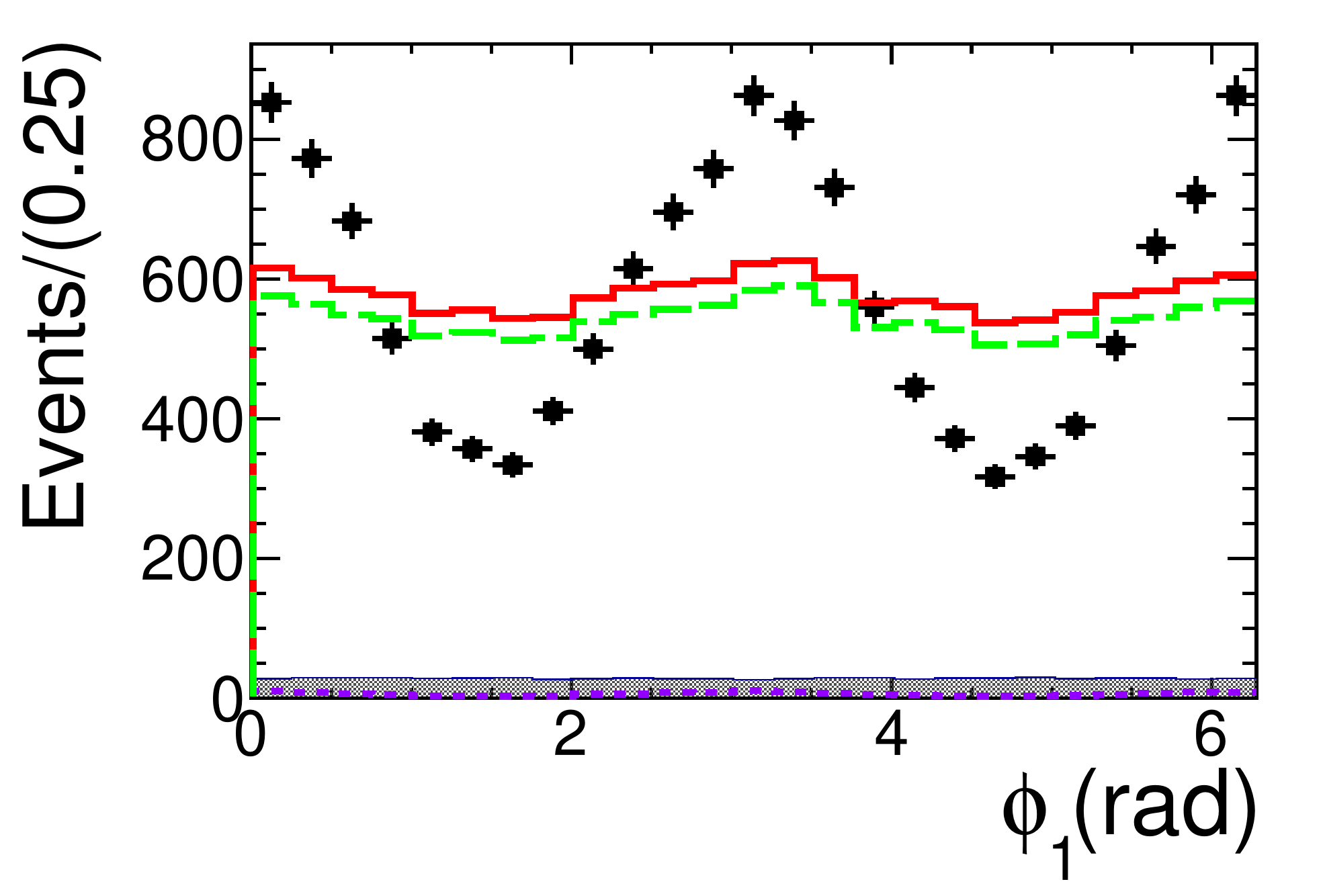}
\caption{
Mass and angular distributions of $D^+$-recoil sample for $J^P=3^-$ hypothesis.
}
\label{fig:angular_rmDp_3}
\end{figure}

Figures~\ref{fig:angular_Dsp_2},~\ref{fig:angular_Dsp_3},~\ref{fig:angular_Dsm_1},~\ref{fig:angular_Dsm_2},~\ref{fig:angular_Dsm_3} show the mass and  angular distributions of the fitting results in $D^{*+}_s$ tag sample and $D^{+}_s$ recoil sample under different hypotheses.
Figures~\ref{fig:angular_D0pi0_1},~\ref{fig:angular_D0pi0_2},~\ref{fig:angular_D0pi0_3},~\ref{fig:angular_rmD0_1},~\ref{fig:angular_rmD0_2},~\ref{fig:angular_rmD0_3} correspond to the fitting results in $D^{*0}$ tag sample and $D^{0}$ recoil sample under different hypotheses.
Figures~\ref{fig:angular_Dppi0_1},~\ref{fig:angular_Dppi0_2},~\ref{fig:angular_Dppi0_3},~\ref{fig:angular_rmDp_1},~\ref{fig:angular_rmDp_2},~\ref{fig:angular_rmDp_3} are the fitting results in $D^{*+}$ tag sample and $D^{+}$ recoil sample under different hypotheses.

\bibliography{basename of .bib file}

%% file: authors.tex
\begin{small}
\begin{center}
M.~Ablikim$^{1}$, M.~N.~Achasov$^{13,b}$, P.~Adlarson$^{73}$, R.~Aliberti$^{34}$, A.~Amoroso$^{72A,72C}$, M.~R.~An$^{38}$, Q.~An$^{69,56}$, Y.~Bai$^{55}$, O.~Bakina$^{35}$, I.~Balossino$^{29A}$
% ...
\\
\vspace{0.2cm}
(BESIII Collaboration)\\
\vspace{0.2cm} {\it
$^{1}$ Institute of High Energy Physics, Beijing 100049, People's Republic of China\\
% ...
$^{13}$ G.I. Budker Institute of Nuclear Physics SB RAS (BINP), Novosibirsk 630090, Russia\\
% ...
$^{29}$ INFN Sezione di Ferrara, (A)INFN Sezione di Ferrara, I-44122, Ferrara, Italy; (B)University of Ferrara, I-44122, Ferrara, Italy\\
% ...
$^{34}$ Johannes Gutenberg University of Mainz, Johann-Joachim-Becher-Weg 45, D-55099 Mainz, Germany\\
$^{35}$ Joint Institute for Nuclear Research, 141980 Dubna, Moscow region, Russia\\
% ...
$^{38}$ Liaoning Normal University, Dalian 116029, People's Republic of China\\
% ...
$^{55}$ Southeast University, Nanjing 211100, People's Republic of China\\
$^{56}$ State Key Laboratory of Particle Detection and Electronics, Beijing 100049, Hefei 230026, People's Republic of China\\
% ...
$^{69}$ University of Science and Technology of China, Hefei 230026, People's Republic of China\\
% ...
$^{72}$ University of Turin and INFN, (A)University of Turin, I-10125, Turin, Italy; (B)University of Eastern Piedmont, I-15121, Alessandria, Italy; (C)INFN, I-10125, Turin, Italy\\
$^{73}$ Uppsala University, Box 516, SE-75120 Uppsala, Sweden\\
% ...
$^{b}$ Also at the Novosibirsk State University, Novosibirsk, 630090, Russia\\
% ...
}\end{center}

\vspace{0.4cm}
\end{small}

%% file: draft_JP.bbl
\begin{thebibliography}{**}



\bibitem{Peruzzi:1976sv}
I.~Peruzzi, M.~Piccolo, G.~J.~Feldman, H.~K.~Nguyen, J.~Wiss, G.~S.~Abrams, M.~S.~Alam, A.~Boyarski, M.~Breidenbach and W.~C.~Carithers \textit{et al.}
Phys. Rev. Lett. \textbf{37}, (1976) 569.

\bibitem{Godfrey:1985xj}
S.~Godfrey and N.~Isgur,
Phys. Rev. D \textbf{32}, (1985) 189.


\bibitem{ParticleDataGroup:2020ssz}
R.~L.~Workman \textit{et al.} (Particle Data Group),
PTEP \textbf{2022}, (2022) 083C01.



\bibitem{Gell-Mann:1964ewy}
M.~Gell-Mann,
Phys. Lett. \textbf{8}, (1964) 214.

\bibitem{Gaillard:1974mw}
M.~K.~Gaillard, B.~W.~Lee and J.~L.~Rosner,
Rev. Mod. Phys. \textbf{47}, (1975) 277.

\bibitem{DeRujula:1975qlm}
A.~De Rujula, H.~Georgi and S.~L.~Glashow,
Phys. Rev. D \textbf{12}, (1975) 147.

\bibitem{Glashow:1970gm}
S.~L.~Glashow, J.~Iliopoulos and L.~Maiani,
Phys. Rev. D \textbf{2}, (1970) 1285.

\bibitem{Nguyen:1977kk}
H.~K.~Nguyen, J.~Wiss, G.~S.~Abrams, M.~S.~Alam, A.~Boyarski, M.~Breidenbach, R.~DeVoe, J.~Dorfan, G.~J.~Feldman and G.~Goldhaber \textit{et al.}
Phys. Rev. Lett. \textbf{39}, (1977) 262.

\bibitem{BaBar:2014omp}
A.~J.~Bevan \textit{et al.} (BaBar and Belle Collaborations),
Eur. Phys. J. C \textbf{74}, (2014) 3026.

\bibitem{BaBar:2010zpy}
P.~del Amo Sanchez \textit{et al.} (BaBar Collaboration),
Phys. Rev. D \textbf{82}, (2010) 111101.

\bibitem{BaBar:2006gme}
B.~Aubert \textit{et al.} (BaBar Collaboration),
Phys. Rev. Lett. \textbf{97}, (2006) 222001.



\bibitem{BaBar:2009rro}
B.~Aubert \textit{et al.} (BaBar Collaboration),
Phys. Rev. D \textbf{80}, (2009) 092003.

\bibitem{BaBar:2014jjr}
J.~P.~Lees \textit{et al.} (BaBar Collaboration),
Phys. Rev. D \textbf{91}, (2015) 052002.

\bibitem{Belle:2007hht}
J.~Brodzicka \textit{et al.} (Belle Collaboration),
Phys. Rev. Lett. \textbf{100}, (2008) 092001.


\bibitem{CLEO:2003ggt}
D.~Besson \textit{et al.} (CLEO Collaboration),
Phys. Rev. D \textbf{68}, (2003) 032002.

\bibitem{Chen:2021ftn}
S.~Chen, Y.~Li, W.~Qian, Y.~Xie, Z.~Yang, L.~Zhang and Y.~Zhang,
[arXiv:2111.14360 [hep-ex]].

\bibitem{BESIII:2020qkh}
M.~Ablikim \textit{et al.} (BESIII Collaboration),
Phys. Rev. Lett. \textbf{126}, (2021) 102001.

\bibitem{BESIII:2022qzr}
M.~Ablikim \textit{et al.} (BESIII Collaboration),
Phys. Rev. Lett. \textbf{129}, (2022) 112003.




\bibitem{BESIII:2009fln}
M.~Ablikim \textit{et al.} (BESIII Collaboration),
Nucl. Instrum. Meth. A \textbf{614}, (2010) 345.

\bibitem{Yu:2016cof}
   C.~H.~Yu {\it et al.},
Proceedings of IPAC2016, Busan, Korea, 2016.
  
  
\bibitem{BESIII:2020nme}
M.~Ablikim \textit{et al.} (BESIII Collaboration),
Chin. Phys. C \textbf{44}, (2020) 040001.
 

\bibitem{etof}
 X.~Li {\it et al.}, Radiat. Detect. Technol. Methods {\bf 1}, (2017) 13 ;
 Y.~X.~Guo {\it et al.}, Radiat. Detect. Technol. Methods {\bf 1}, (2017) 15 ;
 P.~Cao {\it et al.}, Nucl.\ Instrum.\ Meth.\ A {\bf 953}, (2020) 163053.

\bibitem{GEANT4:2002zbu}
S.~Agostinelli \textit{et al.} (GEANT4 Collaboration),
Nucl. Instrum. Meth. A \textbf{506}, (2003) 250-303.

\bibitem{Huang:2022wuo}
K.~X.~Huang, Z.~J.~Li, Z.~Qian, J.~Zhu, H.~Y.~Li, Y.~M.~Zhang, S.~S.~Sun and Z.~Y.~You,
Nucl. Sci. Tech. \textbf{33}, (2022) 142.

\bibitem{ref:kkmc}
  S.~Jadach, B.~F.~L.~Ward and Z.~Was,
  Phys.\ Rev.\ D {\bf 63}, (2001) 113009;
  Comput.\ Phys.\ Commun.\  {\bf 130}, (2000) 260.
  
\bibitem{ref:evtgen}
  D.~J.~Lange,
  Nucl.\ Instrum.\ Meth.\ A {\bf 462}, (2001) 152;
  R.~G.~Ping,
  Chin. Phys. C {\bf 32}, (2008) 599.

\bibitem{ref:lundcharm}
  J.~C.~Chen, G.~S.~Huang, X.~R.~Qi, D.~H.~Zhang and Y.~S.~Zhu,
  Phys.\ Rev.\ D {\bf 62}, (2000) 034003;
  R.~L.~Yang, R.~G.~Ping and H.~Chen,
  Chin.\ Phys.\ Lett.\  {\bf 31}, (2014) 061301.

\bibitem{photos}
  E.~Richter-Was,
  Phys.\ Lett.\ B {\bf 303}, (1993) 163.




\bibitem{Jacob:1959at}
M.~Jacob and G.~C.~Wick,
Annals Phys. \textbf{7},(1959) 404 .

\bibitem{Chung:1971ri}
S.~U.~Chung, (1971), 10.5170/CERN-1971-008.


\bibitem{Chung:1997jn}
S.~U.~Chung,
Phys. Rev. D \textbf{57}, (1998) 431.

\bibitem{Chung:1993da}
S.~U.~Chung,
Phys. Rev. D \textbf{48}, (1993) 1225.

\bibitem{Chung:2007nn}
S.~U.~Chung and J.~Friedrich,
Phys. Rev. D \textbf{78}, (2008) 074027.

\bibitem{Santopinto:2016pkp}
E.~Santopinto and A.~Giachino,
Phys. Rev. D \textbf{96}, (2017) 014014.






\bibitem{James:1994vla}
F.~James,
CERN-D-506(1994).

\bibitem{Langenbruch:2019nwe}
C.~Langenbruch,
Eur. Phys. J. C \textbf{82}, (2022) 393.

\bibitem{Supplemental}
See Supplemental Material at xxx for additional analysis information.


\bibitem{Narsky:1999kt}
I.~V.~Narsky,
Nucl. Instrum. Meth. A \textbf{450}, (2000) 444.



\bibitem{Zhu:2006pfm}
Y.~S.~Zhu,
HEPNP \textbf{30}, (2006) 331.


\bibitem{BESIII:2019kfh}
M.~Ablikim \textit{et al.} (BESIII Collaboration),
Phys. Rev. D \textbf{99},  (2019) 112005.


\bibitem{BESIII:2010ank}
M.~Ablikim \textit{et al.} (BESIII Collaboration),
Phys. Rev. D \textbf{81}, (2010) 052005.


\bibitem{BESIII:2015jmz}
M.~Ablikim \textit{et al.} (BESIII Collaboration),
Phys. Rev. D \textbf{92}, (2015) 112008.



\bibitem{BESIII:2016gbw}
M.~Ablikim \textit{et al.} (BESIII Collaboration),
Eur. Phys. J. C \textbf{76}, (2016) 369.


\bibitem{BESIII:2016hko}
M.~Ablikim \textit{et al.} (BESIII Collaboration),
Chin. Phys. C \textbf{40}, (2016) 113001.



\bibitem{BESIII:2012mpj}
M.~Ablikim \textit{et al.} (BESIII Collaboration),
Phys. Rev. D \textbf{87}, (2013) 012002.

\bibitem{Asner:2008nq} 
  D.~M.~Asner {\it et al.},
  ``Physics at BES-III,''
  Int.\ J.\ Mod.\ Phys.\ A {\bf 24},  (2009) S1-794.

\end{thebibliography}
